\newif\ifdraft
\draftfalse
\newif\ifprintauthors
\printauthorstrue

\ifdefined\DRAFT\drafttrue\fi
\ifdefined\PRINTAUTHORS\printauthorstrue\fi

\ifdraft
\documentclass[twocolumn,tighten,times,linenumbers,trackchanges]{aastex631}
\else
\documentclass[twocolumn,tighten,times]{aastex631}
\fi

\usepackage[T1]{fontenc}

\usepackage{amsmath}

\usepackage{siunitx}
\DeclareSIUnit\Mpc{Mpc}
\usepackage{xspace}
\usepackage{acronym}
\usepackage{array}
\usepackage{xcolor}
\usepackage{makecell}
\usepackage{multirow}
\usepackage{booktabs}

\ifdraft

\let\keeptoday\today
\def\today{\keeptoday, GitID {\normalfont\input{gitID.txt}}}
\fi

\newcommand\gwtc[1][?]{\mbox{GWTC\if#1?\else-#1\fi}}
\newcommand\thisgwtcversionmajor{5}
\newcommand\thisgwtcversionminor{0}
\newcommand\thisgwtcversion{\if\thisgwtcversionminor0\thisgwtcversionmajor\else\thisgwtcversionmajor.\thisgwtcversionminor\fi}
\newcommand\thisgwtcversionfull{\thisgwtcversionmajor.\thisgwtcversionminor}

\newcommand\thisgwtcfull{\gwtc[\thisgwtcversionfull]}

\newcommand{\note}[1]{}
\newcommand{\LVKcollaboration}{The LIGO Scientific Collaboration, the Virgo Collaboration, and the KAGRA Collaboration}

\newcommand{\OfourBEndDate}{{{2025~January~28}}}

\newcommand{\Msun}{\ensuremath{\mathit{M_\odot}}}

\newcommand{\Mc}{\ensuremath{\mathcal{M}}}
\newcommand{\Mtot}{\ensuremath{M}}
\newcommand{\Mf}{\ensuremath{M_\mathrm{f}}}

\newcommand{\mchirp}{\Mc}

\newcommand{\Erad}{\ensuremath{E_\mathrm{rad}}}
\newcommand{\lumpeak}{\ensuremath{\ell_{\text{peak}}}}

\newcommand{\energyrad}{\Erad}

\newcommand{\chieff}{\ensuremath{\chi_\mathrm{eff}}}
\newcommand{\chip}{\ensuremath{\chi_\mathrm{p}}}
\newcommand{\chif}{\ensuremath{\chi_\mathrm{f}}}

\newcommand{\vecspinone}{\ensuremath{\boldsymbol{\chi}_1}}

\newcommand{\vecspintwo}{\ensuremath{\boldsymbol{\chi}_2}}

\newcommand{\ip}[2]{\ensuremath{\langle #1 | #2 \rangle}}

\newcommand\PEpdfp{\ensuremath{p}}
\newcommand\PEdata{\ensuremath{d}}
\newcommand{\PEparameter}{\ensuremath{\boldsymbol{\theta}}}%
\newcommand{\PEparameterIntrinsic}{\ensuremath{\boldsymbol{\theta}_{\rm int}}}
\newcommand{\PEdataTimeDomain}{\ensuremath{{d}}}%
\newcommand{\PEdataFrequencyDomain}{\ensuremath{{\tilde{d}}}}%

\newcommand\PEpdf[2][?]{\ensuremath{\PEpdfp({#2}\ifx#1?\else | {#1}\fi)}}

\newcommand\PEposterior[1][\PEparameter]{\PEpdf[\PEdata]{#1}}
\newcommand\PElikelihood[1][\PEparameter]{\PEpdf[#1]{\PEdata}}
\newcommand\PEevidence[1][?]{\PEpdf[#1]{\PEdata}}

\newcommand\PEpriorpdfpi{\ensuremath{\pi}}
\newcommand\PEpdfprior[1]{\ensuremath{\PEpriorpdfpi({#1})}}
\newcommand\PEprior[1][\PEparameter]{\PEpdfprior{#1}}
\newcommand\PEpriorpe[1][\PEparameter]{{\let\keepPEpriorpdfpi\PEpriorpdfpi\def\PEpriorpdfpi{\keepPEpriorpdfpi_{\text{PE}}}\PEprior[#1]\let\PEpriorpdfpi\keepPEpriorpdfpi}}

\newcommand{\flow}{\ensuremath{f_\mathrm{low}}\xspace}
\newcommand{\fhi}{\ensuremath{f_\mathrm{high}}\xspace}
\newcommand{\fsamp}{\ensuremath{f_\mathrm{s}}}

\newcommand{\alphaRoll}{\ensuremath{\alpha^\mathrm{roll\text{-}off}}}

\newcommand{\PE}[0]{\ac{PE}\xspace}
\newcommand{\PDF}[0]{\ac{PDF}\xspace}
\newcommand{\MCMC}[0]{\ac{MCMC}\xspace}

\newcommand{\HzeroSymbol}{\ensuremath{H_{0}}}
\newcommand{\WmSymbol}{\ensuremath{\Omega_{\mathrm{m}}}}

\newcommand{\PN}[0]{\ac{PN}\xspace}
\newcommand{\BBH}[0]{\ac{BBH}\xspace}
\newcommand{\BNS}[0]{\ac{BNS}\xspace}
\newcommand{\NSBH}[0]{\ac{NSBH}\xspace}

\newcommand{\SNR}[0]{\ac{SNR}\xspace}

\newcommand{\GR}[0]{\ac{GR}\xspace}
\newcommand{\PSD}[0]{\ac{PSD}\xspace}

\newcommand{\VT}{\ensuremath{\langle VT \rangle}\xspace}
\newcommand{\pastro}{\ensuremath{p_{\mathrm{astro}}}\xspace}
\newcommand{\pterr}{\ensuremath{p_{\mathrm{terr}}}\xspace}
\newcommand{\pbbh}{\ensuremath{p_{\mathrm{BBH}}}\xspace}
\newcommand{\pbns}{\ensuremath{p_{\mathrm{BNS}}}\xspace}
\newcommand{\pnsbh}{\ensuremath{p_{\mathrm{NSBH}}}\xspace}

\newcommand{\soft}[1]{\textsc{#1}}

\newcommand{\ASTROPY}{\soft{Astropy}\xspace}
\newcommand{\GSTLAL}{\soft{GstLAL}\xspace}
\newcommand{\BAYESTAR}{\soft{BAYESTAR}\xspace}
\newcommand{\CWB}{\soft{cWB}\xspace}
\newcommand{\CWBTWOG}{\soft{cWB-2G}\xspace}
\newcommand{\CWBBBH}{\soft{cWB-BBH}\xspace}
\newcommand{\PYCBC}{\soft{PyCBC}\xspace}
\newcommand{\MBTA}{\soft{MBTA}\xspace}
\newcommand{\SPIIR}{\soft{SPIIR}\xspace}
\newcommand{\LALINFERENCE}{\soft{LALInference}\xspace}
\newcommand{\BAYESWAVE}{\soft{BayesWave}\xspace}
\newcommand{\BILBY}{\soft{Bilby}\xspace}
\newcommand{\BILBYMCMC}{\soft{BilbyMCMC}\xspace}
\newcommand{\RIFT}{\soft{RIFT}\xspace}
\newcommand{\DINGO}{\soft{Dingo}\xspace}
\newcommand{\LALSUITE}{\soft{LALSuite}\xspace}

\newcommand{\PBILBY}{\soft{ParallelBilby}\xspace}
\newcommand{\ASIMOV}{\soft{Asimov}\xspace}
\newcommand{\PESUMMARY}{\soft{PESummary}\xspace}

\newcommand{\NUMPY}{\soft{NumPy}\xspace}

\newcommand{\MATPLOTLIB}{\soft{Matplotlib}\xspace}

\newcommand{\DYNESTY}{\soft{Dynesty}\xspace}
\newcommand{\CBCFLOW}{\soft{cbcflow}\xspace}
\newcommand{\GRACEDB}{\soft{GraceDB}\xspace}
\newcommand{\MANIFOLD}{\soft{manifold}\xspace}
\newcommand{\DQRBUILD}{\soft{DQRBuild}\xspace}
\newcommand{\VIRGODQR}{\soft{Virgo DQR}\xspace}
\newcommand{\EVENTVALIDATION}{\soft{event-validation}\xspace}

\newcommand{\TAYLOR}{\soft{Taylor}\xspace}
\newcommand{\SPINTAYLOR}{\soft{SpinTaylor}\xspace}
\newcommand{\STTFOUR}{\soft{SpinTaylorT4}\xspace}
\newcommand{\TFTWO}{\soft{TaylorF2}\xspace}

\newcommand{\IMRPhenom}{\soft{IMRPhenom}\xspace}

\newcommand{\IMRPhenomC}{\soft{IMRPhenomC}\xspace}
\newcommand{\IMRPhenomD}{\soft{IMRPhenomD}\xspace}
\newcommand{\IMRPhenomDNRTidal}{\soft{IMRPhenomD\_NRTidal}\xspace}
\newcommand{\IMRPhenomHM}{\soft{IMRPhenomHM}\xspace}
\newcommand{\IMRPhenomP}{\soft{IMRPhenomP}\xspace}
\newcommand{\IMRPhenomPVTWO}{\soft{IMRPhenomPv2}\xspace}

\newcommand{\IMRPhenomPVTHREEHM}{\soft{IMRPhenomPv3HM}\xspace}
\newcommand{\IMRPhenomXAS}{\soft{IMRPhenomXAS}\xspace}
\newcommand{\IMRPhenomXHM}{\soft{IMRPhenomXHM}\xspace}

\newcommand{\IMRPhenomXPHM}{\soft{IMRPhenomXPHM}\xspace}
\newcommand{\IMRPhenomXPHMST}{\soft{IMRPhenomXPHM\_SpinTaylor}\xspace}
\newcommand{\IMRPhenomXOFOURa}{\soft{IMRPhenomXO4a}\xspace}

\newcommand{\TEOB}{\soft{TEOB}\xspace}

\newcommand{\SEOBNR}{\soft{SEOBNR}\xspace}

\newcommand{\SEOBNRTHREE}{\soft{SEOBNRv3}\xspace}
\newcommand{\SEOBNRFOUR}{\soft{SEOBNRv4}\xspace}
\newcommand{\SEOBNRFOUROPT}{\soft{SEOBNRv4\_opt}\xspace}
\newcommand{\SEOBNRFOURROM}{\soft{SEOBNRv4\_ROM}\xspace}
\newcommand{\SEOBNRFOURP}{\soft{SEOBNRv4P}\xspace}
\newcommand{\SEOBNRFOURHM}{\soft{SEOBNRv4HM}\xspace}
\newcommand{\SEOBNRFOURHMROM}{\soft{SEOBNRv4HM\_ROM}\xspace}

\newcommand{\SEOBNRFOURPHM}{\soft{SEOBNRv4PHM}\xspace}
\newcommand{\SEOBNRFOURT}{\soft{SEOBNRv4T}\xspace}
\newcommand{\SEOBNRFOURTSUR}{\soft{SEOBNRv4T\_surrogate}\xspace}
\newcommand{\SEOBNRFIVE}{\soft{SEOBNRv5}\xspace}

\newcommand{\SEOBNRFIVEHM}{\soft{SEOBNRv5HM}\xspace}

\newcommand{\SEOBNRFIVEROM}{\soft{SEOBNRv5\_ROM}\xspace}
\newcommand{\SEOBNRFIVEPHM}{\soft{SEOBNRv5PHM}\xspace}
\newcommand{\SEOBNRFIVEHMROM}{\soft{SEOBNRv5HM\_ROM}\xspace}
\newcommand{\IMRPhenomPTWONRTidal}{\soft{IMRPhenomPv2\_NRTidal}\xspace}
\newcommand{\IMRPhenomPTWONRTidalTWO}{\soft{IMRPhenomPv2\_NRTidalv2}\xspace}
\newcommand{\IMRPhenomXPNR}{\soft{IMRPhenomXPNR}\xspace}
\newcommand{\IMRPhenomXPNRTidalTWO}{\soft{IMRPhenomXP\_NRTidalv2}\xspace}

\newcommand{\IMRPhenomNSBH}{\soft{IMRPhenomNSBH}\xspace}
\newcommand{\SEOBNRFOURNRTidal}{\soft{SEOBNRv4\_ROM\_NRTidal}\xspace}
\newcommand{\SEOBNRFOURNRTidalTWO}{\soft{SEOBNRv4\_ROM\_NRTidalv2}\xspace}
\newcommand{\SEOBNRFOURNRtidalTWONSBH}{\soft{SEOBNRv4\_ROM\_NRTidalv2\_NSBH}\xspace}

\newcommand{\NRSUR}{\soft{NRSurrogate}\xspace}
\newcommand{\SURSEVENDQFOUR}{\soft{NRSur7dq4}\xspace}
\newcommand{\SURSEVENDQTWO}{\soft{NRSur7dq2}\xspace}
\newcommand{\TEOBResumS}{\soft{TEOBResumS}\xspace}

\DeclareSIUnit\parsec{pc}
\DeclareSIUnit\Mpc{\mega\parsec}
\DeclareSIUnit\yr{yr}
\DeclareSIUnit\GpcCubedYear{\giga\parsec\cubed\yr}

\newcommand{\HzeroValue}{{{\qty{67.9}{\km.\second^{-1}.\Mpc^{-1}}}}}
\newcommand{\WmValue}{{{\num{0.3065}}}}

\usepackage{amstext}

\makeatletter
\@ifpackageloaded{cleveref}{
  \AtBeginDocument{
    \def\ltx@label#1{\cref@label{#1}}%
    \def\label@in@display@noarg#1{\cref@old@label@in@display{#1}}%
    \def\label@in@mmeasure@noarg#1{%
      \begingroup%
        \measuring@false%
        \cref@old@label@in@display{#1}%
      \endgroup}%
  }
}{}
\makeatother

\protected\def\protectedacused{\acused}

\acrodef{LIGO}[LIGO]{Laser Interferometer Gravitational-Wave Observatory}
\acrodef{LHO}[LHO]{\ac{LIGO} Hanford Observatory}
\acrodef{LLO}[LLO]{\ac{LIGO} Livingston Observatory}
\acrodef{KAGRA}[KAGRA]{KAGRA}\acused{KAGRA}
\acrodef{iKAGRA}[iKAGRA]{initial \ac{KAGRA}}
\acrodef{bKAGRA}[bKAGRA]{baseline \ac{KAGRA}}
\acrodef{GEO}[GEO\,600]{GEO\,600 \ac{GW} detector}
\acrodef{aLIGO}{Advanced \ac{LIGO}}
\acrodef{A+}{Advanced+ \ac{LIGO}}
\acrodef{Asharp}[\ensuremath{\text{A}^\sharp}]{\ac{LIGO} \acs{Asharp}}
\acrodef{AdV}{Advanced \acl{Virgo}}
\acrodef{AdV+}{Advanced \acl{Virgo}+}
\acrodef{Virgo}{Virgo}\acused{Virgo}
\acrodef{VirgoNEXT}[Virgo\_nEXT]{Virgo\_nEXT}\acused{VirgoNEXT}

\acrodef{LSC}[LSC]{\acs{LIGO} Scientific Collaboration}
\acrodef{LV}[LV]{\acs{LIGO}--\acs{Virgo} Collaboration\protect\protectedacused{LVC}}
\acrodef{LVC}[LV]{\acs{LIGO}--\acs{Virgo} Collaboration\protect\protectedacused{LV}}
\acrodef{LVK}[LVK]{\acs{LIGO}--\ac{Virgo}--\ac{KAGRA} Collaboration}
\acrodef{IGWN}[IGWN]{International \ac{GWH} Observatory Network}

\acrodef{O1}[O1]{first observing run}
\acrodef{O2}[O2]{second observing run}
\acrodef{O3}[O3]{third observing run}
\acrodef{O3a}[O3a]{first half of the third observing run}
\acrodef{O3b}[O3b]{second half of the third observing run}
\acrodef{O3GK}[O3GK]{observing run}
\acrodef{O4}[O4]{fourth observing run}
\acrodef{O4a}[O4a]{first part of the fourth observing run}
\acrodef{O4b}[O4b]{second part of the fourth observing run}
\acrodef{O4c}[O4c]{third part of the fourth observing run}
\acrodef{IR1}[IR1]{intermediate run 1}
\acrodef{O5}[O5]{fifth observing run}

\acrodef{BH}[BH]{black hole}
\acrodef{BBH}[BBH]{binary \ac{BH}}
\acrodef{BNS}[BNS]{binary \ac{NS}}
\acrodef{IMBH}[IMBH]{intermediate-mass \ac{BH}}
\acrodef{NS}[NS]{neutron star}
\acrodef{BHNS}[BHNS]{\ac{BH}--\ac{NS} binary}
\acrodefplural{BHNS}[BHNSs]{\ac{BH}--\ac{NS} binaries}
\acrodef{NSBH}[NSBH]{\ac{NS}--\ac{BH} binary}
\acrodefplural{NSBH}[NSBHs]{\ac{NS}--\ac{BH} binaries}
\acrodef{PBH}[PBH]{primordial \ac{BH}}
\acrodef{CBC}[CBC]{compact binary coalescence}

\acrodef{GW}[GW]{gravitational wave\protect\protectedacused{GWH}}
\acrodef{GWH}[GW]{gravitational-wave\protect\protectedacused{GW}}
\acrodef{IFO}[IFO]{interferometer}
\acrodef{SNR}[SNR]{signal-to-noise ratio}
\acrodef{FAR}[FAR]{false-alarm rate}
\acrodef{IFAR}[IFAR]{inverse false-alarm rate}
\acrodef{FAP}[FAP]{false alarm probability}
\acrodef{PSD}[PSD]{power spectral density}
\acrodef{GR}[GR]{general relativity}
\acrodef{NR}[NR]{numerical relativity}
\acrodef{PN}[PN]{post-Newtonian}
\acrodef{EOB}[EOB]{effective-one-body}
\acrodef{ROM}[ROM]{reduced-order model}
\acrodef{IMR}[IMR]{inspiral--merger--ringdown}
\acrodef{PDF}[pdf]{probability density function}
\acrodef{PE}[PE]{parameter estimation}
\acrodef{CI}[CI]{credible interval}
\acrodef{CL}[CL]{credible level}
\acrodef{EOS}[EoS]{equation of state}
\acrodef{KLD}[KLD]{Kullback--Leibler divergence}
\acrodef{JSD}[JSD]{Jensen--Shannon divergence}
\acrodef{GCN}[GCN]{General Coordinates Network}
\acrodef{GWTC}[GWTC]{Gravitational-Wave Transient Catalog}
\acrodef{GWOSC}[GWOSC]{Gravitational Wave Open Science Center}
\acrodef{WDM}[WDM]{Wilson--Daubechies--Meyer}
\acrodef{DQR}[DQR]{data-quality report}

\acrodef{CWB}[cWB]{coherent WaveBurst}
\acrodef{LAL}[LAL]{\ac{LIGO} algorithm library}

\acrodef{CHRoCC}{central heating radius of curvature correction}
\acrodef{NonSENS}{non-stationary estimation and noise subtraction}
\acrodef{RF}{radio frequency}
\acrodef{PNC}{phase noise cancellation}
\acrodef{ASC}{alignment sensing and control}
\acrodef{WFS}{wave-front sensing}
\acrodef{BPC}{beam position control}
\acrodef{ADS}{alignment dither systems}
\acrodef{OMC}{output mode cleaner}
\acrodef{LVDTs}{linear variable differential transformers}
\acrodef{GAS}{geometrical anti-spring}

\acrodef{PTA}{Pulsar Timing Array}

\newcommand{\GW}[0]{\ac{GW}\xspace}

\newcommand{\PECONFIG}{\soft{peconfigurator}\xspace}
\newcommand{\LIGOSKYMAP}{\soft{ligo.skymap}\xspace}

\newcommand{\PEcalfactor}{\ensuremath{{\tilde{\eta}_R}}}

\newcommand{\GDSStrainFrame}{\texttt{GDS-CALIB\_STRAIN\_CLEAN}\xspace}
\newcommand{\GDSStrainFrameAR}{\texttt{GDS-CALIB\_STRAIN\_CLEAN\_AR}\xspace}
\newcommand{\HrecStrainFrame}{\texttt{Hrec\_hoft\_16384Hz}\xspace}
\newcommand{\HrecStrainFrameAR}{\texttt{Hrec\_hoftRepro1AR\_16384Hz}\xspace}

\newcommand{\PycbcSensImprov}{10\%}

\acrodef{FFT}{fast Fourier transform}
\acrodef{HPD}[HPD]{highest posterior density}
\acrodef{KDE}[KDE]{kernel density estimation}
\acrodef{LR}{likelihood ratio}
\acrodef{MCMC}[MCMC]{Markov chain Monte Carlo}
\acrodef{SPA}{stationary-phase approximation}
\acrodef{SVD}{singular value decomposition}

\newcommand{\GWTC}{GWTC\xspace}

\newcommand{\iDQ}{\soft{iDQ}}

\begin{document}

\title{\thisgwtcfull: Methods for Identifying and Characterizing Gravitational-wave Transients}

\ifprintauthors
\suppressAffiliations

\author[0000-0003-4786-2698]{A.~G.~Abac}
\affiliation{Max Planck Institute for Gravitational Physics (Albert Einstein Institute), D-14476 Potsdam, Germany}
\author{A.~Abe}
\affiliation{Department of Physics, Graduate School of Science, Osaka Metropolitan University, 3-3-138 Sugimoto-cho, Sumiyoshi-ku, Osaka City, Osaka 558-8585, Japan  }
\author{I.~Abouelfettouh}
\affiliation{LIGO Hanford Observatory, Richland, WA 99352, USA}
\author{F.~Acernese}
\affiliation{Dipartimento di Fisica ``E.R. Caianiello'', Universit\`a di Salerno, I-84084 Fisciano, Salerno, Italy}
\affiliation{INFN, Sezione di Napoli, I-80126 Napoli, Italy}
\author[0000-0002-8648-0767]{K.~Ackley}
\affiliation{University of Warwick, Coventry CV4 7AL, United Kingdom}
\author{A.~Adam}
\affiliation{OzGrav, University of Western Australia, Crawley, Western Australia 6009, Australia}
\author[0009-0004-2101-5428]{S.~Adhicary}
\affiliation{The Pennsylvania State University, University Park, PA 16802, USA}
\author{D.~Adhikari}
\affiliation{Max Planck Institute for Gravitational Physics (Albert Einstein Institute), D-30167 Hannover, Germany}
\affiliation{Leibniz Universit\"{a}t Hannover, D-30167 Hannover, Germany}
\author[0000-0002-5731-5076]{R.~X.~Adhikari}
\affiliation{LIGO Laboratory, California Institute of Technology, Pasadena, CA 91125, USA}
\author{V.~K.~Adkins}
\affiliation{Louisiana State University, Baton Rouge, LA 70803, USA}
% ...

\collaboration{0}{\LVKcollaboration\\(See the end matter for the full list of authors)}
\else
\collaboration{0}{\LVKcollaboration}
\fi

\correspondingauthor{LSC P\&P Committee, via LVK Publications as proxy}
\email{lvc.publications@ligo.org}

\begin{abstract}
The Gravitational-Wave Transient Catalog (\GWTC) is a collection of candidate gravitational-wave transient signals identified and characterized by the LIGO--Virgo--KAGRA Collaboration.
Producing the contents of the \GWTC from detector data requires complex analysis methods.
These comprise techniques to model the signal; identify the transients in the data; evaluate the quality of the data and mitigate possible instrumental issues; infer the parameters of each transient; compare the data with the waveform models for compact binary coalescences, and handle the large amount of results associated with all these different analyses.
In this paper, we describe the methods employed to produce the catalog's fifth release, \thisgwtcfull, focusing on the analysis of the second part of the fourth observing run of LIGO, Virgo and KAGRA.

\end{abstract}


\section{Introduction}
\label{sec:intro}

Interferometric gravitational-wave (\acsu{GW}) detectors produce a calibrated discrete digital time
series $d(t)$ known as the \emph{strain} (a dimensionless measure of the relative
difference in arm length of the interferometers).
The \aclu{LIGO} \citep[\acs{LIGO};][]{2015CQGra..32g4001L} and Virgo \citep{2015CQGra..32b4001A} detectors are the most sensitive to date.
Alongside the developing KAGRA detector \citep{2019NatAs...3...35K} the \ac{LVK} has recently undertaken the \ac{O4}.
However, the data produced are dominated by detector noise, with only occasional
occurrences of detectable transient \ac{GW} signals \citep{KAGRA:2013rdx,
2020CQGra..37e5002A}; to date, all such observed signals probably arise from \acp{CBC} involving \acp{BH} and \acp{NS}.
This paper describes the methodology used to analyze
the calibrated strain data up to \OfourBEndDate{}, the end date of the \ac{O4b}, and produce version \thisgwtcversionfull{} of the
Gravitational Wave Transient Catalog (\gwtc), hereafter referred to as
\thisgwtcfull{}. This catalog release also includes updated search results and 
parameter estimation of additional events from the \ac{O4a}. To clarify that these 
\ac{O4a} results supersede those in \gwtc[4.0], we name them \gwtc[4.1].
However, as \gwtc[5.0] is a cumulative catalog, all the results 
labeled \gwtc[4.1] are included in \gwtc[5.0] by definition.
For a general introduction to \thisgwtcfull{}, see \citet{GWTC:Introduction} which
also contains a description of the observed source classes and data-analysis
nomenclature which should be read as a background to this methodology paper.
The scientific results of \thisgwtcfull{} are presented in \citet{GWTC:Results}.
Data-analysis methods not directly related to producing results for \thisgwtcfull{}, such as searches for continuous \acp{GW} or subsolar-mass \acp{CBC}, will be described elsewhere.

\begin{figure*}
\includegraphics[width=\textwidth]{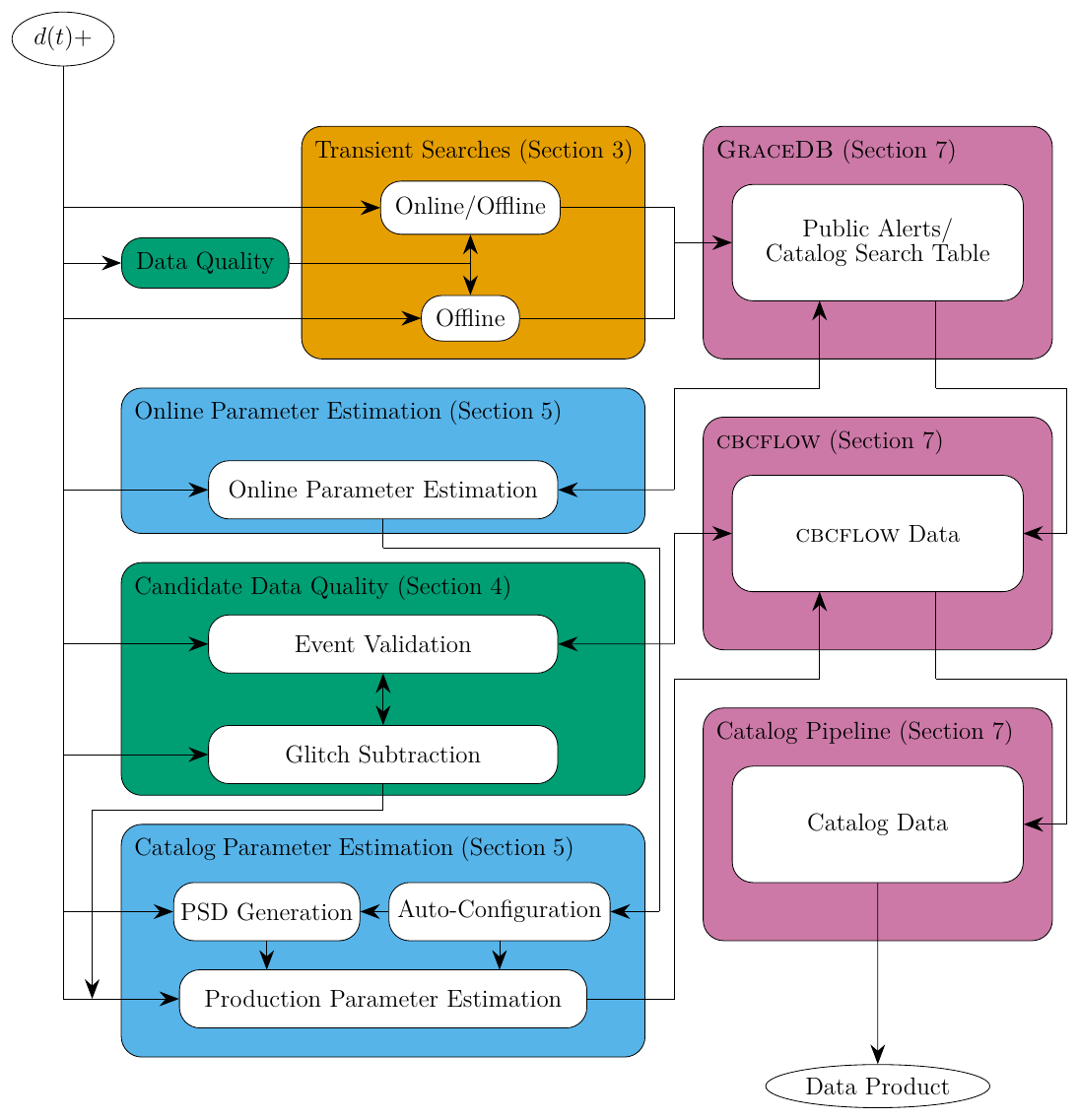}        
\caption{A high-level rendering of the flow of data from the strain and auxiliary data, denoted $d(t)+$, to the production of \thisgwtcfull{} described in this work.
In this work, we use the terms \emph{downstream} and \emph{upstream} to refer to the flow of information in the analysis process (e.g. the parameter estimation is downstream of the compilation of search results since it depends on their outputs).
This complex process enables the use of the most complete set of information for the final results while also leveraging preliminary studies to parallelise and reduce the overall analysis time.
The term \emph{online} (also referred to as \emph{low-latency} in the literature) refers to analyses run on live data with a goal to upload candidates to \GRACEDB immediately and enable rapid public alerts, while \emph{offline} refers to analyses run with the goal to identify candidates for the \gwtc. \acs{PSD} refers to the \acl{PSD}.
In each box, we provide references to the sections of this paper where the methods are explained in more detail.}
\label{fig:data-flow-chart}
\end{figure*}

There are many interconnected elements to the data processing methodology described in this work.
To provide a visual guide and summary, Figure~\ref{fig:data-flow-chart} shows a diagram of the data-processing workflow.
In Section~\ref{sec:waveforms}, we introduce the fundamental concepts behind
modeling \ac{GW} waveforms from \ac{CBC} sources and describe the
waveform approximants used in later analyses. 
The data analysis process then starts with the calibrated strain data $d(t)$ and associated auxiliary data, which is produced by the \ac{LVK} detectors and is the input to the analysis.
The strain data and auxiliary data are inputs to the process of searching for signals and compiling a list of candidates from the strain data
which we describe in Section~\ref{sec:searches}.
Section~\ref{sec:dq} discusses how we
assess data quality around candidates and mitigate the impact of
potential instrumental issues.
The strain data, candidate lists, and data-quality information are then inputs into the methods used to infer the properties of the signals and their sources which we describe in Section~\ref{sec:pe}.
In Section~\ref{sec:wct} we detail consistency tests performed on selected candidates to evaluate how well
\ac{CBC} waveform models match the data.
Section~\ref{sec:data-management} describes the technologies used to manage the flow of
information throughout the analysis process as shown in Figure~\ref{fig:data-flow-chart}.
Finally, we conclude in Section~\ref{sec:conclusions}.

\newcommand{\NRTidal}{\soft{NRTidal}}
\newcommand{\NRTidalvtwo}{\soft{NRTidalv2}}
\newcommand{\NRTidalvthree}{\soft{NRTidalv3}}

\section{Modeling Compact Binary Coalescence Signals}\label{sec:waveforms}

In this section, we will describe the modeling of \ac{CBC} waveform signals, including \ac{BBH}, \ac{BNS}, and \ac{NSBH} systems.
This modeling is a crucial aspect for the detection and astrophysical interpretation of these systems and is used by modeled search algorithms (described in Section~\ref{sec:searches}) and \acl{PE} (\acsu{PE}; described in Section~\ref{sec:pe}).
The focus of this section is on the waveform models used in \thisgwtcfull{}, although we include models employed in analyses of previous \GWTC versions that have been superseded in newer versions, in particular the \PE analyses performed for \gwtc[1.0] and \gwtc[2.0]. 
For an introduction to waveform modeling in general and definitions of our coordinate conventions, see Section~5 of \citet{GWTC:Introduction}.

Several different modeling approaches have been followed in the development of \ac{CBC} waveform models for the complete \ac{IMR} stages of the signal. 
Inspiral-only models have been developed based on \ac{PN} theory \citep{2014LRR....17....2B}, using different \ac{PN}-expansions of the energy balance equations for the two-body dynamics, under the \TAYLOR family \citep{2009PhRvD..80h4043B} for non-spinning systems and the \SPINTAYLOR family \citep{Sturani:2015aaa,Isoyama:2020lls} for spinning systems. 
The \IMRPhenom  approach \citep{Ajith:2007qp} focuses on the description of the \GW signal, traditionally in Fourier domain for an efficient implementation in the data analysis pipelines, combining \ac{PN} and \ac{NR} information into closed-form expressions for describing the inspiral, merger and ringdown stages of the signal.
The \ac{EOB} approach \citep{1999PhRvD..59h4006B,2000PhRvD..62f4015B} focuses on accurately describing the dynamics and resulting waveform of the system in the time-domain, using a combination of resummed analytical information and calibration to \ac{NR}.
Inside this approach, two main different development efforts have been followed, the \SEOBNR \citep{Buonanno:2007pf} and \TEOB \citep{Damour:2014sva,Nagar:2016ayt} approaches.
The \NRSUR approach \citep{2017PhRvD..95j4023B} focuses on producing efficient surrogate models of the \ac{NR} data, with or without hybridization with analytical waveforms, to deliver highly accurate waveforms in a validity region limited by the input numerical data, both in the number of cycles and in the coverage of parameter space.
These modeling approaches provide the waveform models described in this section.

The majority of waveform models that we describe in this section correspond to \ac{CBC} systems on quasi-circular orbits (also called quasi-spherical orbits when spin precession is present), 
therefore neglecting orbital eccentricity effects, since at the moment all models employed in the analysis of \GWTC candidates have this restriction.
One of the main reasons for this limitation is that mature eccentric waveform models have only been developed recently, and reviewed versions were not yet ready at the time of starting the analyses described in this work.
This limitation can lead to biases in the mass parameters \citep{1999PhRvD..60l4008M,2018PhRvD..98h3028L,2020MNRAS.497.1966L,2023PhRvD.108j4018O,2022PhRvD.105b3003F}, and also influence the inferred spins \citep{2020MNRAS.497.1966L,2020ApJ...903L...5R,2023PhRvD.108j4018O,2022PhRvD.105b3003F,Morras:2025nlp,Morras:2025xfu}.
While not described here, eccentric waveform models were used for special candidate analyses~\citep{2025ApJ...993L..21A,2025PhRvL.135k1403A}.
For high-mass systems, where few orbits are observed, neglecting eccentricity can also lead to incorrectly identifying spin-precessing effects due to possible degeneracies of both effects in this regime \citep{Ramos-Buades:2019uvh,CalderonBustillo:2020xms,Romero-Shaw:2022fbf}.
During a binary inspiral, the energy loss from \GW emission rapidly circularizes an eccentric orbit, generally reducing expectation of significant eccentricity \citep{1964PhRv..136.1224P} by the time the signal enters the sensitive frequency band of the interferometers \citep{2021PhRvD.104j4023T}.
Rate estimations \citep{Wen:2002km,Samsing:2017rat,Gupte:2024jfe} constrain the fraction of events with observable eccentricity to a small percentage.

Nevertheless, recent independent analyses have shown evidence of nonzero eccentricity in a few previously detected signals \citep{2020ApJ...903L...5R,Gamba:2021gap,Gayathri:2020coq,Gupte:2024jfe,Planas:2025jny,Morras:2025xfu,Planas:2025plq, Jan:2025fps, Romero-Shaw:2025vbc, Kacanja:2025kpr}.

Therefore, we cannot exclude a priori that a few candidates in \gwtc[5.0] may present nonzero eccentricity.

In the following subsections, we will describe chronologically the relevant waveform modeling efforts that lead to models employed in \GWTC analyses.
For consulting the specific set of waveform models employed in each \GWTC version, we refer the reader to Table~\ref{tab:wf_models}, while a summary of the physics of each model is displayed in Table~\ref{tab:wf_models_physics}. 
Their usage on specific pipelines will be described in the corresponding sections of this work.

\subsection{\ac{BBH} Models} 

Within the \IMRPhenom and \SEOBNR families, the first complete \ac{IMR} models calibrated to \ac{NR} were produced for the dominant spin-weighted spherical harmonic multipole of nonspinning systems \citep{Ajith:2007qp,Buonanno:2007pf}, and then extended to aligned-spin systems \citep{2011PhRvL.106x1101A,2010PhRvD..82f4016S,Taracchini:2013rva}.
Improved versions were developed for \GWTC analyses: \IMRPhenomD \citep{2016PhRvD..93d4006H,2016PhRvD..93d4007K} and \SEOBNRFOUR \citep{2017PhRvD..95d4028B}, increasing the amount of analytical information included in the models, the number and coverage of the calibration dataset, specific details of the model expressions and better accuracy with \ac{NR}.
A highly optimized version of \SEOBNRFOUR was produced to reduce its computational cost in the time domain, \SEOBNRFOUROPT \citep{Devine:2016ovp}, and \ac{ROM} techniques were applied to obtain a fast Fourier-domain version of the original model, \SEOBNRFOURROM \citep{2017PhRvD..95d4028B}.
Within the \TEOB approach, a version of the model for the dominant multipole of \ac{BBH} systems was developed, \TEOBResumS \citep{Nagar:2018zoe}, including a post-adiabatic approximation for the dynamics that increases its computational efficiency \citep{Nagar:2018gnk}.
Additionally, the inspiral-only models \TFTWO \citep{Damour:2000zb,2009PhRvD..80h4043B,2011PhRvD..83h4051V}, which is an analytical Fourier-domain model, and \STTFOUR \citep{Sturani:2015aaa,Isoyama:2020lls}, a time-domain model, have also been employed for searching \GW signals.

Modeling spin-precession effects is crucial for precise measurements of spins \citep{2014PhRvL.112y1101V,2020PhRvR...2d3096P,2022PhRvD.106b3001J,2021PhRvD.104j3018B,Steinle:2022rhj}, providing key information about the formation channels of the observed systems \citep{2016ApJ...832L...2R,2017MNRAS.471.2801S,2017PhRvD..96b3012T,Zhu:2017znf}, and breaking degeneracies in parameter inference \citep{Vecchio:2003tn,Lang:2006bsg,2015ApJ...798L..17C,2022PhRvD.105f4012K}.
Spin-precession can significantly increase the complexity of the signal, and modeling efforts focused first on applying the \emph{twisting-up} approach \citep{2003PhRvD..67j4025B,2011PhRvD..84b4046S,2011PhRvD..84l4011B,OShaughnessy:2012lay,2012PhRvD..86j4063S} to the dominant-multipole models, differing on the description of the spin dynamics of the system.
The \IMRPhenom approach incorporated a closed-form solution of the next-to-next-to-leading order spin-precessing evolution equations for single-spin configurations \citep{Marsat:2013wwa,Bohe:2012mr}, mapping then the expressions effectively to double-spin systems \citep{2015PhRvD..91b4043S}, and employing the twisting-up technique first to \IMRPhenomC \citep{2010PhRvD..82f4016S} to obtain \IMRPhenomP \citep{2014PhRvL.113o1101H}, and then to the more accurate model \IMRPhenomD to obtain \IMRPhenomPVTWO \citep{2014PhRvL.113o1101H,Bohe:PPv2}.
On the \SEOBNR approach, spin-dynamics were incorporated via evolution of the \ac{EOB} equations of motion for a quasi-circular spin-precessing Hamiltonian, constructing the spin-precessing polarizations from the quadrupolar multipoles in the co-precessing frame for producing \SEOBNRTHREE \citep{2014PhRvD..89h4006P}.

Subdominant harmonics in the signal were shown to be important for reducing several degeneracies in the analysis of  signals \citep{Capano:2013raa,2015PhRvD..92b2002G,2021PhRvX..11b1053A,2024PhRvD.109b2001A}, in particular in inclination--distance degeneracy and modeling efforts focused on their inclusion into current models.
\IMRPhenomHM \citep{2018PhRvL.120p1102L} incorporated a set of the most important subdominant harmonics to the \IMRPhenomD model, although without additional calibration of these harmonics to numerical data. 
\SEOBNRFOURHM \citep{2018PhRvD..98h4028C} incorporated a similar list of harmonics to \SEOBNRFOUR, with explicit calibration of the waveform modes to \ac{NR} and test-particle waveforms, and provided an efficient \ac{ROM} version in Fourier-domain, \SEOBNRFOURHMROM \citep{2020PhRvD.101l4040C}.
Spin-precessing versions of these multipolar waveform models were developed in parallel, with some improvements in the spin-dynamics description in the \IMRPhenom approach, resulting in \IMRPhenomPVTHREEHM \citep{2020PhRvD.101b4056K} and \SEOBNRFOURPHM \citep{2020PhRvD.102d4055O}.
Additionally, the \NRSUR approach started producing the first surrogate models for multipolar spin-precessing signals, first with \SURSEVENDQTWO \citep{2017PhRvD..96b4058B} and then extending the parameter space coverage with \SURSEVENDQFOUR \citep{2019PhRvR...1c3015V}.

Inside the \IMRPhenom approach, a new generation of waveform models was developed improving substantially the accuracy of the previous generation, providing a new model for the dominant harmonic of aligned-spin signal, \IMRPhenomXAS \citep{2020PhRvD.102f4001P}, a model for subdominant harmonics explicitly calibrated to \ac{NR} simulations, \IMRPhenomXHM \citep{2020PhRvD.102f4002G} and its extension to spin-precessing signals through the twisting-up technique and the employment of the multiscale expression for the spin dynamics \citep{Klein:2013qda,Chatziioannou:2013dza,Klein:2014bua}, \IMRPhenomXPHM \citep{2021PhRvD.103j4056P}.
This generation also reduced substantially the computational cost of waveform generation via an implementation of the multibanding method \citep{Vinciguerra:2017ngf,2021CQGra..38a5006G}.
On a parallel effort, a new phenomenological family of models was developed in the time domain, IMRPhenomTPHM \citep{2021PhRvD.103l4060E,2022PhRvD.105h4039E,2022PhRvD.105h4040E}, with the aim of overcoming the limitations of the \ac{SPA} in the modeling of precessing signals, and including for the first time in a phenomenological model accurate numerical solutions of the spin-precession equations.

Further improvements for spin-precessing signals were recently introduced in the \IMRPhenomXPHM model, introducing for the first time explicit calibration with spin-precessing \ac{NR} simulations in \IMRPhenomXOFOURa \citep{2021PhRvD.104l4027H,2024PhRvD.109f3012T} as well as the inclusion of the dominant multipole equatorial asymmetry \citep{Ghosh:2023mhc}, a key effect for accurately describing systems with large remnant recoil \citep{Varma:2020nbm,Borchers:2024tdi} and improving accuracy of spin-precessing models \citep{2020PhRvD.101j3014R}.
This effect has been shown in several recent studies to be relevant for the correct inference of spin-precessing signals \citep{Kolitsidou:2024vub, 2026PhRvD.113d4049E}. 
Additionally, improvements in the description of the spin dynamics during the inspiral were incorporated in \IMRPhenomXPHMST \citep{2025PhRvD.111j4019C}, performing numerically the integration of the \ac{PN} spin dynamic equations and enhancing the accuracy for the inspiral stage of the signal. Similar methods to improve the numerical integration were developed, implemented and published independently in \citet{Yu:2023lml} and \citet{Gamba:2021ydi}.
The improvements introduced in \IMRPhenomXOFOURa (including the dominant multipole equatorial asymmetry \citep{Ghosh:2023mhc}) and in \IMRPhenomXPHMST, relative to \IMRPhenomXPHM, have now been combined into a single model, \IMRPhenomXPNR{}~\citep{2026PhRvD.113h4055H}.

The \SEOBNR approach has recently been developed with a new and more accurate generation of \ac{BBH} waveform models, \SEOBNRFIVEHM \citep{2023PhRvD.108l4035P} for aligned-spin systems and \SEOBNRFIVEPHM \citep{2023PhRvD.108l4037R} for spin-precessing systems.
Improvements include more analytical \ac{PN} information \citep{Henry:2022ccf,Khalil:2023kep} and recent developments from second-order gravitational self-force for the flux and gravitational modes \citep{Warburton:2021kwk,vandeMeent:2023ols}, increasing the coverage of parameter space employed in the calibration with \ac{NR} and test-mass limit simulations \citep{Barausse:2011kb,Taracchini:2014zpa}.
Equatorial asymmetry effects for spin-precessing binaries have been incorporated as described in \citep{2026PhRvD.113d4049E}.
There are also substantial improvements to the computational efficiency of the models via the implementation of the post-adiabatic technique \citep{Nagar:2018gnk,Mihaylov:2021bpf} and the release of the modular and highly optimized Python package pySEOBNR \citep{Mihaylov:2023bkc}.
The dominant and multipolar spin-aligned models \SEOBNRFIVE and \SEOBNRFIVEHM also provide efficient \ac{ROM} versions in the Fourier domain, \SEOBNRFIVEROM and \SEOBNRFIVEHMROM \citep{2023PhRvD.108l4035P}.

\subsection{\ac{BNS} Models}
\label{sec:waveformsBNS}

When modeling \ac{BNS} mergers, the inclusion of tidal interactions is critical for accurately describing the two-body dynamics and the corresponding \GW emission.
During the inspiral, these interactions are primarily characterized by the dimensionless tidal deformability parameter $\Lambda$, which encodes the response of a \ac{NS} to the gravitational field of its companion and is directly related to the \ac{EOS} of dense nuclear matter \citep{2008PhRvD..77b1502F,Hinderer:2007mb}.
Measurements of the tidal deformability through \GW observations can constrain the \ac{EOS} of dense nuclear matter \citep{Chatziioannou:2020pqz}. 
Tidal interactions in \ac{BNS} systems are first captured at 5 \ac{PN} order \citep{2008PhRvD..77b1502F,2011PhRvD..83h4051V} in the velocity expansion and have been extended up to 7.5 \ac{PN} order \citep{2012PhRvD..85l3007D,Henry:2020ski}. 
Tidal corrections have been added to the \GW phase of the \ac{PN} inspiral Taylor Fourier-domain model \TFTWO \citep{2011PhRvD..83h4051V}.

Besides being tidally deformed by their companions, rapidly-spinning \acp{NS} acquire an intrinsic oblateness where the \ac{NS} mass-quadrupole moment depends on the dimensionless spin and encodes information about the \ac{EOS}. 
The interaction between this quadrupole and the monopole of the companion generates an additional conservative potential, usually termed quadrupole–monopole coupling, entering the two-body dynamics at the same 2 \ac{PN} order as the spin–spin term and contributing corrections to both the binding energy and the \GW flux \citep{1998PhRvD..57.5287P}.
Neglecting the quadrupole-monopole term can bias tidal-deformability and spin measurements once spins approach millisecond-pulsar values \citep{Agathos:2015uaa,Samajdar:2018dcx}.

In addition to \PN-based tidal corrections and quadrupole-monopole interaction, \ac{NR} simulations play an essential role in accurately modeling the merger and post-merger phases of \BNS coalescence, where nonlinear effects become significant, and several catalogs of \BNS simulations have been produced \citep{Dietrich:2018phi,Gonzalez:2022mgo,Kiuchi:2017pte}. 
The \NRTidal{} approach \citep{2017PhRvD..96l1501D} has been developed to incorporate tidal effects into the frequency-domain \ac{BBH} waveform models, maintaining the computational efficiency of the \ac{BBH} model baseline.
The tidal phase in the frequency domain is modeled by employing the \PN-expanded phase and calibrating it to \BNS \ac{NR} simulations.
The original \NRTidal{} model included \PN-phase corrections augmented with calibration in the time domain using a non-spinning set of equal-mass \BNS \ac{NR} simulations, and then transformed to frequency-domain using the \ac{SPA}.
It was incorporated \citep{2019PhRvD..99b4029D} in the construction of the \BNS models \IMRPhenomDNRTidal and \SEOBNRFOURNRTidal, for aligned-spin systems, \IMRPhenomPTWONRTidal, for precessing systems, which was employed in the analysis of GW170817 \citep{2017PhRvL.119p1101A} together with \SEOBNRFOURNRTidal.

\NRTidalvtwo{} \citep{2019PhRvD.100d4003D} improved upon the previous model by the employment of improved \ac{NR} data, the incorporation of spin effects, and the addition of amplitude corrections to the \GW signal.
It was incorporated to several \ac{BBH} baselines models, including \IMRPhenomPTWONRTidalTWO \citep{2019PhRvD.100d4003D} and more recently \IMRPhenomXPNRTidalTWO \citep{2025PhRvD.111f4025C}, for precessing systems, and \SEOBNRFOURNRTidalTWO for aligned-spin systems.

Within the \ac{EOB} framework, the \SEOBNRFOURT \citep{2016PhRvL.116r1101H, 2016PhRvD..94j4028S} and \TEOBResumS \citep{2015PhRvL.114p1103B,Nagar:2018zoe,Akcay:2018yyh} models aim to provide a more accurate description of the inspiral phase, particularly in regimes where tidal interactions become significant.
\SEOBNRFOURT includes tidally-induced multipole moments (up to $\ell=3$), the effect of dynamical tides from f-mode resonances \citep{2016PhRvL.116r1101H, 2016PhRvD..94j4028S}, and the spin-induced quadrupole moment \citep{1998PhRvD..57.5287P,Harry:2018hke}.
An efficient surrogate version of the model in the frequency domain, based on Gaussian process regression, was developed to reduce the cost of its employment in data-analysis applications \citep{2019PhRvD.100b4002L}.

Similarly, \TEOBResumS incorporates tidally-induced quadrupole moments using a different resummation and new information from gravitational self-force, with the lack of f-mode resonances but the addition of calibration to \ac{NR} \ac{BNS} simulations in a more recent version of the model \citep{Gamba:2023mww}.
In order to improve its efficiency in data-analysis applications, a combination of the post-adiabatic technique and the \ac{SPA} is employed to produce fast frequency-domain inspiral templates \citep{Nagar:2018gnk,Gamba:2020ljo}.

\subsection{\ac{NSBH} Models}

Modeling \ac{NSBH} mergers presents distinct challenges compared to \ac{BNS} systems, primarily because the \ac{NS} may be fully disrupted and accreted by the \ac{BH} without producing a post-merger remnant.
As in \ac{BNS} systems, the tidal response of the \ac{NS} affects the phase evolution of the inspiral, requiring the inclusion of tidal corrections in the waveform phase \citep{2011PhRvD..84j4017P}.
Additionally, in an \ac{EOS}-dependent region of parameter space, typically involving unequal masses or highly-spinning \acp{BH}, the \ac{NS} can be tidally disrupted before reaching the innermost stable circular orbit.
In such cases, the signal amplitude is strongly suppressed at high frequencies \citep{Kyutoku:2011vz,Foucart:2014nda,Kawaguchi:2015bwa}.
While tidal disruption and post-merger remnants can also occur in \ac{BNS} mergers, disrupted \ac{NSBH} systems may lack a distinct merger signature altogether if the \ac{NS} is fully disrupted before crossing the \ac{BH}’s horizon.

Current \ac{NSBH} models employed to analyse the \ac{NSBH} candidate signals from \gwtc[2.0] onwards are the \SEOBNRFOURNRtidalTWONSBH \citep{2020PhRvD.102d3023M} model and \IMRPhenomNSBH \citep{2020PhRvD.101l4059T} model.
Both models incorporate tidal information into the phase of the respective Fourier-domain \ac{BBH} baseline model using the \NRTidalvtwo{} phase model described in Section~\ref{sec:waveformsBNS}.
Disruptive and non-disruptive mergers are handled by adding tidal correction to the \GW amplitude, as well as establishing a parameter-space dependent cut-off frequency for suppressing the \GW amplitude in the case of disruptive events.
In particular, \IMRPhenomNSBH employs an \ac{NR}-calibrated amplitude model \citep{Pannarale:2013uoa,Pannarale:2015jka}, while \SEOBNRFOURNRtidalTWONSBH implements \ac{NR}-calibrated correction factors to the underlying \ac{BNS} \NRTidal{} model.
Both the amplitude model of \SEOBNRFOURNRtidalTWONSBH and the correction factors in \SEOBNRFOURNRtidalTWONSBH incorporate disruptive events by establishing a cut-off frequency and tapering the waveform at frequencies above this cut-off.

One of the main limitations of most current \ac{NSBH} models is that they are restricted to the dominant multipole and assume spin-aligned configurations.
Including only the dominant mode can lead to degeneracies in distance and inclination, as well as to a non-negligible loss in \SNR for unequal-mass \ac{NSBH} systems, since higher-order multipoles can contain a significant fraction of the signal power for asymmetric binaries.
Models are being actively developed to address this limitation~\citep{Gonzalez:2022prs,2025arXiv250700113G}.

\begin{table*}[t]
\startlongtable
\setlength{\tabcolsep}{4pt}
\renewcommand\arraystretch{1.15}
\begin{deluxetable*}{ccccc}

\tabletypesize{\scriptsize}
\tablewidth{0pt}
\tablecaption{
  A summary of waveform models used in each release of the \GWTC.
   Since the catalog is cumulative, later releases (e.g., 
  \gwtc[5.0]) include all previous candidates. This includes the re-analysis
  of \ac{O4a} which we label \gwtc[4.1].
  In most cases, \PE results from earlier releases remain
  unchanged. An exception is \gwtc[2.1], which reanalyzed O1
   and O2 data using updated waveform models, replacing
    the original \PE results from \gwtc[1.0] and \gwtc[2.0].
  As a result, the \PE analyses with the waveform 
  approximants listed for \gwtc[1.0] and \gwtc[2.0] are not
  part of \gwtc[5.0]. The models employed in \gwtc[5.0] 
  are discussed in  Section~\ref{sec:waveforms}, and
   their use within specific pipelines is described in the 
   following sections. Special candidate analyses might employ additional 
   waveform models not listed here, but discussed for those
   particular signals.
  }
  \label{tab:wf_models}

\startdata 
\makecell{Catalog release} &
Data analysed &
\makecell{Search templates} &
\makecell{Sensitivity estimates} &
\makecell{Parameter estimation}
\\ \hline
\makecell{GWTC-1.0\\ \citep{2019PhRvX...9c1040A}} & O1, O2 & \makecell{
    \SEOBNRFOURROM,\\
    \TFTWO}
    & - & \makecell{
    \IMRPhenomPVTWO,
    \IMRPhenomPTWONRTidal, \\
    \SEOBNRTHREE,
    \SEOBNRFOURNRTidal,\\
    \SEOBNRFOURT,
    \TFTWO,
    \TEOBResumS
}\\ \hline
\makecell{GWTC-2.0\\ \citep{2021PhRvX..11b1053A}} & O3a & \makecell{
    \SEOBNRFOURROM,\\
    \TFTWO} & \makecell{
    \SEOBNRFOUROPT
    } & \makecell{
    \IMRPhenomD,
    \IMRPhenomDNRTidal,\\
    \IMRPhenomHM,
    \IMRPhenomPVTWO,\\
    \IMRPhenomPTWONRTidal,\\
    \IMRPhenomPVTHREEHM,\\
    \IMRPhenomNSBH, \SURSEVENDQFOUR,\\
    \SEOBNRFOURNRtidalTWONSBH,\\
    \SEOBNRFOURROM, 
    \SEOBNRFOURHMROM,\\
    \SEOBNRFOURP,
    \SEOBNRFOURPHM,\\
    \SEOBNRFOURTSUR,
    \TFTWO,\\
    \TEOBResumS
}\\
\hline
\makecell{GWTC-2.1\\ \citep{2024PhRvD.109b2001A}} & O1--O3a & \makecell{
    \SEOBNRFOUROPT,\\
    \SEOBNRFOURROM,\\
    \STTFOUR,\\
    \TFTWO
    } & \makecell{
    \SEOBNRFOURP, \\
    \SEOBNRFOURPHM, \\
    \STTFOUR
    } & \makecell{
    \IMRPhenomPTWONRTidal,\\
    \IMRPhenomXPHM,\\
    \SEOBNRFOURPHM,\\
    } \\
\hline
\makecell{GWTC-3.0\\ \citep{2023PhRvX..13d1039A}} & O3b & \makecell{
    \SEOBNRFOUROPT,\\
    \SEOBNRFOURROM,\\
    \STTFOUR,\\
    \TFTWO
    } & \makecell{
    \SEOBNRFOURP, \\
    \SEOBNRFOURPHM, \\
    \STTFOUR
    } & \makecell{
    \IMRPhenomNSBH,\\
    \IMRPhenomXPHM,\\
    \SEOBNRFOURNRtidalTWONSBH,\\
    \SEOBNRFOURPHM\\
    } \\
\hline
\makecell{GWTC-4.0\\ \citep{2025arXiv250818082T}} & O4a & \makecell{
    \IMRPhenomD,\\
    \SEOBNRFOUROPT,\\
    \SEOBNRFOURROM,\\
    \SEOBNRFIVEROM,\\
    \STTFOUR,
    \TFTWO\\
    } & \makecell{
    \IMRPhenomXPHM
    } & \makecell{
    \IMRPhenomNSBH,\\
    \IMRPhenomPTWONRTidalTWO,\\
    \IMRPhenomXOFOURa\\
    \IMRPhenomXPHMST,\\
    \SURSEVENDQFOUR,\\
    \SEOBNRFOURNRtidalTWONSBH,\\
    \SEOBNRFIVEPHM
    } \\
\hline
\makecell{GWTC-5.0\\ \citep{GWTC:Results}} & O4b (and O4a) & \makecell{
    \IMRPhenomD,\\
    \SEOBNRFOUROPT,\\
    \SEOBNRFOURROM,\\
    \SEOBNRFIVEROM,\\
    \STTFOUR,
    \TFTWO\\
    } & \makecell{
    \IMRPhenomXPHM
    } & \makecell{
    \IMRPhenomXPHMST,\\
    \IMRPhenomXPNR,\\
    \SURSEVENDQFOUR,\\
    \SEOBNRFIVEPHM\\
    }

\enddata

\end{deluxetable*}
\end{table*}

\begin{table*}[htbp]
    \caption{Waveform models used in \GWTC analyses. 
    We indicate if the model includes spin-precession, matter effects and which multipoles are included for each model. 
    For spin-precessing models, the multipoles correspond to those available in the coprecessing frame. We indicate the most relevant reference for each model, additional references are given in the main text of Section~\ref{sec:waveforms}. Special candidate analyses might employ additional 
    waveform models not listed here, but discussed for those
    particular signals.\label{tab:wf_models_physics}}
    \small                                      
    \setlength{\tabcolsep}{4pt}
    \renewcommand\arraystretch{1.15}
    \begin{ruledtabular}
        \begin{tabular}{c c c c l}
        Model &
        Precession &
        Multipoles $(\ell,\,|m|)$ &
        Matter &
        Reference \\
        \hline

\STTFOUR                  & $\checkmark$ & $\ell\!\le\!4$                & $\checkmark$ & \citet{Isoyama:2020lls} \\
\TFTWO                    & $\times$     & $(2,2)$                       & $\checkmark$ & \citet{Damour:2000zb} \\[4pt]
\hline

\IMRPhenomD               & $\times$     & $(2,2)$                       & $\times$     & \citet{2016PhRvD..93d4007K} \\
\IMRPhenomDNRTidal        & $\times$     & $(2,2)$                       & $\checkmark$ & \citet{2019PhRvD..99b4029D} \\
\IMRPhenomHM              & $\times$     & $(2,2),(2,1),(3,3),(3,2),(4,4),(4,3)$ & $\times$ & \citet{2018PhRvL.120p1102L} \\
\IMRPhenomNSBH            & $\times$     & $(2,2)$                       & $\checkmark$ & \citet{2020PhRvD.101l4059T} \\
\IMRPhenomPVTWO           & $\checkmark$ & $(2,2)$                       & $\times$     & \citet{Bohe:PPv2} \\
\IMRPhenomPTWONRTidal     & $\checkmark$ & $(2,2)$                       & $\checkmark$ & \citet{2019PhRvD..99b4029D} \\
\IMRPhenomPTWONRTidalTWO     & $\checkmark$ & $(2,2)$                       & $\checkmark$ & \citet{2019PhRvD.100d4003D} \\
\IMRPhenomPVTHREEHM       & $\checkmark$ & $(2,2),(2,1),(3,3),(3,2),(4,4),(4,3)$ & $\times$ & \citet{2020PhRvD.101b4056K} \\
\IMRPhenomXPHM            & $\checkmark$ & $(2,2),(2,1),(3,3),(3,2),(4,4)$ & $\times$   & \citet{2021PhRvD.103j4056P} \\
\IMRPhenomXPHMST          & $\checkmark$ & $(2,2),(2,1),(3,3),(3,2),(4,4)$ & $\times$   & \citet{2025PhRvD.111j4019C} \\
\IMRPhenomXOFOURa         & $\checkmark$ & $(2,2),(2,1),(3,3),(3,2),(4,4)$ & $\times$   & \citet{2024PhRvD.109f3012T} \\
\IMRPhenomXPNR            & $\checkmark$ & $(2,2),(2,1),(3,3),(3,2),(4,4)$ & $\times$   & \citet{2026PhRvD.113h4055H} \\[4pt]
\hline

\SURSEVENDQFOUR           & $\checkmark$ & $\ell\!\le\!4$                & $\times$     & \citet{2019PhRvR...1c3015V} \\[4pt]
\hline

\SEOBNRTHREE              & $\checkmark$ & $(2,2),(2,1)$                 & $\times$     & \citet{2014PhRvD..89h4006P} \\
\SEOBNRFOUROPT            & $\times$  & $(2,2)$                 & $\times$     & \citet{Devine:2016ovp} \\
\SEOBNRFOURROM            & $\times$     & $(2,2)$                       & $\times$     & \citet{2017PhRvD..95d4028B} \\
\SEOBNRFOURHMROM          & $\times$     & $(2,2),(2,1),(3,3),(4,4),(5,5)$ & $\times$   & \citet{2020PhRvD.101l4040C} \\
\SEOBNRFOURNRTidal        & $\times$     & $(2,2)$                       & $\checkmark$ & \citet{2019PhRvD..99b4029D} \\
\SEOBNRFOURNRtidalTWONSBH & $\times$     & $(2,2)$                       & $\checkmark$ & \citet{2020PhRvD.102d3023M} \\
\SEOBNRFOURP              & $\checkmark$ & $(2,2),(2,1)$                 & $\times$     & \citet{2020PhRvD.102d4055O} \\
\SEOBNRFOURPHM            & $\checkmark$ & $(2,2),(2,1),(3,3),(4,4),(5,5)$ & $\times$   & \citet{2020PhRvD.102d4055O} \\
\SEOBNRFOURT              & $\times$     & $(2,2)$                       & $\checkmark$ & \citet{2016PhRvD..94j4028S} \\
\SEOBNRFOURTSUR           & $\times$     & $(2,2)$                       & $\checkmark$ & \citet{2019PhRvD.100b4002L} \\
\SEOBNRFIVEROM            & $\times$     & $(2,2)$ & $\times$ & \citet{2023PhRvD.108l4035P} \\
\SEOBNRFIVEPHM            & $\checkmark$ & $(2,2),(2,1),(3,3),(3,2),(4,4),(4,3),(5,5)$ & $\times$ & \citet{2023PhRvD.108l4037R} \\
                          &              &                                             &          & \citet{2026PhRvD.113d4049E} \\[4pt]
\hline

\TEOBResumS               & $\times$     & $(2,2),(2,1),(3,3),(3,2),(3,1)$                       & $\checkmark$ & \citet{Akcay:2018yyh}

      \end{tabular}
    \end{ruledtabular}
    \normalsize                                
  \end{table*}

\subsection{Luminosity and Remnant Properties}
\label{sec:waveformsfinalproperties}

Besides modeling the \GW waveform, accurate predictions of the mass and spin of the final remnant \ac{BH}, as well as estimations of the peak luminosity, are also crucial for population studies and tests of \GR.
Several analytical fits have been developed using the remnant properties from \ac{NR} \ac{BBH} simulations \citep{2016ApJ...825L..19H,2017PhRvD..96b4006K,2017PhRvD..95b4037H,2017PhRvD..95f4024J}, and recently accurate Gaussian process regression surrogate models have been developed \citep{Varma:2018aht,Islam:2023mob}.
For \ac{NSBH} systems, dependency on the tidal deformability of the \ac{NS} have been included \citep{2019PhRvL.123d1102Z}.
These fitting formulae generally depend on the component masses and spins, typically specified at some reference frequency in the inspiral stage of the signal.
For estimating these quantities from the inferred source properties (see Section~\ref{sec:PECatalogChoices}), the input spin values are evolved forward until a fiducial orbital frequency near merger is achieved, and then results from different fitting formulae are averaged to provide the final estimates.
For precessing systems, these formulae are often augmented with in-plane spin contribution before the average procedure is applied \citep{JohnsonMcDaniel:2016aaa}.

\section{Signal Identification}\label{sec:searches}

The \ac{GW} strain time series produced by advanced-era interferometric detectors can be treated as a linear superposition of continuous non-astrophysical \emph{noise} and occasional transient astrophysical \emph{signals}.
On timescales of tens of seconds, the noise is generally well approximated by colored stationary Gaussian processes, allowing for relatively stable modeling and analysis \citep{2020CQGra..37e5002A}.
However, on longer timescales, the noise becomes non-stationary, exhibiting time-dependent statistical properties \citep{2021PhRvD.104f3034Z, 2020CQGra..37u5014M, Mozzon:2022, Kumar:2025}.
It is frequently contaminated by transient non-Gaussian artifacts, known as \emph{glitches}~\citep{2018RSPTA.37670286N,2023CQGra..40f5004G,2025CQGra..42h5016S}, and is also affected by slowly time-varying broadband disturbances \citep{2020CQGra..37e5002A}.
On the other hand, the signals in the data are presently relatively rare, occurring at a rate of just a few per week.
\ac{GW} \emph{searches} are thus required to perform statistical data reduction, taking in the kilohertz-sampled strain data and producing a list of astrophysical candidates.
The candidate lists are released as part of the \thisgwtcfull{} data release \citep{GWTC-4.0:Search} and also used as input to the downstream data-analysis workflow illustrated in Figure~\ref{fig:data-flow-chart}.

The search for \GW transient candidates is carried out through two distinct phases.
Initially, \emph{online} analyses enable prompt follow-up observations by the global astronomical community \citep{2019ApJ...875..161A,2023PhRvX..13d1039A,OPA}.
Later, \emph{offline} analyses are conducted to produce a more accurate list of candidates by re-evaluating the significance of the initial online candidates and identifying new ones.
Some of the factors responsible for the differences in the online and offline analyses are the use of the final strain data, improved data quality, enhanced algorithms, and a better understanding of the noise background in the data.

Two broad families of detection algorithms are used: search pipelines for minimally-modeled transient sources that do not rely on specific waveform predictions and search pipelines that, conversely, aim to maximize sensitivity to \acp{CBC} by using sets of template waveforms (a subset of the models described in Section~\ref{sec:waveforms}).
For the \gwtc, we use multiple pipelines concurrently and combine their results.
This approach minimizes the risk of missing astrophysical signals and allows us to cover a wider range of the \ac{CBC} parameter space.

Template-based pipelines target specific regions of the \acp{CBC} parameter space, usually defined by the redshifted (detector-frame) masses \citep{1987GReGr..19.1163K} and spins of the two compact objects, which are arranged into \emph{template banks}.
Template waveforms are currently calculated neglecting several physical effects, specifically non-dominant multipole emission, orbital precession, eccentricity and tidal deformability.
The templates thus only model the dominant multipole emission from quasi-circular \BBH coalescences with spins parallel to the orbital angular momentum.

This simplification, which is appropriate for the majority of the \ac{CBC} signals detected so far, reduces the dimensionality of the parameter space to be covered, leading to a beneficial reduction of the computational and operational costs of the analysis.
The price to pay for this simplification is a reduced search sensitivity in regions of the \ac{CBC} parameter space where the neglected physical effects give a large contribution to the signal \citep{DalCanton:2014qjd, 2016PhRvD..93h4019C, Harry:2016ijz, 2017PhRvD..95j4038C, 2019PhRvD..99b4029D, 2020PhRvD.102d3005R, Phukon:2024amh, Chia:2023tle, Mehta:2025jiq}.
Combining template-based with minimally-modeled search methods partially covers such cases, since the latter impose only weak assumptions on the signal.
Moreover, signals that appear to deviate from the simplified templates, such as GW190412 \citep{2020PhRvD.102d3015A} and GW190521 \citep{2020PhRvL.125j1102A}, can still be detected with high significance because the deviation is relatively small.
Still, including the correct physical effects in a template-based search is feasible and has yielded additional marginally-significant candidate events with interesting properties \citep{Dhurkunde:2025, 2024PhRvD.110d4063W}.

In modeled searches, we search the data from each detector by matched filtering each normalized template $\tilde{u}(f| \PEparameterIntrinsic)$ (where $\PEparameterIntrinsic$ is a vector over a subset of intrinsic \ac{CBC} parameters), to produce a \SNR time series defined as \citep{2007gwte.book.....M, 2012PhRvD..85l2006A}

\begin{equation}
  \label{eq:search_snr}
	\rho(t) = 4 \left| \int_{\flow}^{\fhi} \frac{\PEdataFrequencyDomain(f) \tilde{u}^{*}(f| \PEparameterIntrinsic)}{S_n(f)} e^{2i\pi ft} df \right|,
\end{equation}
where $\PEdataFrequencyDomain$ represents the strain data, $S_n$ is the \PSD modeling the detector noise \citep[see Appendix B of][]{GWTC:Introduction}, and the tilde over a symbol indicates the Fourier transform of a time series.

The lower integration limit frequency \flow is determined by considerations of computational cost, data quality, and calibration reliability.
The upper limit \fhi is predominantly based on considerations of computational cost.
Both limits may vary with search implementations and templates.
In practice, $S_n$ is a point estimate of the noise variance at a given time and frequency, and is measured empirically from the data themselves.
This is possible in a straightforward way with current detectors thanks to the sparsity of the transient astrophysical signals and the absence of detectable persistent signals.
This measurement is repeated at regular intervals in order to track the time-dependence of noise properties.

Modeled search pipelines produce triggers by looking for peaks exceeding a given threshold in the \SNR time series.
This same procedure is applied to the calibrated data provided by all available detectors.

All search pipelines, minimally-modeled or templated-based, attempt to pair candidates from one detector with coincident candidates from others, though some pipelines also consider triggers observed in a single interferometer \citep{2019arXiv190108580S, 2025CQGra..42j5009A, 2022CQGra..39u5012C}.
Some pipelines employ an additional \emph{coherent} step, where the signal likelihood is maximized over some of the extrinsic parameters using strain or \SNR time series information from all detectors simultaneously, rather than only from \SNR maxima in each separate detector.
The root-sum-square of the single-detector \acp{SNR}, called the \emph{network \ac{SNR}}, is often used to quantify the overall loudness of a candidate over all available detectors \citep{1994PhRvD..49.2658C,2007gwte.book.....M}.
Each pipeline combines the \SNR information with additional signal-consistency tests to construct its own ranking statistic.

Based on their ranking, search pipelines assign each candidate a \ac{FAR} value, which quantifies the expected rate of non-astrophysical candidates produced with a rank at least as high as the candidate under consideration.

In addition to the measure of detection significance provided by the \ac{FAR} of a candidate, we have also provided the probability \pastro that a candidate is of astrophysical rather than terrestrial (noise) origin.
The calculation of \pastro accounts for both the rate of noise events corresponding to the \ac{FAR} and an estimated rate of astrophysical signals under some prior model of the \ac{CBC} population \citep{2015PhRvD..91b3005F, 2020CQGra..37d5007K}.
It is formally defined as \citep{2016ApJ...833L...1A}
\begin{equation}
	\pastro(x) = \frac{\mathcal{R}_{\mathrm{astro}} f(x)}{\mathcal{R}_{\mathrm{astro}} f(x) + \mathcal{R}_{\mathrm{noise}} b(x)},
\end{equation}
where $\mathcal{R}_{\mathrm{astro}}$ is the total rate of astrophysical candidates in a given pipeline, $\mathcal{R}_{\mathrm{noise}}$ is the total rate of candidates due to noise; $f(x)$ is the \PDF of signal events at the candidate ranking value $x$, and $b(x)$ is the equivalent \PDF for noise events.
Each pipeline estimates these rates and calculates a \pastro value for the candidates it identifies using its own method.

Since \ac{O4a}, most pipelines use a common signal model for $f(x)$ based on a distribution over source masses, spins and distance resembling the astrophysical population inferred from \gwtc[3.0] \citep{2023PhRvX..13a1048A} while supporting possible outliers \citep{2025PhRvD.112j2001E};
 a slightly different model used by one analysis is described in Section~\ref{subsection:GstLAL}.

The total astrophysical probability \pastro is then distributed between three mutually-exclusive source categories: a \BNS class corresponding to both component (source-frame) masses ranging between \qty{1}{\Msun} and \qty{3}{\Msun}; a \NSBH class for one mass between \qty{1}{\Msun} and \qty{3}{\Msun} and the other above; and a \BBH class corresponding to both masses above \qty{3}{\Msun}.
The associated probabilities for each category thus sum to \pastro:
\begin{equation}
	\pbns + \pnsbh + \pbbh = \pastro .
\end{equation}
The complementary probability that the candidate is of terrestrial origin, i.e., caused by noise rather than a \ac{CBC}, is notated as $\pterr = 1 - \pastro$.

Transient search methods used to compile our catalog for data from the \ac{O1} and \ac{O2} are described in Section 3 of \cite{2019PhRvX...9c1040A}, and for data from \ac{O3}, in Section 4 of \cite{2021PhRvX..11b1053A}, Section 3 of \cite{2024PhRvD.109b2001A} and Section 4 of \cite{2023PhRvX..13d1039A}.
We describe here the methods used for data from \ac{O4a} and \ac{O4b}, in the context of this cumulative catalog\note{check if it makes sense to describe methods for O4a and O4b together.}.
In the coming subsections, we present details of the different pipelines used to identify transient \GW candidates, including coherent WaveBurst (\CWB) for minimally-modeled sources \citep{2016PhRvD..93d2004K, 2022PhRvD.105h3018M,2025PhRvD.111b3054M}, and GStreamer LIGO Algorithm Library \citep[\GSTLAL;][]{2012PhRvD..85h1504C, 2017PhRvD..95d2001M, 2019arXiv190108580S, 2020PhRvD.101b2003H, 2021SoftX..1400680C, 2024PhRvD.109d4066S, 2023arXiv230607190R, 2023PhRvD.108d3004T, 2024PhRvD.109d2008E, 2025arXiv250606497J, 2025arXiv250523959J}, Multi-Band Template Analysis \citep[\MBTA;][]{2025CQGra..42j5009A}, \PYCBC \citep{2021ApJ...923..254D}, and Summed Parallel Infinite Impulse Response \citep[\SPIIR;][]{2022PhRvD.105b4023C} for model-based search analyses.
The names we use for the different pipelines typically indicate publicly available and fairly general software packages.
In principle, one may construct many different search algorithms with those packages, and may configure them in arbitrary ways to optimize different figures of merit (e.g.,~sensitivity to specific populations of signals, or computing cost).
For the purposes of \gwtc, each pipeline name identifies the combination of a specific software package with the specific set of configuration choices described in this paper.
In Table~\ref{tab:pipeline_comparison}, we summarize the key features of each search pipeline and the strain data they analyzed.
The table focuses on the configurations used for offline analyses, which are the main scope of this paper, though the online configurations are similar.

Most of these search pipelines also participated in pre-\gwtc[4.0] releases, with some configuration differences.
The main change concerning modeled searches is that some pipelines have refined their search space to reflect past detections \citep{2023PhRvX..13d1039A} and to cover the higher redshift range enabled by improved detector sensitivity.
In addition, all pipelines have implemented new features aimed at better distinguishing non-astrophysical transients and improving sensitivity to \ac{CBC} signals.
In parallel, they continue to improve their computational and operational efficiency.
There is also a continuous effort to improve the population prior used in searches, informed by results on past observing runs \citep{2023PhRvX..13a1048A}.
The main differences for each pipeline are described in the corresponding subsections below.

\begin{deluxetable*}{c|c|c|c|c|c}
\tablecaption{A high-level overview of the similarities and differences between \GW search pipelines used in the analysis of \ac{O4b} data for \thisgwtcfull.
The \GDSStrainFrame and \GDSStrainFrameAR channels are identical other than that the latter contains only data for which the detectors are nominally in a good working state \citep{OpenData}.
}
\label{tab:pipeline_comparison}
\tablehead{
  &
  \colhead{\CWBBBH} &
  \colhead{\GSTLAL} &
  \colhead{\MBTA} &
  \colhead{\PYCBC} &
  \colhead{\SPIIR (online only)}
}
\startdata
Data source (online) & \multicolumn{1}{c|}{\GDSStrainFrame (\ac{LIGO})} & \multicolumn{4}{c}{\GDSStrainFrame (\ac{LIGO})} \\
channel name & \multicolumn{1}{c|}{} & \multicolumn{4}{c}{\HrecStrainFrame (\ac{Virgo})} \\
\hline
Data source (offline) &
  \multicolumn{1}{c|}{\GDSStrainFrame (\ac{LIGO})} &
  \multicolumn{1}{c|}{\GDSStrainFrame (\ac{LIGO})} &
  \multicolumn{2}{c|}{\GDSStrainFrameAR (\ac{LIGO})} &
  N/A \\
channel name  & \multicolumn{1}{c|}{} & 
  \multicolumn{1}{c|}{\HrecStrainFrame (\ac{Virgo})} &
  \multicolumn{2}{c|}{\HrecStrainFrameAR (\ac{Virgo})} &
  N/A \\
\hline
\makecell{Target waveform\\morphology} &
  \makecell{Linear superposition\\of wavelets} &
  \multicolumn{4}{c}{Quasicircular spin-aligned $(2,2)$-mode \acp{CBC}} \\
\hline
\ac{PSD} estimation &
  \makecell{WDM wavelet} &
  \makecell{Welch with geometric-mean\\and median averaging} &
  \multicolumn{3}{c}{Welch with median averaging} \\
\hline
Filtering approach &
  \makecell{Time-frequency\\decomposition} &
  \makecell{Matched filtering via\\time-domain SVD filters} &
  \makecell{Matched filtering\\via multi-band FFT} &
  \makecell{Matched filtering\\via single-band FFT} &
  \makecell{Matched filtering via\\time-domain summed\\parallel IIR filters} \\
\hline
\makecell{Working frequency\\band (Hz)} &
  16--320 &
  \makecell{10--1024\\(template dependent)} &
  \makecell{20--2048\\(template dependent)} &
  \makecell{15--1024\\(template dependent)} &
  15--1024 \\
\hline
Glitch suppression &
  \makecell{Gating, coherence test} &
  \makecell{Mass-based gating,\\autocorrelation $\xi^2$ test} &
  \makecell{Gating,\\autocorrelation $\chi^2$ test,\\trigger-rate penalty} &
  \makecell{Gating,\\time-frequency $\chi^2$ test,\\sine--Gaussian $\chi^2$ test} &
  \makecell{Gating,\\autocorrelation $\chi^2$ test} \\
\hline
\makecell{Multi-detector\\trigger formation} &
  Coherent &
  \multicolumn{3}{c|}{Coincident} &
  Coherent \\
\hline
Single-detector candidates &
  N/A &
  \makecell{\ac{IMBH} excluded;\\Other singles penalized} &
  $\mchirp < 7 \Msun$ &
  \makecell{Template duration $> \qty{0.3}{s}$,\\background-based cuts} &
  N/A \\
\hline
Ranking statistic &
  \makecell{Coherent network \SNR\\weighted by ML classifier} &
  Likelihood ratio &
  \makecell{Reweighted \SNR\\with astrophysical prior} &
  \makecell{Likelihood ratio\\with KDE-based\\template weighting} &
  \makecell{Likelihood ratio,\\extrapolation using\\KNN KDE} \\
\hline
\makecell{Use of external\\data-quality info} &
  \makecell{CAT2 as veto\\for candidates} &
  \multicolumn{2}{c|}{None} &
  \iDQ~included in ranking &
  None \\
\hline
\makecell{Multi-detector\\background estimation} &
  Time shifts &
  \multirow{2}{*}{\makecell{Monte Carlo of single-detector\\trigger distributions,\\ excluding manually-vetted\\online candidates}} &
  \multicolumn{3}{c}{Time shifts} \\
\cline{1-2}
\cline{4-6}
\makecell{Single-detector\\background estimation} &
  N/A &
  &
  \multicolumn{2}{c|}{\makecell{Exponential fit and extrapolation\\of empirical distribution}} &
  N/A \\
\hline
\pastro population model &
  \makecell{Mass and spin distributions\\based on GWTC-3.0 population\\(\ac{BBH} only)} &
  \makecell{Salpeter primary mass function} &
  \multicolumn{3}{c}{Mass and spin distributions based on GWTC-3.0 population} \\
\enddata
\end{deluxetable*}

Each search pipeline provides its own list of candidates, along with corresponding values of the \ac{FAR} and \pastro, which are neither corrected for multiple trials nor combined in any way (see Section~\ref{sec:selection_criteria}).

For offline analyses, a set of \emph{Category 1} (CAT1) flags are determined, which flag periods of severe data-quality or technical issues affecting the interferometers during the observation time.
During CAT1 periods \citep[detailed in Section 4.1 of][]{2025CQGra..42h5016S}, the data are considered too contaminated by noise to be analyzable within realistic computational or human resources.

During the \ac{O1} to \ac{O3} observing runs, short-duration \emph{Category 2} (CAT2) flags were also provided for transient searches \citep{2020CQGra..37e5002A}.
CAT2 flags are based on statistical correlations found between excess noise transients in auxiliary channels and in the strain data.
As of \ac{O4}, these CAT2 flags are no longer used in \ac{CBC} searches, though they continue to be used for minimally-modeled transient searches \citep{2025CQGra..42h5016S}.

As an extra input for \ac{CBC} searches, a time series is generated by the \iDQ{} pipeline \citep{Essick:2020qpo}.
This time series contains statistical data-quality information based on the activity of auxiliary channels deemed safe (i.e., channels that should not be influenced by \acp{GW}).
Observation times marked with \iDQ{} flags are likely to be contaminated by glitches.
Unlike CAT1 flags, which are mandatory vetoes prior to any analysis, \iDQ{} flags are optional and may be used either as vetoes, or as input in the ranking statistic of the candidates.
CAT1 vetoes and the \iDQ{} time series are created from the strain data and auxiliary channels as a pre-processing step and ingested by the offline searches (see Figure~\ref{fig:data-flow-chart}).
Details of the data release of the \iDQ{} time series are provided in \citet{GWTC-4.0:DQ}.

\begin{figure*}[ht]
  \centering
  \begin{minipage}{0.49\textwidth}
      \centering
      \includegraphics[scale=0.38]{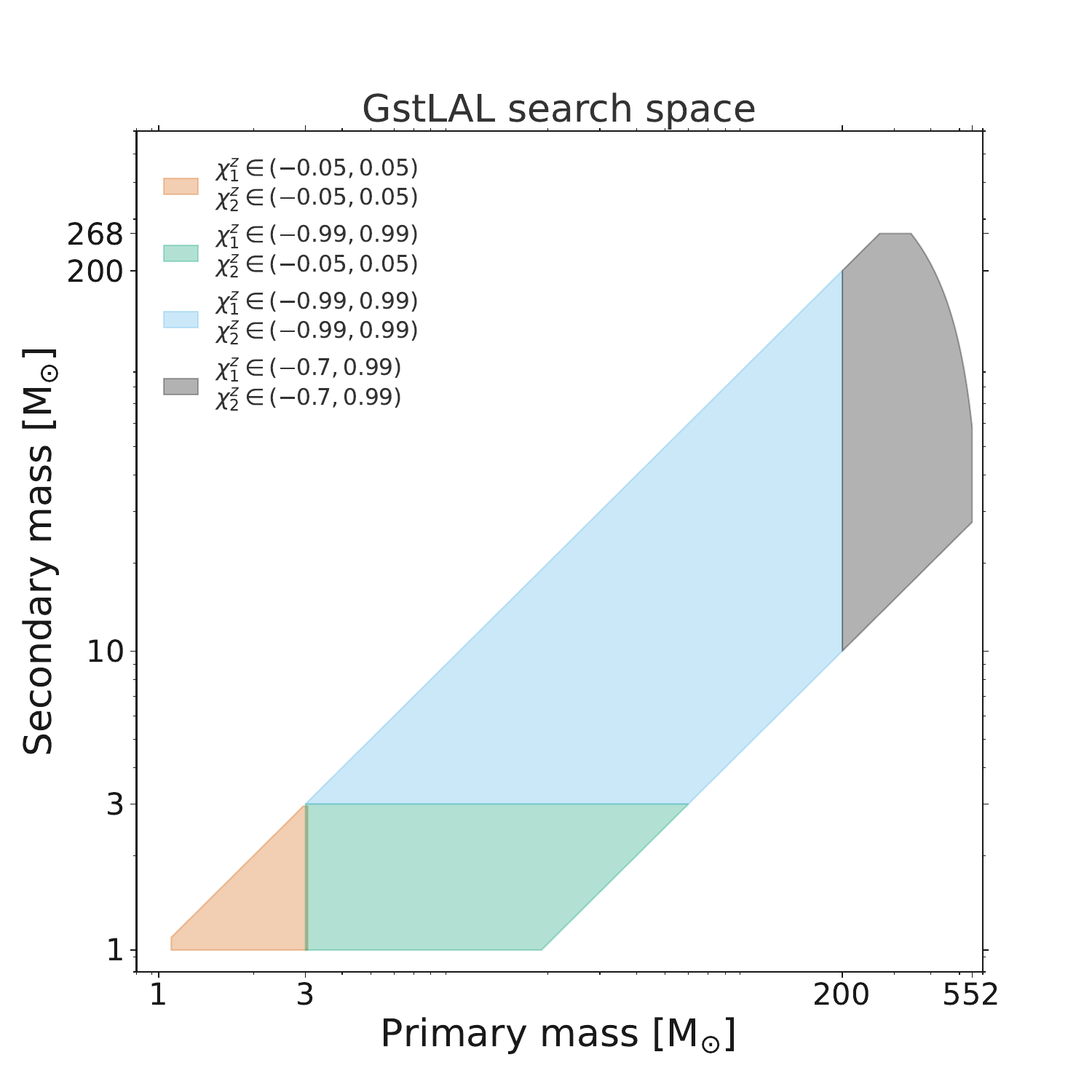}
      \label{fig:search_space_gstlal}
  \end{minipage}
  \begin{minipage}{0.49\textwidth}
    \centering
    \includegraphics[scale=0.38]{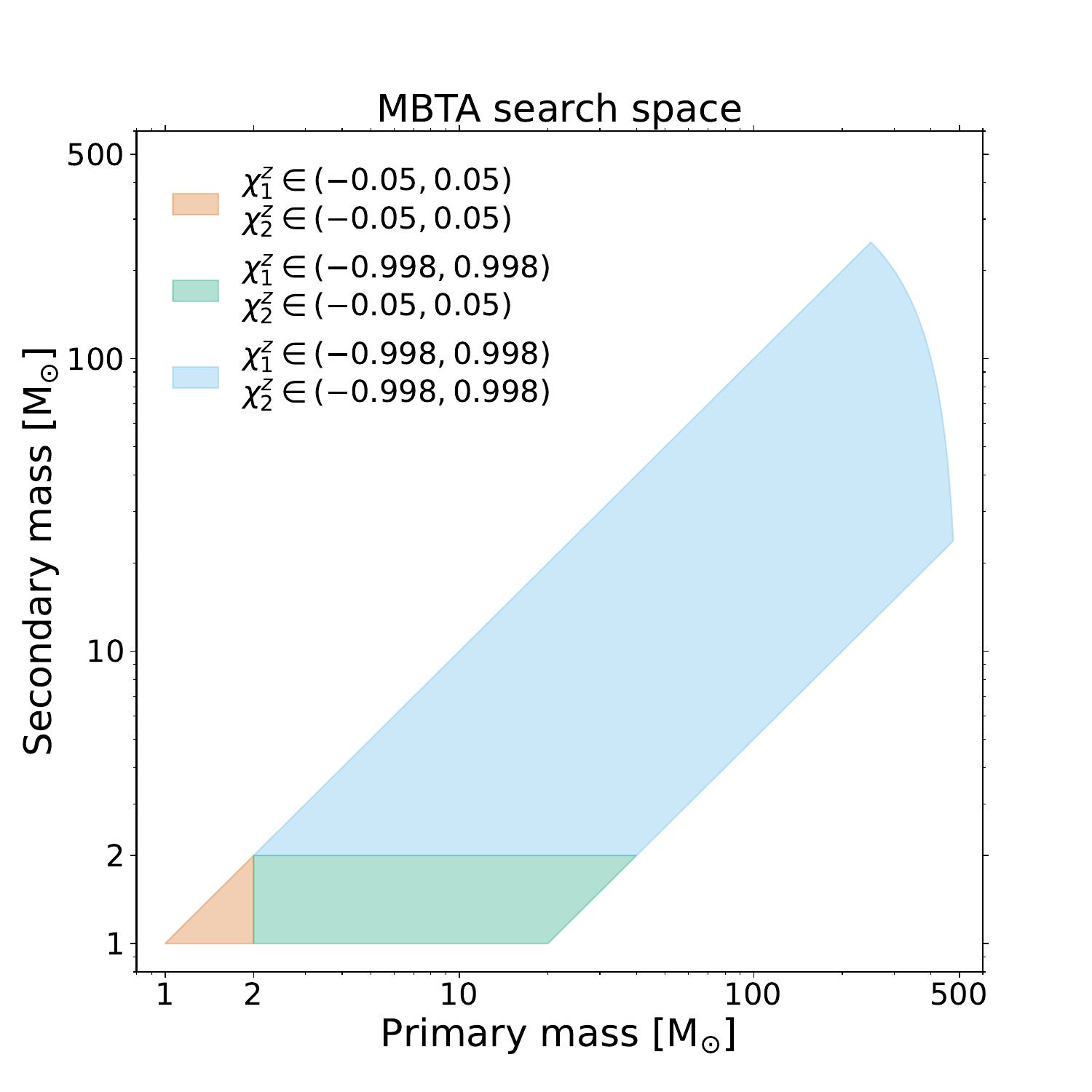}
    \label{fig:search_space_mbta}
  \end{minipage}
  \begin{minipage}{0.49\textwidth}
    \centering
    \includegraphics[scale=0.38]{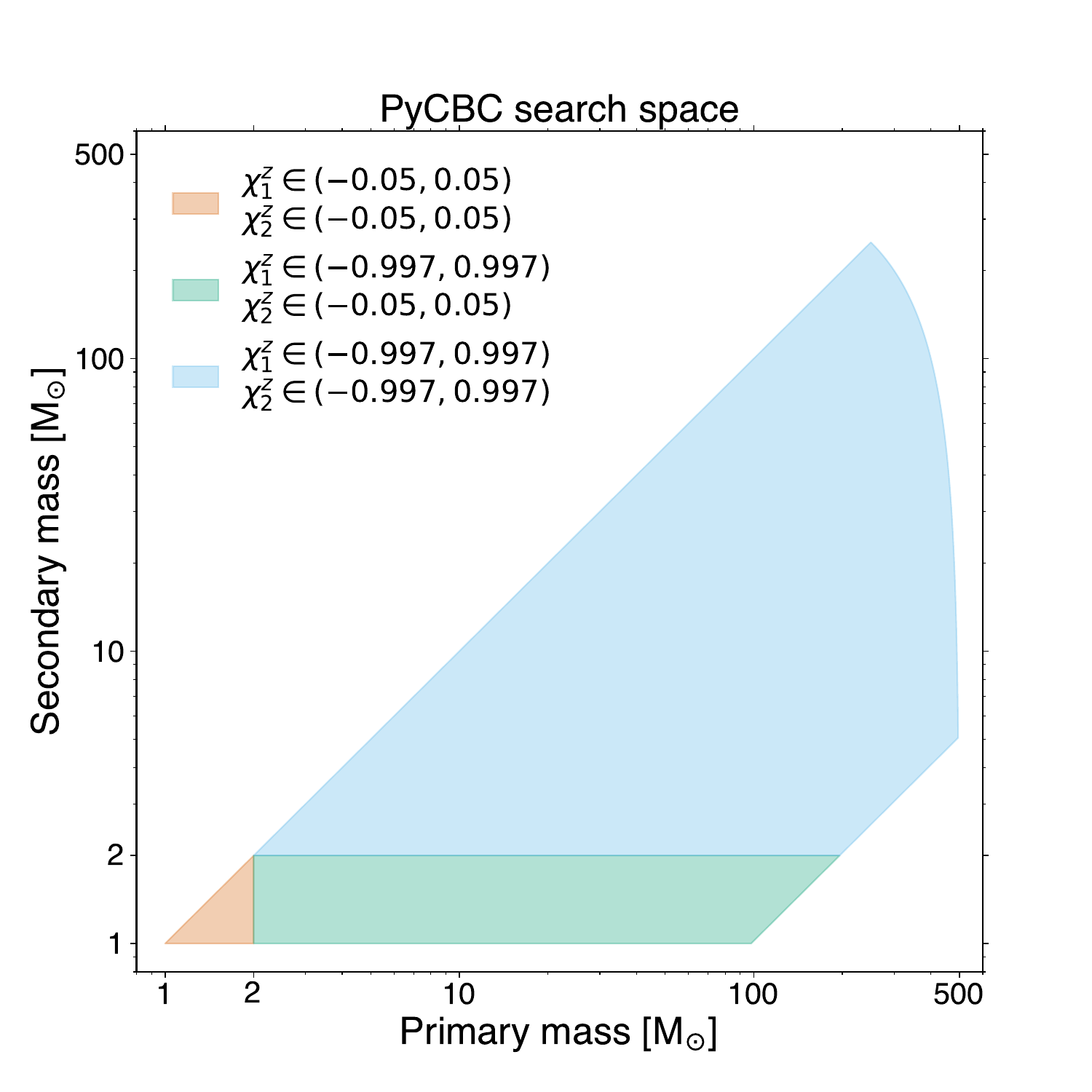}
    \label{fig:search_space_PyCBC}
  \end{minipage}
  \begin{minipage}{0.49\textwidth}
    \centering
    \includegraphics[scale=0.38]{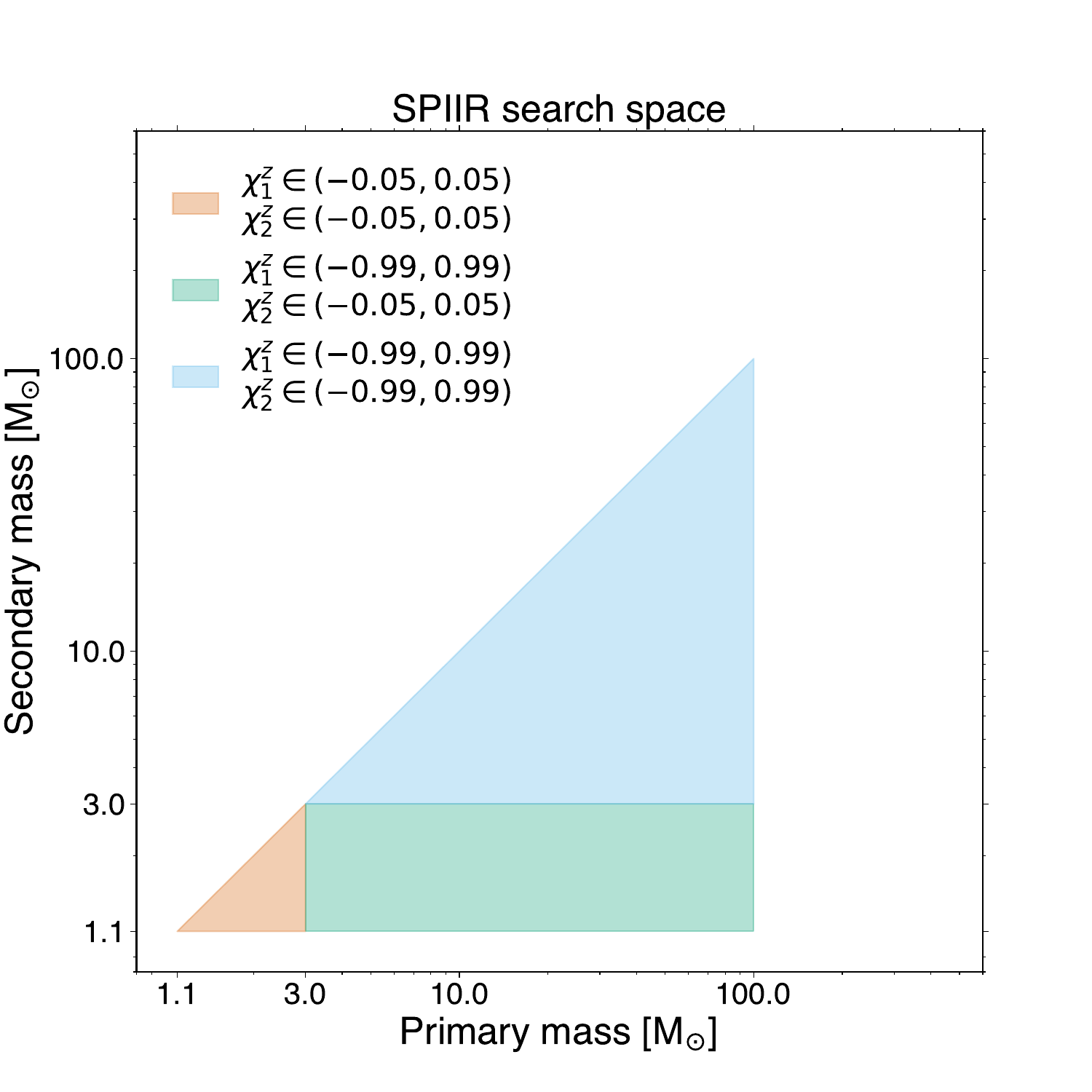}
    \label{fig:search_space_SPIIR}
  \end{minipage}
  \caption{
    Regions of the \ac{CBC} parameter space explored by our different template-based search pipelines during the \ac{O4b} observing run.
    Individual masses are given in the detector frame, i.e.~they are redshifted.
    The spin parameters $\chi^{z}_{1,2}$ are the spin projections along the orbital angular momentum.
  }
  \label{fig:search_space}
\end{figure*}

\subsection{\CWBBBH}
\label{subsection:cWB}

The \CWB algorithm is designed to detect transient \GW signals using networks of \GW detectors without requiring a specific waveform model.

It identifies coincident excess power events by analyzing multi-resolution time\textendash{frequency} (TF) representations of detector data~\citep{2008CQGra..25k4029K,2016PhRvD..93d2004K} whitened in the \ac{WDM} wavelet domain \citep{2012JPhCS.363a2032N}.

Once potential \GW signals or noise events are identified, \CWB reconstructs the source sky location and the signal waveforms recorded by the detectors using the constrained maximum-likelihood method~\citep{2005PhRvD..72l2002K,2016PhRvD..93d2004K}.

This analysis is performed on detector strain data, and can in principle accommodate signal frequencies up to a few kilohertz and durations up to a few seconds, though one might choose a narrower search space as part of the configuration choices (see Table \ref{tab:pipeline_comparison}).

The \CWB detection statistic relies on coherent energy $E_\mathrm{c}$, which is derived from cross-correlating the reconstructed signal waveforms in the detectors and normalizing them by the spectral amplitude of the detector noise.
The square root of $E_\mathrm{c}$ provides a lower bound on the signal network \SNR and is used for the initial selection of candidates identified by \CWB.
As the production rate of \CWB candidates is primarily driven by glitches, additional post-production vetoes are applied to further reduce the background rates.

One primary signal-independent veto statistics is the network correlation coefficient, defined as $c_\mathrm{c} = E_\mathrm{c}/(E_\mathrm{c}+E_\mathrm{n})$, where $E_\mathrm{n}$ represents the normalized residual noise energy after the reconstructed signal has been subtracted from the data.
Typically, for glitches $c_\mathrm{c} \ll 1$.
Consequently, candidates with $c_\mathrm{c} < 0.6$ are rejected as potential glitches.
For authentic \GW signals, the residual energy $E_\mathrm{n}$ is expected to follow the $\chi^2$ distribution with the number of degrees of freedom $N_\mathrm{DoF}$ proportional to the number of TF data samples comprising the signal.
The reduced $\tilde{\chi}^2=\chi^2/N_\mathrm{DoF}$ statistic serves as a powerful signal-independent veto, effectively identifying and removing glitches where $\tilde{\chi}^2$ is significantly greater than 1.
To enhance the search for \ac{CBC} signals, \CWB performs analysis in the frequency band below \qty{512}{\hertz} and also employs weak signal-dependent vetoes.
These vetoes are based on the central frequency $f_\mathrm{c}$ and the detector-frame chirp mass $(1+z)\mchirp$, both of which are estimated from the TF evolution of the signal power~\citep{2016CQGra..33aLT01T}.

All candidates identified by \CWB are ranked by the reduced coherent network \SNR
\begin{eqnarray}
  \label{eq1}
  \rho_\mathrm{cWB} = \sqrt{\frac{E_\mathrm{c}}{1+\tilde{\chi}^2 [ \max(1,\tilde{\chi}^2)-1 ]}} \; .
\end{eqnarray}
To estimate the statistical significance of the \GW candidates, they are ranked against background events generated by the detector noise.
A comprehensive background data sample, equivalent to approximately 1000 years of observation time, is obtained by repeating the \CWB analysis on time-shifted data.
To ensure that astrophysical signals are excluded from the background data, the time shifts are selected to be much larger than the expected signal time delay between the detectors.
The statistical significance of each \CWB candidate is quantified by its \ac{FAR}.
The \ac{FAR} is defined as the rate of background events that exhibit a larger $\rho_\mathrm{cWB}$ value than the \GW candidate in question.
To account for potential long-term variations in detector noise, the candidate significance is estimated using nearby data intervals, typically one to two weeks in length.
For the calculation of \pastro, \CWB combines the time-shifted background analysis and simulated \BBH signals based on the \gwtc[3.0] population \citep{2023PhRvX..13d1039A}.
Only \pbbh is calculated.

Since the publication of \gwtc[3.0], the \CWB algorithm used in the previous observing runs~\citep{2016PhRvD..93d2004K,2021SoftX..1400678D}, i.e., \CWBTWOG, has undergone significant changes, and a new version, \CWBBBH, has been adopted for \ac{CBC} searches: it incorporates two major changes compared to \CWBTWOG.

The first change involves replacing the \ac{WDM} wavelet transform \citep{2012JPhCS.363a2032N} with the multi-resolution \emph{WaveScan} transform, which is based on the Gabor wavelets~\citep{2022arXiv220101096K}.
This update effectively reduces temporal and spectral leakage in the TF data, potentially leading to more accurate signal representation.

Secondly, in addition to the traditional excess-power statistic, the algorithm now incorporates the cross-power statistic \citep{2022arXiv220101096K} for identifying transient signals.
The excess-power represents the total power of a TF data sample (or \emph{WaveScan} pixel) integrated over the detector network.
The cross-power represents the correlation of power in the detectors.
Both statistics are maximized over possible time-of-flight delays of a \GW signal across the detector network.
The excess-power and the cross-power amplitudes, $a_\mathrm{e}$  and $a_\times$, respectively, follow a predictable half-normal distribution with unity variance assuming quasi-stationary detector noise \citep{2022arXiv220101096K}.
For analysis, TF pixels with the excess-power amplitude exceeding 2.3 standard deviations are selected.
Adjacent pixels in the TF plane are then clustered to form the initial \CWBBBH candidates (clusters).

The clustered excess-power $\sum_{i=1}^I{a^2_\mathrm{e}[i]}$ and the cross-power $\sum_{i=1}^I{a^2_\times[i]}$ are used to calculate the upper bounds on the signal network \SNR $\rho_\mathrm{e}$ and the coherent network \SNR $\rho_\times$ respectively~\citep{2025PhRvD.111b3054M}, where $I$ is the number of the TF pixels comprising the cluster.

The $\rho_\mathrm{e}$ is optimized for identification of clusters due to fluctuations of the quasi-stationary detector noise dominating the initial \CWBBBH rate on the order of \qty{100}{\hertz}.

All clusters with $\rho_\mathrm{e}>4$ are accepted for further analysis.

The accepted \CWBBBH clusters are mostly produced by glitches with a typical rate of \qty{0.1}{\hertz}.
After selection, the initial TF clusters are aggregated if they fall within the time and frequency intervals of \qty{0.23}{\second} and \qty{64}{\hertz} respectively.
This improves the energy collection for transient candidates that can be fragmented into clusters by the TF transform.
The glitch rate is further reduced by requiring $\rho_\times>7$, where the upper bound of the coherent network \SNR is calculated for the defragmented candidates.

The above conditions on the $\rho_\mathrm{e}$ and $\rho_\times$ require that the \GW signal clusters with \SNR$>4$ and defragmented \GW candidates with \SNR$>7$ are accepted for the analysis.
The resulting reduced candidates rate, on the order of \qty{1}{\milli\hertz}, makes the subsequent likelihood analysis and reconstruction of the remaining candidates computationally feasible.
At this stage, the sky location of each candidate is determined and the signal waveforms are reconstructed with the inverse \emph{WaveScan} transform \citep{2022arXiv220101096K}.
The \CWBBBH candidates with  $\rho_\mathrm{cWB}>7$ are stored for the post-production analysis.

Moreover, the post-production veto analysis has been replaced by the XGBoost machine-learning algorithm~\citep{PhysRevD.104.023014, 2022PhRvD.105h3018M,2025PhRvD.111b3054M}, based on an ensemble of decision trees.
It performs a classification of \CWBBBH candidates by using a subset of 25 summary statistics including  $\rho_\mathrm{cWB}, c_\mathrm{c}, \tilde{\chi}, \rho_\mathrm{e},  \rho_\times, f_\mathrm{c}$, and $(1+z)\mchirp$.
To improve the distinction between astrophysical \GW signals and glitches, the \CWBBBH detection statistic is modified as
\begin{equation}\label{eq6}
  \rho_\mathrm{r} = \rho_\mathrm{cWB} W_{\mathrm{XGB}},
\end{equation}
where $W_{\mathrm{XGB}}$ is the XGBoost classification factor~\citep{PhysRevD.104.023014} ranging from 0 (indicating a glitch) to 1 (indicating a signal).
The XGBoost response is trained using a subset of background candidates and a representative set of simulated \BBH signals, which are injected into the detector data and recovered with \CWBBBH.
Due to a weak dependence of the \CWBBBH summary statistics on the \ac{CBC} population and signal models, the algorithm is robust to a variety of \ac{CBC} features, including higher multipoles, unequal component masses, misaligned spins, eccentric orbits, and possible deviations from \GR~\citep{2022PhRvD.105h3018M,eBBH:2025sbgt}.

While in principle the \CWB algorithm can run with any number of interferometers, the version used in the \ac{O4b} searches only runs with interferometer pairs.
In particular, for this catalog \CWBBBH only searched data from the \ac{LHO} and \ac{LLO} interferometer pair.

\subsection{\GSTLAL}
\label{subsection:GstLAL}

\GSTLAL is a stream-based time-domain matched-filtering search pipeline \citep{2012PhRvD..85h1504C, 2017PhRvD..95d2001M, 2019arXiv190108580S, 2020PhRvD.101b2003H, 2021SoftX..1400680C, 2024PhRvD.109d4066S, 2023arXiv230607190R, 2023PhRvD.108d3004T, 2024PhRvD.109d2008E, 2025arXiv250606497J, 2025arXiv250523959J}. 
All previous GWTC versions included results produced by \GSTLAL \citep{2019PhRvX...9c1040A, 2021PhRvX..11b1053A, 2024PhRvD.109b2001A, 2023PhRvX..13d1039A, 2025arXiv250818082T}. 
For this analysis, multiple \GSTLAL configurations were used: an online configuration whose results were re-evaluated using the full background, an offline configuration that processed data dropped by the online configuration, and another offline configuration that targeted intermediate mass black hole (\ac{IMBH}) mergers \citep{2025arXiv250606497J}. 
The results published in \gwtc[5.0], include analysis of data from the \ac{LIGO} detectors during \ac{O4a}, and from the \ac{LIGO} and \ac{Virgo} detectors during \ac{O4b}.
The primary configuration was an online search, which identified \ac{CBC} signals in near real-time, enabling rapid alerts and follow-up. 
The online search dropped $\sim$2\% of the observing run data due to, for example, computer downtime. 
The dropped data were processed in an offline configuration, methodologically identical to the online configuration, but executed post \ac{O4b}. 
The online configuration did not cover the high-mass region of the parameter space, so an additional offline configuration was run to target the \ac{IMBH} region, shown in grey in Figure~\ref{fig:search_space}. 
All configurations (online, dropped-data and \ac{IMBH}) were used to evaluate search sensitivity by injecting simulated signals into the detector strain data and analyzing the effectiveness of signal recovery.
Final candidate ranking and significance estimation incorporated the collective results from all configurations.
An additional early-warning configuration, designed to identify \BNS coalescences prior to merger, was also deployed during \ac{O4b}.
However, results from the early-warning configuration are not part of \gwtc[5.0].

The analysis begins by constructing a template bank to cover the targeted astrophysical parameter space.
Strain data are then pre-processed to prepare for matched-filtering, which correlates the pre-processed strain with templates from the bank.
Peaks in the matched-filter output are identified as triggers and subsequently ranked using a \ac{LR} statistic that incorporates signal consistency and detector network properties \citep{2023PhRvD.108d3004T}.
The background \ac{LR} distribution from noise triggers is used to estimate the \ac{FAR} and the significance of each \GW candidate.
In the following paragraphs, we describe each component of the analysis in detail, beginning with the template bank.

For \ac{O4b}, the template bank\note{check if same template bank} targeted \acp{CBC} with detector-frame total masses between \qty{1.95}{\Msun} and \qty{610}{\Msun}, and mass ratios from 0.05 to 1 \citep{2024PhRvD.109d4066S, 2025arXiv250606497J}. 
It was divided into two disjoint regions: stellar-mass (component masses up to \qty{200}{\Msun}) and \ac{IMBH} (primary component mass between \qty{200}{\Msun} and \qty{552}{\Msun} and secondary component mass below \qty{268}{\Msun}). 
While no \acp{BBH} with total mass above \qty{400}{\Msun} have been observed so far, cosmological redshifting of source-frame masses can shift distant stellar-mass binaries into the \ac{IMBH} region in the detector frame, motivating an extended search \citep{2023PhRvX..13d1039A}. 
Templates in both regions were spin aligned, i.e.,~spin components in the orbital plane are zero. 

Aligned spin values were allowed in the range $[-0.99, +0.99]$ for stellar-mass templates, with additional restriction to $[-0.05, +0.05]$ for component masses below \qty{3}{\Msun}, consistent with observations for galactic \acp{BNS} \citep{2018ApJ...854L..22S} and motivated by computational and background considerations.

In the \ac{IMBH} region, spins were restricted to $[-0.70, +0.99]$ to mitigate triggers from noise transients with similar time--frequency structure as high-mass waveforms with large anti-aligned spins \citep{Hanna:2021luk}.
The template bank used is identical to the one used in \ac{O4a}.
The search space is illustrated in Figure~\ref{fig:search_space}.

During template bank creation, templates were placed using a metric based on the \IMRPhenomD approximant \citep{2016PhRvD..93d4006H,2016PhRvD..93d4007K}, with a lower frequency cutoff of \qty{10}{\hertz} and a maximum duration of \qty{128}{\second}, ensuring coverage of the detector's sensitive band while limiting computational cost. 
The duration constraint led to a template-dependent low-frequency cutoff for low-mass systems. 
For example, a \qty{1.4}{\Msun}–\qty{1.4}{\Msun} binary would have a waveform duration of over \qty{1000}{\second} starting from \qty{10}{\hertz}. 
To stay within the \qty{128}{\second} constraint, its corresponding lower frequency cutoff must be raised to $\sim$\qty{22}{\hertz} \citep{1963PhRv..131..435P}. 

Template placement was performed using \MANIFOLD \citep{2023PhRvD.108d2003H}, an algorithm based on a geometric binary tree that tiles the parameter space dictated by the minimal match constraints.
A minimal match of 97\% was applied in the stellar-mass region, corresponding to a ~10\% expected loss in detection rate \citep{1996PhRvD..53.6749O}, while the \ac{IMBH} region used 99\%, resulting in a total of $\sim1.8 \times 10^6$ templates.

To improve fault tolerance of the online configuration, the stellar-mass bank was interleaved into two complementary halves, each independently covering the stellar-mass region and analyzed independently at separate data centers \citep{2020PhDT........28G, 2024PhRvD.109d4066S}.
Each half was subdivided into template bins and each bin was decomposed into filters using \acl{SVD} \citep[\acsu{SVD};][]{2012ApJ...748..136C, 2024PhRvD.109d4066S}.

After template bank construction, the detector strain data were preprocessed to ensure compatibility with the templates and to reduce noise contamination. 
Data were resampled to \qty{2048}{\hertz} and whitened. 
To suppress short-duration high-amplitude noise transients that can mimic astrophysical signals, especially at high masses where template waveforms are short, amplitude-based gating was applied to the whitened data. 
The gating threshold was a linear function of the chirp mass, with values computed per template bin and expressed in units of the standard deviation of the whitened strain data.
This adaptive approach balances glitch suppression with sensitivity to real signals across the mass range \citep{2019arXiv190108580S, 2024PhRvD.109d2008E}.

Along with gating, accurate whitening of strain data and template waveforms required \PSD estimation using Welch's method with geometric-mean and median averaging to characterize frequency-dependent noise \citep{2017PhRvD..95d2001M}.
In the online configuration, \acp{PSD} were estimated using a \qty{4}{\second} \ac{FFT} computed continuously on incoming data and used to whiten the strain in real time.

To account for longer-term noise variations that could impact the \ac{SVD}, template waveforms were re-whitened weekly using updated \acp{PSD} derived from recent data.
For template whitening, the analysis period was divided into continuous segments of up to \qty{8}{\hour}, with each segment producing a \PSD using a \qty{8}{\second} \ac{FFT}.
The final \PSD was constructed by taking the median power in each frequency bin across segments \citep{2017PhRvD..95d2001M}.
The \ac{SVD} basis for each template bin were then recomputed weekly using the same \PSD.
Weekly cadence balanced the need to track gradual detector noise evolution with the high computational cost of recomputing \acp{SVD} \citep{2024PhRvD.109d2008E}.

After whitening the strain data using the \PSD, matched-filtering was performed in the time domain, using the \ac{SVD} basis, starting at \qty{15}{\hertz} for the stellar-mass region (\qty{10}{\hertz} for the \ac{IMBH} region).
The output, an \SNR time series, was scanned for peaks exceeding a threshold of 4, which were defined as triggers.
A signal-consistency statistic, $\xi^2$, was computed for each trigger to quantify the deviation between observed and expected \SNR values across a fixed number of points centered on the peak \citep{2017PhRvD..95d2001M}.
Only the highest \SNR trigger in each \qty{1}{\second} window per template bin was kept.
Templates with chirp masses up to \qty{1.73}{\Msun} used the \TFTWO approximant and 701 samples to compute $\xi^2$.
Higher-mass templates used the \SEOBNRFOURROM approximant and 351 samples to compute $\xi^2$.
The resulting triggers were used to form candidates for further processing.

Candidates were classified as either single-detector or coincident.
Coincident candidates are defined as triggers from the same template in two or more detectors and occurring within the light-travel time between detectors, plus a \qty{5}{\milli\second} window to account for timing uncertainties.
Single-detector candidates had no corresponding trigger in another detector within the window \citep{2019arXiv190108580S}.
Candidates from the \ac{IMBH} region were excluded if they were a single-detector candidate.
Each candidate was ranked using a \ac{LR} quantifying the probability of astrophysical versus noise origin.
The \ac{LR} incorporated per-detector \acp{SNR}, the signal-consistency statistic $\xi^2$, local trigger rates, detector sensitivities, and prior assumptions about the \ac{CBC} population.
For coincident candidates, the relative arrival times and phases at the different detectors were included, with probabilities computed assuming isotropically-distributed sources.
In contrast, single-detector candidates were down-weighted using a \emph{singles-penalty}, empirically tuned via signal simulation campaigns, in order to suppress false positives while retaining sensitivity \citep{2019arXiv190108580S}.

The \ac{LR} was computed independently for each template bin to account for variations in how templates interact with detector noise.
Background distributions for each template bin were built from single-detector triggers occurring during coincident observing time.
Candidates were clustered in an \qty{8}{\second} window, and the candidate with the highest \ac{LR} across all template bins was selected for further processing.
The construction of the ranking statistic, including the modeling of the signal and noise distributions used in the \ac{LR} calculation, is described in \cite{2015arXiv150404632C, 2017PhRvD..95d2001M, 2023PhRvD.108d3004T}.

Candidate significance was quantified using \ac{FAR}, defined as the rate at which detector noise could produce a candidate with equal or higher \ac{LR} than the candidate under consideration.
\ac{FAR} was estimated by populating the background \ac{LR} distribution via Monte Carlo sampling of \SNR and $\xi^2$ values drawn from the empirical background, smoothed via \ac{KDE}.
Samples were assigned coalescence times, phases, and templates from uniform distributions \citep{2013PhRvD..88b4025C, 2025arXiv250606497J}.
To reduce contamination of the background by astrophysical signals, a \qty{10}{\second} window around each \GW candidate is vetted and excluded from the background, but only for candidates identified online by the online configuration \citep{2023PhRvD.108h4032J}.

Dropped data from the online configuration was processed in an offline configuration, yielding additional candidates.
Results from both configurations were merged, and \acp{FAR} recomputed using the full background collected by the online configuration \citep{2025arXiv250523959J}.
The combined set of candidates and background defined the stellar-mass search analysis.
To produce the final \gwtc[5.0] results, candidates and background from the stellar-mass and \ac{IMBH} search-analysis regions were combined into a unified analysis.
Each region was assigned a weight, determined from injection campaigns \citep{2025PhRvD.112j2001E}, reflecting its relative sensitivity and the expected number of detectable astrophysical signals.
The stellar-mass and \ac{IMBH} search analyses were assigned weights of $0.94$ and $0.06$, respectively.
The weights account for the fact that different analyses contribute unequally to the final set of candidates under the assumption of a given prior source population.
\acp{FAR} were rescaled by the inverse of these weights, and \acp{LR} were recomputed to provide a unified, unbiased ranking \citep{2025arXiv250606497J}.

In addition to \ac{FAR}, the total probability of astrophysical origin \pastro, and the classification probability for the mutually-exclusive \BNS, \NSBH, or \BBH source classes, were calculated for each candidate.
Unlike other pipelines, \GSTLAL adopts a signal model that assumes the Salpeter primary mass function \citep{Salpeter:1955it} for each source class, with uniform distributions in mass ratio and spin.
The source-specific probabilities were computed using a Bayesian framework that models the posterior probability distribution of the rates associated with each class.
The framework treats \GW triggers as realizations of independent Poisson processes, with rate estimates derived from previous observing runs and the population properties of each source class \citep{2023arXiv230607190R}.

The \GSTLAL offline analysis of \ac{O4b} included corrections from \ac{O4a} which affected the ranking statistic and the calculation of \pastro{}. 
The first change revised the handling of the horizon distance for each operating detector which is dependent on the detector configuration while observing \citep{2023PhRvD.108d3004T}. 
This corrected our unintended loss of sensitivity in the construction of the horizon distance. 
The second correction excluded the online vetted \ac{GW} events from the background estimation during the re-ranking steps \citep{2025arXiv250523959J}. 
This prevents the contamination of the background and bias FAR assignments ~\citep{2023PhRvD.108h4032J}. 
Finally, the normalization used to map ranking information to \pastro{} was revised.
This revision increases the odds-ratio, ${\pastro}/{(1-\pastro)}$, by a factor of $\sim 3.6$, indicating that the previous normalization systematically underestimated \pastro{} values.
The \GSTLAL search configuration, template bank design, and candidate ranking used for \ac{O4a} and \ac{O4b} differ from those used for O3. 
While \gwtc[4.0] and \gwtc[5.0] results were produced by using online results supplemented with offline analyses to avoid redundant processing, the \gwtc[3.0] results were produced entirely from a post-run offline analysis \citep{2023PhRvX..13d1039A}. 
The O3 template bank extended to detector-frame total masses of \qty{758}{\Msun}, with aligned-spin parameters in $[-0.999, 0.999]$ for components more massive than \qty{3}{\Msun}, and was constructed using a stochastic placement algorithm \citep{2021PhRvX..11b1053A}. 
No duration constraint was imposed during matched filtering with the O3 template bank. 
Ranking of candidates identified by matched-filtering used an \ac{LR} similar to \ac{O4a} and \ac{O4b}, but also incorporated the \iDQ{} glitch likelihood \citep{2020arXiv201015282G, 2021PhRvX..11b1053A}. 
The \iDQ{} glitch likelihood was used only for single-detector candidates prior to \gwtc[2.1], and the singles-penalty parameter, used to down-weight such triggers, was tuned differently than in \ac{O4a} and \ac{O4b}. 
Additionally, the \ac{LR} term modeling the distribution of $\xi^2$ used an empirically-tuned analytical function in O3, rather than being directly informed by the statistical properties of $\xi^2$ as done in \ac{O4a} and \ac{O4b}. 
Furthermore, data near vetted \GW candidates were not excluded from background estimation before \ac{O4a}.

The \GSTLAL pipeline used during O2 differed from O3 primarily in the template bank construction and candidate ranking methodology.
The O2 template bank covered detector-frame total masses up to \qty{400}{\Msun} and mass ratios from 1/98 to 1, with aligned-spin parameters in $[-0.999, 0.999]$ for components above \qty{2}{\Msun} \citep{2019PhRvX...9c1040A, 2021PhRvD.103h4047M}.
The \ac{LR} in O2 assumed uniform signal recovery across the template bank, ignoring the non-uniform template density and any astrophysical prior on the source population \citep{2021PhRvX..11b1053A, Fong:2018elx}.

The O2 \GSTLAL pipeline differed from O1 in its template bank construction, candidate ranking methodology and data conditioning.
The O1 template bank covered detector-frame total masses from \qty{2}{\Msun} to \qty{100}{\Msun} \citep{2021PhRvD.103h4047M}.
In O1, the \ac{LR} was computed only for coincident candidates and excluded time and phase differences between detectors \citep{2020PhRvD.101b2003H}.
The \ac{LR} included the joint probability of observed \acp{SNR} but assumed equal horizon distances across detectors: an approximation that neglected how detector-specific noise characteristics impact the horizon distance.
It was improved in O2 by pre-computing joint \SNR distributions for a discrete set of horizon-distance ratios.
Additionally, the O1 pipeline also lacked a template-mass-dependent glitch excision threshold \citep{2019arXiv190108580S}.

\subsection{\MBTA}
\label{subsection:MBTA}
\MBTA \citep{2012ApJ...760...12A, 2016CQGra..33q5012A, 2021CQGra..38i5004A, 2025CQGra..42j5009A} has performed online searches for \acp{CBC} since the initial-detector era science runs \citep{LIGOScientific:2011jth}.
\MBTA analyzes each detector independently using matched-filtering, before searching for candidates seen in coincidence.
To reduce the computational cost, the pipeline splits the matched-filtering process over several frequency bands (typically two).
Since \ac{O3}, \MBTA runs offline search analyses as well, with the aim of more accurately assessing the significance of candidates and contributing to the \GWTC \citep{2024PhRvD.109b2001A, 2023PhRvX..13d1039A}.

The parameter space explored by \MBTA's template bank is entirely defined by the detector-frame masses and spins (assumed parallel to the orbital angular momentum) of the binary components.
The \ac{O4b} analysis covers individual masses greater than \qty{1}{\Msun}, total masses up to \qty{500}{\Msun}, and mass ratios ranging from $1/20$ to $1$.

In line with astrophysical expectations for galactic \acp{NS} \citep{2018ApJ...854L..22S}, spin magnitudes are restricted to $0.05$ for objects with masses below \qty{2}{\Msun}, while more massive objects are allowed to have spin magnitudes of up to $0.998$.

Very short templates, which tend to generate a background excess, are removed by introducing a cut-off of \qty{200}{\milli\second} on the template duration, calculated starting from \qty{18}{\hertz}.
A visual representation of the parameter space covered by the \ac{O4b} \MBTA stellar-mass bank is given in the upper-right part of Figure~\ref{fig:search_space}.

The template banks used for \ac{O4} are generated with different algorithms.
The stellar-mass search space is first divided into two regions, corresponding to \BNS and symmetrical \BBH (no more than a factor $3$ between the two component masses).
Templates in the \BNS region are geometrically placed according to a local metric estimate \citep{2013PhRvD..87h2004B}, aiming for \SNR recovery of at least $98\%$.
Symmetrical \BBH template placement uses a hybrid algorithm \citep{2017PhRvD..95j4045R, 2019PhRvD..99b4048R}, with a similar expected \SNR recovery.
The rest of the bank is then completed using the same algorithm, with the objective of limiting the maximum \SNR loss to $3.5\%$.
As \MBTA filters data in two disjoint frequency bands, a bank has been created for each of them, as well as one for the whole frequency range.
Each template from the latter bank is associated with a template from other banks, maximizing their overlap in their common frequency band.

\MBTA uses a total of about $942000$ templates.
The shortest $3000$ templates are processed in the full-frequency band due to their intrinsically narrow bandwidth.
For the longer templates, the two-band decomposition leads to approximately $50000$ low-frequency-band templates and $20000$ high-frequency-band templates.
Filtering with such a decomposition leads to an \SNR loss that is much smaller than the minimal match of the bank.

Templates use the \STTFOUR approximant \citep{Klein:2014bua} for total masses below \qty{4}{\Msun} and \SEOBNRFOUROPT \citep{2017PhRvD..95d4028B} above.

Before matched filtering, \MBTA applies multiple pre-processing steps to the calibrated strain data delivered by the detectors.
Data are first resampled to \qty{4096}{\hertz}.
Next, a gating procedure is applied, designed to remove short and high-amplitude glitches.
The gating used for \ac{O4} is similar to the version deployed for \ac{O3} and is triggered by drops in the detector sensitivity \citep{2021CQGra..38i5004A}.
It is also applied to periods of poor quality data identified by data-quality vetoes \citep{2020CQGra..37e5002A}.
Less than $0.1\%$ of the data provided by each detector was removed by \MBTA gating during \ac{O4b} \citep{2025CQGra..42j5009A}.
An ungated search analysis is also performed for part of the parameter space, with the aim of finding the massive, intense \acp{CBC} that are likely to trigger the gating \citep{2021CQGra..38i5004A}.

\MBTA regularly re-estimates the noise \PSD of each detector by taking the median over thousands of seconds of data from several \acp{FFT}, whose lengths range from seconds to hundreds of seconds, depending on the targeted region of the parameter space \citep{Aubin:2025TDS}.

For most templates, the matched-filter calculation is carried out in two disjoint frequency bands, starting at \qty{24}{\hertz} and separated at \qty{80}{\hertz}.
When a sufficiently high \SNR is detected, bands are then coherently recombined to produce a unified single-detector trigger from the two bands.
The shortest templates, however, are processed using a single band starting at \qty{20}{\hertz} due to their intrinsically narrower frequency band.

Triggers with an \SNR above $4.4$ (or $9$ for the ungated analysis) are recorded.
\MBTA then applies several tools to reject triggers caused by transient noise.
A basic $\chi^2$ test discards any trigger whose \SNR is not distributed as expected across the filtered bands \citep{2016CQGra..33q5012A}.
A second $\chi^2$ statistic is used to quantify the discrepancy between the measured \SNR time series and the template autocorrelation, and to reweight the \SNR, thereby defining a single-detector ranking statistic \citep{2021CQGra..38i5004A}.
\MBTA also downgrades the ranking statistics of triggers occurring during periods when detectors show signs of poor data quality.
It builds a $\chi^2$ statistic from the differences between the time series of the maximum \SNR observed around each trigger and models based on a large population of simulated signals \citep{2025CQGra..42j5009A}.

With the addition of \ac{Virgo} in \ac{O4b} (absent during \ac{O4a}), \MBTA now looks for coincident triggers between the two \ac{LIGO} detectors and \ac{Virgo} detector sharing the same templates, within a limited time window.
A combined ranking statistic is computed as the quadrature sum of single detector ranking statistics.
This statistic also includes a term measuring the consistency of arrival times and phases across the detectors \citep{2021CQGra..38i5004A}.
To mitigate the impact of loud glitches on a single detector, paired with random fluctuations from other detectors, the pipeline rejects high-mass coincidences resulting from a heavily reweighted trigger and a low SNR trigger.
In addition to the search for coincident triggers between the two \ac{LIGO} and \ac{Virgo} detectors, since the start of \ac{O4}, \MBTA now also produces single-Hanford and single-Livingston candidates. In \ac{O4b}, Virgo is only used for coincidences.
In order to limit the risk of false alarms, while maximizing the chances of detecting electromagnetically-bright sources, the \MBTA search for single-Hanford and single-Livingston candidates is restricted to chirp masses below \qty{7}{\Msun}.

The significance of the candidates obtained by the stellar-mass search analysis is evaluated based on their probable origin, quantified by \pastro.
These probabilities are obtained using a method similar to the \ac{O3} method \citep{2022CQGra..39e5002A}.
For a given astrophysical source, the ratio of expected number of foreground triggers to the total number of triggers (foreground + background) is calculated.
The foreground is estimated using population models similar to those used to estimate search sensitivity (see Section~\ref{sec:search_sensitivity}) and by fitting the rate observed by \MBTA.
The background is obtained by fitting the distribution of the ranking statistic measured by \MBTA, after eliminating confidently-detected astrophysical signals.
In the case of coincidences, the statistic is increased by considering fake coincidences between noise triggers, regardless of the time of arrival.
These operations are performed over several bins in the mass and spin parameter space covered by the search.

The pipeline associates a \ac{FAR} to candidates using a method similar to the \pastro background estimation.
In order to make the values of \pastro and \ac{FAR} consistent with each other, and to include an astrophysical prior in the \ac{FAR} definition, \pastro is used since \ac{O4a} as a new ranking statistic.

\MBTA performed several online search analyses during \ac{O4b} using a pipeline version similar to the \ac{O4a} offline analysis \citep{2025arXiv250818081T}, in addition to searching for coincidences with Virgo data.
All coincident candidates, as well as single-Hanford and single-Livingston candidates with chirp mass below \qty{7}{\Msun}, from the full-bandwidth analysis with a \acp{FAR} below $\qty{2}{\hour^{-1}}$ were submitted to \GRACEDB with a median latency of \qty{21}{\second}.
The online \pastro calculation performed by the pipeline during \ac{O4b} used the same model as for the offline \ac{O4a} analysis \citep{2023PhRvX..13d1039A}, adjusting the signal rates to account for the increased sensitivity.
\MBTA was also conducting \emph{early-warning} searches for \acp{CBC} with component masses between \qty{1}{\Msun} and \qty{2.5}{\Msun} and aligned spins below \num{0.05}, with the aim of rapidly identifying high-\SNR signals.

\MBTA also conducted offline search analyses during the \ac{O3} and \ac{O4a} period, and thus participated in the production of the previous catalog versions \gwtc[2.1] \citep{2024PhRvD.109b2001A}, \gwtc[3.0] \citep{2023PhRvX..13d1039A} and \gwtc[4.0] \citep{2025ApJ...995L..18A}.
During \ac{O4a}, \MBTA was operated in a similar configuration than for \ac{O4b}, with the main difference of searching for mass ratio down to $1/50$.
During \ac{O3}, the pipeline processing was slightly different.
The \ac{O3} stellar-mass search template bank was created using a purely stochastic method \citep{2014PhRvD..89b4003P}, guaranteeing a \SNR loss below $3\%$, compared with $2\%$ for parts of the \ac{O4} banks.
During \ac{O3}, the pipeline only searched for \acp{CBC} with (detector-frame) total mass below \qty{200}{\Msun}, but no constraints on waveform duration were applied.
Re-ranking of triggers occurring during periods of poor data quality was handled by comparing the trigger rate before and after the application of various other rejection criteria.
Before \ac{O4}, \MBTA only calculated significance for candidates seen in at least two detectors.
The \ac{FAR} values were based on a ranking that assumed all templates were equiprobable, which was responsible for some inconsistencies with the associated \pastro values.
Trials factors were used to account for the various coincidence types and parameter space regions \citep{2021CQGra..38i5004A}.

\subsection{\PYCBC}
\label{subsection:PyCBC}

The \PYCBC search pipeline~\citep{2014PhRvD..90h2004D,2016CQGra..33u5004U,2017ApJ...849..118N} is a descendant of the IHOPE pipeline~\citep{2012PhRvD..85l2006A,2013PhRvD..87b4033B}, which was used to search LIGO--Virgo data during the initial detector era up to $2010$~\citep{LIGOScientific:2011jth, LIGOScientific:2012vij}.
The input to \PYCBC is the calibrated strain data from the detectors, which undergo a series of conditioning steps to prepare for analysis.
After removing invalid or contaminated strain data flagged by CAT1 vetoes, we then apply high-pass filtering to suppress low-frequency noise (below \qty{15}{\hertz}), downsampling from \qty{16}{\kilo\hertz} to \qty{2}{\kilo\hertz} to reduce data volume, and gating (windowing out) to remove loud glitches~\citep{2016CQGra..33u5004U}.
The pipeline is also limited to analyze data segments with a minimum length of \qty{500}{\second}, to ensure a sufficiently precise \PSD estimate.

After data selection and input conditioning, data from each detector are filtered using a fixed template bank.
The templates are placed in a four-dimensional parameter space of component detector-frame masses and orbit-aligned spins.
For \ac{O4a}, the template bank was generated with a hybrid geometric-random placement algorithm~\citep{2017PhRvD..95j4045R}, imposing a minimum template waveform duration of \qty{70}{ms} which allows coverage of systems with \acp{IMBH}, and applying a dynamic minimal match criterion to ensure smooth template density across the mass range~\citep{2017PhRvD..95j4045R, 2019PhRvD..99b4048R}. The \ac{O4b} analysis reused the template bank generated during \ac{O4a}.
The match was calculated assuming the projected \ac{LIGO} \ac{O4} sensitivity that predicts a \qty{160}{Mpc} \ac{BNS} range \citep{LIGO-T2000012}.
The waveform model used is \SEOBNRFIVEROM \citep{2023PhRvD.108l4035P} for total mass above \qty{4}{\Msun}, and \TFTWO \citep{2011PhRvD..83h4051V} below.
A fixed lower cutoff frequency of \qty{15}{\hertz} is applied for systems with total mass above \qty{100}{\Msun}.
The bank covers a total mass range of \qty{2}{\Msun} to \qty{500}{\Msun}, with mass ratios from $1/97.989$ to $1$, and aligned spin ranges from $-0.997$ to $0.997$ for \ac{BH} components and from $-0.05$ to $0.05$ for \ac{NS} components, as shown in Figure~\ref{fig:search_space}.

Data from each detector are independently matched filtered against the template bank, producing an \SNR time series for each template: triggers, corresponding to potential signals, are then identified as local \SNR maxima above a given threshold within predefined time windows.
A $\chi^2$ test is applied to assess whether the time--frequency distribution of power in the data matches the expected distribution from the template waveform~\citep{2005PhRvD..71f2001A}.
The pipeline re-weights the matched-filter \SNR using the reduced $\chi^2$ normalized such that the expected value in Gaussian noise for a signal matching the template is unity~\citep{LIGOScientific:2011jth, 2016CQGra..33u5004U}.
Triggers with re-weighted \SNR below a threshold of \num{4} are discarded as they are likely to correspond to noise transients.
Since short-duration blip glitches~\citep{2019CQGra..36o5010C} may pass the time--frequency $\chi^2$ test for some templates, a high-frequency sine--Gaussian $\chi^2$ test is used to further identify such glitches~\citep{2018CQGra..35c5016N}.

A single-detector ranking statistic $\hat{\rho}$ is then calculated, incorporating the time--frequency and sine--Gaussian $\chi^2$ tests, as well as a correction to the \SNR due to short-term \PSD variability~\citep{2021PhRvD.104f3034Z, 2020CQGra..37u5014M}.

The next step of the pipeline is to identify candidates consistent with potential \GW signals by comparing the coalescence times and template parameters of triggers from multiple detectors.
Coincident candidates are found if triggers from two or more detectors, associated with the same template, occur within a time window that accounts for the light travel time between detectors and a small margin for timing errors.
If data from more than two detectors are analyzed, coincident candidates may be formed across multiple combinations of two and three detectors via the same process.

Starting in O4, candidates consisting of triggers in only a single detector are also included~\citep{2022CQGra..39u5012C}, though only from templates with duration larger than \qty{0.3}{\second}
and satisfying additional cuts related to the background distribution.

Following this, the pipeline calculates the statistical significance of candidates by estimating their \ac{FAR}.
A ranking statistic is assigned to each candidate, reflecting its likelihood of being a true \GW signal versus background noise.
For coincident candidates seen in two or three detectors, the \ac{FAR} is computed by comparing their ranking statistics to a set of artificially generated background candidates created by time-shifting triggers in one of the participating detectors. 
In \ac{O4a}, where only \ac{LHO} and \ac{LLO} data were available, background candidates were generated by repeatedly shifting triggers in one detector with respect to the other by a fixed interval. 
In \ac{O4b}, incorporating data from \ac{Virgo}, background candidates are generated by extending the time-shifting procedure to a multi-detector scheme \citep{2020PhRvD.102b2004D}. Artificial triple-detector background coincidences are generated by identifying coincident triggers in any two detectors and shifting the triggers in the third detector by fixed intervals. For three-detector analyses, the \ac{FAR} is computed by summing the \ac{FAR} contributions from various coincidence types at a given ranking statistic threshold: triple-detector (HLV) and the three possible double-detector combinations (HL, HV, and LV) \citep{2020PhRvD.102b2004D}.

To ensure that candidates are statistically independent, clustering is performed on both coincident (un-shifted) and time-shifted analyses: only the candidate with the highest detection value within a set time window is kept.
To reduce background contamination due to loud signals, any candidate detected with \ac{FAR} below a threshold of $1$ per \qty{100}{\yr} is removed from the background estimation for less significant candidates \citep{2016PhRvX...6d1015A,2019ApJ...872..195N}.

For single-detector candidates in \ac{O4}, since their \acp{FAR} cannot be estimated from time-shifted analysis, the ranking statistic distribution is fitted with a falling exponential, and extrapolated with a maximum possible \ac{IFAR} assignment of \qty{1000}{\yr} \citep{2022CQGra..39u5012C}.
In times when multiple detectors are active, \ac{FAR} estimates from all candidate types at the ranking statistic value of the candidate are added to give the final \ac{FAR} estimate.

The ranking statistic assigned to both coincident and single-detector candidates is designed to optimize the detection rate by reflecting the relative densities of signal vs.\ noise over the binary source parameter space.
To account for noise variations across different templates, the single-detector background distributions are fitted, with the fit parameters allowed to vary over the template parameter space~\citep{2017ApJ...849..118N}.
We define an intermediate expression $R_3$, which is the basis for both the multi-detector and single-detector statistic:
\begin{equation}
\label{eq:pycbc-R3-stat}
\begin{split}
    R_3
    &= \ln \left( \frac{\sigma_{\mathrm{min},i}^3 / \bar{\sigma}_{\mathrm{HL},i}^3}{A_{\mathrm{N} \{ d \}} \prod_{d} r_{d,i}(\hat{\rho}_{d})}
       \frac{p(\boldsymbol{\Omega}|\mathrm{S})}{p(\boldsymbol{\Omega}|\mathrm{N})} \right) \\
    &= \ln p(\boldsymbol{\Omega}|\mathrm{S}) - \ln p(\boldsymbol{\Omega}|\mathrm{N}) - \ln A_{\mathrm{N}\{d\}} \\
    &\phantom{=}\ -\sum_d \ln r_{d,i}(\hat{\rho}_{d}) + 3 (\ln \sigma_{\mathrm{min},i} - \ln \bar{\sigma}_{\mathrm{HL},i}).
\end{split}
\end{equation}
Here $p(\boldsymbol{\Omega}|\mathrm{S})$ and $p(\boldsymbol{\Omega}|\mathrm{N})$ are the probabilities of a signal or noise candidate, respectively, to have extrinsic parameters $\boldsymbol{\Omega}$, comprising relative phases, arrival times and amplitudes between detectors \citep{2017ApJ...849..118N}.
We consider the noise distribution $p(\boldsymbol{\Omega}|\mathrm{N})$ as uniform for a given combination of detectors, absorbing this constant factor into the normalization of $p(\boldsymbol{\Omega}|\mathrm{S})$.
The rate of noise events is predicted from the allowed time window $A_{\mathrm{N}\{d\}}$ for coincident events involving detectors $\{d\}$ and the expected rate of noise triggers $r_{d, i}(\hat{\rho}_d)$ in template $i$ at re-weighted \SNR $\hat{\rho}_d$.
The last term accounts for the time-dependent rate of signals, which is proportional to the network sensitive volume for a given template, and scales with the cube of the minimum sensitive distance $\sigma_{\mathrm{min},i}$ over triggering detectors, normalized to an average LIGO sensitive distance $\bar{\sigma}_{\mathrm{HL},i}$.

The ranking statistic used in O4 includes additional terms beyond Equation~\eqref{eq:pycbc-R3-stat}.
First, an explicit model of the signal distribution over binary masses and spins is incorporated, intended to optimize the detection rate of the known \ac{CBC} population while remaining sensitive to so far unpopulated parameter regions~\citep{2024PhRvD.110d3036K}. 
Second, the noise event rate model is updated to account for variations in the rate of noise artifacts, correlated with auxiliary channels that monitor the detector state or environmental disturbances, as measured by \iDQ{} \citep{Essick:2020qpo}.
We then have:
\begin{equation}
\label{eq:pycbc-o4-stat}
\begin{split}
    R_\textrm{O4} &= R_3 + \ln d_\mathrm{S}(\PEparameterIntrinsic) - \ln d_\mathrm{N}(\PEparameterIntrinsic) \\
      &\phantom{=}\ - \sum_d \ln \delta_d\left(\Delta_d(t),B_d(\PEparameterIntrinsic)\right) \\
      & \phantom{=}\ + \sum_D \ln C_{D},
\end{split}
\end{equation}
where $d_\mathrm{S}(\PEparameterIntrinsic)$ and $d_\mathrm{N}(\PEparameterIntrinsic)$ represent a \ac{KDE}-based model distribution of signals over template parameter space $\PEparameterIntrinsic$, and the distribution of template points, respectively.
The time-dependent excess rate of non-Gaussian triggers is notated as $\delta_d(\Delta_d(t),B_d(\PEparameterIntrinsic)) \geq 1$ for a detector $d$.
Our estimate of $\delta_d$ depends on a discrete detector state variable $\Delta_d(t)$ and on an index $B_d(\PEparameterIntrinsic)$ labelling template bins.
We consider three possible detector states, corresponding to (i) auxiliary channels indicating significant likelihood of noise artefacts; (ii) presence of loud gated glitches nearby in time, or (iii) neither.
In the case of single-detector candidates, the ranking statistic omits terms that depend on multi-detector coincidence quantities~\citep{2022CQGra..39u5012C}; specifically, the terms $A_{\mathrm{N}\{d\}}$, $p(\boldsymbol{\Omega}|\mathrm{S})$, and $p(\boldsymbol{\Omega}|\mathrm{N})$ present in Equation~\eqref{eq:pycbc-R3-stat} are omitted.

The remaining term $C_{D}$ is a constant correction dependent on the candidate type $D$ (a single detector or combination of detectors).  It accounts for the different rates of triggers due to detected signals and noise over the available combinations of detectors, after other terms in Eq.~\eqref{eq:pycbc-o4-stat} are applied, with the aim of improving the overall search sensitivity.
For the \gwtc[4.0] analysis of O4a data \cite{2025arXiv250818081T}, this term was not included.  For the analysis of \ac{O4b} data and a revised O4a analysis, both included in \gwtc[5.0] \citep{GWTC:Results}, corrections are applied to all candidates, apart from \ac{LHO}\textendash\ac{LLO} double coincidences (i.e., we impose $C_\mathrm{HL} = 0$).  The values of $C_{D}$ are empirically tuned using simulated signals (injections) over representative periods of O4 data.  Inclusion of the correction for O4a reanalysis resulted in an increase of up to \PycbcSensImprov{} in PyCBC's sensitivity (at a threshold of \ac{FAR} $< 1 \mathrm{yr}^{-1}$).

The probability of astrophysical origin \pastro is estimated by a Poisson mixture inference \citep{2016ApJ...833L...1A,2016ApJS..227...14A}, for which the distribution of signal events over the ranking statistic is estimated via histograms from a set of simulated signals and distributions of noise events are estimated via histograms of time-shifted background for coincident candidates, or via exponential fits for single-detector candidates~\citep{2022CQGra..39u5012C}.
Such histograms are generated separately for each type of (single-detector or coincident) candidate in single-detector or multi-detector observing time to allow for differences in event rates and distributions~\citep{Dent:multi_pastro}.
Marginalization over the unknown rate of astrophysical signals is performed by generalized Gauss--Laguerre quadrature~\citep{Creighton:fgmc_identities}.
The resulting candidate \pastro values are apportioned between different astrophysical categories using an estimate of the source chirp mass based on search outputs~\citep{2021ApJ...923..254D,Villa-Ortega:2022qdo}.

Several features of the \PYCBC online search analysis \citep{2021ApJ...923..254D, 2018PhRvD..98b4050N} differ from the offline search analysis, these are detailed in Section 3.4 of \cite{2025arXiv250818081T}.
The main differences between \ac{O4a} and \ac{O4b} \PYCBC online analysis, are:
\begin{itemize}
\item Improved gating: to improve glitch mitigation, the gating window was extended to a data segment significantly longer than the analysis chunk (16~s). This ensures that extended clusters of multiple glitches are detected and gated across multiple chunks, preventing them from persisting through the duration of the longest signal templates.
\item \ac{Virgo} data was used in the end-to-end offline analysis. For online analysis \ac{Virgo} data was not used to perform low-latency matched-filtering but only for localization when at least one other detector in the network registered a trigger.
\item Deploying an improved version of the SNR optimizer used for \ac{O3} analysis \citep{2023PhRvX..13d1039A}: for every trigger a refined search near to the triggering template is performed to recover the potentially lost SNR due to the discreteness of the template bank. The O4b SNR optimizer is tuned to improve upon the settings (in particular the range of allowed masses) used in the previous O3 analysis.
\end{itemize}

The above descriptions concern \PYCBC analyses of \ac{O4a} and \ac{O4b} data.
Here, for completeness, we summarize significant changes in search methods used for archival catalog (offline) results since the \gwtc[1.0] release.
The O3 template bank was constructed using the same hybrid geometric-random method as in O4 to ensure efficient coverage of the parameter space~\citep{2017PhRvD..95j4045R}.
Three different independent search analyses (PyCBC-BBH, PyCBC-Broad, and PyCBC-IMBH) were used, with each covering distinct regions of the parameter space~\citep{2023PhRvX..13d1039A, 2021PhRvD.104d2004C}.
The waveform model \SEOBNRFOURROM was used for systems above \qty{4}{\Msun}, and \TFTWO for lower-mass systems~\citep{2019PhRvD..99b4048R}.
Templates with durations below \qty{0.15}{\second} were excluded to reduce false alarms from short glitches, thereby also placing an upper limit on the mass of the systems, unlike in O4, where shorter-duration templates are included to improve coverage of high-mass systems.
The template bank used in the earlier O2 and O1 analyses \citep{2017arXiv170501845D} spanned total masses from \qty{2}{\Msun} to \qty{500}{\Msun}, with mass ratios from $1/98$ to $1$, and spin limits based on component mass: $0.05$ for masses below \qty{2}{\Msun} and up to $0.998$ for higher masses~\citep{2019PhRvX...9c1040A}.
For O3, the background estimation was extended to a three-detector network by using fixed relative time shifts between detectors~\citep{2020PhRvD.102b2004D}, whereas in earlier runs, a two-detector method (also used in O4a) was employed.
Unlike in O4, single-detector candidates were not included in O3 or previous runs.
The O3 broad search statistic is given by Equation~\eqref{eq:pycbc-R3-stat}, i.e., the O4 statistic without the additional \ac{KDE} and data-quality terms.
For the O3-BBH search, an alternate ranking statistic was adopted, incorporating an additional term in the broad search statistic dependent on the template chirp mass to enhance sensitivity to the observed \BBH population~\citep{2020ApJ...891..123N}.
In earlier runs, only one type of event (\ac{LHO}\textendash\ac{LLO}) was considered, and sensitivity variations across the parameter space, such as those described by $\sigma_{i}$, were neglected.
As a result, many terms in the O3 statistic in Equation~\eqref{eq:pycbc-R3-stat} either disappeared or became constants and were omitted~\citep{2017ApJ...849..118N}.

\subsection{\SPIIR}
\label{subsection:SPIIR}

\SPIIR is a coherent search pipeline developed primarily for the online detection of \GW signals from \acp{CBC} \citep{2012PhRvD..86b4012H, 2012PhRvD..85j2002L, 2022PhRvD.105b4023C}.
It became operational during O3, during which it registered \num{38} out of the \num{56} non-retracted public alerts \citep{2021PhRvX..11b1053A, 2024PhRvD.109b2001A, 2022PhRvD.105b4023C}.
The design of \SPIIR focuses on efficiency and rapid processing, enabling it to provide detections with latencies around \qty{11}{\second}.
This capability makes \SPIIR particularly useful for issuing early warning alerts for electromagnetic follow-ups of \GW candidates \citep{2022ApJ...927L...9K}.
While an offline version of \SPIIR is under development, it has not yet participated in any of the offline searches through the end of \ac{O4b}\note{check}.

The \SPIIR pipeline employs infinite impulse-response (IIR) filters as template banks to perform matched filtering directly in the time domain \citep{2012PhRvD..86b4012H}.
Unlike traditional methods, which can be time-consuming, the IIR filtering approach allows \SPIIR to execute matched filtering rapidly.
The primary advantage of IIR filters is their ability to approximate matched-filtering \SNR in a way that can be efficiently parallelized and executed on graphics processing units (GPUs), which significantly accelerate high-performance computing tasks \citep{2012CQGra..29w5018L, 2018CoPhC.231...62G}.
The \SPIIR template-bank generation is a two-step process.
The first step is to generate the template points in the \ac{CBC} parameter space using the stochastic placement algorithm \citep{2014PhRvD..89b4003P, 2009PhRvD..80j4014H} with a minimum match of \num{0.97}.
The approximant \STTFOUR \citep{Klein:2014bua} is used for chirp mass below \qty{1.73}{\Msun} and \SEOBNRFOURROM for larger masses.
In the second step, each corresponding waveform is decomposed into approximately 350 IIR filters \citep{2012PhRvD..86b4012H, 2012CQGra..29w5018L, 2018CoPhC.231...62G}.
These IIR filters are then utilized for matched filtering using GPU parallelization.

The real-time data analysis process begins by retrieving strain data for all detectors in observing mode.

The data are then conditioned, applying data-quality vetoes, down-sampling from \qty{16384}{\hertz} to \qty{2048}{\hertz}, and whitening, which requires real-time \PSD calculation.
The \PSD calculation takes place using the Welch method with tens of \qty{4}{\second} overlapping data blocks.
This method prevents sporadic transient signals from biasing the \PSD estimate.
The data are also subjected to gating to mitigate the effects of loud transient noises.
Once conditioned, the whitened data from each detector is passed through a set of IIR filters that span the parameter space of \ac{CBC} templates, generating corresponding \SNR time series for each template.

A key distinguishing feature of the \SPIIR pipeline is its coherent search algorithm \citep{Bose:2011chs, Harry:2011chs}.
The coherent search algorithm looks for time and phase consistency across all detectors.
\SPIIR calculates a maximum network likelihood ratio statistic to evaluate the likelihood of a coherent \GW signal being present in the data.
This approach enhances the pipeline's sensitivity, particularly to weaker signals, and improves localization accuracy by combining data from multiple detectors.

The coherent analysis can be computationally demanding.
To address this challenge, \SPIIR uses \ac{SVD} \citep{Wen:2008chs}, which allows for efficient calculation of the coherent statistic by focusing on the most significant signal components.
This optimization reduces the complexity of the search and enables rapid scanning across different sky positions to localize the source.

After determining the coherent \SNR, \SPIIR applies a $\xi^2$ signal-consistency test \citep{2017PhRvD..95d2001M} to asses the morphological consistency between data and template and combines it with the coherent \SNR into a multidimensional ranking statistic. \acp{FAR} are assigned by comparing candidates with background distributions generated from time-shifted analyses.
To ensure robust estimation, the pipeline collects at least one million background candidates, typically accumulated over several hours.

\acp{FAR} are assigned using a minimum of one week of background data.
The background
 is extrapolated using a $k$-nearest-neighbor (KNN) \ac{KDE} method with $k = 11$, smoothing the probability density using neighboring bins.
This approach allows accurate \ac{FAR} estimation even for rare, high-significance candidates.
To account for non-stationary noise, \SPIIR additionally maintains shorter-term background sets over \qty{2}{\hour} and \qty{1}{\day}.
Furthermore, it stores individual single-detector \SNR and $\xi^2$ values and applies the same KNN-based extrapolation method to estimate single-detector \acp{FAR} used in veto stages before making the final decision about the trigger's validity.

In preparation for O4, \SPIIR has undergone several upgrades, with some already implemented and others still in testing and review.
These improvements include: automated signal trigger removal from the background estimation, preventing astrophysical triggers from affecting the background calculation;
fine tuning of the signal consistency test $\xi^2$, aiming to improve sensitivity to \BNS and \NSBH; and a dynamic approach to background collection.
Furthermore, \SPIIR has implemented the necessary infrastructure to process \ac{KAGRA} data.

\subsection{Criteria for Inclusion in the Catalog}
\label{sec:selection_criteria}

Search pipelines can in principle produce a rate of candidates as high as one every few seconds, depending on their configuration choices.
With the sensitivity of current detectors, however, the vast majority of such candidates would have low significance and would be unlikely to have an astrophysical origin.
For practical consideration, we therefore select a smaller number of candidates to report in the \GWTC.

The criterion for selecting which candidates to include has evolved with the observing runs and \GWTC versions, due to the rapid evolution of the field and the computational and human cost required to retroactively apply the latest analysis methods to previous observing runs.
In \gwtc[1.0] \citep{2019PhRvX...9c1040A}, we published all candidates from \ac{O1} and \ac{O2} for which at least one matched-filtering pipeline (\PYCBC or \GSTLAL at the time) estimated a \ac{FAR} below $1$ per \qty{30}{\day}.

\gwtc[2.1] \citep{2024PhRvD.109b2001A} and \gwtc[3.0] \citep{2023PhRvX..13d1039A} respectively added candidates from the \ac{O3a} and \ac{O3b} with a \ac{FAR} below $\qty{2}{\day^{-1}}$ in any pipeline, with the exception of \SPIIR.
Following this logic, candidates from \ac{O4a} whose \ac{FAR} is less than $\qty{2}{\day^{-1}}$ in any of the pipelines described in Section \ref{sec:searches}, with the exception of \SPIIR, were added to \gwtc[4.0], and we now do the same for \ac{O4b} and \gwtc[5.0].
In general, we define a combined candidate list (denoted as the \textit{Any} pipeline results) for a given \ac{FAR} threshold by including candidates whose \ac{FAR} is below that threshold in at least one pipeline.

When candidates from multiple pipelines are reported with times within a given clustering window \citep{OPA}, then they are grouped under a single event name.
\acp{FAR} from individual pipelines are presented side-by-side and are neither corrected for multiple trials, nor combined in any way.
Downstream analyses that need a single clustered list of candidates are then free to apply any appropriate combination method, e.g.,~choosing the minimum \ac{FAR} across the available pipelines.

Stricter candidate selection criteria are typically applied when performing downstream analyses and for the purpose of displaying lists and properties of the candidates.
Stricter criteria might, for example, be motivated by computational or person-power considerations.
Furthermore, downstream analyses might not necessarily have the same requirements; for instance, a sufficiently large \SNR might be necessary, as opposed to a sufficiently low \ac{FAR}.
Such considerations, and consequently the corresponding selection criteria, might in turn evolve as data-analysis methods become more efficient.
Stricter criteria will therefore be described wherever necessary.

\subsection{Search Sensitivity}
\label{sec:search_sensitivity}

Apart from the candidates themselves, an important product of search pipelines is a measurement of their sensitivity, which is necessary to diagnose the behavior of the pipelines and to infer the rate density of the astrophysical sources.
We quantify the sensitivity of the search pipelines described above by their estimated time--volume product or hypervolume \VT ~\citep{2023PhRvX..13d1039A}.
The hypervolume represents the sensitivity of a given search to a set of sources assumed uniformly distributed in both comoving volume and source-frame time.
An estimate of the total number of signals $\hat{N}$ that a pipeline is likely to detect during a period of observation is given by
\begin{equation}
	\hat{N} = \VT \mathcal{R} ,
\end{equation}
with $\mathcal{R}$ the rate of \acp{CBC} per unit comoving volume and source-frame time \citep{2016ApJ...833L...1A}.
In practice, hypervolumes are obtained through weighted Monte Carlo simulations,
which use simulated \ac{GW} signals referred to as \emph{injections}.

The injections are added to the strain data, which are then processed by the search pipelines to determine which injections are detectable according to a given set of selection criteria.
The resulting \VT estimate is a function of the selection criteria, which for \thisgwtcfull{} have evolved over the observing runs as described in Section \ref{sec:selection_criteria}.
Downstream analyses that use the entirety of \gwtc{}, such as rate density inferences, can and should account for this evolution by applying the appropriate time-dependent selection criteria when considering which injections are detectable.

Our current population of injected signals includes only \acp{CBC}, as no transient \ac{GW} signal so far detected is inconsistent with a \ac{CBC} source at high confidence. 
The parameter ranges and distribution models were chosen to represent \ac{CBC} signals the detectors were possibly sensitive to, based on previous releases of \GWTC \citep{2023PhRvX..13a1048A}.
The rate of injections was chosen to be much higher than the astrophysical rate in order to reach a sufficient precision in \VT.
As the production of injections required detector \acp{PSD} to be measured over each month of observation, injection analysis was performed offline.
To ensure consistency in the \VT measurements, all pipelines contributing \GWTC candidates (\GSTLAL, \MBTA, \PYCBC, and \CWBBBH) analyzed the same sets of injections.
The pipelines processed data with injections using methods equivalent to those used to produce candidates from injection-free data, described earlier in this section.
All pipelines used background data collected by their injection-free analyses to assign significances to the recovered injections.
The choice of population priors and the injection generation process are detailed in \citet{2025PhRvD.112j2001E}, while results from the injection analysis are described in \citet{GWTC:Results}.

The sensitive \VT differs between pipelines, reflecting their abilities to detect \acp{CBC} within a given range of parameters as well as their different effective live times.
As \CWBBBH is designed to search for \BBH signals as described in Section~\ref{subsection:cWB}, for this pipeline we calculate \VT only for masses corresponding to \BBH sources.

There are differences in how certain pipelines process injections compared to injection-free data, which arise from computational efficiency or technical considerations and do not significantly affect the estimated sensitivity. 
The \CWBBBH pipeline performs injection analysis in a $\pm$\qty{5}{\second} time window around the time of each injection.
The \GSTLAL and \PYCBC pipelines process data using subsets of their template banks based on the known chirp mass of the injections, in order to limit the computational cost. 
In addition, \GSTLAL's injection processing has two other major differences compared to its injection-free data processing.
First, injections are processed entirely in an offline configuration, as opposed to reassigning significances to triggers obtained in online analysis, as injection data were not available during the online analysis. 
This is a difference from other pipelines, which process both injection-free and injection data in their offline configurations to produce the \gwtc[5.0] results and \VT estimations.
Second, \GSTLAL computes \SNR time series only over a short time window (${\sim}\qty{6}{\second}$) around the time of the injections, to limit the computational cost, as opposed to computing the \SNR time series for all available data in \ac{O4b}.

The search sensitivities may be calculated at different significance thresholds, which
enables us to self-consistently estimate the number of astrophysical signals
among the subthreshold candidates. If the source population distribution assumed
for \pastro calculations is close to the true astrophysical distribution, then
for each individual pipeline the sum of \pastro values below the given threshold provides the expected
number of true signals among the candidates produced by the pipeline. This expected signal count is
itself a realization of a Poisson process with a mean proportional to the pipeline's \VT for the true
signal population. This reasoning also applies to the total set of candidates identified by at least one pipeline,
i.e., to the catalog as a whole, if we consider the \textit{Any} pipeline sensitivity. Therefore,
following a similar approach in \gwtc[3.0]~\citep{2025arXiv250818082T}, we scale the sum of \pastro
values for each pipeline by the ratios of the \textit{Any} pipeline \VT to the individual pipeline's, each of
which is measured by total number of recovered injections without any weighting in the mass distribution. 
While these ratios in principle depend on the \ac{FAR} threshold at which we count injections as recovered, the dependence is small and we typically impose a threshold of $\qty{2}{\day^{-1}}$.
We here make use of the injected distribution~\citep{2025PhRvD.112j2001E} designed to broadly follow
the astrophysical \ac{CBC} distribution inferred from \ac{O3} observations described in \citet{2023PhRvX..13a1048A}.

We thus obtain an estimate of the signal count over subthreshold candidates for the \textit{Any} pipeline
by scaling the sum of \pastro from each of the four pipelines; finally, we take the average of these estimates
across the four pipelines.
The purity of the subthreshold candidate set is this estimated signal count divided by the overall number of subthreshold candidates.

\section{Data-quality Vetting Around Candidates}
\label{sec:dq}

A subset of the candidates produced by the search algorithms described in
Section~\ref{sec:searches} undergo studies to understand the quality of the
interferometric data surrounding the candidates with a goal to characterise and, if
required, mitigate against any departures from stationary Gaussian noise such as glitches.
This is vital to ensure that downstream analyses, particularly \PE (see Figure~\ref{fig:data-flow-chart} and Section~\ref{sec:pe}), which typically assume stationary Gaussian noise, produce unbiased results \citep{2018PhRvD..98h4016P, 2018CQGra..35o5017P, Ghonge:2023ksb}.
These methods are all applied after candidates are identified by search algorithms and to short sections of data.
A discussion of the data-quality studies that feed into the search algorithms is given in Section~\ref{sec:searches}.

Event validation is the process of identifying and examining potential data-quality issues in the vicinity of a \GW candidate.
Several distinct tests are used, due to the variety of data-quality issues that may impact the detection or analysis of \GW candidates. 
For each \GW candidate, test results are collected into a report referred to as a \ac{DQR}.
The creation of a single report simplifies the process of interpreting the results of multiple tests together and allows for summary statistics about individual \GW candidates to be produced based on the results of multiple aggregated analyses.
We apply this process to all candidates reported by online search pipelines and labeled as significant, with an internationally-coordinated team to provide rapid vetting aided by an automated \ac{DQR} framework \citep{2025CQGra..42h5016S, 2026JPhCS3177a2067A}.
Manual data-quality vetting of these online candidates was only conducted when at least one test identified a potential data-quality issue; if no issues were identified in this automated step, no additional vetting was conducted at this stage. 
The manual validation process is also completed for all candidates identified by offline search pipelines that meet the requirements for \PE analyses.  

In contrast to \ac{O3}, where \ac{LIGO} and Virgo conducted separate validation processes \citep{2022CQGra..39x5013D, 2023CQGra..40r5006A}, \ac{O4} employs a unified infrastructure
across all active detectors.
This uniformity in validation improves efficiency, streamlines information flow between different analysis groups, and reduces the demand on human resources.
The first step in event validation is the creation of a \ac{DQR}, which includes numerous tests of strain data-quality as well as analyses of auxiliary sensor information. 
The auxiliary sensor data from each observatory is generally only available at the observatory site, requiring these \ac{DQR} analyses to be conducted at multiple locations across the network. 
In O4b, there were four \ac{DQR} nodes that rely on a combination of the \DQRBUILD \citep{2026arXiv260516183D} toolkit and \VIRGODQR \citep{2023CQGra..40r5005F}: one at each of the LIGO Hanford, LIGO Livingston, and Virgo observatories, and a central node that hosts a database tracking results from all nodes. 

Each \ac{DQR} node is launched in response to an \soft{igwn-alert};\footnote{\url{https://igwn-alert.readthedocs.io}} the corresponding data-quality analyses are managed by a \DQRBUILD workflow configurator at the central node and at LIGO Hanford and LIGO Livingston, and by \VIRGODQR at Virgo. 
Once results are generated, they are then uploaded to the central node to be tracked by a \DQRBUILD SQL database. 
This database is used to generate a centralized result page that tracks the global set of data-quality analyses conducted for each event to simplify the manual event validation process.  

The data-quality analyses hosted by each \ac{DQR} node include
tests of the Gaussianity and stationarity of the strain data residual \citep{2023CQGra..40c5008V, 2020CQGra..37u5014M, 2024arXiv240115392D},
machine-learning signal versus noise classification \citep{2024CQGra..41h5007A},
estimates of the environmental coupling \citep{2024CQGra..41n5003H},
statistical correlations with auxiliary sensors \citep{2020arXiv200512761E, 2011CQGra..28w5005S, 2022PhRvD.105j2005V, 2010JPhCS.243a2005I},
tests of anomalies in the long-term behavior of the detectors,
checks of known data-quality issues,
and investigations of the detector observing state.
Analyses that are statistical in nature automatically produce conclusions about the data quality of a candidate; however, the results from many tasks still require manual investigation by data-quality experts.

Once the initial online vetting is completed, all significant candidates undergo further scrutiny to determine a final assessment of the data-quality status.
In this stage, tracked with the \EVENTVALIDATION package \citep{2025CQGra..42h5016S,2023chep.confE.110D}, a data-quality expert reaches a conclusion based on the preponderance of evidence in the automated \ac{DQR} analysis results. 
If data-quality issues are identified in this stage of the process, additional investigation and noise mitigation efforts are conducted to ensure that these issues have minimal impact on estimates of the astrophysical parameters of candidates. 

If data-quality issues, such as the presence of glitches, are identified around a candidate and these issues are not severe enough to warrant a retraction, we quantify if further action is required by comparing the distribution of excess power in the relevant region with expected Gaussian noise and computing a $p$-value \citep{2023CQGra..40c5008V}.
Until November 2023, a conservative threshold of 0.1 was adopted.
This threshold proved to result in an unsustainable rate of false positives, and was therefore changed to $0.05$.
For $p$-values greater than this threshold, we conclude the data are consistent with Gaussian noise, and the candidates are ready for downstream analyses.
On the other hand, if $p \le 0.05$, we attempt to subtract the noise using either a modeled Bayesian inference approach implemented in \BAYESWAVE \citep{2018PhRvD..98h4016P, 2021PhRvD.103d4006C, 2021PhRvD.103d4013C, 2022PhRvD.106d2006H, 2024PhRvD.109f4040G}, or a linear noise subtraction based on auxiliary witness channels \citep{2022CQGra..39x5013D}.
For all candidates in \ac{O4b}, \BAYESWAVE was used and applied to the \GDSStrainFrameAR channel as described in \citet{OpenData}; no correction is made for calibration uncertainty in this process\note{check}.
For past observing runs, \BAYESWAVE has predominantly been applied, except in cases where a witness channel was available \citep{2022CQGra..39x5013D, 2023PhRvX..13d1039A}.
However, if the noise is extended in time or frequency, as is often the case for certain glitch classes \citep[e.g.,][]{2022PhRvD.106d2006H, 2021CQGra..38b5016S, Soni:2023kqq},
making subtraction difficult, or if the data remains insufficiently stationary
after subtraction, we instead restrict the time and frequency analysis window
to avoid the afflicted region.
For badly afflicted candidates, preliminary \PE studies are routinely performed to validate the mitigation technique and understand the impact on parameter estimates \citep{2022CQGra..39x5013D}.
Once a satisfactory mitigation approach has been identified, the candidates are ready for downstream analyses.
In \citet{GWTC:Results}, we provide a summary of new \thisgwtcfull{} candidates that required mitigation and details of the mitigation approach that was applied.
We also release the glitch-mitigated strain data as part of the \thisgwtcfull{} data release \citep{GWTC-5.0:Deglitch}.

\section{Parameter Estimation for \acp{CBC}}
\label{sec:pe}

For a selection of candidates identified by the search algorithms (see Section~\ref{sec:searches}) and either having passed validation or with appropriate mitigation (see Section~\ref{sec:dq}), we use Bayesian inference to estimate the source parameters $\PEparameter$ of the \GW signal (see Figure~\ref{fig:data-flow-chart}), which includes both
the intrinsic parameters of the \ac{CBC} source and the extrinsic parameters that localize and orient the source in spacetime \citep{GWTC:Introduction}.
These inferences come in the form of posterior distributions $\PEposterior$ over the source parameters.
The posteriors represent our best understanding of the properties of the source and its location, including all correlations and degeneracies in the measured parameters for a given set of prior assumptions.
The posteriors for the selected \GW candidates from \ac{O4b} are reported for the first time in~\cite{GWTC:Results} 
and can be obtained from the \thisgwtcfull{} data release~\citep{GWTC-4.0:PE}.
We have not updated our inferences for candidates identified in previous observing runs.
Thus, our preferred parameter inferences for candidates identified before \ac{O4b} remain the same as previously reported in~\citet{2024PhRvD.109b2001A,2023PhRvX..13d1039A,2025arXiv250818082T}, with the exception of the \BNS candidate GW170817 whose preferred parameter inferences are reported in~\citet{2019PhRvX...9c1040A}\note{check}.

In Sections~\ref{sec:preliminary} to \ref{sec:AppxLocalization}, we describe the broad elements of \GW \PE and discuss the particular choices made in each version of the catalog
in Section~\ref{sec:PECatalogChoices}, noting that configuration files to reproduce the analysis of individual candidates are provided with the data release.

\subsection{Bayesian Formalism}

To carry out Bayesian \PE, we assume the data $\PEdata$ contains colored Gaussian noise 
and an astrophysical \GW signal which is well-approximated by a quasi-circular \ac{CBC} 
waveform model (see Section~\ref{sec:waveforms}), consistent with \GR
and absent of environmental effects.
These assumptions determine the form of our likelihood $\PElikelihood$, which for a single detector is Gaussian in the residuals between data and waveform~\citep{1992PhRvD..46.5236F,1994PhRvD..49.2658C,GWTC:Introduction}. 
We use the standard formulation for this likelihood and the associated frequency-domain 
noise-weighted inner product~\citep[e.g.,][]{2015PhRvD..91d2003V,2019PASA...36...10T}.
Given the likelihood and a prior distribution $\PEprior$, we
compute the posterior distribution from Bayes' theorem
\begin{align}
\PEposterior & = \frac{\PElikelihood \PEprior}{\PEevidence} \,,
\label{eqn:posterior}
\end{align}
where $\PEevidence$ is the normalizing evidence:
\begin{align}
\PEevidence  = \int \PElikelihood \PEprior \, d \PEparameter\,.
\label{eqn:evidence}
\end{align}
The evidence is not required in most approaches to \PE, but can be used for model comparison.

In our baseline analyses the joint likelihood also includes additional parameters to account for the calibration state of the detector and the uncertainty in this calibration (Section~\ref{sec:calibration}).
The full likelihood is evaluated coherently across the detector network by multiplying the individual interferometer likelihoods, under the assumption that the noise is independent in each.

\subsection{Preliminary \PE}
\label{sec:preliminary}

After initial identification of a candidate by a search algorithm (Section~\ref{sec:searches}), we carry out preliminary \PE in order to understand the properties of the source and guide our analysis settings.
We choose our initial settings and waveform model using the output of the search pipelines, e.g., the point estimates of the source masses.
As needed, we carry out further analyses with modified priors or configuration settings, in some cases utilizing multiple waveform models to understand systematic modeling errors.
If there are data-quality issues, this preliminary estimation is iterated alongside glitch-mitigation studies, as described in Section~\ref{sec:dq}.

Preliminary \PE also guides our choice of waveform models for inference (Section \ref{sec:waveforms}).

\subsection{Likelihood Calculation: Elements of the Inner Product}
\label{sec:PEInnerProduct}

The integration of the inner product that defines the likelihood is carried out in the frequency 
domain~\citep{GWTC:Introduction,2015PhRvD..91d2003V,2019PASA...36...10T}.
Following our standard conventions~\citep{GWTC:Introduction}, 
given a discrete raw time-series data $\PEdataTimeDomain(t)$ with a duration $T$ and
sampling frequency $\fsamp$, we first apply a \ac{FFT} and divide
by the sampling frequency to produce the discrete frequency series $\tilde{\PEdata}$.
Similarly, time-domain waveforms are transformed into frequency series, while frequency-domain models
can be evaluated directly for use in the inner product.

The amount of data analyzed for each candidate depends on the estimated mass of the system
as determined by preliminary \PE analyses,
with lower mass signals in the sensitive band for longer periods of time.
The segment of data used in \PE is selected with the estimated time of coalescence set \qty{2}{\second} before
the end of the segment, with the total duration set to powers of $2$ ranging from \qty{4}{\second} up to \qty{128}{\second}
such that the evolution of the signal starting from $\flow$ is included.
This choice sets the discrete frequency spacing $\Delta f = 1/T$ in the inner product.
Before transforming time-domain data to the frequency domain to evaluate the likelihood, we apply a symmetric Tukey window function~\citep{Harris:1978} with a roll-on time that balances how much data in the segment is affected by the windowing and the impact of spectral leakage of the noise, especially at the lowest frequencies analyzed where the noise is steeply varying.

We compute the required integral from a lower frequency $\flow$ to $\fhi$ and discard the data outside this range.
Generally, $\flow$ is set to \qty{20}{\hertz}, below which the detector noise steeply increases.
The upper frequency $\fhi$ is chosen to be the Nyquist frequency, half the discrete sampling rate $\fsamp$
of the data, with further modifications to account for power lost due to applying a low-pass Butterworth filter 
when downsampling the data to the desired sampling rate~\citep{2015PhRvL.115n1101V,2020MNRAS.499.3295R}.
The result is $\fhi = \alphaRoll \fsamp/2$, with the particular choice of $\alphaRoll = 0.875$ selected to 
limit power loss to 1\%~\citep{2023PhRvX..13d1039A}.
The data are downsampled to limit the computational cost of the likelihood evaluation, and $\fsamp$ is chosen for 
each candidate to ensure that selected higher multipolar modes are resolved for candidates which coalesce in 
the sensitive frequency band of the detectors, or else set to a high enough value where noise 
dominates the signal power for low-mass candidates that coalesce out of the sensitive band.
This upper limit is generally $\fsamp = 4096$ Hz or $\fsamp = 8192$ Hz.

A key ingredient in the likelihood evaluation is an estimate of the noise \PSD for the analyzed data.

In order to mitigate the effects of the nonstationary nature of the instrumental noise, 
for \PE results presented in \thisgwtcfull{}, we use an on-source estimate for the \PSD generated with the \BAYESWAVE algorithm~\citep{2015CQGra..32m5012C,2015PhRvD..91h4034L,2021PhRvD.103d4006C,2024PhRvD.109f4040G}.

This approach uses Bayesian inference to construct a posterior distribution over \PSD realizations, and
we take the median \PSD value at each frequency~\citep{2019PhRvD.100j4004C} for the point-estimate used in the inner product.

\subsection{Calibration Marginalization}
\label{sec:calibration}

In order to account for the uncertain calibration of the strain data, 
we allow for independent frequency-dependent shifts of the 
amplitude and phase of the waveform in each interferometer~\citep{Farr:2014aab}.
The frequency-dependent shifts are modeled using additional free parameters in our waveform model, 
which we call calibration parameters. 
We marginalize over the calibration parameters when reporting \PE results~\citep{2016PhRvL.116x1102A}.

The strain data are produced in near-real time using an initial calibration model, 
which may be subsequently updated in order to recalibrate the data \citep{OpenData}.
For the \ac{LIGO} detectors the methods used to calibrate the strain data provide 
a measured distribution of frequency-dependent correction factors
$\PEcalfactor (f)$ \citep{2017PhRvD..96j2001C,2020CQGra..37v5008S,2021arXiv210700129S,O4aCalPaper}\note{check, add calibration O4b reference if available}. 
These factors correct the calibrated strain data $\PEdata$ in each detector to the strain that would be measured with perfect calibration $\PEdata_\star$,
\begin{align}
\label{eq:PECalCorrection}
\tilde{\PEdata}_\star & =\PEcalfactor \tilde{\PEdata}  \,.
\end{align}
For the Virgo detector the methods used for calibration give an estimate of the inverse of the
correction factor $1/\PEcalfactor$ (which has been unity through \ac{O3}) and the corresponding 
frequency-dependent uncertainties~\citep{2014CQGra..31p5013A,2018CQGra..35t5004A,2022CQGra..39d5006A}.

The likelihood used in \PE relies on knowledge of the noise properties of the imperfectly calibrated data $\PEdata$ through the estimated \PSD.
The likelihood $\PElikelihood$ is evaluated in the frequency domain using $\tilde{\PEdata}$,
\begin{align}
\tilde{\PEdata} & = \frac{1}{\PEcalfactor}[\tilde{h}(\PEparameter) + \tilde{n}_\star] =  \frac{\tilde{h}(\PEparameter) }{\PEcalfactor} + \tilde{n} \,,
\end{align}
where $\tilde{n}_\star$ is the true noise in the detector, assumed to be Gaussian, and $\tilde n = \tilde{n}_\star/\PEcalfactor$ is the noise in the calibrated strain, also assumed to be Gaussian.
The \PSD describes the properties of $\tilde n$, so the correction factors $\PEcalfactor$ must be applied to correct the \GW waveform $\tilde{h}$ to account for imperfect calibration.
The corrections $1/\PEcalfactor$ are approximated by small frequency-dependent amplitude and phase corrections, which are modeled using splines~\citep{Farr:2014aab}.
The values of these splines at fixed frequency nodes are the calibration parameters, and for \PE we use Gaussian priors
on these additional parameters.
For \ac{LIGO} data we use the median and standard deviation of the measured distribution of  $\PEcalfactor$~\citep{2017PhRvD..96j2001C,2020CQGra..37v5008S,2021arXiv210700129S,O4aCalPaper} to set the mean and 
standard deviation of the priors over the calibration parameters describing $1/\PEcalfactor$\note{check, add calibration O4b reference if available}. 
For Virgo data, the priors on the calibration parameters through \ac{O3} are zero-mean Gaussians with standard deviations corresponding to the directly measured uncertainties in $1/\PEcalfactor$. For O4, frequency-dependent error is also provided, even if it was mainly negligible since the error was corrected in low-latency to provide an (almost) unbiased strain.

Up to \ac{O3}, the data from \ac{LHO} and \ac{LLO} were recalibrated prior to final analysis in order to approximately remove the systematic miscalibration arising from the initial model.
For \PE of these candidates, we used final calibrated data as described in Table 4 of \citet{OpenData}.
Because of this recalibration, the errors and uncertainty on the calibration parameters are small, such that the priors for the calibration parameters are nearly centered at zero (corresponding to no calibration correction).
However, in \ac{O4} the data in \ac{LHO}, \ac{LLO} and \ac{Virgo} are no longer recalibrated in this way by default, except for an approximately two-month period in \ac{O4b} when recalibration was needed~\citep{OpenData}\note{check}.
Nevertheless, recalibration during \PE can be effectively carried out using calibration priors centered away from zero.
For this reason the priors associated with the new \GW candidates presented in \gwtc[5.0] may lie more than a standard 
deviation away from zero for \gwtc[5.0] at various frequency points.

\subsection{Priors}
\label{sec:Priors}

To estimate the posterior distribution, Equation~\eqref{eqn:posterior}, we must select an appropriate prior distribution over the binary parameters $\PEparameter$.
Since our inferences are made over a catalog of \ac{CBC} systems with a wide range of properties and using a variety of waveform models, our prior ranges are selected to be appropriate for each candidate.
The priors chosen are agnostic and wide enough to cover the region of parameter space where the posteriors have support, while ensuring a reasonable amount of analysis time and accounting for the parameter ranges over which our waveform models are calibrated.
Prior ranges (e.g., on the component masses) are selected during preliminary \PE analysis and adjusted afterward as needed, for example to ensure that 
an arbitrary prior boundary does not affect the posterior.

For all candidates we use priors that are uniform over the (redshifted) detector-frame component masses $(1+z)m_i$, with boundaries in detector-frame total mass, detector-frame chirp mass, and mass ratio appropriate for each candidate and applied model (see Section~\ref{sec:waveforms}).
The priors are uniform on the spin magnitudes and isotropic in spin orientations.
We use isotropic priors on the binary orientation, and priors that are uniform in comoving volume and comoving time so that the priors on the sky location are isotropic.
For this we carry out sampling in a reference cosmology~\citep{GWTC:Introduction} and reweight our final samples to priors appropriate for the cosmological model of~\citet{2016A&A...594A..13P}, which is a flat-$\Lambda$CDM model with $\HzeroSymbol = \HzeroValue$ and $\WmSymbol = \WmValue$, following our approach in past versions of the \gwtc.

\subsection{Sampling from the Posterior}
\label{sec:PESampling}

Due to the high dimensionality of the source parameters $\PEparameter$, brute-force evaluation of the posteriors is intractable. 
Instead we represent the posteriors for each candidate with discrete samples $\PEparameter_i$ which represent fair draws from 
the posterior distribution.
The challenge then reduces to the problem of stochastically sampling from the posterior distribution $\PEposterior$.

A number of algorithms exist to tackle this sampling challenge.
These include nested sampling~\citep{Skilling:2006gxv}, a Monte Carlo technique to estimate the evidence and which produces samples as a byproduct, 
and \acl{MCMC}~\citep[\acsu{MCMC};][]{1970Bimka..57...97H}, which samples through random walks and forms the core of many sampling algorithms.
The evidence can be computed with \MCMC methods by e.g., thermodynamic integration~\citep{2004AIPC..707...59G,2009PhRvD..80f3007L} via parallel tempering~\citep{2005PCCP....7.3910E,2008CQGra..25r4011V,2015PhRvD..91d2003V}.
For many analyses, we use the \DYNESTY~\citep{2020MNRAS.493.3132S} nested-sampling algorithm, and accessed through the \BILBY package~\citep{2019ApJS..241...27A,2020MNRAS.499.3295R} which includes a custom stepping method optimized for use on parallelized computing resources, specifically large-core-count CPUs. While for older results parameter inference was carried out using \LALINFERENCE \citep{2015PhRvD..91d2003V}, which provides an implementation of \MCMC and nested sampling, as well as tailored methods for proposing new sample points.

We also make use of additional, highly parallelized sampling techniques to tackle computationally expensive analyses.
These include \PBILBY~\citep{2020MNRAS.498.4492S} and the \RIFT algorithm~\citep{2015PhRvD..92b3002P,2017PhRvD..96j4041L,2019PhRvD..99h4026W}.
\PBILBY is optimized to use distributed computing to perform nested sampling.
\RIFT is a highly parallel, iterative sampling algorithm that uses adaptive, grid-based explorations of the likelihood while marginalizing over the extrinsic parameters, followed by a final stage of Monte Carlo sampling of the intrinsic and extrinsic parameters.

For the new candidates presented in \gwtc[5.0], we employ nested sampling in \BILBY as well as the \RIFT algorithm to perform sampling. 
Additionally, \BILBYMCMC~\citep{2021MNRAS.507.2037A}, which is a parallel-tempered ensemble Metropolis--Hastings Monte Carlo sampling algorithm, and a new machine-learning based pipeline, \DINGO~\citep{2020PhRvD.102j4057G,2021PhRvL.127x1103D,2023PhRvL.130q1403D} trained using \SEOBNRFIVEPHM, were used for cross-validation. 
\DINGO employs neural posterior estimation to train a normalizing flow-based neural density estimator that directly approximates the posterior distribution. 
Once fully trained using simulated data, the model enables amortised inference, reducing parameter estimation times by several orders of magnitude relative to conventional stochastic samplers. A subsequent importance sampling step is used to correct for residual discrepancies between the learned posterior and the true likelihood. 

To improve sampling performance, certain parameters in the waveform model can be marginalized out of the posterior during sampling.
Depending on the waveform model employed and the \GW candidate, any of the coalescence phase~\citep{2015PhRvD..91d2003V,Farr:2014aaa}, coalescence time~\citep{Farr:2014aaa,2020MNRAS.499.3295R}, or luminosity distance~\citep{2016PhRvD..93b4013S} have been marginalized over, as discussed in Section~\ref{sec:PECatalogChoices}.
This can be done analytically or numerically depending on the parameter and the details of the waveform model.
When marginalization is carried out, the full posterior distributions are reconstructed in postprocessing~\citep{2019PASA...36...10T}.

The method of marginalizing over calibration uncertainties varies depending on the sampling method used. 
For inferences carried out using \RIFT, the marginalization over the calibration model parameters is carried out in post-processing, by reweighting the likelihood of samples produced first without utilizing the spline calibration model~\citep{2020PhRvD.102l2004P,2023PhRvX..13d1039A}.
Meanwhile, for \PE using \BILBY, \PBILBY, and \LALINFERENCE, the parameters of the calibration model are inferred alongside the intrinsic and extrinsic binary parameters during sampling.

A large number of \PE analyses and steps within each analysis is required to produce the inferences included in \gwtc[5.0].
These analyses were managed using the \ASIMOV software library~\citep{2023JOSS....8.4170W}.
Postprocessing of the samples is carried out with the \PESUMMARY software library~\citep{2021SoftX..1500765H}.
This includes the computation of derived parameters such as the remnant properties following \BBH merger and the evolution of the spin components to large binary separations as described in Section~\ref{sec:waveformsfinalproperties}.
These are implemented with routines in \LALSUITE~\citep{lalsuite, 2020SoftX..1200634W} and used by \PESUMMARY.
For reweighting of samples, e.g., to our preferred cosmological model or to incorporate calibration marginalization with \RIFT, either routines in \PESUMMARY or simple custom routines (as in the case for new candidates in \gwtc[5.0]) are used and provided with our data release.

\subsection{Posterior Samples}
\label{sec:PEposterior}

Following sampling, our measurements of the binary parameters $\PEparameter$ for each candidate are represented by a discrete set of parameter values $\PEparameter_i$ sampled from the posterior distribution. 
These samples can be used to compute expectation values over quantities of interest, e.g., through Monte Carlo integration.
Marginalization is achieved by only considering the dependence of the samples on the quantities that are not marginalized out.
Each sample includes the fifteen intrinsic and extrinsic parameters required to describe the \GW strain from quasi-circular \BBH{s} as well as the parameters required to describe the inferred calibration state.
In addition to these a number of quantities can be derived from the intrinsic parameters of the binary using the methods described in Section~\ref{sec:waveformsfinalproperties}.
These include the final mass $\Mf$ and final dimensionless spin $\chif$ of the remnant compact object following coalescence, the peak luminosity $\lumpeak$ and total energy radiated $\energyrad$, as well as the final kick velocity of the remnant $v_{\rm k}$ when the relevant waveform model supports such estimates.

Some intrinsic properties of the binaries vary over the course of the binary coalescence, i.e., the orientations of the dimensionless spins $\vecspinone$ and $\vecspintwo$ in precessing systems.
Our initial inferences report these quantities at a reference frequency $f_{\rm ref}$, taken to be \qty{20}{\hertz}, corresponding to a variable reference time before coalescence.
The spin tilt angles with respect to the orbital angular momentum, $\theta_1$ and $\theta_2$, in particular carry valuable information about the formation mechanism of the binary.
These angles approach well-defined limits at infinite binary separation~\citep{2015PhRvD..92f4016G}, where they can be used in population inferences \citep[e.g.,][]{Mould:2021xst}, and so we report the spin tilts and derived quantities such as $\chieff$ and $\chip$ in this limit.
The derived spins are evolved from their reference values~\citep{2022PhRvD.106b3001J} using \PN expressions for the precession-averaged dynamics~\citep{2015PhRvD..92f4016G,2017PhRvD..95j4004C} to formally infinite separations \citep[cf.][]{Gerosa:2023xsx}.
Wherever possible, the more accurate hybrid evolution using a combination of orbit-averaged precession followed by precession-average evolution is employed, but for a small number candidates only the computationally faster precession-averaging approach is used.
We report these evolved spin quantities in our final samples for \ac{BBH} candidates, and for \ac{NSBH} candidates when analyzing 
them with waveform models that neglect the imprint of matter onto the \ac{GW} signal.

In reporting and plotting the inferred parameters for each candidate, we marginalize out all but one or two parameters at a time, and report the median value and \acp{CI} of the marginalized posteriors for those parameters.
Unless otherwise noted, we report the median and 90\% \ac{CI} of each binary parameter, marginalizing over the others.
Generally we use symmetric \acp{CI} for single parameters, so that 5\% of the posterior density lies below the lower bound of the 90\% \ac{CI} and 5\% of the density lies above the upper bound.
In some instances where the posteriors have support near a physical prior boundary, the symmetric \ac{CI} gives the appearance of excluding the boundary value even if there is high probability there.
In such cases we may report the 90\% \ac{HPD}, the smallest interval containing 90\% of the posterior density.
When plotting marginalized densities we make use of one- and two-dimensional \ac{KDE} to produce continuous densities from samples. 
Our two-dimensional credible regions are constructed using the \ac{HPD} method.

Our inferences on the location of \GW candidates are available in two formats. 
The samples themselves represent our full posterior over the source parameters, including the sky location and distance of the detected \GW sources.
In addition, we provide three-dimensional localizations in the same format as the localizations included in our public \GW alerts,
by applying \ac{KDE} to the samples.
These localizations are created using \LIGOSKYMAP \citep{2016PhRvD..93b4013S,2016ApJ...829L..15S,Singer:2016erz}, which includes the \BAYESTAR package used to localize \GW candidates from modeled online and offline searches.

\subsection{Approximate Spatial Localization}
\label{sec:AppxLocalization}

We do not carry out full \PE for all candidates which meet the criteria for inclusion in \gwtc[5.0] (Section \ref{sec:selection_criteria}).
Our criteria for full \PE analysis have evolved with \gwtc{} versions 
(Section \ref{sec:PECatalogChoices}) due to the increased number of detection candidates and resource limitations.
However, we provide approximate spatial localization for all candidates in \gwtc[5.0] to enable multimessenger analyses of large samples of weak \ac{GW} candidates using the same methods as for our public \GW alerts.
Such methods are computationally much cheaper than full \PE, and already integrated into the candidate management system.
In particular, for candidates produced by modeled \ac{CBC} searches, the
approximate localization is carried out with \BAYESTAR \citep{2016PhRvD..93b4013S,
2016ApJ...829L..15S, Singer:2016erz}.
For candidates produced by \CWB-BBH, the localization method is directly part of \CWB
and described in Section~\ref{subsection:cWB}.

\subsection{Analysis Settings and Details}
\label{sec:PECatalogChoices}

Here we describe the particular analysis settings which vary across the inferences carried out for each candidate in \gwtc[5.0].

\subsubsection{Candidates Found in \ac{O4}}

For the new candidates from \ac{O4a} and \ac{O4b} presented in \gwtc[5.0], we perform full Bayesian \PE on a high-purity subset of the candidates (Section \ref{sec:selection_criteria}), namely those with \acp{FAR} $< \qty{1}{\yr^{-1}}$ and $\pastro > 0.5$ and not determined to be of instrumental origin (Section \ref{sec:dq}).

For this subset, the analysis settings depend on the nature of the binary as inferred using preliminary \PE and confirmed in our final analysis.
As described in \citet{GWTC:Results} these candidates include a large number of \acp{BBH} but no significant \ac{NSBH} or \ac{BNS} candidates.

Based on preliminary \PE and any input from data validation (cf.\ Section~\ref{sec:dq}), we categorize the signals and determine an appropriate prior, waveform model, and data segment to analyze (in all cases, the data product is based on the \GDSStrainFrameAR channel, as described in \citet{OpenData}).
For nearly all cases, the durations are selected so that the $\ell =3$, $|m|=3$ higher harmonic is fully captured within the frequency range integrated over in our likelihood.
In higher-mass systems $\ell =4$ modes may make a non-negligible contribution in our sensitive frequency band~\citep[e.g.,][]{2021PhRvD.103b4042M}, and in such cases we set the starting frequency so that the $\ell =4$, $|m|=4$ mode begins in the range of integration.
Specifically, we do so for events whose source properties have support for redshifted mass $(1+z)\Mtot > 200 \Msun$.
In each case it is possible that 
at  the starting frequency of the waveform even higher harmonics will then begin in band, which may cause aliasing in the case of time-domain waveform models.
Due to the small amplitude of such higher harmonics relative to the lower-frequency multipole moments, the effect is expected to be negligible.
For all events, we adopt the additional prior on the binary parameters during sampling such that the $\ell = 3$ mode is resolved in band, given the chosen sampling rate. 
This only modifies the prior for a small number of candidates, and we have checked that in these cases, the prior makes no discernible difference in the final posterior samples.

For all candidates the Tukey window applied before transforming time-domain data to the frequency domain has a roll-on of \qty{1}{\second}.
This is longer than the \qty{0.4}{\second} roll-on used in prior \PE analyses.
The longer window is chosen to reduce spectral leakage, because for the first time the instrumental noise near $\flow$ is sufficiently low that the small amount of leakage from even lower frequencies into frequency range integrated over in our likelihood made noticeable impact on our analyses.

The analysis of the initial \PE results to determine our final settings is carried out using the \PECONFIG software package.
The resulting information is also used to determine the appropriate waveform models to use to capture systematic uncertainties across waveform models.

For \BBH candidates, we use both the phenomenological models \IMRPhenomXPHMST~\citep{2020PhRvD.102f4002G,2020PhRvD.102f4001P,2025PhRvD.111j4019C} and \IMRPhenomXPNR{}~\citep{2026PhRvD.113h4055H}  and the effective one body model \SEOBNRFIVEPHM~\citep{Khalil:2023kep,2023PhRvD.108l4035P,2023PhRvD.108l4037R,2026PhRvD.113d4049E} for all candidates.
In addition, we use \SURSEVENDQFOUR~\citep{2019PhRvR...1c3015V} for those candidates whose parameters lie within the range of total mass and mass ratio values supported by the model, as determined by preliminary \PE.
In the case of \SURSEVENDQFOUR we use a fixed duration of $10000(G/c^3)\Mtot(1+z)$ for the waveform model when evaluating the likelihood, with $\Mtot(1+z)$ the total detector-frame mass of the source.

For all new candidates presented in \gwtc[5.0], the luminosity distance is marginalized over during sampling.
For sampling we use a distance prior uniform in comoving volume and comoving time with the default cosmological model of the \ASTROPY software package, a flat-$\Lambda$ cold dark matter (CDM) cosmology with $\HzeroSymbol = \ensuremath{67.66~\mathrm{km\,s^{-1}\,Mpc^{-1}}}$  and $\WmSymbol = 0.30966$.
The cosmology assumed during sampling is not the default cosmology we present our results in.
Instead, the final samples are reweighted to our preferred cosmology~\citep{2016A&A...594A..13P} as described in Section~\ref{sec:Priors}.

The analysis of the candidates from \ac{O4a} first presented in \gwtc[4.0] used the same methods and settings as described above, except for the choice of waveform models as per Table~\ref{tab:wf_models}, and the use of the \DYNESTY sampling algorithm through the \BILBY package for all \ac{PE} results.

\subsubsection{Candidates found in \ac{O1}, \ac{O2}, \ac{O3}}

As a cumulative catalog, 
\gwtc[5.0] includes candidates detected during the first three observing runs and the first part of the fourth observing run of the advanced detector network, as discussed in Section~\ref{sec:selection_criteria}.
For \GW candidates identified in \ac{O1}, \ac{O2}, \ac{O3} and \ac{O4a} we have previously performed full \PE for the subset of these candidates identified with $\pastro > 0.5$ plus the \NSBH candidate GW200105\_162426~\citep{2021ApJ...915L...5A}, and our \PE results for these candidates remain the same in \gwtc[5.0].
These inferences were first presented in \gwtc[2.1]~(\citealt{2024PhRvD.109b2001A}, for candidates found in \ac{O1}, \ac{O2}, and \ac{O3a}) and \gwtc[3.0]~(\citealt{2023PhRvX..13d1039A}, for candidates found in \ac{O3b}).
The exception is the \BNS candidate GW170817, for which the \PE results were not updated in \gwtc[2.1], and so remain the same as those presented in \gwtc[1.0]~\citep{2019PhRvX...9c1040A}.
Both sets of analyses from \gwtc[2.1] and \gwtc[3.0] used the same methods and settings which we now summarize.

For each candidate, \acp{PSD} were generated using \BAYESWAVE~\citep{2015CQGra..32m5012C,2015PhRvD..91h4034L,2021PhRvD.103d4006C,2024PhRvD.109f4040G} as described in Section~\ref{sec:PEInnerProduct}.

For these candidates the roll-on of Tukey window applied to the time domain-data before transforming to the frequency domain was \qty{0.4}{\second}.
Multiple waveform models were used in each case in order to understand systematic uncertainties in the inferences.

For \BBH candidates, \PE was carried out using two waveform models which include higher harmonics and the effects of orbital precession.
The phenomenological model \IMRPhenomXPHM~\citep{2020PhRvD.102f4002G,2020PhRvD.102f4001P} was used with a multiscale prescription for the precession dynamics~\citep{2017PhRvD..95j4004C} and sampling was carried out with \BILBY.
The \ac{EOB} model \SEOBNRFOURPHM~\citep{2020PhRvD.102d4055O} was also used for each \BBH candidate, using \RIFT to carry out \PE. 

For the \NSBH candidates, multiple waveform models were also used for \PE.
Our baseline results are drawn from the same models as for \BBH candidates, namely \IMRPhenomXPHM  and \SEOBNRFOURPHM.
These models include higher harmonics and precession but neglect matter effects such as the tidal deformation or disruption of the \ac{NS} on the waveform.
This is because such effects are negligible for the \NSBH candidates identified in \ac{O3b}~\citep{2021ApJ...915L...5A}.
In order to assess the importance of matter effects on the \GW signal, models which include the effects to tidal deformation and models which are specifically tailored for \NSBH systems were also used for inference.
These were \IMRPhenomNSBH and \SEOBNRFOURNRtidalTWONSBH.
In addition, \IMRPhenomPTWONRTidalTWO is employed to approximately incorporate both matter and precession effects. There were no new \NSBH candidates detected in \ac{O4b}.

Up to \gwtc[4.0], two \BNS candidates have been identified satisfying the \PE threshold, GW170817 and GW190425.
Multiple models are used for \PE of GW170817~\citep{2019PhRvX...9c1040A}. 
GW170817 was analyzed with three frequency-domain models using \LALINFERENCE: \TFTWO including tidal effects \citep{Sturani:2015aaa,Isoyama:2020lls,2008PhRvD..77b1502F,2011PhRvD..83h4051V}, \IMRPhenomPTWONRTidal, and \SEOBNRFOURNRTidal, and with two time-domain models using \RIFT: \SEOBNRFOURT \citep{2016PhRvL.116r1101H,2016PhRvD..94j4028S}, and \TEOBResumS \citep{Nagar:2018zoe}.
For this candidate we marginalize over the phase analytically for models that assume spins aligned with the orbital plane.
As with our other analyses on candidates including \acp{NS}, multiple analyses are carried out allowing only for relatively small spin magnitudes $\chi_i \leq 0.05$ and allowing for spin magnitudes up to the maximum allowed for a given model, as large as $\chi_i \leq 0.99$, 
which allows us to infer an upper bound on the component spin magnitudes from the data without prior constraints~\citep{2019PhRvX...9a1001A}.
The second \BNS candidate, GW190425~\citep{2020ApJ...892L...3A}, was detected during \ac{O3a} and its \PE results updated in \gwtc[2.1]~\citep{2024PhRvD.109b2001A}.
For this candidate we present results using the precessing, tidal approximant \IMRPhenomPTWONRTidal using \DYNESTY through \BILBY, and both relatively low- and high-spin prior limits on the component spins.
To accelerate inference we employ reduced-order quadrature~\citep{2015PhRvL.114g1104C,2016PhRvD..94d4031S,2023PhRvD.108l3040M} 

When using \BILBY or \PBILBY to analyze candidates from the first three observing runs, the posteriors are marginalized over luminosity distance and geocenter time, with the exception of the \BILBY analysis of GW190425 which only used distance marginalization.
For these candidates, initial sampling was carried out with a distance prior uniform in Euclidean volume.
During postprocessing, the posterior samples were then reweighted to the cosmological model described in Section~\ref{sec:Priors}~\citep{2016A&A...594A..13P}.

\subsubsection{Calibration Prior Settings for Candidates from \ac{O1}, \ac{O2} and \ac{O3}}

We marginalized over the calibration uncertainties when producing
the \PE results for \ac{O1}, \ac{O2}, and \ac{O3}, as discussed in Section~\ref{sec:calibration}.
Due to an error in implementation, an incorrect prior on the calibration parameters for \ac{LHO} and \ac{LLO} was used for these results.
The priors were set using the median and \acp{CI} of $\PEcalfactor$ from Equation \eqref{eq:PECalCorrection} for data from the \ac{LIGO} detectors, rather than those of $1/\PEcalfactor$ as required by the method. The inconsistency was discovered in the analysis of GW240925\_005809, where the \ac{LHO} calibration was inferred directly from the signal~\citep{gzrj-mwv3}.
In the limit of small calibration uncertainties, this amounts to a sign error in the means of the Gaussian priors for the calibration parameters.
In the case that the means of the priors are zero the error has no effect.
However, the calibration uncertainties for \ac{LHO} and \ac{LLO} have nonzero means.
Meanwhile, the priors on the calibration parameters were set correctly for Virgo data. 

For candidates detected in \ac{O1}, \ac{O2}, and \ac{O3} the means are generally small relative to the standard deviation of the priors on the calibration parameters.
Further the absolute sizes of the standard deviations of these parameters are small 
in the sensitive band of the detectors, of the order of a few percent in amplitude and a few degrees in phase for the two \ac{LIGO} interferometers~\citep{2019PhRvX...9c1040A,2021PhRvX..11b1053A,2024PhRvD.109b2001A,2023PhRvX..13d1039A}, 
and so \emph{a priori} the impact of this error on our inferences is expected to be small.
We have verified this expectation through preliminary re-analysis of the potentially impacted candidates, 
carried out by repeating \PE with corrected priors and by reweighting the likelihood values of 
existing \PE samples in order to correct for the erroneous calibration priors \citep{Baka:2025bbb}.
None of the scientific conclusions reported in previous studies are affected by the error.
In addition, the error does not impact the significance of any of our candidates, since the \GW searches described in Section \ref{sec:searches} do not incorporate the effect of uncertain strain calibration.

The typical change in the posteriors following preliminary re-analysis is within statistical sampling error, 
as quantified by the Jensen--Shannon divergence \citep{1991ITIT...37..145L} between one-dimensional marginal distributions 
before and after correction \citep[cf.][Appendix A]{2020MNRAS.499.3295R,2021PhRvX..11b1053A}.
As a particular case, the localization of the source of GW170817 receives only a small correction when re-analyzed, 
and its association with the electromagnetic counterpart emission from AT 2017gfo \citep{2017ApJ...848L..12A} is unaffected by the error.

Another case is GW150914, which displays visible differences in the sky location posteriors when re-analyzed (although the bounds of the $90\%$ \ac{CI} of the right ascension and declination remain nearly unchanged).
Meanwhile, our inferences of the intrinsic parameters of GW150914 remain unchanged to within sampling errors.

While the impact on individual candidates is small, a possible concern is that this error can bias analyses 
that aggregate data from multiple candidates, such as population studies and cosmological inferences.
We are investigating the impact of the calibration marginalization error on these analyses, 
but the effects are expected to be negligible compared to other sources of systematic error.
This error does not impact the \PE of new candidates observed in \ac{O4b}~\citep{GWTC:Results}.

\subsection{On the Likelihood used for Inference in \ac{O1}, \ac{O2} and \ac{O3}}
\label{sec:likelihood_issue}

During the production stage for \ac{O4a}, we discovered a normalization error in the likelihood used for the inference codes \BAYESWAVE, \BILBY, \LALINFERENCE, \PBILBY, and \RIFT.
The error arises due to the incorrect application of a window factor that was intended to account for the power lost in the noise residuals due to the Tukey window applied to the data before transforming them to the frequency domain.
This error causes the likelihood to be overly constrained by a factor depending on the window function used to mitigate spectral leakage.
The incorrect likelihood $\hat{{p}}(\PEdata | \PEparameter) $ is related to the correct likelihood ${\PElikelihood}$ via the average power in the Tukey window $\hat{{p}}(\PEdata | \PEparameter)=\PElikelihood^{\beta} $ with
\begin{equation}
\beta = \left( 1 - \frac{5 T_\mathrm{w}}{4 T} \right)^{-1} \,,
\end{equation}
where $T_\mathrm{w}$ is the roll-off time of the Tukey window to one side, and $T$ is the segment duration.
The window factor is applied when computing the \ac{PSD} via standard methods, but should not be applied to
the data when the signal has support only where the window function is unity, as is the case in our analyses.
Although accounting for the windowing using a single window factor in \ac{PSD} estimation is still an approximation when computing
the likelihood, multiple investigations and the use of 
probability--probability tests \citep{2015PhRvD..91d2003V,2020MNRAS.499.3295R} have confirmed its accuracy.
Further, these tests have confirmed the normalization error in our inference codes and the correctness of our updated likelihood.
More details on the error and validation of the updated likelihood can be found in \citet{2025CQGra..42w5023T}.

For the new candidates from \ac{O4}, we used the correct likelihood $\PElikelihood$.
Comparisons with preliminary results for \ac{O4a} generated with the incorrect likelihood $\hat{{p}}(\PEdata | \PEparameter)$ show that the differences in the posteriors between the two methods are small, but correcting the error systematically widens the posterior distributions.
This is the expected behavior, since the effect of the error corresponds to systematically overestimating the \ac{SNR} inferred from \ac{PE}.\note{update and check}

While the error is not present in the source properties reported in \gwtc[4.0] \citep{2025arXiv250818082T}, 
due to resource constraints, posteriors from earlier analyses that included
this error were used in the waveform consistency tests \citep{2025arXiv250818081T} 
for the new candidates from \ac{O4a}. 
The error in the likelihood remains present in the \BAYESWAVE analyses used in those tests.
For the worst cases, $T_\mathrm{w} = \qty{1}{\second}$ and $T = \qty{4}{\second}$, yielding $\beta =  1.45$.
This corresponds to overestimated \acp{SNR} from \ac{PE} by a factor $1.21$.
The impact is less for lower-mass candidates whose durations are larger, so that for \ac{BBH} candidates with $T_\mathrm{w} = \qty{1}{\second}$ and $T = \qty{8}{\second}$, the factor is $\beta = 1.19$, and the \acp{SNR} are overestimated by a factor of $1.09$.
To check the impact of this error on the waveform consistency tests, 
we reran the consistency tests for a subset of events using the correct posterior distributions, 
and find no significant changes in the results reported in~\cite{2025arXiv250818082T}.\note{update and check}

For \ac{O4a} candidates, we released our final posteriors generated using the correct likelihood $\PElikelihood$;
posteriors generated using the incorrect likelihood $\hat{{p}}(\PEdata | \PEparameter)$;
and a set of reweighted posteriors that correct the error by applying rejection sampling with replacement
to posteriors generated with $\hat{{p}}(\PEdata | \PEparameter)$, 
with an acceptance ratio given by the ratio of the correct to incorrect likelihood.\note{update and check}

The error does affect previous analyses from \ac{O1} through \ac{O3}, but the impact is smaller than for \ac{O4} because of the smaller Tukey window applied
when performing \ac{PE} for candidates from those observing runs, and we have not corrected these past inferences.
For results from these previous runs, the worst cases have $T_\mathrm{w} = \qty{0.4}{\second}$ and $T = \qty{4}{\second}$, yielding $\beta =  1.14$.
In this case, the \acp{SNR} are overestimated by a factor of $1.07$.

This normalization error in the likelihood does not impact the \PE results for the new candidates presented in \gwtc[5.0], since the correct likelihood $\PElikelihood$ was used for those analyses.

\section{Waveform Consistency Tests}
\label{sec:wct}

As we have seen in the earlier sections, for our modeled analyses we use several waveforms based on different approximations.
Events like GW190521 \citep{2020PhRvL.125j1102A}, GW200105\_042309 \citep{2021ApJ...915L...5A}, and GW231123\_135430~\citep{2025ApJ...993L..25A} highlight the importance of continuously checking the validity of our assumptions, in order to ensure the reliability of astrophysical interpretations.
To this end, we perform \emph{waveform consistency tests} that compare different waveform reconstruction techniques to assess their agreement and identify any unexpected mismatch between the reconstructed signals. Potential mismatches can arise from assumptions at the pipeline level such as the noise properties, the effects of the subtraction of glitches from the data stream, and inadequate waveform choice. Mismatches can also originate from physical or astrophysical properties, such as deviations from \GR, the presence of eccentricity unaccounted for in waveforms, and gravitational lensing-induced deformations of the measured waveforms.

Waveform reconstruction techniques can be minimally modeled (as seen for \CWB in Section \ref{subsection:cWB}) or template based (Sections \ref{sec:searches} and \ref{sec:pe}).
Minimally-modeled techniques use time--frequency representations, based on wavelets or other time--frequency techniques, to identify coherent features in the data of a network of multiple detectors.
This generic approach enables the discovery of unexpected phenomena but does not provide a direct mapping between the reconstructed waveform and the source's physical properties, such as the masses and spins of a binary system and the distance to the source.

To evaluate the consistency between template-based and minimally-modeled reconstructions, we implement a systematic injection study~\citep{2019PhRvX...9c1040A,2021PhRvX..11b1053A,2023PhRvX..13d1039A,2019PhRvD.100d2003S,2020PhRvD.102f4056G,2022PhRvD.105d4020J}.
This involves injecting \ac{CBC} waveform samples from the posterior parameter distributions (Section \ref{sec:pe}) into detector data at times near but distinct from the candidate (\emph{off-source} injections).
These injections are then reconstructed using minimally-modeled methods.
By comparing the reconstructed waveforms from off-source injections ($w_i$) with the reconstruction of the candidate ($\hat{w}$), we can assess how well the \ac{CBC} \ac{PE} posteriors align with the minimally-modeled reconstruction.
We quantify the agreement between waveform reconstructions using the \emph{overlap}, defined as
\begin{equation}
  \mathcal{O}( h_1,h_2  ) = \frac{\ip{h_1}{h_2}}{\sqrt{\ip{h_1}{h_1}\ip{h_2}{h_2}}}\,, 
  \label{eq:overlap}
\end{equation}
where $h_1$ and $h_2$ are the waveforms being compared and $\langle \cdot | \cdot \rangle$ denotes the noise-weighted inner product~\citep[][Appendix B]{GWTC:Introduction}.
The overlap, $\mathcal{O}( h_1,h_2  )$, is bounded between $-1$ and $+1$. 

For each candidate, we compute the empirical distribution $\mathcal{O}$ of \emph{off-source} overlaps $\mathcal{O}(w_i,h_i)$ between injected waveforms and their minimally-modeled reconstructions. Next, we use the \emph{on-source} overlap $\mathcal{O}(\hat{w},h_\mathrm{maxL})$ between the maximum-likelihood posterior sample and the actual minimally-modeled reconstruction to compute a $p$-value on the left tail of the overlap distribution (i.e., for low overlap values).

This comparison has an inherent asymmetry: for off-source cases, we calculate the overlap between known waveforms and their minimally-modeled reconstructions, while for on-source cases, we compare the maximum-likelihood template from \PE with the minimally-modeled reconstruction.
This asymmetry typically results in off-source overlaps being systematically lower than what would be expected from a true null distribution~\citep{2023PhRvX..13d1039A}.
Consequently, the derived $p$-values are conservative by construction.
Despite this limitation, the on-source $p$-value remains a useful indicator for identifying unexpected signal features, though with reduced statistical power compared to an unbiased test.

To ensure reliable waveform consistency tests, we selected all \ac{O4b} candidates that meet four key criteria\note{check}:
(i) data availability from both \ac{LIGO} detectors,
(ii) presence of \ac{PE} results using the \IMRPhenomXPHMST{} waveform family,
(iii) detector-frame chirp mass $(1+z)\Mc >15\Msun$, and
(iv) a network $\SNR > 10$.
These criteria reflect that minimally-modeled methods require multi-detector data, and that they are particularly effective for high-mass \acp{CBC} with high \SNR signals.

The candidates satisfying these criteria are analyzed in a companion paper \citep{GWTC:Results} using three minimally-modeled waveform reconstruction methods: \BAYESWAVE~\citep{2015CQGra..32m5012C,2021PhRvD.103d4006C,2020PhRvD.102f4056G,2024PhRvD.109f4040G}, and two configurations of the \CWB pipeline.
The two \CWB configurations are \CWBTWOG~\citep{2016PhRvD..93d2004K,2021SoftX..1400678D} which employs the \ac{WDM} wavelet transform and an excess power statistic to identify coherent features, and \CWB-BBH~\citep{2022arXiv220101096K}, also described in Section~\ref{subsection:cWB}, which uses the WaveScan transform and a cross-power statistic.
Both \BAYESWAVE and \CWBTWOG are designed for generic \GW transients, whereas \CWB-BBH is optimized for \ac{CBC} signals, also adopting specialized frequency bands and time--frequency resolutions.

For each selected candidate, our injection campaign used for
\begin{itemize}
    \item \BAYESWAVE approximately \num{200} random posterior samples injected within $\pm \qty{8192}{\second}$ of the candidate time,
    \item \CWBTWOG and \CWB-BBH approximately \num{2000} random posterior samples injected across a 2--3 week period surrounding the candidate time.
\end{itemize}

This analysis provides a robust statistical baseline to quantify the degree of agreement between template-based and minimally-modeled reconstructions for a relatively large subset of the candidates from \gwtc[5.0].
By comparing the on-source overlap to the distribution from off-source injections, we can verify their consistency with the corresponding \ac{CBC} reconstructions, and possibly identify those that may exhibit unexpected features.

\section{Data Management}
\label{sec:data-management}

The workflow of data analyses described in this paper outlines a complex chain of disparate analyses required to find and characterize \GW transients (see Figure~\ref{fig:data-flow-chart}).
In addition to the internal complexity of each analysis, coordinating each stage of this analysis and effectively tracking that input and output data is a significant challenge.
This challenge grows with the number of \GW candidates observed, necessitating the development of tools to manage these tasks with little to no human intervention.
For \ac{O4}, this development included the augmentation of existing infrastructure and the development of new software packages including \CBCFLOW and the catalog data-product pipeline\note{check}.

The online and offline analysis results from the search pipelines described in Section~\ref{sec:searches} are stored in \GRACEDB \citep{GraceDB2014}.
For offline results, \GRACEDB is also used to generate the final candidate list for the catalog.
To facilitate the tracking of these offline results, \GRACEDB has been augmented to provide version-controlled snapshots of the state of the catalog during the progression of the offline analyses.

As for catalogs since \gwtc[2.0], the \PE analyses described in Section~\ref{sec:pe} are managed with the \ASIMOV software package \citep{2023JOSS....8.4170W}.
\ASIMOV ingests data-quality recommendations and preliminary \PE analyses, and uses this to automatically determine appropriate configuration for \PE analyses, as well as automating the production of \PSD estimates.
Once all desired \PE is complete, \ASIMOV packages results into a standard format including all inputs and configuration required for reproduction using \PESUMMARY \citep{2021SoftX..1500765H} .

The \CBCFLOW \citep{cbcflow} software package is used to manage the flow of data between the various analyses and to track metadata about each stage of the analysis.
A monitor process fetches search metadata from \GRACEDB and preliminary \PE results from a shared directory on the Caltech computing cluster.
Data-quality information about candidates is updated following the studies described in Section~\ref{sec:dq}.
\ASIMOV reads \CBCFLOW for search, data quality, and preliminary-\PE metadata, which is used to configure production analyses.
Upon completion of those analyses, metadata about them are written back to \CBCFLOW.
All other downstream analyses such as searches for lensed pairs of candidates \citep{GWTC:Lensing} and tests of \GR \citep{GWTC:TGR} also utilize \CBCFLOW to track their progress and results.

Finally, the release data product is assembled by a collection of scripts.
These read \CBCFLOW to identify the preferred search and \PE results, as well as tracking their finalization status.
These are then collated into the data product itself, and tables of summary information are generated to facilitate downstream use.

\section{Conclusion}
\label{sec:conclusions}

Leading on from the introduction presented in \citet{GWTC:Introduction}, this article describes the analysis methods used to transform the interferometric strain data from the \ac{LVK} detectors into version~\thisgwtcversionfull{} of the \gwtc; the results of these analyses are presented in \citet{GWTC:Results}.
We began in Section~\ref{sec:waveforms} with a description of the waveform models used to describe the \GW signals from \acp{CBC} involving \acp{BH} and \acp{NS}.
In Section~\ref{sec:searches}, we described the search methods we use to filter the strain data to identify candidate transient \GW signals (sensitive to both \ac{CBC} sources as well as minimally-modeled bursts of \GW radiation), and how these candidates are then ranked to identify the most significant ones.
Likely \GW candidates are then studied to understand the quality of data and identify any transient non-Gaussian noise that may bias later analyses (see Section~\ref{sec:dq}).
Section~\ref{sec:pe} described how a subset of likely \GW candidates are then characterized using computational Bayesian inference methods to estimate the parameters of the \GW source, such as the masses and spins of the compact objects involved.
To check the assumptions made in the PE analysis, we described in Section~\ref{sec:wct} how waveform consistency tests compare different waveform reconstruction techniques to assess their agreement and identify any unexpected features in the reconstructed signals.
Finally, in Section~\ref{sec:data-management}, we described the data management and workflow tools used to coordinate the various analyses and track the input and output data.

The methods described in this work are continually developed to improve the capabilities of the \ac{LVK} network to detect and characterize \GW transients.
Specifically, as the detectors evolve towards higher sensitivity, and more detectors are added to the network, we continue to see a greater number of signals and more high-fidelity signals at larger \ac{SNR} \citep{KAGRA:2013rdx}.
This necessitates efficiency improvements across all the methods to avoid computational bottlenecks and to ensure that the \GW transient candidates can be processed in a timely manner.
Moreover, the increased \ac{SNR} requires refinements to the waveform models to reduce systematic biases as much as possible.
The evolution of the detectors also necessitates continual improvements to the data analysis methods to ensure optimal data processing (e.g., as the noise floor of the detectors lowers, new glitch classes may become relevant, requiring development to searches and data-quality studies).
Finally, as the time--volume explored increases, we hope to further expand the range of astrophysical sources that can be detected and characterized, which in turn requires the development of new models, search, and \PE methods.

\emph{Data Availability:} The data products generated by the methods described within this work are openly available in the \thisgwtcfull{} online catalog, which is hosted at \url{https://gwosc.org/GWTC-5.0}, documented further in \citet{OpenData} and the results described in \citet{GWTC:Results}.

\section*{Acknowledgements}
This material is based upon work supported by NSF's LIGO Laboratory, which is a
major facility fully funded by the National Science Foundation.
The authors also gratefully acknowledge the support of
the Science and Technology Facilities Council (STFC) of the
United Kingdom, the Max-Planck-Society (MPS), and the State of
Niedersachsen/Germany for support of the construction of Advanced LIGO 
and construction and operation of the GEO\,600 detector. 
Additional support for Advanced LIGO was provided by the Australian Research Council.
The authors gratefully acknowledge the Italian Istituto Nazionale di Fisica Nucleare (INFN),  
the French Centre National de la Recherche Scientifique (CNRS) and
the Netherlands Organization for Scientific Research (NWO)
for the construction and operation of the Virgo detector
and the creation and support  of the EGO consortium. 
The authors also gratefully acknowledge research support from these agencies as well as by 
the Council of Scientific and Industrial Research of India, 
the Department of Science and Technology, India,
the Science \& Engineering Research Board (SERB), India,
the Ministry of Human Resource Development, India,
the Spanish Agencia Estatal de Investigaci\'on (AEI),
the Spanish Ministerio de Ciencia, Innovaci\'on y Universidades,
the European Union NextGenerationEU/PRTR (PRTR-C17.I1),
the ICSC - CentroNazionale di Ricerca in High Performance Computing, Big Data
and Quantum Computing, funded by the European Union NextGenerationEU,
the Comunitat Auton\`oma de les Illes Balears through the Conselleria d'Educaci\'o i Universitats,
the Conselleria d'Innovaci\'o, Universitats, Ci\`encia i Societat Digital de la Generalitat Valenciana and
the CERCA Programme Generalitat de Catalunya, Spain,
the Polish National Agency for Academic Exchange,
the National Science Centre of Poland and the European Union - European Regional
Development Fund;
the Foundation for Polish Science (FNP),
the Polish Ministry of Science and Higher Education,
the Swiss National Science Foundation (SNSF),
the Russian Science Foundation,
the European Commission,
the European Social Funds (ESF),
the European Regional Development Funds (ERDF),
the Royal Society, 
the Scottish Funding Council, 
the Scottish Universities Physics Alliance, 
the Hungarian Scientific Research Fund (OTKA),
the French Lyon Institute of Origins (LIO),
the Belgian Fonds de la Recherche Scientifique (FRS-FNRS), 
Actions de Recherche Concert\'ees (ARC) and
Fonds Wetenschappelijk Onderzoek - Vlaanderen (FWO), Belgium,
the Paris \^{I}le-de-France Region, 
the National Research, Development and Innovation Office of Hungary (NKFIH), 
the National Research Foundation of Korea,
the Natural Sciences and Engineering Research Council of Canada (NSERC),
the Canadian Foundation for Innovation (CFI),
the Brazilian Ministry of Science, Technology, and Innovations,
the International Center for Theoretical Physics South American Institute for Fundamental Research (ICTP-SAIFR), 
the Research Grants Council of Hong Kong,
the National Natural Science Foundation of China (NSFC),
the Israel Science Foundation (ISF),
the US-Israel Binational Science Fund (BSF),
the Leverhulme Trust, 
the Research Corporation,
the National Science and Technology Council (NSTC), Taiwan,
the United States Department of Energy,
and
the Kavli Foundation.
The authors gratefully acknowledge the support of the NSF, STFC, INFN and CNRS for provision of computational resources.

This work was supported by MEXT,
the JSPS Leading-edge Research Infrastructure Program,
JSPS Grant-in-Aid for Specially Promoted Research 26000005,
JSPS Grant-in-Aid for Scientific Research on Innovative Areas 2402: 24103006,
24103005, and 2905: JP17H06358, JP17H06361 and JP17H06364,
JSPS Core-to-Core Program A.\ Advanced Research Networks,
JSPS Grants-in-Aid for Scientific Research (S) 17H06133 and 20H05639,
JSPS Grant-in-Aid for Transformative Research Areas (A) 20A203: JP20H05854,
the joint research program of the Institute for Cosmic Ray Research,
University of Tokyo,
the National Research Foundation (NRF),
the Computing Infrastructure Project of the Global Science experimental Data hub
Center (GSDC) at KISTI,
the Korea Astronomy and Space Science Institute (KASI),
the Ministry of Science and ICT (MSIT) in Korea,
Academia Sinica (AS),
the AS Grid Center (ASGC) and the National Science and Technology Council (NSTC)
in Taiwan under grants including the Science Vanguard Research Program,
the Advanced Technology Center (ATC) of NAOJ, 
the Mechanical Engineering Center of KEK
and Vietnam National Foundation for Science and Technology Development 
(NAFOSTED) 103.01-2025.147.

Additional acknowledgements for support of individual authors may be found in the following document: \\
\texttt{https://dcc.ligo.org/LIGO-M2300033/public}.
For the purpose of open access, the authors have applied a Creative Commons Attribution (CC BY)
license to any Author Accepted Manuscript version arising.
We request that citations to this article use 'A. G. Abac {\it et al.} (LIGO-Virgo-KAGRA Collaboration), ...' or similar phrasing, depending on journal convention.

\software{Plots were prepared with \MATPLOTLIB{}~\citep{2007CSE.....9...90H}, \NUMPY~\citep{2020Natur.585..357H}, and \soft{TikZ}~\citep{tantau:2023}.}

\bibliography{}

\begin{thebibliography}{}
\expandafter\ifx\csname natexlab\endcsname\relax\def\natexlab#1{#1}\fi
\providecommand{\url}[1]{\href{#1}{#1}}
\providecommand{\dodoi}[1]{doi:~\href{http://doi.org/#1}{\nolinkurl{#1}}}
\providecommand{\doeprint}[1]{\href{http://ascl.net/#1}{\nolinkurl{http://ascl.net/#1}}}
\providecommand{\doarXiv}[1]{\href{https://arxiv.org/abs/#1}{\nolinkurl{https://arxiv.org/abs/#1}}}

\bibitem[{Aasi {et~al.}(2013)}]{LIGOScientific:2012vij}
Aasi, J., {et~al.} 2013, Phys. Rev. D, 87, 022002,
  \dodoi{10.1103/PhysRevD.87.022002}

\bibitem[{{Aasi} {et~al.}(2015){Aasi}, {Abbott}, {Abbott}, {Abbott},
  {Abernathy}, {Ackley}, {Adams}, {Adams}, {Addesso}, {Adhikari}, {Adya},
  {Affeldt}, {Aggarwal}, {Aguiar}, {Ain}, {Ajith}, {Alemic}, {Allen},
  {Amariutei}, {Anderson}, {Anderson}, {Arai}, {Araya}, {Arceneaux}, {Areeda},
  {Ashton}, {Ast}, {Aston}, {Aufmuth}, {Aulbert}, {Aylott}, {Babak}, {Baker},
  {Ballmer}, {Barayoga}, {Barbet}, {Barclay}, {Barish}, {Barker}, {Barr},
  {Barsotti}, {Bartlett}, {Barton}, {Bartos}, {Bassiri}, {Batch}, {Baune},
  {Behnke}, {Bell}, {Bell}, {Benacquista}, {Bergman}, {Bergmann}, {Berry},
  {Betzwieser}, {Bhagwat}, {Bhandare}, {Bilenko}, {Billingsley}, {Birch},
  {Biscans}, {Biwer}, {Blackburn}, {Blackburn}, {Blair}, {Blair}, {Bock},
  {Bodiya}, {Bojtos}, {Bond}, {Bork}, {Born}, {Bose}, {Brady}, {Braginsky},
  {Brau}, {Bridges}, {Brinkmann}, {Brooks}, {Brown}, {Brown}, {Brown},
  {Buchman}, {Buikema}, {Buonanno}, {Cadonati}, {Calder{\'o}n Bustillo},
  {Camp}, {Cannon}, {Cao}, {Capano}, {Caride}, {Caudill}, {Cavagli{\`a}},
  {Cepeda}, {Chakraborty}, {Chalermsongsak}, {Chamberlin}, {Chao}, {Charlton},
  {Chen}, {Cho}, {Cho}, {Chow}, {Christensen}, {Chu}, {Chung}, {Ciani},
  {Clara}, {Clark}, {Collette}, {Cominsky}, {Constancio}, {Cook}, {Corbitt},
  {Cornish}, {Corsi}, {Costa}, {Coughlin}, {Countryman}, {Couvares}, {Coward},
  {Cowart}, {Coyne}, {Coyne}, {Craig}, {Creighton}, {Creighton}, {Cripe},
  {Crowder}, {Cumming}, {Cunningham}, {Cutler}, {Dahl}, {Dal Canton},
  {Damjanic}, {Danilishin}, {Danzmann}, {Dartez}, {Dave}, {Daveloza}, {Davies},
  {Daw}, {DeBra}, {Del Pozzo}, {Denker}, {Dent}, {Dergachev}, {DeRosa},
  {DeSalvo}, {Dhurandhar}, {D́{\i}az}, {Di Palma}, {Dojcinoski}, {Dominguez},
  {Donovan}, {Dooley}, {Doravari}, {Douglas}, {Downes}, {Driggers}, {Du},
  {Dwyer}, {Eberle}, {Edo}, {Edwards}, {Edwards}, {Effler}, {Eggenstein},
  {Ehrens}, {Eichholz}, {Eikenberry}, {Essick}, {Etzel}, {Evans}, {Evans},
  {Factourovich}, {Fairhurst}, {Fan}, {Fang}, {Farr}, {Farr}, {Favata}, {Fays},
  {Fehrmann}, {Fejer}, {Feldbaum}, {Ferreira}, {Fisher}, {Frei}, {Freise},
  {Frey}, {Fricke}, {Fritschel}, {Frolov}, {Fuentes-Tapia}, {Fulda}, {Fyffe},
  \& {Gair}}]{2015CQGra..32g4001L}
{Aasi}, J., {Abbott}, B.~P., {Abbott}, R., {et~al.} 2015, Classical and Quantum
  Gravity, 32, 074001, \dodoi{10.1088/0264-9381/32/7/074001}

\bibitem[{Abac {et~al.}(2026{\natexlab{a}})Abac, Abouelfettouh, Acernese,
  {et~al.}}]{GWTC:Introduction}
Abac, A.~G., Abouelfettouh, I., Acernese, F., {et~al.} 2026{\natexlab{a}}, To
  be published in this issue.
\newblock \url{https://dcc.ligo.org/LIGO-P2500701/public}

\bibitem[{Abac {et~al.}(2026{\natexlab{b}})Abac, Abouelfettouh, Acernese,
  {et~al.}}]{GWTC:Results}
---. 2026{\natexlab{b}}, To be published in this issue.
\newblock \url{https://dcc.ligo.org/LIGO-P2600152/public}

\bibitem[{Abac {et~al.}(2026{\natexlab{c}})Abac, Abouelfettouh, Acernese,
  {et~al.}}]{OpenData}
---. 2026{\natexlab{c}}, To be published in this issue.
\newblock \url{https://dcc.ligo.org/LIGO-P2600085/public}

\bibitem[{Abac {et~al.}(2026{\natexlab{d}})Abac, Abouelfettouh, Acernese,
  {et~al.}}]{GWTC:Lensing}
---. 2026{\natexlab{d}}, To be published in this issue

\bibitem[{Abac {et~al.}(2026{\natexlab{e}})Abac, Abouelfettouh, Acernese,
  {et~al.}}]{GWTC:TGR}
---. 2026{\natexlab{e}}, To be published in this issue

\bibitem[{{Abac} {et~al.}(2025{\natexlab{a}}){Abac}, {Abouelfettouh},
  {Acernese}, {Ackley}, {Adamcewicz}, {Adhicary}, {Adhikari}, {Adhikari},
  {Adhikari}, {Adkins}, {Afroz}, {Agapito}, {Agarwal}, {Agathos}, {Aggarwal},
  {Aggarwal}, {Aguiar}, {Ahrend}, {Aiello}, {Ain}, {Ajith}, {Akutsu},
  {Albanesi}, {Ali}, {Al-Kershi}, {All{\'e}n{\'e}}, {Allocca}, {Al-Shammari},
  {Altin}, {Alvarez-Lopez}, {Amar}, {Amarasinghe}, {Amato}, {Amicucci}, {Amra},
  {Ananyeva}, {Anderson}, {Anderson}, {Andia}, {Ando}, {Andr{\'e}s-Carcasona},
  {Andri{\'c}}, {Anglin}, {Ansoldi}, {Antelis}, {Antier}, {Antonini}, {Aoumi},
  {Appavuravther}, {Appert}, {Apple}, {Arai}, {Ara{\'u}jo-{\'A}lvarez},
  {Araya}, {Araya}, {Arca Sedda}, {Areeda}, {Aritomi}, {Armato}, {Armstrong},
  {Arnaud}, {Arogeti}, {Aronson}, {Arun}, {Ashton}, {Aso}, {Asprea}, {Assiduo},
  {Assis de Souza Melo}, {Aston}, {Astone}, {Aswathi}, {Attadio}, {Aubin},
  {Aultoneal}, {Avallone}, {Avila}, {Babak}, {Badger}, {Bae}, {Bagnasco},
  {Baiotti}, {Bajpai}, {Baka}, {Baker}, {Baker}, {Baker}, {Baldi}, {Baldicchi},
  {Ball}, {Ballardin}, {Ballmer}, {Banagiri}, {Banerjee}, {Bankar}, {Baptiste},
  {Baral}, {Baratti}, {Barayoga}, {Barish}, {Barker}, {Barman}, {Barneo},
  {Barone}, {Barr}, {Barsotti}, {Barsuglia}, {Barta}, {Bartoletti}, {Barton},
  {Bartos}, {Basalaev}, {Bassiri}, {Basti}, {Bawaj}, {Baxi}, {Bayley},
  {Baylor}, {Baynard}, {Bazzan}, {Bedakihale}, {Beirnaert}, {Bejger},
  {Belardinelli}, {Bell}, {Bellie}, {Bellizzi}, {Benoit}, {Bentara}, {Bentley},
  {Ben Yaala}, {Bera}, {Bergamin}, {Berger}, {Bernuzzi}, {Beroiz}, {Berry},
  {Bersanetti}, {Bertheas}, {Bertolini}, {Betzwieser}, {Beveridge},
  {Bevilacqua}, {Bevins}, {Bhandare}, {Bhatt}, {Bhattacharjee},
  {Bhattacharyya}, {Bhaumik}, {Biancalana}, {Bianchi}, {Bilenko},
  {Billingsley}, {Binetti}, {Bini}, {Binu}, {Biot}, {Birnholtz}, {Biscoveanu},
  {Bisht}, {Bitossi}, {Bizouard}, {Blaber}, {Blackburn}, {Blagg}, {Blair},
  {Blair}, {Bode}, {Boettner}, {Boileau}, {Boldrini}, {Bolingbroke},
  {Bolliand}, {Bonavena}, {Bondarescu}, {Bondu}, {Bonilla}, {Bonilla},
  {Bonino}, {Bonnand}, {Borchers}, {Borhanian}, {Boschi}, {Bose}, {Bossilkov},
  {Bothra}, {Boudon}, {Bourg}, {Boyle}, {Bozzi}, {Bradaschia}, {Brady},
  {Branch}, {Branchesi}, {Braun}, {Briant}, {Brillet}, {Brinkmann}, {Brockill},
  \& {Brockmueller}}]{2025ApJ...993L..21A}
{Abac}, A.~G., {Abouelfettouh}, I., {Acernese}, F., {et~al.}
  2025{\natexlab{a}}, \apjl, 993, L21, \dodoi{10.3847/2041-8213/ae0d54}

\bibitem[{{Abac} {et~al.}(2025{\natexlab{b}}){Abac}, {Abouelfettouh},
  {Acernese}, {Ackley}, {Adamcewicz}, {Adhicary}, {Adhikari}, {Adhikari},
  {Adhikari}, {Adkins}, {Afroz}, {Agapito}, {Agarwal}, {Agathos}, {Aggarwal},
  {Aggarwal}, {Aguiar}, {Ahrend}, {Aiello}, {Ain}, {Ajith}, {Akutsu},
  {Al-Kershi}, {Al-Shammari}, {Albanesi}, {Ali}, {All{\'e}n{\'e}}, {Allocca},
  {Altin}, {Alvarez-Lopez}, {Amar}, {Amarasinghe}, {Amato}, {Amicucci}, {Amra},
  {Ananyeva}, {Anderson}, {Anderson}, {Andia}, {Ando}, {Andr{\'e}s-Carcasona},
  {Andri{\'c}}, {Anglin}, {Ansoldi}, {Antelis}, {Antier}, {Aoumi},
  {Appavuravther}, {Appert}, {Apple}, {Arai}, {Araya}, {Araya}, {Arca Sedda},
  {Areeda}, {Aritomi}, {Armato}, {Armstrong}, {Arnaud}, {Arogeti}, {Aronson},
  {Ashton}, {Aso}, {Asprea}, {Assiduo}, {Assis de Souza Melo}, {Aston},
  {Astone}, {Attadio}, {Aubin}, {Aultoneal}, {Avallone}, {Avila}, {Babak},
  {Badger}, {Bae}, {Bagnasco}, {Baiotti}, {Bajpai}, {Baka}, {Baker}, {Baker},
  {Baker}, {Baldi}, {Baldicchi}, {Ball}, {Ballardin}, {Ballmer}, {Banagiri},
  {Banerjee}, {Bankar}, {Baptiste}, {Baral}, {Baratti}, {Barayoga}, {Barish},
  {Barker}, {Barman}, {Barneo}, {Barone}, {Barr}, {Barsotti}, {Barsuglia},
  {Barta}, {Bartoletti}, {Barton}, {Bartos}, {Basalaev}, {Bassiri}, {Basti},
  {Bawaj}, {Baxi}, {Bayley}, {Baylor}, {Baynard}, {Bazzan}, {Bedakihale},
  {Beirnaert}, {Bejger}, {Belardinelli}, {Bell}, {Bellie}, {Bellizzi},
  {Benoit}, {Bentara}, {Bentley}, {Ben Yaala}, {Bera}, {Bergamin}, {Berger},
  {Bernuzzi}, {Beroiz}, {Berry}, {Bersanetti}, {Bertheas}, {Bertolini},
  {Betzwieser}, {Beveridge}, {Bevilacqua}, {Bevins}, {Bhagwat}, {Bhandare},
  {Bhatt}, {Bhattacharjee}, {Bhattacharyya}, {Bhaumik}, {Biancalana},
  {Bianchi}, {Bilenko}, {Billingsley}, {Binetti}, {Bini}, {Binu}, {Biot},
  {Birnholtz}, {Biscoveanu}, {Bisht}, {Bitossi}, {Bizouard}, {Blaber},
  {Blackburn}, {Blagg}, {Blair}, {Blair}, {Bode}, {Boettner}, {Boileau},
  {Boldrini}, {Bolingbroke}, {Bolliand}, {Bonavena}, {Bondarescu}, {Bondu},
  {Bonilla}, {Bonilla}, {Bonino}, {Bonnand}, {Borchers}, {Borhanian}, {Boschi},
  {Bose}, {Bossilkov}, {Bothra}, {Boudon}, {Bourg}, {Boyle}, {Bozzi},
  {Bradaschia}, {Brady}, {Branch}, {Branchesi}, {Braun}, {Briant}, {Brillet},
  {Brinkmann}, {Brockill}, {Brockmueller}, {Brooks}, {Brown}, \&
  {Brown}}]{2025PhRvL.135k1403A}
---. 2025{\natexlab{b}}, \prl, 135, 111403, \dodoi{10.1103/kw5g-d732}

\bibitem[{{Abac} {et~al.}(2025{\natexlab{c}}){Abac}, {Abouelfettouh},
  {Acernese}, {Ackley}, {Adamcewicz}, {Adhicary}, {Adhikari}, {Adhikari},
  {Adhikari}, {Adkins}, {Afroz}, {Agapito}, {Agarwal}, {Agathos}, {Aggarwal},
  {Aggarwal}, {Aguiar}, {Ahrend}, {Aiello}, {Ain}, {Ajith}, {Akutsu},
  {Albanesi}, {Ali}, {Al-Kershi}, {All{\'e}n{\'e}}, {Allocca}, {Al-Shammari},
  {Altin}, {Alvarez-Lopez}, {Amar}, {Amarasinghe}, {Amato}, {Amicucci}, {Amra},
  {Ananyeva}, {Anderson}, {Anderson}, {Andia}, {Ando}, {Andr{\'e}s-Carcasona},
  {Andri{\'c}}, {Anglin}, {Ansoldi}, {Antelis}, {Antier}, {Aoumi},
  {Appavuravther}, {Appert}, {Apple}, {Arai}, {Araya}, {Araya}, {Arca Sedda},
  {Areeda}, {Aritomi}, {Armato}, {Armstrong}, {Arnaud}, {Arogeti}, {Aronson},
  {Arun}, {Ashton}, {Aso}, {Asprea}, {Assiduo}, {Assis de Souza Melo}, {Aston},
  {Astone}, {Attadio}, {Aubin}, {AultONeal}, {Avallone}, {Avila}, {Babak},
  {Badger}, {Bae}, {Bagnasco}, {Baiotti}, {Bajpai}, {Baka}, {Baker}, {Baker},
  {Baker}, {Baldi}, {Baldicchi}, {Ball}, {Ballardin}, {Ballmer}, {Banagiri},
  {Banerjee}, {Bankar}, {Baptiste}, {Baral}, {Baratti}, {Barayoga}, {Barish},
  {Barker}, {Barman}, {Barneo}, {Barone}, {Barr}, {Barsotti}, {Barsuglia},
  {Barta}, {Bartoletti}, {Barton}, {Bartos}, {Basalaev}, {Bassiri}, {Basti},
  {Bawaj}, {Baxi}, {Bayley}, {Baylor}, {Baynard}, {Bazzan}, {Bedakihale},
  {Beirnaert}, {Bejger}, {Belardinelli}, {Bell}, {Bellie}, {Bellizzi},
  {Benoit}, {Bentara}, {Bentley}, {Ben Yaala}, {Bera}, {Bergamin}, {Berger},
  {Bernuzzi}, {Beroiz}, {Berry}, {Bersanetti}, {Bertheas}, {Bertolini},
  {Betzwieser}, {Beveridge}, {Bevilacqua}, {Bevins}, {Bhandare}, {Bhatt},
  {Bhattacharjee}, {Bhattacharyya}, {Bhaumik}, {Biancalana}, {Bianchi},
  {Bilenko}, {Billingsley}, {Binetti}, {Bini}, {Binu}, {Biot}, {Birnholtz},
  {Biscoveanu}, {Bisht}, {Bitossi}, {Bizouard}, {Blaber}, {Blackburn}, {Blagg},
  {Blair}, {Blair}, {Bode}, {Boettner}, {Boileau}, {Boldrini}, {Bolingbroke},
  {Bolliand}, {Bonavena}, {Bondarescu}, {Bondu}, {Bonilla}, {Bonilla},
  {Bonino}, {Bonnand}, {Borchers}, {Borhanian}, {Boschi}, {Bose}, {Bossilkov},
  {Bothra}, {Boudon}, {Bourg}, {Boyle}, {Bozzi}, {Bradaschia}, {Brady},
  {Branch}, {Branchesi}, {Braun}, {Briant}, {Brillet}, {Brinkmann}, {Brockill},
  \& {Brockmueller}}]{2025arXiv250818082T}
---. 2025{\natexlab{c}}, arXiv e-prints, arXiv:2508.18082,
  \dodoi{10.48550/arXiv.2508.18082}

\bibitem[{{Abac} {et~al.}(2025{\natexlab{d}}){Abac}, {Abouelfettouh},
  {Acernese}, {Ackley}, {Adhicary}, {Adhikari}, {Adhikari}, {Adhikari},
  {Adkins}, {Afroz}, {Agarwal}, {Agathos}, {Aghaei Abchouyeh}, {Aguiar},
  {Ahmadzadeh}, {Aiello}, {Ain}, {Ajith}, {Akcay}, {Akutsu}, {Albanesi},
  {Alfaidi}, {Al-Jodah}, {All{\'e}n{\'e}}, {Allocca}, {Al-Shammari}, {Altin},
  {Alvarez-Lopez}, {Amarasinghe}, {Amato}, {Amra}, {Ananyeva}, {Anderson},
  {Anderson}, {Andia}, {Ando}, {Andrade}, {Andr{\'e}s-Carcasona}, {Andri{\'c}},
  {Anglin}, {Ansoldi}, {Antelis}, {Antier}, {Aoumi}, {Appavuravther}, {Appert},
  {Apple}, {Arai}, {Araya}, {Araya}, {Arca Sedda}, {Areeda}, {Argianas},
  {Aritomi}, {Armato}, {Armstrong}, {Arnaud}, {Arogeti}, {Aronson}, {Ashton},
  {Aso}, {Assiduo}, {Assis de Souza Melo}, {Aston}, {Astone}, {Attadio},
  {Aubin}, {AultONeal}, {Avallone}, {Babak}, {Badaracco}, {Badger}, {Bae},
  {Bagnasco}, {Bagui}, {Baiotti}, {Bajpai}, {Baka}, {Baker}, {Ball},
  {Ballardin}, {Ballmer}, {Banagiri}, {Banerjee}, {Bankar}, {Baptiste},
  {Baral}, {Barayoga}, {Barish}, {Barker}, {Barman}, {Barneo}, {Barone},
  {Barr}, {Barsotti}, {Barsuglia}, {Barta}, {Bartoletti}, {Barton}, {Bartos},
  {Basak}, {Basalaev}, {Bassiri}, {Basti}, {Bates}, {Bawaj}, {Baxi}, {Bayley},
  {Baylor}, {Baynard}, {Bazzan}, {Bedakihale}, {Beirnaert}, {Bejger},
  {Belardinelli}, {Bell}, {Bellie}, {Bellizzi}, {Benoit}, {Bentara}, {Bentley},
  {Ben Yaala}, {Bera}, {Bergamin}, {Berger}, {Bernuzzi}, {Beroiz}, {Berry},
  {Bersanetti}, {Bertolini}, {Betzwieser}, {Beveridge}, {Bevilacqua}, {Bevins},
  {Bhandare}, {Bhat}, {Bhatt}, {Bhattacharjee}, {Bhaumik}, {Bhowmick},
  {Biancalana}, {Bianchi}, {Bilenko}, {Billingsley}, {Binetti}, {Bini}, {Binu},
  {Birnholtz}, {Biscoveanu}, {Bisht}, {Bitossi}, {Bizouard}, {Blaber},
  {Blackburn}, {Blagg}, {Blair}, {Blair}, {Bobba}, {Bode}, {Boileau},
  {Boldrini}, {Bolingbroke}, {Bolliand}, {Bonavena}, {Bondarescu}, {Bondu},
  {Bonilla}, {Bonilla}, {Bonino}, {Bonnand}, {Booker}, {Borchers}, {Borhanian},
  {Boschi}, {Bose}, {Bossilkov}, {Boudon}, {Bozzi}, {Bradaschia}, {Brady},
  {Branch}, {Branchesi}, {Braun}, {Briant}, {Brillet}, {Brinkmann}, {Brockill},
  {Brockmueller}, {Brooks}, {Brown}, {Brown}, {Brozzetti}, {Brunett}, {Bruno},
  {Bruntz}, {Bryant}, \& {Bu}}]{2025arXiv250818081T}
---. 2025{\natexlab{d}}, arXiv e-prints, arXiv:2508.18081,
  \dodoi{10.48550/arXiv.2508.18081}

\bibitem[{{Abac} {et~al.}(2025{\natexlab{e}}){Abac}, {Abouelfettouh},
  {Acernese}, {Ackley}, {Adhicary}, {Adhikari}, {Adhikari}, {Adhikari},
  {Adkins}, {Afroz}, {Agarwal}, {Agathos}, {Aghaei Abchouyeh}, {Aguiar},
  {Ahmadzadeh}, {Aiello}, {Ain}, {Ajith}, {Akcay}, {Akutsu}, {Albanesi},
  {Alfaidi}, {Al-Jodah}, {All{\'e}n{\'e}}, {Allocca}, {Al-Shammari}, {Altin},
  {Alvarez-Lopez}, {Amarasinghe}, {Amato}, {Amra}, {Ananyeva}, {Anderson},
  {Anderson}, {Andia}, {Ando}, {Andrade}, {Andr{\'e}s-Carcasona}, {Andri{\'c}},
  {Anglin}, {Ansoldi}, {Antelis}, {Antier}, {Aoumi}, {Appavuravther}, {Appert},
  {Apple}, {Arai}, {Araya}, {Araya}, {Arca Sedda}, {Areeda}, {Argianas},
  {Aritomi}, {Armato}, {Armstrong}, {Arnaud}, {Arogeti}, {Aronson}, {Ashton},
  {Aso}, {Assiduo}, {Assis de Souza Melo}, {Aston}, {Astone}, {Attadio},
  {Aubin}, {Aultoneal}, {Avallone}, {Babak}, {Badaracco}, {Badger}, {Bae},
  {Bagnasco}, {Bagui}, {Baiotti}, {Bajpai}, {Baka}, {Baker}, {Ball},
  {Ballardin}, {Ballmer}, {Banagiri}, {Banerjee}, {Bankar}, {Baptiste},
  {Baral}, {Barayoga}, {Barish}, {Barker}, {Barman}, {Barneo}, {Barone},
  {Barr}, {Barsotti}, {Barsuglia}, {Barta}, {Bartoletti}, {Barton}, {Bartos},
  {Basak}, {Basalaev}, {Bassiri}, {Basti}, {Bates}, {Bawaj}, {Baxi}, {Bayley},
  {Baylor}, {Baynard}, {Bazzan}, {Bedakihale}, {Beirnaert}, {Bejger},
  {Belardinelli}, {Bell}, {Bellie}, {Bellizzi}, {Benoit}, {Bentara}, {Bentley},
  {Ben Yaala}, {Bera}, {Bergamin}, {Berger}, {Bernuzzi}, {Beroiz}, {Berry},
  {Bersanetti}, {Bertolini}, {Betzwieser}, {Beveridge}, {Bevilacqua}, {Bevins},
  {Bhandare}, {Bhat}, {Bhatt}, {Bhattacharjee}, {Bhaumik}, {Bhowmick},
  {Biancalana}, {Bianchi}, {Bilenko}, {Billingsley}, {Binetti}, {Bini}, {Binu},
  {Birnholtz}, {Biscoveanu}, {Bisht}, {Bitossi}, {Bizouard}, {Blaber},
  {Blackburn}, {Blagg}, {Blair}, {Blair}, {Bobba}, {Bode}, {Boileau},
  {Boldrini}, {Bolingbroke}, {Bolliand}, {Bonavena}, {Bondarescu}, {Bondu},
  {Bonilla}, {Bonilla}, {Bonino}, {Bonnand}, {Booker}, {Borchers}, {Borhanian},
  {Boschi}, {Bose}, {Bossilkov}, {Boudon}, {Bozzi}, {Bradaschia}, {Brady},
  {Branch}, {Branchesi}, {Braun}, {Briant}, {Brillet}, {Brinkmann}, {Brockill},
  {Brockmueller}, {Brooks}, {Brown}, {Brown}, {Brozzetti}, {Brunett}, {Bruno},
  {Bruntz}, {Bryant}, {Bu}, {Bucci}, {Buchanan}, \&
  {Bulashenko}}]{2025ApJ...995L..18A}
---. 2025{\natexlab{e}}, \apjl, 995, L18, \dodoi{10.3847/2041-8213/ae0c06}

\bibitem[{{Abac} {et~al.}(2025{\natexlab{f}}){Abac}, {Abouelfettouh},
  {Acernese}, {Ackley}, {Adamcewicz}, {Adhicary}, {Adhikari}, {Adhikari},
  {Adhikari}, {Adkins}, {Afroz}, {Agapito}, {Agarwal}, {Agathos}, {Aggarwal},
  {Aggarwal}, {Aguiar}, {Ahrend}, {Aiello}, {Ain}, {Ajith}, {Akutsu},
  {Albanesi}, {Ali}, {Al-Kershi}, {All{\'e}n{\'e}}, {Allocca}, {Al-Shammari},
  {Altin}, {Alvarez-Lopez}, {Amar}, {Amarasinghe}, {Amato}, {Amicucci}, {Amra},
  {Ananyeva}, {Anderson}, {Anderson}, {Andia}, {Ando}, {Andr{\'e}s-Carcasona},
  {Andri{\'c}}, {Anglin}, {Ansoldi}, {Antelis}, {Antier}, {Aoumi},
  {Appavuravther}, {Appert}, {Apple}, {Arai}, {Alvarez}, {Araya}, {Araya},
  {Arca Sedda}, {Areeda}, {Aritomi}, {Armato}, {Armstrong}, {Arnaud},
  {Arogeti}, {Aronson}, {Arun}, {Ashton}, {Aso}, {Asprea}, {Assiduo}, {Assis de
  Souza Melo}, {Aston}, {Astone}, {Attadio}, {Aubin}, {AultONeal}, {Avallone},
  {Avila}, {Babak}, {Badger}, {Bae}, {Bagnasco}, {Baiotti}, {Bajpai}, {Baka},
  {Baker}, {Baker}, {Baker}, {Baldi}, {Baldicchi}, {Ball}, {Ballardin},
  {Ballmer}, {Banagiri}, {Banerjee}, {Bankar}, {Baptiste}, {Baral}, {Baratti},
  {Barayoga}, {Barish}, {Barker}, {Barman}, {Barneo}, {Barone}, {Barr},
  {Barsotti}, {Barsuglia}, {Barta}, {Bartoletti}, {Barton}, {Bartos},
  {Basalaev}, {Bassiri}, {Basti}, {Bawaj}, {Baxi}, {Bayley}, {Baylor},
  {Baynard}, {Bazzan}, {Bedakihale}, {Beirnaert}, {Bejger}, {Belardinelli},
  {Bell}, {Bellie}, {Bellizzi}, {Benoit}, {Bentara}, {Bentley}, {Ben Yaala},
  {Bera}, {Bergamin}, {Berger}, {Bernuzzi}, {Beroiz}, {Berry}, {Bersanetti},
  {Bertheas}, {Bertolini}, {Betzwieser}, {Beveridge}, {Bevilacqua}, {Bevins},
  {Bhandare}, {Bhatt}, {Bhattacharjee}, {Bhattacharyya}, {Bhaumik}, {Bhagwat},
  {Biancalana}, {Bianchi}, {Bilenko}, {Billingsley}, {Binetti}, {Bini}, {Binu},
  {Biot}, {Birnholtz}, {Biscoveanu}, {Bisht}, {Bitossi}, {Bizouard}, {Blaber},
  {Blackburn}, {Blagg}, {Blair}, {Blair}, {Bode}, {Boettner}, {Boileau},
  {Boldrini}, {Bolingbroke}, {Bolliand}, {Bonavena}, {Bondarescu}, {Bondu},
  {Bonilla}, {Bonilla}, {Bonino}, {Bonnand}, {Borchers}, {Borhanian}, {Boschi},
  {Bose}, {Bossilkov}, {Bothra}, {Boudon}, {Bourg}, {Boyle}, {Bozzi},
  {Bradaschia}, {Brady}, {Branch}, {Branchesi}, {Braun}, {Briant}, {Brillet},
  {Brinkmann}, {Brockill}, {Brockmueller}, \& {Brooks}}]{2025ApJ...993L..25A}
---. 2025{\natexlab{f}}, \apjl, 993, L25, \dodoi{10.3847/2041-8213/ae0c9c}

\bibitem[{Abac {et~al.}(2026{\natexlab{f}})Abac, Abouelfettouh, Acernese,
  Ackley, Adam, Adamcewicz, Adhicary, Adhikari, Adhikari, Adhikari, Adkins,
  Afroz, Agapito, Agarwal, Agathos, Aggarwal, Aggarwal, Aguiar, Ahrend, Aiello,
  Ain, Ajith, Akutsu, Albanesi, Albers, Ali, Al-Kershi, Alléné, Allocca,
  Al-Shammari, Altin, Alvarez-Lopez, Amar, Amarasinghe, Amato, Amicucci, Amra,
  Anand, Ananyeva, Anderson, Anderson, Andia, Ando, Andrés-Carcasona, Andrey,
  Andrić, Anglin, Anna, Ansoldi, Antelis, Antier, Aoumi, Appavuravther,
  Appert, Apple, Arai, Araya, Araya, Sedda, Arciprete, Areeda, Aritomi, Armato,
  Armstrong, Arnaud, Arogeti, Aronson, Arun, Ashton, Aso, Asprea, Assiduo,
  Melo, Aston, Astone, Attadio, Aubin, AultONeal, Avallone, Avila, Babak,
  Badger, Bae, Bagnasco, Baiotti, Bajpai, Baka, Baker, Baker, Balbi, Baldi,
  Baldicchi, Ball, Ballardin, Ballmer, Banagiri, Banerjee, Bankar, Baptiste,
  Baral, Baratti, Barayoga, Baric, Barish, Barker, Barman, Barneo, Barone,
  Barr, Barrios, Barsotti, Barsuglia, Barta, Barton, Bartos, Basalaev, Bassiri,
  Basti, Bawaj, Baxi, Bayley, Baylor, II, Bazzan, Bedakihale, Beirnaert,
  Bejger, Belardinelli, Bell, Bellani, Bellizzi, Beltran-Martinez, Benoit,
  Bentara, Yaala, Bera, Bergamin, Berger, Bernuzzi, Beroiz, Berry, Berry,
  Bersanetti, Bertheas, Bertolini, Betzwieser, Beveridge, Bevilacqua, Bevins,
  Bhandare, Bhatt, Bhattacharjee, Bhattacharjee, Bhattacharyya, Bhaumik,
  Biancalana, Bianchi, Bianchi, Bilenko, Billingsley, Binetti, Bini, Binu,
  Biot, Birnholtz, Biscoveanu, Bisht, Bitossi, Bizouard, Blaber, Blackburn,
  Blagg, Blair, Blair, Bode, Boettner, Bogdan, Boileau, Boldrini, Bolingbroke,
  Bolliand, Bonavena, Bondarescu, Bondu, Bonhomme, Bonilla, Bonilla, Bonino,
  Bonnand, Borchers, Borghi, Boschi, Bose, Bossilkov, Bothra, Boudon, Boyle,
  Bozzi, Bradaschia, Brady, Brady, Branch, Branchesi, Briant, Brillet,
  Brinkmann, Brockill, Brockmueller, Brooks, Brown, Brown, Brozzetti, Brunett,
  Bruno, Bruntz, Bryant, Bu, Bucci, Buchanan, Bulashenko, Bulik, Bulten,
  Buonanno, Burtnyk, Buscicchio, Buskulic, Buy, Byer, Cabrita,
  Cáceres-Barbosa, Cadonati, Cagnoli, Cahillane, Calafat, Callister, Calloni,
  Callos, Canepa, Santoro, Cannon, Cao, Capistran, Capocasa, Capoccia, Capote,
  Capurri, Carapella, Carbognani, Cardona-Martínez, Carlassara, Carlin,
  Carlson, Carney, Carpinelli, Carrillo, Carter, Carullo, Casallas-Lagos, Diaz,
  Casentini, Caudill, Cavaglià, Cavalieri, Cella, Cepic, Cerdá-Durán,
  Cesarini, Chabbra, Chaibi, Chakraborty, Chakraborty, Chakraborty,
  Subrahmanya, Chalmers, Chan, Chan, Chan, Chang, Charlton, Chassande-Mottin,
  Chatterjee, Chatterjee, Chatterjee, Chaturvedi, Chaty, Chatziioannou, Chen,
  Chen, Chen, Chen, Chen, Chen, Chen, Cheng, Cheng, Chessa, Cheunchitra,
  Cheung, Cheung, Chiadini, Chiarini, Chiba, Chincarini, Chintala, Chiofalo,
  Chiummo, Chou, Choudhary, Christensen, Chua, Ciani, Ciecielag, Cieślar,
  Cifaldi, Cirok, Clara, Clark, Clarke, Clearwater, Clesse, Cleva, Clyne,
  Coccia, Codazzo, Cohadon, Cohen, Colace, Colangeli, Cole, Colleoni, Collette,
  Collins, Colloms, Colombo, Compton, Connolly, Conti, Corbitt,
  Cordero-Carrión, Corezzi, Cornish, Coronado, Corsi, Corubolo, Cotnoir,
  Cottingham, Coughlin, Couvares, Coward, Coyne, Coyne, Cozzumbo, Creighton,
  Creighton, Crook, Crouch, Csizmazia, Cudell, Cullen, Cumming, Cuoco,
  Cusinato, Conceição, Canton, Dall’Osso, Pra, Dálya, Dan, Dang,
  D’Angelo, Danilishin, D’Antonio, Danzmann, Darroch, Dartez, Das,
  Dasgupta, Dattilo, Daumas, Dave, Davenport, Davier, Davies, Davis, Davis,
  Davis, Davis, Daw, Dax, Bolle, Deenadayalan, Degallaix, Laurentis, Mendez,
  Lillo, Torre, Pozzo, Rio, Demagny, Marco, Demasi, Matteis, Demos, Dent,
  Depasse, DePergola, Pietri, Rosa, Rossi, Desai, Deshmukh, Simone, Determan,
  Dhani, Dhurkunde, Diab, Diaz, Díaz, Cesare, Dideron, Dietrich, Fiore,
  Fronzo, Giovanni, Girolamo, Diksha, Ding, Pace, Palma, Piero, Renzo,
  Divyajyoti, Dmitriev, Docherty, Doctor, Doerksen, Dohmen, Doke, Souza,
  D’Onofrio, Donovan, Dooley, Dooney, Doravari, Dorosh, Dosopoulou, Doyle,
  Drago, Driggers, Dubois, Dumbreck, Dunn, Dupletsa, D’Urso, Roy, Duval,
  Duverne, Dwyer, Eassa, Eberhardt, Ebersold, Eckhardt, Eddolls, Effler,
  Eichholz, Einsle, Eisenmann, Eisenstein, Emma, Endo, Enficiaud, Errico,
  Espinosa, Esposito, Essick, Estellés, Etzel, Evans, Evstafyeva, Ewing,
  Ezquiaga, Fabrizi, Fafone, Fairhurst, Fan, Farah, Farr, Farr, Favata, Fays,
  Fazio, Feicht, Fejer, Feldhusen, Fenyvesi, Fernandes, Fernandes, Fernando,
  Ferraiuolo, Ferreira, Ferrer, Fidecaro, Figura, Finch, Fiori, Fiori,
  Fishbach, Fisher, Fittipaldi, Fiumara, Flaminio, Fleischer, Fleming, Floden,
  Fong, Font, Fontinele-Nunes, Foo, Fornal, Forsyth, Franceschetti,
  Franco-Ordovas, Frappez, Frasca, Frasconi, Freed, Frei, Freise, Freitas,
  Frey, Frischhertz, Fritschel, Frolov, Fuentes-Garcia, Fujii, Fujimori, Fulda,
  Fyffe, Gadre, Gair, Galaudage, Galdi, Gamba, Gamboa, Gamoji, Ganguly,
  Garaventa, Abia, García-Bellido, García-Quirós, Gardner, Garg, Gargiulo,
  Garrido, Garron, Garufi, Garver, Gasbarra, Gateley, Gautier, Gayathri, Gayer,
  Gemme, Gennai, Gennari, George, George, Gerberding, Gergely, Ghosh, Ghosh,
  Ghosh, Ghosh, Ghosh, Ghosh, Giaime, Giardina, Gibson, Gier, Gkaitatzis,
  Glanzer, Glotin, Godfrey, Godley, Godwin, Goettel, Goetz, Golomb, Lopez,
  González, Goodarzi, Goode, Goodwin-Jones, Gosselin, Gostiaux, Gouaty, Gould,
  Govorkova, Grado, Granados, Granata, Granata, Gras, Grassia, Gray, Gray,
  Greco, Green, Green, Green, Green, Gretarsson, Gretarsson, Griffin, Griffith,
  Griggs, Grignani, Grimaud, Grote, Grunewald, Guerra, Guerrero, Guetta, Guidi,
  Guidry, Gulati, Gulminelli, Gunny, Guo, Guo, Guo, Gupta, Gupta, Gupta, Gupta,
  Gupta, Gupte, Gurs, Gutierrez, Guttman, Guzman, Haba, Haberland, Haino, Hall,
  Hamilton, Hammond, Haney, Hanks, Hanna, Hannam, Hannuksela, Hansen, Hanson,
  Harada, Hardison, Harikumar, Haris, Harley-Trochimczyk, Harmark, Harms,
  Harry, Harry, Hart, Hartman, Haskell, Haster, Haughian, Hayakawa, Hayama,
  Heffernan, Hegde, Heintze, Heinze, Heinzel, Heitmann, Hellman,
  Helmling-Cornell, Hemming, Henderson-Sapir, Hendry, Heng, Hennig, Henshaw,
  Heurs, Hewitt, Heynen, Heyns, Higginbotham, Hild, Hill, Himemoto, Hirata,
  Hirose, Hofman, Hogan, Holland, Holley-Bockelmann, Hollows, Holz, Honet,
  Hoops, Hoque, Horton-Bailey, Hough, Hourihane, Howard, Howell, Hoy,
  Hrishikesh, Hsi, Hsieh, Hsieh, Hsiung, Hsu, Hsu, Hu, Huang, Huang, Huang,
  Huddart, Hughey, Hui, Husa, Iampieri, Iandolo, Ianni, Iannone, Iascau, Ide,
  Iden, Ierardi, Ikeda, Imafuku, Inoue, Iorio, Iosif, Irwin, Ishikawa,
  Ishikawa, Isi, Isleif, Itoh, Iwaguchi, Iwaya, Iyer, Jackson, Jacquet,
  Jacquet, Jacquot, Jadhav, Jadhav, Jain, Jain, James, Jani, Janquart,
  Janthalur, Jaraba, Jaranowski, Jaume, Javed, Jensen, Jia, Jiang, Jin, Johns,
  Johnson, Johnson-McDaniel, Johnston, Johny, Jones, Jones, Jones, Jose, Joshi,
  Joshi, Joubert, Ju, Ju, Juarez-Reyes, Jung, Junker, Juste, Kabagoz, Kajita,
  Kaku, Kalogera, Kalomenopoulos, Kamiizumi, Kanda, Kandhasamy, Kang, Kanner,
  KantiMahanty, Kapadia, Kapasi, Karthikeyan, Kasprzack, Kato, Kato,
  Katsavounidis, Katzman, Kaushik, Kawabe, Kawamoto, Keitel, Kemper, Kemperman,
  Kennington, Kerkow, Kesharwani, Key, Khadela, Khadka, Khadkikar, Khalili,
  Khan, Khanam, Khursheed, Khusid, Kiendrebeogo, Kijbunchoo, Kim, Kim, Kim,
  Kim, Kim, Kim, Kimball, Kimes, Kinnear, Kissel, Klimenko, Knee, Knox, Knust,
  Kobayashi, Koehlenbeck, Koekoek, Kohri, Kokeyama, Koley, Kolitsidou,
  Koloniari, Komori, Kompanets, Kong, Kontos, Kopczuk, Koponen, Korobko, Kou,
  Koushik, Kouvatsos, Kovalam, Koyama, Kozak, Kraja, Kranzhoff, Kringel,
  Krishnendu, Kroker, Królak, Kruska, Kubisz, Kuehn, Ramamohan, Kumar, Kumar,
  Kumar, Kumar, Kumar, Kumar, Kumar, Kume, Kuns, Kuntimaddi, Kuroyanagi,
  Kuwahara, Kwak, Kwan, Kwon, Lacaille, Laghi, Laity, Lakhal, Lalande,
  Lalleman, Lalvani, Landry, Lang, Lange, Langgin, Lantz, Rosa,
  Lartaux-Vollard, Lasky, Lavezzi, Lawrence, Laxen, Lazarte, Lazzarini,
  Lazzaro, Leaci, Leali, Lecoeuche, Lee, Lee, Lee, Lee, Lee, Lee, Lee, Lee,
  Lee, Legred, Lehmann, Lehner, Jean, Lemaître, Lenti, Leonardi, Lequime,
  Leroy, Lesovsky, Letendre, Lethuillier, Levin, Levin, Lexmond, Leyde, Li, Li,
  Li, Li, Li, Liang, Lihos, Lin, Lin, Lin, Lin, Lindsay, Linker, Liu, Liu, Liu,
  Liu, Villarreal, Llobera-Querol, Lo, Locquet, Loggins, Loizou, London, Longo,
  Lopez, Portilla, Lorenzo-Medina, Loriette, Lormand, Losurdo, Lotti, IV,
  Lough, Loughlin, Lousto, Low, Lu, Lucchesi, Lück, Lukina, Lumaca, Lundgren,
  Lunghini, Lussier, Ma, Ma, Macleod, MacMillan, Macquet, Madekar, Maeda,
  Maenaut, Magare, Magee, Maggio, Maggiore, Magnozzi, Mahapatra, Mahesh, Majhi,
  Majorana, Makarem, Makelele, Malakar, Malaquias-Reis, Mali, Maliakal, Malik,
  Mallick, Malz, Man, Mancarella, Mandic, Mangano, Mannix, Mansell, Manske,
  Mantovani, Mapelli, Marchetti, Marinelli, Marion, Markosyan, Markowitz,
  Maros, Marsat, Martelli, Martin, Martin, Martinez, Martinez, Martinez,
  Martinez, Martini, Martins, Martynov, Marx, Massaro, Masserot, Masso-Reid,
  Masters, Mastrogiovanni, Mastropasqua, Matcovich, Matiushechkina,
  Matte-Landry, Maurin, Mavalvala, Maxwell, McCarrol, McCarthy, McClelland,
  McCormick, McCuller, McDermott, McEachin, McElhenny, McGhee, McGowan, McIver,
  McLeod, McRae, McTeague, Meacher, Meagher, Mechum, Meijer, Melatos, Menoni,
  Mera, Mercer, Mereni, Merfeld, Merilh, Merino, Mérou, Merritt, Merzougui,
  Messick, Mestichelli, Meyer-Conde, Meylahn, Mhaske, Miani, Miao,
  Michaloliakos, Michel, Michimura, Middleton, Mihaylov, Miller, Millhouse,
  Milotti, Milotti, Minenkov, Minihan, Mir, Mirasola, Miritescu, Mishra,
  Mishra, Mishra, Mitchell, Mitchell, Mitchem, Mitra, Mitrofanov, Mitsuhashi,
  Mittleman, Miyakawa, Miyoki, Mo, Mobilia, Mohapatra, Mohite, Molina-Ruiz,
  Mondin, Montani, Moore, Moraru, More, More, Moreno, Moreno, Moreno, Serra,
  Morgan, Morisaki, Moriwaki, Morras, Moscatello, Mould, Mours, Mow-Lowry,
  Muccillo, Muciaccia, Mukherjee, Mukherjee, Mukherjee, Mukherjee, Mukherjee,
  Mukherjee, Mukund, Mullavey, Muller, Mungioli, Murakoshi, Murray, Nabari,
  Nadji, Nadji, Nagar, Nagarajan, Nakagaki, Nakamura, Nakano, Nakano,
  Nanadoumgar-Lacroze, Nandi, Napolano, Naqvi, Narayan, Nardecchia, Narikawa,
  Narola, Naticchioni, Nayak, Neeson, Negri, Nela, Nelle, Nelson, Nelson,
  Nemmani, Nery, Neunzert, Newell, Ng, Quynh, Nielsen, Nishino, Nishizawa,
  Nissanke, Niu, Nocera, Noller, Norman, North, Novak, Nowicki, Siles, Nurbek,
  Nuttall, Obayashi, Oberling, Ochoa, O’Dell, Oertel, Oganesyan, O’Hanlon,
  Ohashi, Ohme, Oke, Omer, O’Neal, Onishi, Oohara, O’Reilly, Orselli,
  O’Shaughnessy, Oshino, Osthelder, Ota, Othman, Ottaway, Ouzriat, Overmier,
  Owen, Ozaki, Pace, Pagano, Page, Pai, Paiella, Pal, Pal, Palaia, Pálfi,
  Palma, Palomba, Palud, Pan, Pan, Pan, Panda, Pandey, Pandey, Pang, Pannarale,
  Pannone, Pant, Panther, Panzeri, Paoletti, Paolone, Papadopoulos,
  Papalexakis, Papalini, Papigkiotis, Paquis, Parisi, Park, Park, Parker,
  Pascale, Pascucci, Pasqualetti, Passaquieti, Passenger, Passuello, Patane,
  Patel, Pathak, Patra, Patricelli, Patterson, Paul, Paul, Payne, Pearce,
  Pedraza, Pele, Arellano, Peng, Peng, Penn, Penuliar, Perego, Pereira,
  Périgois, Perna, Perreca, Perret, Perriès, Perry, Peters, Petracca,
  Petrillo, Pfeiffer, Pham, Pham, Phukon, Phurailatpam, Piarulli, Piccari,
  Piccinni, Pichot, Pied, Piendibene, Piergiovanni, Pierini, Pierra, Pierro,
  Pietrzak, Pillas, Pinard, Pinto, Pinto, Piotrzkowski, Pirello, Pitkin,
  Placidi, Placidi, Planas, Plastino, Plunkett, Poggiani, Polini, Pomper,
  Pompili, Poon, Porcelli, Porter, Porter, Posnansky, Poulton, Powell, Prabhu,
  Pracchia, Pradhan, Pradier, Prajapati, Prasai, Prasanna, Prasia, Pratten,
  Praveen, Principe, Prodi, Prosperi, Prosposito, Puecher, Pullin, Puppo,
  Pürrer, Qi, Qiao, Qin, Quéméner, Quetschke, Quinonez, Rading, Rainho,
  Raja, Rajan, Rajbhandari, Ramirez, Vidal, Arevalo, Ramos-Buades, Ranjan,
  Ranjbar, Ransom, Rapagnani, Ratto, Ravichandran, Ray, Raymond, Razzano, Read,
  Regan, Regimbau, Reichardt, Reid, Reissel, Reitze, Renzini, Revenu, Peña,
  Ricca, Ricci, Ricci, Ricciardone, Rice, Richardson, Richardson, Rijal, Riles,
  Riley, Rinaldi, Rittmeyer, Robertson, Robinet, Robinson, Rocchi, Rolland,
  Rollins, Romano, Romano, Romero-Rodríguez, Romero-Shaw, Romie, Ronchini,
  Roocke, Rosa, Rosauer, Rose, Rosińska, Ross, Rossello-Sastre, Rowan,
  Rowlands, Roy, Roy, Rozza, Ruggi, Ruhama, Ruiz, Morales, Ruiz-Rocha, Russ,
  Sachdev, Sadecki, Saffarieh, Safi-Harb, Sah, Saha, Sainrat, Menon, Sakai,
  Sakai, Sakellariadou, Sakon, Salafia, Salces-Carcoba, Salconi, Saleem,
  Salemi, Sallé, Salunkhe, Salvador, Salvarese, Samajdar, Sanchez, Sanchez,
  Sanchis-Gual, Sanders, Sänger, Santoliquido, Sarandrea, Saravanan, Sarin,
  Sarkar, Sasli, Sassi, Sassolas, Sathyaprakash, Sato, Sato, Sato, Sato,
  Sauter, Savage, Sawada, Sawant, Sayah, Scacco, Schaetzl, Scheel,
  Schiebelbein, Schiworski, Schmidt, Schmidt, Schnabel, Schneewind, Schofield,
  Schouteden, Schulte, Schulz, Schutz, Schwartz, Scialpi, Scott, Scott, Sedas,
  Seetharamu, Seglar-Arroyo, Sekiguchi, Sellers, Sembo, Sengupta, Seo, Seo,
  Sequino, Serra, Sevrin, Shaffer, Shah, Shaikh, Shao, Sharkey, Sharma, Sharma,
  Sharma, Sharma, Sharma-Chaudhary, Shawhan, Shcheblanov, Sheridan, Shi,
  Shimomura, Shinkai, Shirke, Shoemaker, Shoemaker, Short, ShyamSundar, Sider,
  Siegel, Sierra, Sigg, Silenzi, Silvestri, Simmonds, Singer, Singh, Singh,
  Singh, Singh, Singh, Singh, Sinha, Sintes, Sipala, Skliris, Slagmolen,
  Slaven-Blair, Smetana, Smith, Smith, Smith, Smith, Smith, Filho, Somiya,
  Song, Soni, Sordini, Sorrentino, Sotani, Spada, Spagnuolo, Spencer,
  Spinicelli, Srivastava, Stachurski, Stark, Steer, Steinle, Steinlechner,
  Steinlechner, Stergioulas, Stevens, StPierre, Strong, Strunk, Stuver,
  Suchenek, Sudhagar, Sudo, Sueltmann, Suleiman, Sullivan, Sun, Sun, Sunil,
  Suresh, Sutton, Sutton, Suzuki, Suzuki, Svizzeretto, Swain, Swinkels, Syx,
  Szczepańczyk, Szewczyk, Tacca, Tagliazucchi, Tagoshi, Tait, Takada,
  Takahashi, Takahashi, Takamori, Takano, Takeda, Takeshita, Schmiegelow,
  Takou-Ayaoh, Talbot, Tamaki, Tamanini, Tanabe, Tanaka, Tanaka, Tanioka,
  Tanner, Tanner, Tao, Tapia, Martín, Taranto, Taruya, Tasson, Tau, Tejera,
  Tenorio, Themann, Theodoropoulos, Thirugnanasambandam, Thomas, Thomas,
  Thomas, Thompson, Thondapu, Thorne, Thrane, Tissino, Tiwari, Tiwari, Tiwari,
  Tiwari, Tiwari, Todd, Todd, Tofani, Toffano, Toivonen, Toland, Tolley,
  Tomaru, Tommasini, Tomura, Tong, Tong-Yu, Torres-Forné, Torrie, Melo,
  Tournefier, Nery, Trapananti, Travaglini, Travasso, Traylor, Trevor,
  Tringali, Tripathee, Troian, Trovato, Trozzo, Trudeau, Tsang, Tsuchida,
  Tsuji, Tsukada, Turbang, Turconi, Turski, Ubach, Ubhi, Uchikata, Uchiyama,
  Udall, Uehara, Ueno, Undheim, Uronen, Ushiba, Vacatello, Vahlbruch, Vajente,
  Valencia, Valentini, Vallejo-Pagès, Vallejo-Peña, Vallero, Dael, Bossche,
  Brand, Broeck, Kolk, Sluys, Walle, Dongen, Vandra, VanDyke, Haevermaet,
  Heijningen, Hove, Vanier, Vanosky, Remortel, Vardaro, Vargas, Varma, Vecchio,
  Vedovato, Veitch, Veitch, Venikoudis, Venneberg, Venterea, Verdier,
  Vereecken, Verkindt, Verma, Verma, Vermeulen, Vetrano, Veutro, Viceré,
  Vidyant, Viets, Vijaykumar, Vilkha, Espinosa, Villa-Ortega, Vincent, Vinet,
  Viret, Vitale, Vives, Vizmeg, Vocca, Voigt, Reis, Wrangel, Vossius, Vujeva,
  Vyatchanin, Wack, Wade, Wade, Wagner, Wallace, Wang, Wang, Wang, Wang, Wang,
  Waratkar, Ward, Warner, Was, Washimi, Washington, Weaver, Webster,
  Weickhardt, Weinert, Weinstein, Weiss, Wen, Wette, Wheeler, Whelan, Whiting,
  Whittall, Wickens, Wilken, Williams, Williams, Williams, Williams, Willis,
  Willke, Wils, Wilson, Winborn, Winterflood, Wipf, Woan, Woehler, Wolfe, Wong,
  Wong, Wong, Wouters, Wright, Wright, Wu, Wu, Wu, Wu, Wu, Wu, Wu, Wuchner,
  Wysocki, Xu, Xu, Yadav, Yamamoto, Yamamoto, Yamamoto, Yamamoto, Yamazaki,
  Yan, Yang, Yang, Yang, Yarbrough, Yebana, Yeh, Yelikar, Yin, Yokoyama,
  Yokozawa, Yuan, Yuzurihara, Zanolin, Zeeshan, Zelenova, Zendri, Zeoli,
  Zerrad, Zevin, Zhang, Zhang, Zhang, Zhang, Zhang, Zhao, Zhao, Zhao, Zhao,
  Zheng, Zhong, Zhou, Zhu, Zhu, Zhu, Zimmerman, Zimmermann, \&
  Zucker}]{gzrj-mwv3}
Abac, A.~G., Abouelfettouh, I., Acernese, F., {et~al.} 2026{\natexlab{f}},
  Phys. Rev. Lett., \dodoi{10.1103/gzrj-mwv3}

\bibitem[{{Abadie} {et~al.}(2012){Abadie}, {Abbott}, {Abbott}, {Abbott},
  {Abernathy}, {Accadia}, {Acernese}, {Adams}, {Adhikari}, {Affeldt},
  {Agathos}, {Agatsuma}, {Ajith}, {Allen}, {Amador Ceron}, {Amariutei},
  {Anderson}, {Anderson}, {Arai}, {Arain}, {Araya}, {Aston}, {Astone},
  {Atkinson}, {Aufmuth}, {Aulbert}, {Aylott}, {Babak}, {Baker}, {Ballardin},
  {Ballmer}, {Barayoga}, {Barker}, {Barone}, {Barr}, {Barsotti}, {Barsuglia},
  {Barton}, {Bartos}, {Bassiri}, {Bastarrika}, {Basti}, {Batch}, {Bauchrowitz},
  {Bauer}, {Bebronne}, {Beck}, {Behnke}, {Bejger}, {Beker}, {Bell},
  {Belopolski}, {Benacquista}, {Berliner}, {Bertolini}, {Betzwieser},
  {Beveridge}, {Beyersdorf}, {Bilenko}, {Billingsley}, {Birch}, {Biswas},
  {Bitossi}, {Bizouard}, {Black}, {Blackburn}, {Blackburn}, {Blair}, {Bland},
  {Blom}, {Bock}, {Bodiya}, {Bogan}, {Bondarescu}, {Bondu}, {Bonelli},
  {Bonnand}, {Bork}, {Born}, {Boschi}, {Bose}, {Bosi}, {Bouhou}, {Braccini},
  {Bradaschia}, {Brady}, {Braginsky}, {Branchesi}, {Brau}, {Breyer}, {Briant},
  {Bridges}, {Brillet}, {Brinkmann}, {Brisson}, {Britzger}, {Brooks}, {Brown},
  {Bulik}, {Bulten}, {Buonanno}, {Burguet-Castell}, {Buskulic}, {Buy}, {Byer},
  {Cadonati}, {Calloni}, {Camp}, {Campsie}, {Cannizzo}, {Cannon}, {Canuel},
  {Cao}, {Capano}, {Carbognani}, {Carbone}, {Caride}, {Caudill},
  {Cavagli{\`a}}, {Cavalier}, {Cavalieri}, {Cella}, {Cepeda}, {Cesarini},
  {Chaibi}, {Chalermsongsak}, {Charlton}, {Chassande-Mottin}, {Chelkowski},
  {Chen}, {Chen}, {Chen}, {Chincarini}, {Chiummo}, {Cho}, {Chow},
  {Christensen}, {Chua}, {Chung}, {Chung}, {Ciani}, {Clara}, {Clark}, {Clark},
  {Clayton}, {Cleva}, {Coccia}, {Cohadon}, {Colacino}, {Colas}, {Colla},
  {Colombini}, {Conte}, {Conte}, {Cook}, {Corbitt}, {Cordier}, {Cornish},
  {Corsi}, {Costa}, {Coughlin}, {Coulon}, {Couvares}, {Coward}, {Cowart},
  {Coyne}, {Creighton}, {Creighton}, {Cruise}, {Cumming}, {Cunningham},
  {Cuoco}, {Cutler}, {Dahl}, {Danilishin}, {Dannenberg}, {D'Antonio},
  {Danzmann}, {Dattilo}, {Daudert}, {Daveloza}, {Davier}, {Daw}, {Day},
  {Dayanga}, {De Rosa}, {DeBra}, {Debreczeni}, {Degallaix}, {Del Pozzo}, {del
  Prete}, {Dent}, {Dergachev}, {DeRosa}, {DeSalvo}, {Dhurandhar}, {Di Fiore},
  {Di Lieto}, {Di Palma}, \& {Emilio}}]{2012ApJ...760...12A}
{Abadie}, J., {Abbott}, B.~P., {Abbott}, R., {et~al.} 2012, \apj, 760, 12,
  \dodoi{10.1088/0004-637X/760/1/12}

\bibitem[{Abadie {et~al.}(2012)}]{LIGOScientific:2011jth}
Abadie, J., {et~al.} 2012, Phys. Rev. D, 85, 082002,
  \dodoi{10.1103/PhysRevD.85.082002}

\bibitem[{{Abbott} {et~al.}(2016{\natexlab{a}}){Abbott}, {Abbott}, {Abbott},
  {Abernathy}, {Acernese}, {Ackley}, {Adams}, {Adams}, {Addesso}, {Adhikari},
  {Adya}, {Affeldt}, {Agathos}, {Agatsuma}, {Aggarwal}, {Aguiar}, {Aiello},
  {Ain}, {Ajith}, {Allen}, {Allocca}, {Altin}, {Anderson}, {Anderson}, {Arai},
  {Araya}, {Arceneaux}, {Areeda}, {Arnaud}, {Arun}, {Ascenzi}, {Ashton}, {Ast},
  {Aston}, {Astone}, {Aufmuth}, {Aulbert}, {Babak}, {Bacon}, {Bader}, {Baker},
  {Baldaccini}, {Ballardin}, {Ballmer}, {Barayoga}, {Barclay}, {Barish},
  {Barker}, {Barone}, {Barr}, {Barsotti}, {Barsuglia}, {Barta}, {Bartlett},
  {Bartos}, {Bassiri}, {Basti}, {Batch}, {Baune}, {Bavigadda}, {Bazzan},
  {Behnke}, {Bejger}, {Bell}, {Bell}, {Berger}, {Bergman}, {Bergmann}, {Berry},
  {Bersanetti}, {Bertolini}, {Betzwieser}, {Bhagwat}, {Bhandare}, {Bilenko},
  {Billingsley}, {Birch}, {Birney}, {Biscans}, {Bisht}, {Bitossi}, {Biwer},
  {Bizouard}, {Blackburn}, {Blair}, {Blair}, {Blair}, {Bloemen}, {Bock},
  {Bodiya}, {Boer}, {Bogaert}, {Bogan}, {Bohe}, {Bojtos}, {Bond}, {Bondu},
  {Bonnand}, {Boom}, {Bork}, {Boschi}, {Bose}, {Bouffanais}, {Bozzi},
  {Bradaschia}, {Brady}, {Braginsky}, {Branchesi}, {Brau}, {Briant}, {Brillet},
  {Brinkmann}, {Brisson}, {Brockill}, {Brooks}, {Brown}, {Brown}, {Brown},
  {Buchanan}, {Buikema}, {Bulik}, {Bulten}, {Buonanno}, {Buskulic}, {Buy},
  {Byer}, {Cadonati}, {Cagnoli}, {Cahillane}, {Calder{\'o}n Bustillo},
  {Callister}, {Calloni}, {Camp}, {Cannon}, {Cao}, {Capano}, {Capocasa},
  {Carbognani}, {Caride}, {Casanueva Diaz}, {Casentini}, {Caudill},
  {Cavagli{\`a}}, {Cavalier}, {Cavalieri}, {Cella}, {Cepeda}, {Cerboni
  Baiardi}, {Cerretani}, {Cesarini}, {Chakraborty}, {Chalermsongsak},
  {Chamberlin}, {Chan}, {Chao}, {Charlton}, {Chassande-Mottin}, {Chen}, {Chen},
  {Cheng}, {Chincarini}, {Chiummo}, {Cho}, {Cho}, {Chow}, {Christensen}, {Chu},
  {Chua}, {Chung}, {Ciani}, {Clara}, {Clark}, {Cleva}, {Coccia}, {Cohadon},
  {Colla}, {Collette}, {Cominsky}, {Constancio}, {Conte}, {Conti}, {Cook},
  {Corbitt}, {Cornish}, {Corsi}, {Cortese}, {Costa}, {Coughlin}, {Coughlin},
  {Coulon}, {Countryman}, {Couvares}, {Cowan}, {Coward}, {Cowart}, {Coyne},
  {Coyne}, {Craig}, {Creighton}, \& {Cripe}}]{2016ApJ...833L...1A}
{Abbott}, B.~P., {Abbott}, R., {Abbott}, T.~D., {et~al.} 2016{\natexlab{a}},
  \apjl, 833, L1, \dodoi{10.3847/2041-8205/833/1/L1}

\bibitem[{{Abbott} {et~al.}(2016{\natexlab{b}}){Abbott}, {Abbott}, {Abbott},
  {Abernathy}, {Acernese}, {Ackley}, {Adams}, {Adams}, {Addesso}, {Adhikari},
  {Adya}, {Affeldt}, {Agathos}, {Agatsuma}, {Aggarwal}, {Aguiar}, {Aiello},
  {Ain}, {Ajith}, {Allen}, {Allocca}, {Altin}, {Anderson}, {Anderson}, {Arai},
  {Araya}, {Arceneaux}, {Areeda}, {Arnaud}, {Arun}, {Ascenzi}, {Ashton}, {Ast},
  {Aston}, {Astone}, {Aufmuth}, {Aulbert}, {Babak}, {Bacon}, {Bader}, {Baker},
  {Baldaccini}, {Ballardin}, {Ballmer}, {Barayoga}, {Barclay}, {Barish},
  {Barker}, {Barone}, {Barr}, {Barsotti}, {Barsuglia}, {Barta}, {Bartlett},
  {Bartos}, {Bassiri}, {Basti}, {Batch}, {Baune}, {Bavigadda}, {Bazzan},
  {Bejger}, {Bell}, {Berger}, {Bergmann}, {Berry}, {Bersanetti}, {Bertolini},
  {Betzwieser}, {Bhagwat}, {Bhandare}, {Bilenko}, {Billingsley}, {Birch},
  {Birney}, {Birnholtz}, {Biscans}, {Bisht}, {Bitossi}, {Biwer}, {Bizouard},
  {Blackburn}, {Blair}, {Blair}, {Blair}, {Bloemen}, {Bock}, {Boer}, {Bogaert},
  {Bogan}, {Bohe}, {Bond}, {Bondu}, {Bonnand}, {Boom}, {Bork}, {Boschi},
  {Bose}, {Bouffanais}, {Bozzi}, {Bradaschia}, {Brady}, {Braginsky},
  {Branchesi}, {Brau}, {Briant}, {Brillet}, {Brinkmann}, {Brisson}, {Brockill},
  {Broida}, {Brooks}, {Brown}, {Brown}, {Brown}, {Brunett}, {Buchanan},
  {Buikema}, {Bulik}, {Bulten}, {Buonanno}, {Buskulic}, {Buy}, {Byer},
  {Cabero}, {Cadonati}, {Cagnoli}, {Cahillane}, {Calder{\'o}n Bustillo},
  {Callister}, {Calloni}, {Camp}, {Cannon}, {Cao}, {Capano}, {Capocasa},
  {Carbognani}, {Caride}, {Casanueva Diaz}, {Casentini}, {Caudill},
  {Cavagli{\`a}}, {Cavalier}, {Cavalieri}, {Cella}, {Cepeda}, {Cerboni
  Baiardi}, {Cerretani}, {Cesarini}, {Chamberlin}, {Chan}, {Chao}, {Charlton},
  {Chassande-Mottin}, {Cheeseboro}, {Chen}, {Chen}, {Cheng}, {Chincarini},
  {Chiummo}, {Cho}, {Cho}, {Chow}, {Christensen}, {Chu}, {Chua}, {Chung},
  {Ciani}, {Clara}, {Clark}, {Cleva}, {Coccia}, {Cohadon}, {Colla}, {Collette},
  {Cominsky}, {Constancio}, {Conte}, {Conti}, {Cook}, {Corbitt}, {Cornish},
  {Corsi}, {Cortese}, {Costa}, {Coughlin}, {Coughlin}, {Coulon}, {Countryman},
  {Couvares}, {Cowan}, {Coward}, {Cowart}, {Coyne}, {Coyne}, {Craig},
  {Creighton}, {Cripe}, {Crowder}, \& {Cumming}}]{2016PhRvX...6d1015A}
---. 2016{\natexlab{b}}, Physical Review X, 6, 041015,
  \dodoi{10.1103/PhysRevX.6.041015}

\bibitem[{{Abbott} {et~al.}(2016{\natexlab{c}}){Abbott}, {Abbott}, {Abbott},
  {Abernathy}, {Acernese}, {Ackley}, {Adams}, {Adams}, {Addesso}, {Adhikari},
  {Adya}, {Affeldt}, {Agathos}, {Agatsuma}, {Aggarwal}, {Aguiar}, {Aiello},
  {Ain}, {Ajith}, {Allen}, {Allocca}, {Altin}, {Anderson}, {Anderson}, {Arai},
  {Araya}, {Arceneaux}, {Areeda}, {Arnaud}, {Arun}, {Ascenzi}, {Ashton}, {Ast},
  {Aston}, {Astone}, {Aufmuth}, {Aulbert}, {Babak}, {Bacon}, {Bader}, {Baker},
  {Baldaccini}, {Ballardin}, {Ballmer}, {Barayoga}, {Barclay}, {Barish},
  {Barker}, {Barone}, {Barr}, {Barsotti}, {Barsuglia}, {Barta}, {Bartlett},
  {Bartos}, {Bassiri}, {Basti}, {Batch}, {Baune}, {Bavigadda}, {Bazzan},
  {Behnke}, {Bejger}, {Bell}, {Bell}, {Berger}, {Bergman}, {Bergmann}, {Berry},
  {Bersanetti}, {Bertolini}, {Betzwieser}, {Bhagwat}, {Bhandare}, {Bilenko},
  {Billingsley}, {Birch}, {Birney}, {Biscans}, {Bisht}, {Bitossi}, {Biwer},
  {Bizouard}, {Blackburn}, {Blair}, {Blair}, {Blair}, {Bloemen}, {Bock},
  {Bodiya}, {Boer}, {Bogaert}, {Bogan}, {Bohe}, {Bojtos}, {Bond}, {Bondu},
  {Bonnand}, {Boom}, {Bork}, {Boschi}, {Bose}, {Bouffanais}, {Bozzi},
  {Bradaschia}, {Brady}, {Braginsky}, {Branchesi}, {Brau}, {Briant}, {Brillet},
  {Brinkmann}, {Brisson}, {Brockill}, {Brooks}, {Brown}, {Brown}, {Brown},
  {Buchanan}, {Buikema}, {Bulik}, {Bulten}, {Buonanno}, {Buskulic}, {Buy},
  {Byer}, {Cadonati}, {Cagnoli}, {Cahillane}, {Calder{\'o}n Bustillo},
  {Callister}, {Calloni}, {Camp}, {Cannon}, {Cao}, {Capano}, {Capocasa},
  {Carbognani}, {Caride}, {Casanueva Diaz}, {Casentini}, {Caudill},
  {Cavagli{\`a}}, {Cavalier}, {Cavalieri}, {Cella}, {Cepeda}, {Cerboni
  Baiardi}, {Cerretani}, {Cesarini}, {Chakraborty}, {Chalermsongsak},
  {Chamberlin}, {Chan}, {Chao}, {Charlton}, {Chassande-Mottin}, {Chen}, {Chen},
  {Cheng}, {Chincarini}, {Chiummo}, {Cho}, {Cho}, {Chow}, {Christensen}, {Chu},
  {Chua}, {Chung}, {Ciani}, {Clara}, {Clark}, {Cleva}, {Coccia}, {Cohadon},
  {Colla}, {Collette}, {Cominsky}, {Constancio}, {Conte}, {Conti}, {Cook},
  {Corbitt}, {Cornish}, {Corsi}, {Cortese}, {Costa}, {Coughlin}, {Coughlin},
  {Coulon}, {Countryman}, {Couvares}, {Cowan}, {Coward}, {Cowart}, {Coyne},
  {Coyne}, {Craig}, {Creighton}, \& {Cripe}}]{2016ApJS..227...14A}
---. 2016{\natexlab{c}}, \apjs, 227, 14, \dodoi{10.3847/0067-0049/227/2/14}

\bibitem[{{Abbott} {et~al.}(2016{\natexlab{d}}){Abbott}, {Abbott}, {Abbott},
  {Abernathy}, {Acernese}, {Ackley}, {Adams}, {Adams}, {Addesso}, {Adhikari},
  {Adya}, {Affeldt}, {Agathos}, {Agatsuma}, {Aggarwal}, {Aguiar}, {Aiello},
  {Ain}, {Ajith}, {Allen}, {Allocca}, {Altin}, {Anderson}, {Anderson}, {Arai},
  {Araya}, {Arceneaux}, {Areeda}, {Arnaud}, {Arun}, {Ascenzi}, {Ashton}, {Ast},
  {Aston}, {Astone}, {Aufmuth}, {Aulbert}, {Babak}, {Bacon}, {Bader}, {Baker},
  {Baldaccini}, {Ballardin}, {Ballmer}, {Barayoga}, {Barclay}, {Barish},
  {Barker}, {Barone}, {Barr}, {Barsotti}, {Barsuglia}, {Barta}, {Bartlett},
  {Bartos}, {Bassiri}, {Basti}, {Batch}, {Baune}, {Bavigadda}, {Bazzan},
  {Behnke}, {Bejger}, {Bell}, {Bell}, {Berger}, {Bergman}, {Bergmann}, {Berry},
  {Bersanetti}, {Bertolini}, {Betzwieser}, {Bhagwat}, {Bhandare}, {Bilenko},
  {Billingsley}, {Birch}, {Birney}, {Birnholtz}, {Biscans}, {Bisht}, {Bitossi},
  {Biwer}, {Bizouard}, {Blackburn}, {Blair}, {Blair}, {Blair}, {Bloemen},
  {Bock}, {Bodiya}, {Boer}, {Bogaert}, {Bogan}, {Bohe}, {Bojtos}, {Bond},
  {Bondu}, {Bonnand}, {Boom}, {Bork}, {Boschi}, {Bose}, {Bouffanais}, {Bozzi},
  {Bradaschia}, {Brady}, {Braginsky}, {Branchesi}, {Brau}, {Briant}, {Brillet},
  {Brinkmann}, {Brisson}, {Brockill}, {Brooks}, {Brown}, {Brown}, {Brown},
  {Buchanan}, {Buikema}, {Bulik}, {Bulten}, {Buonanno}, {Buskulic}, {Buy},
  {Byer}, {Cadonati}, {Cagnoli}, {Cahillane}, {Calder{\'o}n Bustillo},
  {Callister}, {Calloni}, {Camp}, {Cannon}, {Cao}, {Capano}, {Capocasa},
  {Carbognani}, {Caride}, {Casanueva Diaz}, {Casentini}, {Caudill},
  {Cavagli{\`a}}, {Cavalier}, {Cavalieri}, {Cella}, {Cepeda}, {Cerboni
  Baiardi}, {Cerretani}, {Cesarini}, {Chakraborty}, {Chalermsongsak},
  {Chamberlin}, {Chan}, {Chao}, {Charlton}, {Chassande-Mottin}, {Chen}, {Chen},
  {Cheng}, {Chincarini}, {Chiummo}, {Cho}, {Cho}, {Chow}, {Christensen}, {Chu},
  {Chua}, {Chung}, {Ciani}, {Clara}, {Clark}, {Cleva}, {Coccia}, {Cohadon},
  {Colla}, {Collette}, {Cominsky}, {Constancio}, {Conte}, {Conti}, {Cook},
  {Corbitt}, {Cornish}, {Corsi}, {Cortese}, {Costa}, {Coughlin}, {Coughlin},
  {Coulon}, {Countryman}, {Couvares}, {Cowan}, {Coward}, {Cowart}, {Coyne},
  {Coyne}, {Craig}, \& {Creighton}}]{2016PhRvL.116x1102A}
---. 2016{\natexlab{d}}, \prl, 116, 241102,
  \dodoi{10.1103/PhysRevLett.116.241102}

\bibitem[{{Abbott} {et~al.}(2017{\natexlab{a}}){Abbott}, {Abbott}, {Abbott},
  {Acernese}, {Ackley}, {Adams}, {Adams}, {Addesso}, {Adhikari}, {Adya},
  {Affeldt}, {Afrough}, {Agarwal}, {Agathos}, {Agatsuma}, {Aggarwal}, {Aguiar},
  {Aiello}, {Ain}, {Ajith}, {Allen}, {Allen}, {Allocca}, {Altin}, {Amato},
  {Ananyeva}, {Anderson}, {Anderson}, {Angelova}, {Antier}, {Appert}, {Arai},
  {Araya}, {Areeda}, {Arnaud}, {Arun}, {Ascenzi}, {Ashton}, {Ast}, {Aston},
  {Astone}, {Atallah}, {Aufmuth}, {Aulbert}, {AultONeal}, {Austin},
  {Avila-Alvarez}, {Babak}, {Bacon}, {Bader}, {Bae}, {Bailes}, {Baker},
  {Baldaccini}, {Ballardin}, {Ballmer}, {Banagiri}, {Barayoga}, {Barclay},
  {Barish}, {Barker}, {Barkett}, {Barone}, {Barr}, {Barsotti}, {Barsuglia},
  {Barta}, {Barthelmy}, {Bartlett}, {Bartos}, {Bassiri}, {Basti}, {Batch},
  {Bawaj}, {Bayley}, {Bazzan}, {B{\'e}csy}, {Beer}, {Bejger}, {Belahcene},
  {Bell}, {Berger}, {Bergmann}, {Bernuzzi}, {Bero}, {Berry}, {Bersanetti},
  {Bertolini}, {Betzwieser}, {Bhagwat}, {Bhandare}, {Bilenko}, {Billingsley},
  {Billman}, {Birch}, {Birney}, {Birnholtz}, {Biscans}, {Biscoveanu}, {Bisht},
  {Bitossi}, {Biwer}, {Bizouard}, {Blackburn}, {Blackman}, {Blair}, {Blair},
  {Blair}, {Bloemen}, {Bock}, {Bode}, {Boer}, {Bogaert}, {Bohe}, {Bondu},
  {Bonilla}, {Bonnand}, {Boom}, {Bork}, {Boschi}, {Bose}, {Bossie},
  {Bouffanais}, {Bozzi}, {Bradaschia}, {Brady}, {Branchesi}, {Brau}, {Briant},
  {Brillet}, {Brinkmann}, {Brisson}, {Brockill}, {Broida}, {Brooks}, {Brown},
  {Brown}, {Brunett}, {Buchanan}, {Buikema}, {Bulik}, {Bulten}, {Buonanno},
  {Buskulic}, {Buy}, {Byer}, {Cabero}, {Cadonati}, {Cagnoli}, {Cahillane},
  {Calder{\'o}n Bustillo}, {Callister}, {Calloni}, {Camp}, {Canepa},
  {Canizares}, {Cannon}, {Cao}, {Cao}, {Capano}, {Capocasa}, {Carbognani},
  {Caride}, {Carney}, {Carullo}, {Casanueva Diaz}, {Casentini}, {Caudill},
  {Cavagli{\`a}}, {Cavalier}, {Cavalieri}, {Cella}, {Cepeda},
  {Cerd{\'a}-Dur{\'a}n}, {Cerretani}, {Cesarini}, {Chamberlin}, {Chan}, {Chao},
  {Charlton}, {Chase}, {Chassande-Mottin}, {Chatterjee}, {Chatziioannou},
  {Cheeseboro}, {Chen}, {Chen}, {Chen}, {Cheng}, {Chia}, {Chincarini},
  {Chiummo}, {Chmiel}, {Cho}, {Cho}, {Chow}, {Christensen}, {Chu}, {Chua}, \&
  {Chua}}]{2017PhRvL.119p1101A}
---. 2017{\natexlab{a}}, \prl, 119, 161101,
  \dodoi{10.1103/PhysRevLett.119.161101}

\bibitem[{{Abbott} {et~al.}(2017{\natexlab{b}}){Abbott}, {Abbott}, {Abbott},
  {Acernese}, {Ackley}, {Adams}, {Adams}, {Addesso}, {Adhikari}, {Adya},
  {Affeldt}, {Afrough}, {Agarwal}, {Agathos}, {Agatsuma}, {Aggarwal}, {Aguiar},
  {Aiello}, {Ain}, {Ajith}, {Allen}, {Allen}, {Allocca}, {Altin}, {Amato},
  {Ananyeva}, {Anderson}, {Anderson}, {Angelova}, {Antier}, {Appert}, {Arai},
  {Araya}, {Areeda}, {Arnaud}, {Arun}, {Ascenzi}, {Ashton}, {Ast}, {Aston},
  {Astone}, {Atallah}, {Aufmuth}, {Aulbert}, {AultONeal}, {Austin},
  {Avila-Alvarez}, {Babak}, {Bacon}, {Bader}, {Bae}, {Baker}, {Baldaccini},
  {Ballardin}, {Ballmer}, {Banagiri}, {Barayoga}, {Barclay}, {Barish},
  {Barker}, {Barkett}, {Barone}, {Barr}, {Barsotti}, {Barsuglia}, {Barta},
  {Barthelmy}, {Bartlett}, {Bartos}, {Bassiri}, {Basti}, {Batch}, {Bawaj},
  {Bayley}, {Bazzan}, {B{\'e}csy}, {Beer}, {Bejger}, {Belahcene}, {Bell},
  {Berger}, {Bergmann}, {Bero}, {Berry}, {Bersanetti}, {Bertolini},
  {Betzwieser}, {Bhagwat}, {Bhandare}, {Bilenko}, {Billingsley}, {Billman},
  {Birch}, {Birney}, {Birnholtz}, {Biscans}, {Biscoveanu}, {Bisht}, {Bitossi},
  {Biwer}, {Bizouard}, {Blackburn}, {Blackman}, {Blair}, {Blair}, {Blair},
  {Bloemen}, {Bock}, {Bode}, {Boer}, {Bogaert}, {Bohe}, {Bondu}, {Bonilla},
  {Bonnand}, {Boom}, {Bork}, {Boschi}, {Bose}, {Bossie}, {Bouffanais}, {Bozzi},
  {Bradaschia}, {Brady}, {Branchesi}, {Brau}, {Briant}, {Brillet}, {Brinkmann},
  {Brisson}, {Brockill}, {Broida}, {Brooks}, {Brown}, {Brown}, {Brunett},
  {Buchanan}, {Buikema}, {Bulik}, {Bulten}, {Buonanno}, {Buskulic}, {Buy},
  {Byer}, {Cabero}, {Cadonati}, {Cagnoli}, {Cahillane}, {Calder{\'o}n
  Bustillo}, {Callister}, {Calloni}, {Camp}, {Canepa}, {Canizares}, {Cannon},
  {Cao}, {Cao}, {Capano}, {Capocasa}, {Carbognani}, {Caride}, {Carney},
  {Casanueva Diaz}, {Casentini}, {Caudill}, {Cavagli{\`a}}, {Cavalier},
  {Cavalieri}, {Cella}, {Cepeda}, {Cerd{\'a}-Dur{\'a}n}, {Cerretani},
  {Cesarini}, {Chamberlin}, {Chan}, {Chao}, {Charlton}, {Chase},
  {Chassande-Mottin}, {Chatterjee}, {Chatziioannou}, {Cheeseboro}, {Chen},
  {Chen}, {Chen}, {Cheng}, {Chia}, {Chincarini}, {Chiummo}, {Chmiel}, {Cho},
  {Cho}, {Chow}, {Christensen}, {Chu}, {Chua}, {Chua}, {Chung}, {Chung}, \&
  {Ciani}}]{2017ApJ...848L..12A}
---. 2017{\natexlab{b}}, \apjl, 848, L12, \dodoi{10.3847/2041-8213/aa91c9}

\bibitem[{{Abbott} {et~al.}(2019{\natexlab{a}}){Abbott}, {Abbott}, {Abbott},
  {Abraham}, {Acernese}, {Ackley}, {Adams}, {Adhikari}, {Adya}, {Affeldt},
  {Agathos}, {Agatsuma}, {Aggarwal}, {Aguiar}, {Aiello}, {Ain}, {Ajith},
  {Allen}, {Allocca}, {Aloy}, {Altin}, {Amato}, {Ananyeva}, {Anderson},
  {Anderson}, {Angelova}, {Antier}, {Appert}, {Arai}, {Araya}, {Areeda},
  {Ar{\`e}ne}, {Arnaud}, {Arun}, {Ascenzi}, {Ashton}, {Aston}, {Astone},
  {Aubin}, {Aufmuth}, {AultONeal}, {Austin}, {Avendano}, {Avila-Alvarez},
  {Babak}, {Bacon}, {Badaracco}, {Bader}, {Bae}, {Baker}, {Baldaccini},
  {Ballardin}, {Ballmer}, {Banagiri}, {Barayoga}, {Barclay}, {Barish},
  {Barker}, {Barkett}, {Barnum}, {Barone}, {Barr}, {Barsotti}, {Barsuglia},
  {Barta}, {Bartlett}, {Bartos}, {Bassiri}, {Basti}, {Bawaj}, {Bayley},
  {Bazzan}, {B{\'e}csy}, {Bejger}, {Belahcene}, {Bell}, {Beniwal}, {Berger},
  {Bergmann}, {Bernuzzi}, {Bero}, {Berry}, {Bersanetti}, {Bertolini},
  {Betzwieser}, {Bhandare}, {Bidler}, {Bilenko}, {Bilgili}, {Billingsley},
  {Birch}, {Birney}, {Birnholtz}, {Biscans}, {Biscoveanu}, {Bisht}, {Bitossi},
  {Bizouard}, {Blackburn}, {Blackman}, {Blair}, {Blair}, {Blair}, {Bloemen},
  {Bode}, {Boer}, {Boetzel}, {Bogaert}, {Bondu}, {Bonilla}, {Bonnand},
  {Booker}, {Boom}, {Booth}, {Bork}, {Boschi}, {Bose}, {Bossie}, {Bossilkov},
  {Bosveld}, {Bouffanais}, {Bozzi}, {Bradaschia}, {Brady}, {Bramley},
  {Branchesi}, {Brau}, {Briant}, {Briggs}, {Brighenti}, {Brillet}, {Brinkmann},
  {Brisson}, {Brockill}, {Brooks}, {Brown}, {Brunett}, {Buikema}, {Bulik},
  {Bulten}, {Buonanno}, {Buskulic}, {Bustamante Rosell}, {Buy}, {Byer},
  {Cabero}, {Cadonati}, {Cagnoli}, {Cahillane}, {Calder{\'o}n Bustillo},
  {Callister}, {Calloni}, {Camp}, {Campbell}, {Canepa}, {Cannon}, {Cao}, {Cao},
  {Capocasa}, {Carbognani}, {Caride}, {Carney}, {Carullo}, {Casanueva Diaz},
  {Casentini}, {Caudill}, {Cavagli{\`a}}, {Cavalier}, {Cavalieri}, {Cella},
  {Cerd{\'a}-Dur{\'a}n}, {Cerretani}, {Cesarini}, {Chaibi}, {Chakravarti},
  {Chamberlin}, {Chan}, {Chao}, {Charlton}, {Chase}, {Chassande-Mottin},
  {Chatterjee}, {Chaturvedi}, {Chatziioannou}, {Cheeseboro}, {Chen}, {Chen},
  {Chen}, {Cheng}, {Cheong}, {Chia}, {Chincarini}, {Chiummo}, {Cho}, {Cho},
  {Cho}, {Christensen}, {Chu}, {Chua}, \& {Chung}}]{2019PhRvX...9c1040A}
---. 2019{\natexlab{a}}, Physical Review X, 9, 031040,
  \dodoi{10.1103/PhysRevX.9.031040}

\bibitem[{{Abbott} {et~al.}(2019{\natexlab{b}}){Abbott}, {Abbott}, {Abbott},
  {Abraham}, {Acernese}, {Ackley}, {Adams}, {Adhikari}, {Adya}, {Affeldt},
  {Agathos}, {Agatsuma}, {Aggarwal}, {Aguiar}, {Aiello}, {Ain}, {Ajith},
  {Allen}, {Allocca}, {Aloy}, {Altin}, {Amato}, {Ananyeva}, {Anderson},
  {Anderson}, {Angelova}, {Antier}, {Appert}, {Arai}, {Araya}, {Areeda},
  {Ar{\`e}ne}, {Arnaud}, {Ascenzi}, {Ashton}, {Aston}, {Astone}, {Aubin},
  {Aufmuth}, {AultONeal}, {Austin}, {Avendano}, {Avila-Alvarez}, {Babak},
  {Bacon}, {Badaracco}, {Bader}, {Bae}, {Baker}, {Baldaccini}, {Ballardin},
  {Ballmer}, {Banagiri}, {Barayoga}, {Barclay}, {Barish}, {Barker}, {Barkett},
  {Barnum}, {Barone}, {Barr}, {Barsotti}, {Barsuglia}, {Barta}, {Bartlett},
  {Bartos}, {Bassiri}, {Basti}, {Bawaj}, {Bayley}, {Bazzan}, {B{\'e}csy},
  {Bejger}, {Belahcene}, {Bell}, {Beniwal}, {Berger}, {Bergmann}, {Bernuzzi},
  {Bero}, {Berry}, {Bersanetti}, {Bertolini}, {Betzwieser}, {Bhandare},
  {Bidler}, {Bilenko}, {Bilgili}, {Billingsley}, {Birch}, {Birney},
  {Birnholtz}, {Biscans}, {Biscoveanu}, {Bisht}, {Bitossi}, {Bizouard},
  {Blackburn}, {Blair}, {Blair}, {Blair}, {Bloemen}, {Bode}, {Boer}, {Boetzel},
  {Bogaert}, {Bondu}, {Bonilla}, {Bonnand}, {Booker}, {Boom}, {Booth}, {Bork},
  {Boschi}, {Bose}, {Bossie}, {Bossilkov}, {Bosveld}, {Bouffanais}, {Bozzi},
  {Bradaschia}, {Brady}, {Bramley}, {Branchesi}, {Brau}, {Briant}, {Briggs},
  {Brighenti}, {Brillet}, {Brinkmann}, {Brisson}, {Brockill}, {Brooks},
  {Brown}, {Brunett}, {Buikema}, {Bulik}, {Bulten}, {Buonanno}, {Buskulic},
  {Buy}, {Byer}, {Cabero}, {Cadonati}, {Cagnoli}, {Cahillane}, {Calder{\'o}n
  Bustillo}, {Callister}, {Calloni}, {Camp}, {Campbell}, {Canepa}, {Cannon},
  {Cao}, {Cao}, {Capocasa}, {Carbognani}, {Caride}, {Carney}, {Carullo},
  {Casanueva Diaz}, {Casentini}, {Caudill}, {Cavagli{\`a}}, {Cavalier},
  {Cavalieri}, {Cella}, {Cerd{\'a}-Dur{\'a}n}, {Cerretani}, {Cesarini},
  {Chaibi}, {Chakravarti}, {Chamberlin}, {Chan}, {Chao}, {Charlton}, {Chase},
  {Chassande-Mottin}, {Chatterjee}, {Chaturvedi}, {Cheeseboro}, {Chen}, {Chen},
  {Chen}, {Cheng}, {Cheong}, {Chia}, {Chincarini}, {Chiummo}, {Cho}, {Cho},
  {Cho}, {Christensen}, {Chu}, {Chua}, {Chung}, {Chung}, {Ciani}, {Ciobanu}, \&
  {Ciolfi}}]{2019ApJ...875..161A}
---. 2019{\natexlab{b}}, \apj, 875, 161, \dodoi{10.3847/1538-4357/ab0e8f}

\bibitem[{{Abbott} {et~al.}(2019{\natexlab{c}}){Abbott}, {Abbott}, {Abbott},
  {Acernese}, {Ackley}, {Adams}, {Adams}, {Addesso}, {Adhikari}, {Adya},
  {Affeldt}, {Agarwal}, {Agathos}, {Agatsuma}, {Aggarwal}, {Aguiar}, {Aiello},
  {Ain}, {Ajith}, {Allen}, {Allen}, {Allocca}, {Aloy}, {Altin}, {Amato},
  {Ananyeva}, {Anderson}, {Anderson}, {Angelova}, {Antier}, {Appert}, {Arai},
  {Araya}, {Areeda}, {Ar{\`e}ne}, {Arnaud}, {Arun}, {Ascenzi}, {Ashton}, {Ast},
  {Aston}, {Astone}, {Atallah}, {Aubin}, {Aufmuth}, {Aulbert}, {AultONeal},
  {Austin}, {Avila-Alvarez}, {Babak}, {Bacon}, {Badaracco}, {Bader}, {Bae},
  {Baker}, {Baldaccini}, {Ballardin}, {Ballmer}, {Banagiri}, {Barayoga},
  {Barclay}, {Barish}, {Barker}, {Barkett}, {Barnum}, {Barone}, {Barr},
  {Barsotti}, {Barsuglia}, {Barta}, {Bartlett}, {Bartos}, {Bassiri}, {Basti},
  {Batch}, {Bawaj}, {Bayley}, {Bazzan}, {B{\'e}csy}, {Beer}, {Bejger},
  {Belahcene}, {Bell}, {Beniwal}, {Bensch}, {Berger}, {Bergmann}, {Bernuzzi},
  {Bero}, {Berry}, {Bersanetti}, {Bertolini}, {Betzwieser}, {Bhandare},
  {Bilenko}, {Bilgili}, {Billingsley}, {Billman}, {Birch}, {Birney},
  {Birnholtz}, {Biscans}, {Biscoveanu}, {Bisht}, {Bitossi}, {Bizouard},
  {Blackburn}, {Blackman}, {Blair}, {Blair}, {Blair}, {Bloemen}, {Bock},
  {Bode}, {Boer}, {Boetzel}, {Bogaert}, {Bohe}, {Bondu}, {Bonilla}, {Bonnand},
  {Booker}, {Boom}, {Booth}, {Bork}, {Boschi}, {Bose}, {Bossie}, {Bossilkov},
  {Bosveld}, {Bouffanais}, {Bozzi}, {Bradaschia}, {Brady}, {Bramley},
  {Branchesi}, {Brau}, {Briant}, {Brighenti}, {Brillet}, {Brinkmann},
  {Brisson}, {Brockill}, {Brooks}, {Brown}, {Brunett}, {Buchanan}, {Buikema},
  {Bulik}, {Bulten}, {Buonanno}, {Buskulic}, {Buy}, {Byer}, {Cabero},
  {Cadonati}, {Cagnoli}, {Cahillane}, {Bustillo}, {Callister}, {Calloni},
  {Camp}, {Canepa}, {Canizares}, {Cannon}, {Cao}, {Cao}, {Capano}, {Capocasa},
  {Carbognani}, {Caride}, {Carney}, {Carullo}, {Diaz}, {Casentini}, {Caudill},
  {Cavagli{\`a}}, {Cavalier}, {Cavalieri}, {Cella}, {Cepeda},
  {Cerd{\'a}-Dur{\'a}n}, {Cerretani}, {Cesarini}, {Chaibi}, {Chamberlin},
  {Chan}, {Chao}, {Charlton}, {Chase}, {Chassande-Mottin}, {Chatterjee},
  {Chatziioannou}, {Cheeseboro}, {Chen}, {Chen}, {Chen}, {Cheng}, {Chia}, \&
  {Chincarini}}]{2019PhRvX...9a1001A}
---. 2019{\natexlab{c}}, Physical Review X, 9, 011001,
  \dodoi{10.1103/PhysRevX.9.011001}

\bibitem[{Abbott {et~al.}(2020)}]{KAGRA:2013rdx}
Abbott, B.~P., {et~al.} 2020, Living Rev. Rel., 23, 3,
  \dodoi{10.1007/s41114-020-00026-9}

\bibitem[{{Abbott} {et~al.}(2020{\natexlab{a}}){Abbott}, {Abbott}, {Abbott},
  {Abraham}, {Acernese}, {Ackley}, {Adams}, {Adya}, {Affeldt}, {Agathos},
  {Agatsuma}, {Aggarwal}, {Aguiar}, {Aiello}, {Ain}, {Ajith}, {Alford},
  {Allen}, {Allocca}, {Aloy}, {Altin}, {Amato}, {Ananyeva}, {Anderson},
  {Anderson}, {Angelova}, {Antier}, {Appert}, {Arai}, {Araya}, {Areeda},
  {Ar{\`e}ne}, {Arnaud}, {Arun}, {Ascenzi}, {Ashton}, {Aston}, {Astone},
  {Aubin}, {Aufmuth}, {AultONeal}, {Austin}, {Avendano}, {Avila-Alvarez},
  {Babak}, {Bacon}, {Badaracco}, {Bader}, {Bae}, {Baker}, {Baldaccini},
  {Ballardin}, {Ballmer}, {Banagiri}, {Barayoga}, {Barclay}, {Barish},
  {Barker}, {Barkett}, {Barnum}, {Barone}, {Barr}, {Barsotti}, {Barsuglia},
  {Barta}, {Bartlett}, {Bartos}, {Bassiri}, {Basti}, {Bawaj}, {Bayley},
  {Bazzan}, {B{\'e}csy}, {Bejger}, {Belahcene}, {Bell}, {Beniwal}, {Berger},
  {Bergmann}, {Bernuzzi}, {Bero}, {Berry}, {Bersanetti}, {Bertolini},
  {Betzwieser}, {Bhandare}, {Bidler}, {Bilenko}, {Bilgili}, {Billingsley},
  {Birch}, {Birney}, {Birnholtz}, {Biscans}, {Biscoveanu}, {Bisht}, {Bitossi},
  {Bizouard}, {Blackburn}, {Blair}, {Blair}, {Blair}, {Bloemen}, {Bode},
  {Boer}, {Boetzel}, {Bogaert}, {Bondu}, {Bonilla}, {Bonnand}, {Booker},
  {Boom}, {Booth}, {Bork}, {Boschi}, {Bose}, {Bossie}, {Bossilkov}, {Bosveld},
  {Bouffanais}, {Bozzi}, {Bradaschia}, {Brady}, {Bramley}, {Branchesi}, {Brau},
  {Briant}, {Briggs}, {Brighenti}, {Brillet}, {Brinkmann}, {Brisson},
  {Brockill}, {Brooks}, {Brown}, {Brunett}, {Buikema}, {Bulik}, {Bulten},
  {Buonanno}, {Buskulic}, {Buy}, {Byer}, {Cabero}, {Cadonati}, {Cagnoli},
  {Cahillane}, {Calder{\'o}n Bustillo}, {Callister}, {Calloni}, {Camp},
  {Campbell}, {Canepa}, {Cannon}, {Cao}, {Cao}, {Capocasa}, {Carbognani},
  {Caride}, {Carney}, {Carullo}, {Casanueva Diaz}, {Casentini}, {Caudill},
  {Cavagli{\`a}}, {Cavalier}, {Cavalieri}, {Cella}, {Cerd{\'a}-Dur{\'a}n},
  {Cerretani}, {Cesarini}, {Chaibi}, {Chakravarti}, {Chamberlin}, {Chan},
  {Chao}, {Charlton}, {Chase}, {Chassande-Mottin}, {Chatterjee}, {Chaturvedi},
  {Chatziioannou}, {Cheeseboro}, {Chen}, {Chen}, {Chen}, {Cheng}, {Cheong},
  {Chia}, {Chincarini}, {Chiummo}, {Cho}, {Cho}, {Cho}, {Christensen}, {Chu},
  {Chua}, {Chung}, {Chung}, \& {Ciani}}]{2020CQGra..37e5002A}
{Abbott}, B.~P., {Abbott}, R., {Abbott}, T.~D., {et~al.} 2020{\natexlab{a}},
  Classical and Quantum Gravity, 37, 055002, \dodoi{10.1088/1361-6382/ab685e}

\bibitem[{{Abbott} {et~al.}(2020{\natexlab{b}}){Abbott}, {Abbott}, {Abbott},
  {Abraham}, {Acernese}, {Ackley}, {Adams}, {Adhikari}, {Adya}, {Affeldt},
  {Agathos}, {Agatsuma}, {Aggarwal}, {Aguiar}, {Aiello}, {Ain}, {Ajith},
  {Allen}, {Allocca}, {Aloy}, {Altin}, {Amato}, {Anand}, {Ananyeva},
  {Anderson}, {Anderson}, {Angelova}, {Antier}, {Appert}, {Arai}, {Araya},
  {Areeda}, {Ar{\`e}ne}, {Arnaud}, {Aronson}, {Arun}, {Ascenzi}, {Ashton},
  {Aston}, {Astone}, {Aubin}, {Aufmuth}, {AultONeal}, {Austin}, {Avendano},
  {Avila-Alvarez}, {Babak}, {Bacon}, {Badaracco}, {Bader}, {Bae}, {Baird},
  {Baker}, {Baldaccini}, {Ballardin}, {Ballmer}, {Bals}, {Banagiri},
  {Barayoga}, {Barbieri}, {Barclay}, {Barish}, {Barker}, {Barkett}, {Barnum},
  {Barone}, {Barr}, {Barsotti}, {Barsuglia}, {Barta}, {Bartlett}, {Bartos},
  {Bassiri}, {Basti}, {Bawaj}, {Bayley}, {Baylor}, {Bazzan}, {B{\'e}csy},
  {Bejger}, {Belahcene}, {Bell}, {Beniwal}, {Benjamin}, {Berger}, {Bergmann},
  {Bernuzzi}, {Berry}, {Bersanetti}, {Bertolini}, {Betzwieser}, {Bhandare},
  {Bidler}, {Biggs}, {Bilenko}, {Bilgili}, {Billingsley}, {Birney},
  {Birnholtz}, {Biscans}, {Bischi}, {Biscoveanu}, {Bisht}, {Bitossi},
  {Bizouard}, {Blackburn}, {Blackman}, {Blair}, {Blair}, {Blair}, {Bloemen},
  {Bobba}, {Bode}, {Boer}, {Boetzel}, {Bogaert}, {Bondu}, {Bonnand}, {Booker},
  {Boom}, {Bork}, {Boschi}, {Bose}, {Bossilkov}, {Bosveld}, {Bouffanais},
  {Bozzi}, {Bradaschia}, {Brady}, {Bramley}, {Branchesi}, {Brau}, {Breschi},
  {Briant}, {Briggs}, {Brighenti}, {Brillet}, {Brinkmann}, {Brockill},
  {Brooks}, {Brooks}, {Brown}, {Brunett}, {Buikema}, {Bulik}, {Bulten},
  {Buonanno}, {Buskulic}, {Buy}, {Byer}, {Cabero}, {Cadonati}, {Cagnoli},
  {Cahillane}, {Calder{\'o}n Bustillo}, {Callister}, {Calloni}, {Camp},
  {Campbell}, {Canepa}, {Cannon}, {Cao}, {Cao}, {Carapella}, {Carbognani},
  {Caride}, {Carney}, {Carullo}, {Casanueva Diaz}, {Casentini}, {Caudill},
  {Cavagli{\`a}}, {Cavalier}, {Cavalieri}, {Cella}, {Cerd{\'a}-Dur{\'a}n},
  {Cesarini}, {Chaibi}, {Chakravarti}, {Chamberlin}, {Chan}, {Chao},
  {Charlton}, {Chase}, {Chassande-Mottin}, {Chatterjee}, {Chaturvedi},
  {Chatziioannou}, {Cheeseboro}, {Chen}, {Chen}, {Chen}, {Cheng}, {Cheong},
  {Chia}, {Chiadini}, {Chincarini}, {Chiummo}, {Cho}, \&
  {Cho}}]{2020ApJ...892L...3A}
---. 2020{\natexlab{b}}, \apjl, 892, L3, \dodoi{10.3847/2041-8213/ab75f5}

\bibitem[{{Abbott} {et~al.}(2020{\natexlab{c}}){Abbott}, {Abbott}, {Abraham},
  {Acernese}, {Ackley}, {Adams}, {Adhikari}, {Adya}, {Affeldt}, {Agathos},
  {Agatsuma}, {Aggarwal}, {Aguiar}, {Aich}, {Aiello}, {Ain}, {Ajith}, {Akcay},
  {Allen}, {Allocca}, {Altin}, {Amato}, {Anand}, {Ananyeva}, {Anderson},
  {Anderson}, {Angelova}, {Ansoldi}, {Antier}, {Appert}, {Arai}, {Araya},
  {Areeda}, {Ar{\`e}ne}, {Arnaud}, {Aronson}, {Arun}, {Asali}, {Ascenzi},
  {Ashton}, {Aston}, {Astone}, {Aubin}, {Aufmuth}, {AultONeal}, {Austin},
  {Avendano}, {Babak}, {Bacon}, {Badaracco}, {Bader}, {Bae}, {Baer}, {Baird},
  {Baldaccini}, {Ballardin}, {Ballmer}, {Bals}, {Balsamo}, {Baltus},
  {Banagiri}, {Bankar}, {Bankar}, {Barayoga}, {Barbieri}, {Barish}, {Barker},
  {Barkett}, {Barneo}, {Barone}, {Barr}, {Barsotti}, {Barsuglia}, {Barta},
  {Bartlett}, {Bartos}, {Bassiri}, {Basti}, {Bawaj}, {Bayley}, {Bazzan},
  {B{\'e}csy}, {Bejger}, {Belahcene}, {Bell}, {Beniwal}, {Benjamin}, {Benkel},
  {Bentley}, {Bergamin}, {Berger}, {Bergmann}, {Bernuzzi}, {Berry},
  {Bersanetti}, {Bertolini}, {Betzwieser}, {Bhandare}, {Bhandari}, {Bidler},
  {Biggs}, {Bilenko}, {Billingsley}, {Birney}, {Birnholtz}, {Biscans},
  {Bischi}, {Biscoveanu}, {Bisht}, {Bissenbayeva}, {Bitossi}, {Bizouard},
  {Blackburn}, {Blackman}, {Blair}, {Blair}, {Blair}, {Bobba}, {Bode}, {Boer},
  {Boetzel}, {Bogaert}, {Bondu}, {Bonilla}, {Bonnand}, {Booker}, {Boom},
  {Bork}, {Boschi}, {Bose}, {Bossilkov}, {Bosveld}, {Bouffanais}, {Bozzi},
  {Bradaschia}, {Brady}, {Bramley}, {Branchesi}, {Brau}, {Breschi}, {Briant},
  {Briggs}, {Brighenti}, {Brillet}, {Brinkmann}, {Brito}, {Brockill}, {Brooks},
  {Brooks}, {Brown}, {Brunett}, {Bruno}, {Bruntz}, {Buikema}, {Bulik},
  {Bulten}, {Buonanno}, {Buskulic}, {Byer}, {Cabero}, {Cadonati}, {Cagnoli},
  {Cahillane}, {Calder{\'o}n Bustillo}, {Callaghan}, {Callister}, {Calloni},
  {Camp}, {Canepa}, {Cannon}, {Cao}, {Cao}, {Carapella}, {Carbognani},
  {Caride}, {Carney}, {Carullo}, {Casanueva Diaz}, {Casentini},
  {Casta{\~n}eda}, {Caudill}, {Cavagli{\`a}}, {Cavalier}, {Cavalieri}, {Cella},
  {Cerd{\'a}-Dur{\'a}n}, {Cesarini}, {Chaibi}, {Chakravarti}, {Chan}, {Chan},
  {Chao}, {Charlton}, {Chase}, {Chassande-Mottin}, {Chatterjee}, {Chaturvedi},
  {Chatziioannou}, {Chen}, \& {Chen}}]{2020PhRvD.102d3015A}
{Abbott}, R., {Abbott}, T.~D., {Abraham}, S., {et~al.} 2020{\natexlab{c}},
  \prd, 102, 043015, \dodoi{10.1103/PhysRevD.102.043015}

\bibitem[{{Abbott} {et~al.}(2020{\natexlab{d}}){Abbott}, {Abbott}, {Abraham},
  {Acernese}, {Ackley}, {Adams}, {Adhikari}, {Adya}, {Affeldt}, {Agathos},
  {Agatsuma}, {Aggarwal}, {Aguiar}, {Aich}, {Aiello}, {Ain}, {Ajith}, {Akcay},
  {Allen}, {Allocca}, {Altin}, {Amato}, {Anand}, {Ananyeva}, {Anderson},
  {Anderson}, {Angelova}, {Ansoldi}, {Antier}, {Appert}, {Arai}, {Araya},
  {Areeda}, {Ar{\`e}ne}, {Arnaud}, {Aronson}, {Arun}, {Asali}, {Ascenzi},
  {Ashton}, {Aston}, {Astone}, {Aubin}, {Aufmuth}, {AultONeal}, {Austin},
  {Avendano}, {Babak}, {Bacon}, {Badaracco}, {Bader}, {Bae}, {Baer}, {Baird},
  {Baldaccini}, {Ballardin}, {Ballmer}, {Bals}, {Balsamo}, {Baltus},
  {Banagiri}, {Bankar}, {Bankar}, {Barayoga}, {Barbieri}, {Barish}, {Barker},
  {Barkett}, {Barneo}, {Barone}, {Barr}, {Barsotti}, {Barsuglia}, {Barta},
  {Bartlett}, {Bartos}, {Bassiri}, {Basti}, {Bawaj}, {Bayley}, {Bazzan},
  {B{\'e}csy}, {Bejger}, {Belahcene}, {Bell}, {Beniwal}, {Benjamin}, {Bentley},
  {Bergamin}, {Berger}, {Bergmann}, {Bernuzzi}, {Berry}, {Bersanetti},
  {Bertolini}, {Betzwieser}, {Bhandare}, {Bhandari}, {Bidler}, {Biggs},
  {Bilenko}, {Billingsley}, {Birney}, {Birnholtz}, {Biscans}, {Bischi},
  {Biscoveanu}, {Bisht}, {Bissenbayeva}, {Bitossi}, {Bizouard}, {Blackburn},
  {Blackman}, {Blair}, {Blair}, {Blair}, {Bobba}, {Bode}, {Boer}, {Boetzel},
  {Bogaert}, {Bondu}, {Bonilla}, {Bonnand}, {Booker}, {Boom}, {Bork}, {Boschi},
  {Bose}, {Bossilkov}, {Bosveld}, {Bouffanais}, {Bozzi}, {Bradaschia}, {Brady},
  {Bramley}, {Branchesi}, {Brau}, {Breschi}, {Briant}, {Briggs}, {Brighenti},
  {Brillet}, {Brinkmann}, {Brockill}, {Brooks}, {Brooks}, {Brown}, {Brunett},
  {Bruno}, {Bruntz}, {Buikema}, {Bulik}, {Bulten}, {Buonanno}, {Buscicchio},
  {Buskulic}, {Byer}, {Cabero}, {Cadonati}, {Cagnoli}, {Cahillane},
  {Calder{\'o}n Bustillo}, {Callaghan}, {Callister}, {Calloni}, {Camp},
  {Canepa}, {Cannon}, {Cao}, {Cao}, {Carapella}, {Carbognani}, {Caride},
  {Carney}, {Carullo}, {Casanueva Diaz}, {Casentini}, {Casta{\~n}eda},
  {Caudill}, {Cavagli{\`a}}, {Cavalier}, {Cavalieri}, {Cella},
  {Cerd{\'a}-Dur{\'a}n}, {Cesarini}, {Chaibi}, {Chakravarti}, {Chan}, {Chan},
  {Chandra}, {Chao}, {Charlton}, {Chase}, {Chassande-Mottin}, {Chatterjee},
  {Chaturvedi}, {Chatziioannou}, {Chen}, \& {Chen}}]{2020PhRvL.125j1102A}
---. 2020{\natexlab{d}}, \prl, 125, 101102,
  \dodoi{10.1103/PhysRevLett.125.101102}

\bibitem[{{Abbott} {et~al.}(2021{\natexlab{a}}){Abbott}, {Abbott}, {Abraham},
  {Acernese}, {Ackley}, {Adams}, {Adams}, {Adhikari}, {Adya}, {Affeldt},
  {Agathos}, {Agatsuma}, {Aggarwal}, {Aguiar}, {Aiello}, {Ain}, {Ajith},
  {Akcay}, {Allen}, {Allocca}, {Altin}, {Amato}, {Anand}, {Ananyeva},
  {Anderson}, {Anderson}, {Angelova}, {Ansoldi}, {Antelis}, {Antier}, {Appert},
  {Arai}, {Araya}, {Areeda}, {Ar{\`e}ne}, {Arnaud}, {Aronson}, {Arun}, {Asali},
  {Ascenzi}, {Ashton}, {Aston}, {Astone}, {Aubin}, {Aufmuth}, {AultONeal},
  {Austin}, {Avendano}, {Babak}, {Badaracco}, {Bader}, {Bae}, {Baer},
  {Bagnasco}, {Baird}, {Ball}, {Ballardin}, {Ballmer}, {Bals}, {Balsamo},
  {Baltus}, {Banagiri}, {Bankar}, {Bankar}, {Barayoga}, {Barbieri}, {Barish},
  {Barker}, {Barneo}, {Barnum}, {Barone}, {Barr}, {Barsotti}, {Barsuglia},
  {Barta}, {Bartlett}, {Bartos}, {Bassiri}, {Basti}, {Bawaj}, {Bayley},
  {Bazzan}, {Becher}, {B{\'e}csy}, {Bedakihale}, {Bejger}, {Belahcene},
  {Beniwal}, {Benjamin}, {Bennett}, {Bentley}, {Bergamin}, {Berger},
  {Bergmann}, {Bernuzzi}, {Berry}, {Bersanetti}, {Bertolini}, {Betzwieser},
  {Bhandare}, {Bhandari}, {Bhattacharjee}, {Bidler}, {Bilenko}, {Billingsley},
  {Birney}, {Birnholtz}, {Biscans}, {Bischi}, {Biscoveanu}, {Bisht}, {Bitossi},
  {Bizouard}, {Blackburn}, {Blackman}, {Blair}, {Blair}, {Blair}, {Blanch},
  {Bobba}, {Bode}, {Boer}, {Boetzel}, {Bogaert}, {Boldrini}, {Bondu},
  {Bonilla}, {Bonnand}, {Booker}, {Boom}, {Bork}, {Boschi}, {Bose},
  {Bossilkov}, {Boudart}, {Bouffanais}, {Bozzi}, {Bradaschia}, {Brady},
  {Bramley}, {Branchesi}, {Brau}, {Breschi}, {Briant}, {Briggs}, {Brighenti},
  {Brillet}, {Brinkmann}, {Brockill}, {Brooks}, {Brooks}, {Brown}, {Brunett},
  {Bruno}, {Bruntz}, {Buikema}, {Bulik}, {Bulten}, {Buonanno}, {Buscicchio},
  {Buskulic}, {Byer}, {Cabero}, {Cadonati}, {Caesar}, {Cagnoli}, {Cahillane},
  {Calder{\'o}n Bustillo}, {Callaghan}, {Callister}, {Calloni}, {Camp},
  {Canepa}, {Cannon}, {Cao}, {Cao}, {Carapella}, {Carbognani}, {Carney},
  {Carpinelli}, {Carullo}, {Carver}, {Casanueva Diaz}, {Casentini}, {Caudill},
  {Cavagli{\`a}}, {Cavalier}, {Cavalieri}, {Cella}, {Cerd{\'a}-Dur{\'a}n},
  {Cesarini}, {Chaibi}, {Chakravarti}, {Chan}, {Chan}, {Chandra}, {Chanial},
  {Chao}, {Charlton}, \& {Chase}}]{2021PhRvX..11b1053A}
---. 2021{\natexlab{a}}, Physical Review X, 11, 021053,
  \dodoi{10.1103/PhysRevX.11.021053}

\bibitem[{{Abbott} {et~al.}(2021{\natexlab{b}}){Abbott}, {Abbott}, {Abraham},
  {Acernese}, {Ackley}, {Adams}, {Adams}, {Adhikari}, {Adya}, {Affeldt},
  {Agarwal}, {Agathos}, {Agatsuma}, {Aggarwal}, {Aguiar}, {Aiello}, {Ain},
  {Ajith}, {Akutsu}, {Aleman}, {Allen}, {Allocca}, {Altin}, {Amato}, {Anand},
  {Ananyeva}, {Anderson}, {Anderson}, {Ando}, {Angelova}, {Ansoldi}, {Antelis},
  {Antier}, {Appert}, {Arai}, {Arai}, {Arai}, {Araki}, {Araya}, {Araya},
  {Areeda}, {Ar{\`e}ne}, {Aritomi}, {Arnaud}, {Aronson}, {Arun}, {Asada},
  {Asali}, {Ashton}, {Aso}, {Aston}, {Astone}, {Aubin}, {Aufmuth}, {Aultoneal},
  {Austin}, {Babak}, {Badaracco}, {Bader}, {Bae}, {Bae}, {Baer}, {Bagnasco},
  {Bai}, {Baiotti}, {Baird}, {Bajpai}, {Ball}, {Ballardin}, {Ballmer}, {Bals},
  {Balsamo}, {Baltus}, {Banagiri}, {Bankar}, {Bankar}, {Barayoga}, {Barbieri},
  {Barish}, {Barker}, {Barneo}, {Barone}, {Barr}, {Barsotti}, {Barsuglia},
  {Barta}, {Bartlett}, {Barton}, {Bartos}, {Bassiri}, {Basti}, {Bawaj},
  {Bayley}, {Baylor}, {Bazzan}, {B{\'e}csy}, {Bedakihale}, {Bejger},
  {Belahcene}, {Benedetto}, {Beniwal}, {Benjamin}, {Benkel}, {Bennett},
  {Bentley}, {Benyaala}, {Bergamin}, {Berger}, {Bernuzzi}, {Berry},
  {Bersanetti}, {Bertolini}, {Betzwieser}, {Bhandare}, {Bhandari},
  {Bhattacharjee}, {Bhaumik}, {Bidler}, {Bilenko}, {Billingsley}, {Birney},
  {Birnholtz}, {Biscans}, {Bischi}, {Biscoveanu}, {Bisht}, {Biswas}, {Bitossi},
  {Bizouard}, {Blackburn}, {Blackman}, {Blair}, {Blair}, {Blair}, {Bobba},
  {Bode}, {Boer}, {Bogaert}, {Boldrini}, {Bondu}, {Bonilla}, {Bonnand},
  {Booker}, {Boom}, {Bork}, {Boschi}, {Bose}, {Bose}, {Bossilkov}, {Boudart},
  {Bouffanais}, {Bozzi}, {Bradaschia}, {Brady}, {Bramley}, {Branch},
  {Branchesi}, {Brau}, {Breschi}, {Briant}, {Briggs}, {Brillet}, {Brinkmann},
  {Brockill}, {Brooks}, {Brooks}, {Brown}, {Brunett}, {Bruno}, {Bruntz},
  {Bryant}, {Buikema}, {Bulik}, {Bulten}, {Buonanno}, {Buscicchio}, {Buskulic},
  {Byer}, {Cadonati}, {Caesar}, {Cagnoli}, {Cahillane}, {Cain}, {Calder{\'o}n
  Bustillo}, {Callaghan}, {Callister}, {Calloni}, {Camp}, {Canepa},
  {Cannavacciuolo}, {Cannon}, {Cao}, {Cao}, {Cao}, {Capocasa}, {Capote},
  {Carapella}, {Carbognani}, {Carlin}, \& {Carney}}]{2021ApJ...915L...5A}
---. 2021{\natexlab{b}}, \apjl, 915, L5, \dodoi{10.3847/2041-8213/ac082e}

\bibitem[{{Abbott} {et~al.}(2023{\natexlab{a}}){Abbott}, {Abbott}, {Acernese},
  {Ackley}, {Adams}, {Adhikari}, {Adhikari}, {Adya}, {Affeldt}, {Agarwal},
  {Agathos}, {Agatsuma}, {Aggarwal}, {Aguiar}, {Aiello}, {Ain}, {Ajith},
  {Akcay}, {Akutsu}, {Albanesi}, {Allocca}, {Altin}, {Amato}, {Anand}, {Anand},
  {Ananyeva}, {Anderson}, {Anderson}, {Ando}, {Andrade}, {Andres},
  {Andri{\'c}}, {Angelova}, {Ansoldi}, {Antelis}, {Antier}, {Appert}, {Arai},
  {Arai}, {Arai}, {Araki}, {Araya}, {Araya}, {Areeda}, {Ar{\`e}ne}, {Aritomi},
  {Arnaud}, {Arogeti}, {Aronson}, {Arun}, {Asada}, {Asali}, {Ashton}, {Aso},
  {Assiduo}, {Aston}, {Astone}, {Aubin}, {Austin}, {Babak}, {Badaracco},
  {Bader}, {Badger}, {Bae}, {Bae}, {Baer}, {Bagnasco}, {Bai}, {Baiotti},
  {Baird}, {Bajpai}, {Ball}, {Ballardin}, {Ballmer}, {Balsamo}, {Baltus},
  {Banagiri}, {Bankar}, {Barayoga}, {Barbieri}, {Barish}, {Barker}, {Barneo},
  {Barone}, {Barr}, {Barsotti}, {Barsuglia}, {Barta}, {Bartlett}, {Barton},
  {Bartos}, {Bassiri}, {Basti}, {Bawaj}, {Bayley}, {Baylor}, {Bazzan},
  {B{\'e}csy}, {Bedakihale}, {Bejger}, {Belahcene}, {Benedetto}, {Beniwal},
  {Bennett}, {Bentley}, {Benyaala}, {Bergamin}, {Berger}, {Bernuzzi}, {Berry},
  {Bersanetti}, {Bertolini}, {Betzwieser}, {Beveridge}, {Bhandare}, {Bhardwaj},
  {Bhattacharjee}, {Bhaumik}, {Bilenko}, {Billingsley}, {Bini}, {Birney},
  {Birnholtz}, {Biscans}, {Bischi}, {Biscoveanu}, {Bisht}, {Biswas}, {Bitossi},
  {Bizouard}, {Blackburn}, {Blair}, {Blair}, {Blair}, {Bobba}, {Bode}, {Boer},
  {Bogaert}, {Boldrini}, {Bonavena}, {Bondu}, {Bonilla}, {Bonnand}, {Booker},
  {Boom}, {Bork}, {Boschi}, {Bose}, {Bose}, {Bossilkov}, {Boudart},
  {Bouffanais}, {Bozzi}, {Bradaschia}, {Brady}, {Bramley}, {Branch},
  {Branchesi}, {Brandt}, {Brau}, {Breschi}, {Briant}, {Briggs}, {Brillet},
  {Brinkmann}, {Brockill}, {Brooks}, {Brooks}, {Brown}, {Brunett}, {Bruno},
  {Bruntz}, {Bryant}, {Bulik}, {Bulten}, {Buonanno}, {Buscicchio}, {Buskulic},
  {Buy}, {Byer}, {Davies}, {Cadonati}, {Cagnoli}, {Cahillane}, {Bustillo},
  {Callaghan}, {Callister}, {Calloni}, {Cameron}, {Camp}, {Canepa},
  {Canevarolo}, {Cannavacciuolo}, {Cannon}, {Cao}, {Cao}, {Capocasa}, {Capote},
  {Carapella}, \& {Carbognani}}]{2023PhRvX..13d1039A}
{Abbott}, R., {Abbott}, T.~D., {Acernese}, F., {et~al.} 2023{\natexlab{a}},
  Physical Review X, 13, 041039, \dodoi{10.1103/PhysRevX.13.041039}

\bibitem[{{Abbott} {et~al.}(2023{\natexlab{b}}){Abbott}, {Abbott}, {Acernese},
  {Ackley}, {Adams}, {Adhikari}, {Adhikari}, {Adya}, {Affeldt}, {Agarwal},
  {Agathos}, {Agatsuma}, {Aggarwal}, {Aguiar}, {Aiello}, {Ain}, {Ajith},
  {Akutsu}, {de Alarc{\'o}n}, {Akcay}, {Albanesi}, {Allocca}, {Altin}, {Amato},
  {Anand}, {Anand}, {Ananyeva}, {Anderson}, {Anderson}, {Ando}, {Andrade},
  {Andres}, {Andri{\'c}}, {Angelova}, {Ansoldi}, {Antelis}, {Antier},
  {Antonini}, {Appert}, {Arai}, {Arai}, {Arai}, {Araki}, {Araya}, {Araya},
  {Areeda}, {Ar{\`e}ne}, {Aritomi}, {Arnaud}, {Arogeti}, {Aronson}, {Arun},
  {Asada}, {Asali}, {Ashton}, {Aso}, {Assiduo}, {Aston}, {Astone}, {Aubin},
  {Austin}, {Babak}, {Badaracco}, {Bader}, {Badger}, {Bae}, {Bae}, {Baer},
  {Bagnasco}, {Bai}, {Baiotti}, {Baird}, {Bajpai}, {Ball}, {Ballardin},
  {Ballmer}, {Balsamo}, {Baltus}, {Banagiri}, {Bankar}, {Barayoga}, {Barbieri},
  {Barish}, {Barker}, {Barneo}, {Barone}, {Barr}, {Barsotti}, {Barsuglia},
  {Barta}, {Bartlett}, {Barton}, {Bartos}, {Bassiri}, {Basti}, {Bawaj},
  {Bayley}, {Baylor}, {Bazzan}, {B{\'e}csy}, {Bedakihale}, {Bejger},
  {Belahcene}, {Benedetto}, {Beniwal}, {Bennett}, {Bentley}, {Benyaala},
  {Bergamin}, {Berger}, {Bernuzzi}, {Berry}, {Bersanetti}, {Bertolini},
  {Betzwieser}, {Beveridge}, {Bhandare}, {Bhardwaj}, {Bhattacharjee},
  {Bhaumik}, {Bilenko}, {Billingsley}, {Bini}, {Birney}, {Birnholtz},
  {Biscans}, {Bischi}, {Biscoveanu}, {Bisht}, {Biswas}, {Bitossi}, {Bizouard},
  {Blackburn}, {Blair}, {Blair}, {Blair}, {Bobba}, {Bode}, {Boer}, {Bogaert},
  {Boldrini}, {Bonavena}, {Bondu}, {Bonilla}, {Bonnand}, {Booker}, {Boom},
  {Bork}, {Boschi}, {Bose}, {Bose}, {Bossilkov}, {Boudart}, {Bouffanais},
  {Bozzi}, {Bradaschia}, {Brady}, {Bramley}, {Branch}, {Branchesi}, {Brandt},
  {Brau}, {Breschi}, {Briant}, {Briggs}, {Brillet}, {Brinkmann}, {Brockill},
  {Brooks}, {Brooks}, {Brown}, {Brunett}, {Bruno}, {Bruntz}, {Bryant}, {Bulik},
  {Bulten}, {Buonanno}, {Buscicchio}, {Buskulic}, {Buy}, {Byer}, {Cadonati},
  {Cagnoli}, {Cahillane}, {Bustillo}, {Callaghan}, {Callister}, {Calloni},
  {Cameron}, {Camp}, {Canepa}, {Canevarolo}, {Cannavacciuolo}, {Cannon}, {Cao},
  {Cao}, {Capocasa}, {Capote}, \& {Carapella}}]{2023PhRvX..13a1048A}
---. 2023{\natexlab{b}}, Physical Review X, 13, 011048,
  \dodoi{10.1103/PhysRevX.13.011048}

\bibitem[{{Abbott} {et~al.}(2024){Abbott}, {Abbott}, {Acernese}, {Ackley},
  {Adams}, {Adhikari}, {Adhikari}, {Adya}, {Affeldt}, {Agarwal}, {Agathos},
  {Agatsuma}, {Aggarwal}, {Aguiar}, {Aiello}, {Ain}, {Ajith}, {Albanesi},
  {Allocca}, {Altin}, {Amato}, {Anand}, {Anand}, {Ananyeva}, {Anderson},
  {Anderson}, {Andrade}, {Andres}, {Andri{\'c}}, {Angelova}, {Ansoldi},
  {Antelis}, {Antier}, {Appert}, {Arai}, {Araya}, {Areeda}, {Ar{\`e}ne},
  {Arnaud}, {Aronson}, {Arun}, {Asali}, {Ashton}, {Assiduo}, {Aston}, {Astone},
  {Aubin}, {Austin}, {Babak}, {Badaracco}, {Bader}, {Badger}, {Bae}, {Baer},
  {Bagnasco}, {Bai}, {Baird}, {Ball}, {Ballardin}, {Ballmer}, {Balsamo},
  {Baltus}, {Banagiri}, {Bankar}, {Barayoga}, {Barbieri}, {Barish}, {Barker},
  {Barneo}, {Barone}, {Barr}, {Barsotti}, {Barsuglia}, {Barta}, {Bartlett},
  {Barton}, {Bartos}, {Bassiri}, {Basti}, {Bawaj}, {Bayley}, {Baylor},
  {Bazzan}, {B{\'e}csy}, {Bedakihale}, {Bejger}, {Belahcene}, {Benedetto},
  {Beniwal}, {Bennett}, {Bentley}, {BenYaala}, {Bergamin}, {Berger},
  {Bernuzzi}, {Berry}, {Bersanetti}, {Bertolini}, {Betzwieser}, {Beveridge},
  {Bhandare}, {Bhardwaj}, {Bhattacharjee}, {Bhaumik}, {Bilenko}, {Billingsley},
  {Bini}, {Birney}, {Birnholtz}, {Biscans}, {Bischi}, {Biscoveanu}, {Bisht},
  {Biswas}, {Bitossi}, {Bizouard}, {Blackburn}, {Blair}, {Blair}, {Blair},
  {Bobba}, {Bode}, {Boer}, {Bogaert}, {Boldrini}, {Bonavena}, {Bondu},
  {Bonilla}, {Bonnand}, {Booker}, {Boom}, {Bork}, {Boschi}, {Bose}, {Bose},
  {Bossilkov}, {Boudart}, {Bouffanais}, {Bozzi}, {Bradaschia}, {Brady},
  {Bramley}, {Branch}, {Branchesi}, {Brau}, {Breschi}, {Briant}, {Briggs},
  {Brillet}, {Brinkmann}, {Brockill}, {Brooks}, {Brooks}, {Brown}, {Brunett},
  {Bruno}, {Bruntz}, {Bryant}, {Bulik}, {Bulten}, {Buonanno}, {Buscicchio},
  {Buskulic}, {Buy}, {Byer}, {Cadonati}, {Cagnoli}, {Cahillane}, {Bustillo},
  {Callaghan}, {Callister}, {Calloni}, {Cameron}, {Camp}, {Canepa},
  {Canevarolo}, {Cannavacciuolo}, {Cannon}, {Cao}, {Capote}, {Carapella},
  {Carbognani}, {Carlin}, {Carney}, {Carpinelli}, {Carrillo}, {Carullo},
  {Carver}, {Diaz}, {Casentini}, {Castaldi}, {Caudill}, {Cavagli{\`a}},
  {Cavalier}, {Cavalieri}, {Ceasar}, {Cella}, {Cerd{\'a}-Dur{\'a}n},
  {Cesarini}, \& {Chaibi}}]{2024PhRvD.109b2001A}
---. 2024, \prd, 109, 022001, \dodoi{10.1103/PhysRevD.109.022001}

\bibitem[{{Accadia} {et~al.}(2014){Accadia}, {Acernese}, {Agathos}, {Allocca},
  {Astone}, {Ballardin}, {Barone}, {Barsuglia}, {Basti}, {Bauer}, {Bejger},
  {Beker}, {Belczynski}, {Bersanetti}, {Bertolini}, {Bitossi}, {Bizouard},
  {Blom}, {Boer}, {Bondu}, {Bonelli}, {Bonnand}, {Boschi}, {Bosi},
  {Bradaschia}, {Branchesi}, {Briant}, {Brillet}, {Brisson}, {Bulik}, {Bulten},
  {Buskulic}, {Buy}, {Cagnoli}, {Calloni}, {Canuel}, {Carbognani}, {Cavalier},
  {Cavalieri}, {Cella}, {Cesarini}, {Chassande-Mottin}, {Chincarini},
  {Chiummo}, {Cleva}, {Coccia}, {Cohadon}, {Colla}, {Colombini}, {Conte},
  {Coulon}, {Cuoco}, {D'Antonio}, {Dattilo}, {Davier}, {Day}, {Debreczeni},
  {Degallaix}, {Del{\'e}glise}, {Del Pozzo}, {Dereli}, {De Rosa}, {Di Fiore},
  {Di Lieto}, {Di Virgilio}, {Drago}, {Endr{\H{o}}czi}, {Fafone}, {Farinon},
  {Ferrante}, {Ferrini}, {Fidecaro}, {Fiori}, {Flaminio}, {Fournier}, {Franco},
  {Frasca}, {Frasconi}, {Gammaitoni}, {Garufi}, {Gemme}, {Genin}, {Gennai},
  {Giazotto}, {Gouaty}, {Granata}, {Groot}, {Guidi}, {Heidmann}, {Heitmann},
  {Hello}, {Hemming}, {Jaranowski}, {Jonker}, {Kasprzack}, {K{\'e}f{\'e}lian},
  {Kowalska}, {Kr{\'o}lak}, {Kutynia}, {Lazzaro}, {Leonardi}, {Leroy},
  {Letendre}, {Li}, {Lorenzini}, {Loriette}, {Losurdo}, {Majorana},
  {Maksimovic}, {Malvezzi}, {Man}, {Mangano}, {Mantovani}, {Marchesoni},
  {Marion}, {Marque}, {Martelli}, {Martinelli}, {Masserot}, {Meacher},
  {Meidam}, {Michel}, {Milano}, {Minenkov}, {Mohan}, {Morgado}, {Mours},
  {Nagy}, {Nardecchia}, {Naticchioni}, {Nelemans}, {Neri}, {Neri}, {Nocera},
  {Palomba}, {Paoletti}, {Paoletti}, {Pasqualetti}, {Passaquieti}, {Passuello},
  {Pichot}, {Piergiovanni}, {Pinard}, {Poggiani}, {Prijatelj}, {Prodi},
  {Punturo}, {Puppo}, {Rabeling}, {R{\'a}cz}, {Rapagnani}, {Re}, {Regimbau},
  {Ricci}, {Robinet}, {Rocchi}, {Rolland}, {Romano}, {Rosi{\'n}ska}, {Ruggi},
  {Saracco}, {Sassolas}, {Sentenac}, {Sequino}, {Shah}, {Siellez}, {Sperandio},
  {Straniero}, {Sturani}, {Swinkels}, {Tacca}, {ter Braack}, {Toncelli},
  {Tonelli}, {Torre}, {Travasso}, {Vajente}, {van den Brand}, {Van Den Broeck},
  {van der Putten}, {van der Sluys}, {van Heijningen}, {Vas{\'u}th},
  {Vedovato}, {Veitch}, {Verkindt}, {Vetrano}, {Vicer{\'e}}, {Vinet}, {Vitale},
  {Vocca}, {Wei}, {Yvert}, {Zadrozny}, \& {Zendri}}]{2014CQGra..31p5013A}
{Accadia}, T., {Acernese}, F., {Agathos}, M., {et~al.} 2014, Classical and
  Quantum Gravity, 31, 165013, \dodoi{10.1088/0264-9381/31/16/165013}

\bibitem[{{Acernese} {et~al.}(2015){Acernese}, {Agathos}, {Agatsuma}, {Aisa},
  {Allemandou}, {Allocca}, {Amarni}, {Astone}, {Balestri}, {Ballardin},
  {Barone}, {Baronick}, {Barsuglia}, {Basti}, {Basti}, {Bauer}, {Bavigadda},
  {Bejger}, {Beker}, {Belczynski}, {Bersanetti}, {Bertolini}, {Bitossi},
  {Bizouard}, {Bloemen}, {Blom}, {Boer}, {Bogaert}, {Bondi}, {Bondu},
  {Bonelli}, {Bonnand}, {Boschi}, {Bosi}, {Bouedo}, {Bradaschia}, {Branchesi},
  {Briant}, {Brillet}, {Brisson}, {Bulik}, {Bulten}, {Buskulic}, {Buy},
  {Cagnoli}, {Calloni}, {Campeggi}, {Canuel}, {Carbognani}, {Cavalier},
  {Cavalieri}, {Cella}, {Cesarini}, {Mottin}, {Chincarini}, {Chiummo}, {Chua},
  {Cleva}, {Coccia}, {Cohadon}, {Colla}, {Colombini}, {Conte}, {Coulon},
  {Cuoco}, {Dalmaz}, {D'Antonio}, {Dattilo}, {Davier}, {Day}, {Debreczeni},
  {Degallaix}, {Del{\'e}glise}, {Pozzo}, {Dereli}, {Rosa}, {Fiore}, {Lieto},
  {Virgilio}, {Doets}, {Dolique}, {Drago}, {Ducrot}, {Endr{\H{o}}czi},
  {Fafone}, {Farinon}, {Ferrante}, {Ferrini}, {Fidecaro}, {Fiori}, {Flaminio},
  {Fournier}, {Franco}, {Frasca}, {Frasconi}, {Gammaitoni}, {Garufi},
  {Gaspard}, {Gatto}, {Gemme}, {Gendre}, {Genin}, {Gennai}, {Ghosh},
  {Giacobone}, {Giazotto}, {Gouaty}, {Granata}, {Greco}, {Groot}, {Guidi},
  {Harms}, {Heidmann}, {Heitmann}, {Hello}, {Hemming}, {Hennes}, {Hofman},
  {Jaranowski}, {Jonker}, {Kasprzack}, {K{\'e}f{\'e}lian}, {Kowalska}, {Kraan},
  {Kr{\'o}lak}, {Kutynia}, {Lazzaro}, {Leonardi}, {Leroy}, {Letendre}, {Li},
  {Lieunard}, {Lorenzini}, {Loriette}, {Losurdo}, {Magazz{\`u}}, {Majorana},
  {Maksimovic}, {Malvezzi}, {Man}, {Mangano}, {Mantovani}, {Marchesoni},
  {Marion}, {Marque}, {Martelli}, {Martellini}, {Masserot}, {Meacher},
  {Meidam}, {Mezzani}, {Michel}, {Milano}, {Minenkov}, {Moggi}, {Mohan},
  {Montani}, {Morgado}, {Mours}, {Mul}, {Nagy}, {Nardecchia}, {Naticchioni},
  {Nelemans}, {Neri}, {Neri}, {Nocera}, {Pacaud}, {Palomba}, {Paoletti},
  {Paoli}, {Pasqualetti}, {Passaquieti}, {Passuello}, {Perciballi}, {Petit},
  {Pichot}, {Piergiovanni}, {Pillant}, {Piluso}, {Pinard}, {Poggiani},
  {Prijatelj}, {Prodi}, {Punturo}, {Puppo}, {Rabeling}, {R{\'a}cz},
  {Rapagnani}, {Razzano}, {Re}, {Regimbau}, {Ricci}, {Robinet}, {Rocchi},
  {Rolland}, {Romano}, {Rosi{\'n}ska}, {Ruggi}, \&
  {Saracco}}]{2015CQGra..32b4001A}
{Acernese}, F., {Agathos}, M., {Agatsuma}, K., {et~al.} 2015, Classical and
  Quantum Gravity, 32, 024001, \dodoi{10.1088/0264-9381/32/2/024001}

\bibitem[{{Acernese} {et~al.}(2018){Acernese}, {Adams}, {Agatsuma}, {Aiello},
  {Allocca}, {Aloy}, {Amato}, {Antier}, {Ar{\`e}ne}, {Arnaud}, {Ascenzi},
  {Astone}, {Aubin}, {Babak}, {Bacon}, {Badaracco}, {Bader}, {Baldaccini},
  {Ballardin}, {Barone}, {Barsuglia}, {Barta}, {Basti}, {Bawaj}, {Bazzan},
  {Bejger}, {Belahcene}, {Bernuzzi}, {Bersanetti}, {Bertolini}, {Bitossi},
  {Bizouard}, {Bloemen}, {Boer}, {Bogaert}, {Bondu}, {Bonnand}, {Boom},
  {Boschi}, {Bouffanais}, {Bozzi}, {Bradaschia}, {Branchesi}, {Briant},
  {Brighenti}, {Brillet}, {Brisson}, {Bulik}, {Bulten}, {Buskulic}, {Buy},
  {Cagnoli}, {Calloni}, {Canepa}, {Canizares}, {Capocasa}, {Carbognani},
  {Casanueva Diaz}, {Casentini}, {Caudill}, {Cavalier}, {Cavalieri}, {Cella},
  {Cerd{\'a}-Dur{\'a}n}, {Cerretani}, {Cesarini}, {Chaibi}, {Chassande-Mottin},
  {Chincarini}, {Chiummo}, {Christensen}, {Chua}, {Ciani}, {Ciolfi},
  {Cipriano}, {Cirone}, {Cleva}, {Coccia}, {Cohadon}, {Cohen}, {Colla},
  {Conti}, {Cordero-Carri{\'o}n}, {Cortese}, {Coulon}, {Cuoco}, {D'Antonio},
  {Dattilo}, {Davier}, {De Rossi}, {Degallaix}, {De Laurentis},
  {Del{\'e}glise}, {Del Pozzo}, {De Pietri}, {De Rosa}, {Di Fiore}, {Di
  Giovanni}, {Di Girolamo}, {Di Lieto}, {Di Pace}, {Di Palma}, {Di Renzo},
  {Dolique}, {Drago}, {Eisenmann}, {Estevez}, {Fafone}, {Farinon}, {Feng},
  {Ferrante}, {Ferrini}, {Fidecaro}, {Fiori}, {Fiorucci}, {Flaminio}, {Font},
  {Fournier}, {Frasca}, {Frasconi}, {Frey}, {Gammaitoni}, {Garufi}, {Gemme},
  {Genin}, {Gennai}, {Germain}, {Ghosh}, {Giacomazzo}, {Giazotto}, {Giordano},
  {Gonzalez Castro}, {Gosselin}, {Gouaty}, {Grado}, {Granata}, {Greco},
  {Groot}, {Gruning}, {Guidi}, {Halim}, {Harms}, {Heidmann}, {Heitmann},
  {Hello}, {Hemming}, {Hinderer}, {Hoak}, {Hofman}, {Hreibi}, {Huet}, {Iess},
  {Intini}, {Isac}, {Jacqmin}, {Jaranowski}, {Jonker}, {Katsanevas},
  {K{\'e}f{\'e}lian}, {Khan}, {Koley}, {Kowalska}, {Kr{\'o}lak}, {Kutynia},
  {Lartaux-Vollard}, {Lazzaro}, {Leaci}, {Leonardi}, {Leroy}, {Letendre},
  {Longo}, {Lorenzini}, {Loriette}, {Losurdo}, {Lumaca}, {Majorana},
  {Maksimovic}, {Man}, {Mantovani}, {Marchesoni}, {Marion}, {Marquina},
  {Martelli}, {Martellini}, {Masserot}, {Mastrogiovanni}, {Meidam}, {Mereni},
  {Merzougui}, {Metzdorff}, {Michel}, {Milano}, {Miller}, {Minazzoli},
  {Minenkov}, {Mohan}, {Montani}, {Mours}, {Nardecchia}, \&
  {Naticchioni}}]{2018CQGra..35t5004A}
{Acernese}, F., {Adams}, T., {Agatsuma}, K., {et~al.} 2018, Classical and
  Quantum Gravity, 35, 205004, \dodoi{10.1088/1361-6382/aadf1a}

\bibitem[{{Acernese} {et~al.}(2022){Acernese}, {Agathos}, {Ain}, {Albanesi},
  {Allocca}, {Amato}, {Andrade}, {Andres}, {Andri{\'c}}, {Ansoldi}, {Antier},
  {Ar{\`e}ne}, {Arnaud}, {Assiduo}, {Astone}, {Aubin}, {Babak}, {Badaracco},
  {Bader}, {Bagnasco}, {Baird}, {Ballardin}, {Baltus}, {Barbieri}, {Barneo},
  {Barone}, {Barsuglia}, {Barta}, {Basti}, {Bawaj}, {Bazzan}, {Bejger},
  {Belahcene}, {Benedetto}, {Bernuzzi}, {Bersanetti}, {Bertolini}, {Bhardwaj},
  {Bini}, {Bischi}, {Bitossi}, {Bizouard}, {Bobba}, {Boer}, {Bogaert},
  {Boldrini}, {Bonavena}, {Bondu}, {Bonnand}, {Boom}, {Boschi}, {Boudart},
  {Bouffanais}, {Bozzi}, {Bradaschia}, {Branchesi}, {Breschi}, {Briant},
  {Brillet}, {Brooks}, {Bruno}, {Bulik}, {Bulten}, {Buskulic}, {Buy},
  {Cagnoli}, {Calloni}, {Canepa}, {Canevarolo}, {Cannavacciuolo}, {Carapella},
  {Carbognani}, {Carpinelli}, {Carullo}, {Diaz}, {Casentini}, {Caudill},
  {Cavalier}, {Cavalieri}, {Cella}, {Cerd{\'a}-Dur{\'a}n}, {Cesarini},
  {Chaibi}, {Chanial}, {Chassande-Mottin}, {Chaty}, {Chiadini}, {Chiarini},
  {Chierici}, {Chincarini}, {Chiofalo}, {Chiummo}, {Christensen}, {Ciani},
  {Cie{\'s}lar}, {Ciecielag}, {Cifaldi}, {Ciolfi}, {Cipriano}, {Cirone},
  {Clesse}, {Cleva}, {Coccia}, {Codazzo}, {Cohadon}, {Cohen}, {Colombo},
  {Colpi}, {Conti}, {Cordero-Carri{\'o}n}, {Corezzi}, {Corre}, {Cortese},
  {Coulon}, {Croquette}, {Cudell}, {Cuoco}, {Cury{\l}o}, {Dabadie}, {Canton},
  {Dall'Osso}, {D'Angelo}, {Danilishin}, {D'Antonio}, {Dattilo}, {Davier}, {De
  Laurentis}, {De Lillo}, {De Matteis}, {De Pietri}, {De Rosa}, {De Rossi}, {De
  Simone}, {Degallaix}, {Del{\'e}glise}, {Del Pozzo}, {Depasse}, {Di Fiore},
  {Di Giorgio}, {Di Giovanni}, {Di Giovanni}, {Di Girolamo}, {Di Lieto}, {Di
  Pace}, {Di Palma}, {Di Renzo}, {Dietrich}, {D'Onofrio}, {Drago}, {Ducoin},
  {Durante}, {D'Urso}, {Duverne}, {Eisenmann}, {Errico}, {Estevez}, {Fafone},
  {Farinon}, {Favaro}, {Fays}, {Fenyvesi}, {Ferrante}, {Fidecaro}, {Figura},
  {Fiori}, {Fittipaldi}, {Fiumara}, {Flaminio}, {Font}, {Frasca}, {Frasconi},
  {Fronz{\'e}}, {Gamba}, {Garaventa}, {Garufi}, {Gemme}, {Gennai}, {Ghosh},
  {Giacomazzo}, {Giacoppo}, {Giri}, {Gissi}, {Goncharov}, {Gosselin}, {Gouaty},
  {Grado}, {Granata}, {Granata}, {Greco}, {Grignani}, {Grimaldi}, {Grimm},
  {Gruning}, {Guerra}, {Guidi}, {Guix{\'e}}, {Guo}, {Gupta}, {Haegel}, \&
  {Halim}}]{2022CQGra..39d5006A}
{Acernese}, F., {Agathos}, M., {Ain}, A., {et~al.} 2022, Classical and Quantum
  Gravity, 39, 045006, \dodoi{10.1088/1361-6382/ac3c8e}

\bibitem[{{Acernese} {et~al.}(2023{\natexlab{a}}){Acernese}, {Agathos}, {Ain},
  {Albanesi}, {Allocca}, {Amato}, {Andrade}, {Andres}, {Andr{\'e}s-Carcasona},
  {Andri{\'c}}, {Ansoldi}, {Antier}, {Apostolatos}, {Appavuravther},
  {Ar{\`e}ne}, {Arnaud}, {Assiduo}, {Assis de Souza Melo}, {Astone}, {Aubin},
  {Babak}, {Badaracco}, {M Bader}, {Bagnasco}, {Baird}, {Baka}, {Ballardin},
  {Baltus}, {Banerjee}, {Barbieri}, {Barneo}, {Barone}, {Barsuglia}, {Barta},
  {Basti}, {Bawaj}, {Bazzan}, {Beirnaert}, {Bejger}, {Belahcene}, {Benedetto},
  {Berbel}, {Bernuzzi}, {Bersanetti}, {Bertolini}, {Bhardwaj}, {Bianchi},
  {Bini}, {Bischi}, {Bitossi}, {Bizouard}, {Bobba}, {Bo{\"e}r}, {Bogaert},
  {Boldrini}, {Bonavena}, {Bondu}, {Bonnand}, {Boom}, {Boschi}, {Boudart},
  {Bouffanais}, {Bozzi}, {Bradaschia}, {Branchesi}, {Breschi}, {Briant},
  {Brillet}, {Brooks}, {Bruno}, {Bucci}, {Bulik}, {Bulten}, {Buskulic}, {Buy},
  {Cabourn Davies}, {Cabras}, {Cabrita}, {Cagnoli}, {Calloni}, {Canepa},
  {Canevarolo}, {Cannavacciuolo}, {Capocasa}, {Carapella}, {Carbognani},
  {Carpinelli}, {Carullo}, {Casanueva Diaz}, {Casentini}, {Caudill},
  {Cavalier}, {Cavalieri}, {Cella}, {Cerd{\'a}-Dur{\'a}n}, {Cesarini},
  {Chaibi}, {Chanial}, {Chassande-Mottin}, {Chaty}, {Chiadini}, {Chiarini},
  {Chierici}, {Chincarini}, {Chiofalo}, {Chiummo}, {Choudhary}, {Christensen},
  {Ciani}, {Ciecielag}, {Cie{\'s}lar}, {Cifaldi}, {Ciolfi}, {Cipriano},
  {Clesse}, {Cleva}, {Coccia}, {Codazzo}, {Cohadon}, {Cohen}, {Colombo},
  {Colpi}, {Conti}, {Cordero-Carri{\'o}n}, {Corezzi}, {Corre}, {Cortese},
  {Coulon}, {Croquette}, {Cudell}, {Cuoco}, {Cury{\l}o}, {Dabadie}, {Dal
  Canton}, {Dall'Osso}, {D{\'a}lya}, {D'Angelo}, {Danilishin}, {D'Antonio},
  {Dattilo}, {Davier}, {Davis}, {Degallaix}, {De Laurentis}, {Del{\'e}glise},
  {De Lillo}, {Dell'Aquila}, {Del Pozzo}, {De Matteis}, {Depasse}, {De Pietri},
  {De Rosa}, {De Rossi}, {De Simone}, {Di Fiore}, {Di Giorgio}, {Di Giovanni},
  {Di Giovanni}, {Di Girolamo}, {Di Lieto}, {Di Michele}, {Di Pace}, {Di
  Palma}, {Di Renzo}, {D'Onofrio}, {Drago}, {Ducoin}, {Dupletsa}, {Durante},
  {D'Urso}, {Duverne}, {Eisenmann}, {Errico}, {Estevez}, {Fabrizi}, {Faedi},
  {Fafone}, {Farinon}, {Favaro}, {Fays}, {Fenyvesi}, {Ferrante}, {Fidecaro},
  {Figura}, {Fiori}, {Fiori}, {Fittipaldi}, {Fiumara}, {Flaminio}, {Font},
  {Frasca}, {Frasconi}, {Freise}, {Freitas}, {Fronz{\'e}}, {Gadre}, {Gamba},
  {Garaventa}, {Garufi}, \& {Gemme}}]{2023CQGra..40r5006A}
---. 2023{\natexlab{a}}, Classical and Quantum Gravity, 40, 185006,
  \dodoi{10.1088/1361-6382/acd92d}

\bibitem[{{Acernese} {et~al.}(2023{\natexlab{b}}){Acernese}, {Agathos}, {Ain},
  {Albanesi}, {Allocca}, {Amato}, {Andrade}, {Andres}, {Andr{\'e}s-Carcasona},
  {Andri{\'c}}, {Ansoldi}, {Antier}, {Apostolatos}, {Appavuravther},
  {Ar{\`e}ne}, {Arnaud}, {Assiduo}, {Assis de Souza Melo}, {Astone}, {Aubin},
  {Babak}, {Badaracco}, {M Bader}, {Bagnasco}, {Baird}, {Baka}, {Ballardin},
  {Baltus}, {Banerjee}, {Barbieri}, {Barneo}, {Barone}, {Barsuglia}, {Barta},
  {Basti}, {Bawaj}, {Bazzan}, {Beirnaert}, {Bejger}, {Belahcene}, {Benedetto},
  {Berbel}, {Bernuzzi}, {Bersanetti}, {Bertolini}, {Bhardwaj}, {Bianchi},
  {Bini}, {Bischi}, {Bitossi}, {Bizouard}, {Bobba}, {Bo{\"e}r}, {Bogaert},
  {Boldrini}, {Bonavena}, {Bondu}, {Bonnand}, {Boom}, {Boschi}, {Boudart},
  {Bouffanais}, {Bozzi}, {Bradaschia}, {Branchesi}, {Breschi}, {Briant},
  {Brillet}, {Brooks}, {Bruno}, {Bucci}, {Bulik}, {Bulten}, {Buskulic}, {Buy},
  {Cabourn Davies}, {Cabras}, {Cabrita}, {Cagnoli}, {Calloni}, {Canepa},
  {Canevarolo}, {Cannavacciuolo}, {Capocasa}, {Carapella}, {Carbognani},
  {Carpinelli}, {Carullo}, {Casanueva Diaz}, {Casentini}, {Caudill},
  {Cavalier}, {Cavalieri}, {Cella}, {Cerd{\'a}-Dur{\'a}n}, {Cesarini},
  {Chaibi}, {Chanial}, {Chassande-Mottin}, {Chaty}, {Chiadini}, {Chiarini},
  {Chierici}, {Chincarini}, {Chiofalo}, {Chiummo}, {Choudhary}, {Christensen},
  {Ciani}, {Ciecielag}, {Cie{\'s}lar}, {Cifaldi}, {Ciolfi}, {Cipriano},
  {Clesse}, {Cleva}, {Coccia}, {Codazzo}, {Cohadon}, {Cohen}, {Colombo},
  {Colpi}, {Conti}, {Cordero-Carri{\'o}n}, {Corezzi}, {Corre}, {Cortese},
  {Coulon}, {Croquette}, {Cudell}, {Cuoco}, {Cury{\l}o}, {Dabadie}, {Dal
  Canton}, {Dall'Osso}, {D{\'a}lya}, {D'Angelo}, {Danilishin}, {D'Antonio},
  {Dattilo}, {Davier}, {Davis}, {Degallaix}, {De Laurentis}, {Del{\'e}glise},
  {De Lillo}, {Dell'Aquila}, {Del Pozzo}, {De Matteis}, {Depasse}, {De Pietri},
  {De Rosa}, {De Rossi}, {De Simone}, {Di Fiore}, {Di Giorgio}, {Di Giovanni},
  {Di Giovanni}, {Di Girolamo}, {Di Lieto}, {Di Michele}, {Di Pace}, {Di
  Palma}, {Di Renzo}, {D'Onofrio}, {Drago}, {Ducoin}, {Dupletsa}, {Durante},
  {D'Urso}, {Duverne}, {Eisenmann}, {Errico}, {Estevez}, {Fabrizi}, {Faedi},
  {Fafone}, {Farinon}, {Favaro}, {Fays}, {Fenyvesi}, {Ferrante}, {Fidecaro},
  {Figura}, {Fiori}, {Fiori}, {Fittipaldi}, {Fiumara}, {Flaminio}, {Font},
  {Frasca}, {Frasconi}, {Freise}, {Freitas}, {Fronz{\'e}}, {Gadre}, {Gamba},
  {Garaventa}, {Garufi}, \& {Gemme}}]{2023CQGra..40r5005F}
---. 2023{\natexlab{b}}, Classical and Quantum Gravity, 40, 185005,
  \dodoi{10.1088/1361-6382/acdf36}

\bibitem[{{Adams} {et~al.}(2016){Adams}, {Buskulic}, {Germain}, {Guidi},
  {Marion}, {Montani}, {Mours}, {Piergiovanni}, \&
  {Wang}}]{2016CQGra..33q5012A}
{Adams}, T., {Buskulic}, D., {Germain}, V., {et~al.} 2016, Classical and
  Quantum Gravity, 33, 175012, \dodoi{10.1088/0264-9381/33/17/175012}

\bibitem[{{Ade} {et~al.}(2016){Ade}, {Aghanim}, {Arnaud}, {Ashdown}, {Aumont},
  {Baccigalupi}, {Banday}, {Barreiro}, {Bartlett}, {Bartolo}, {Battaner},
  {Battye}, {Benabed}, {Beno{\^\i}t}, {Benoit-L{\'e}vy}, {Bernard},
  {Bersanelli}, {Bielewicz}, {Bock}, {Bonaldi}, {Bonavera}, {Bond}, {Borrill},
  {Bouchet}, {Boulanger}, {Bucher}, {Burigana}, {Butler}, {Calabrese},
  {Cardoso}, {Catalano}, {Challinor}, {Chamballu}, {Chary}, {Chiang}, {Chluba},
  {Christensen}, {Church}, {Clements}, {Colombi}, {Colombo}, {Combet},
  {Coulais}, {Crill}, {Curto}, {Cuttaia}, {Danese}, {Davies}, {Davis}, {de
  Bernardis}, {de Rosa}, {de Zotti}, {Delabrouille}, {D{\'e}sert}, {Di
  Valentino}, {Dickinson}, {Diego}, {Dolag}, {Dole}, {Donzelli}, {Dor{\'e}},
  {Douspis}, {Ducout}, {Dunkley}, {Dupac}, {Efstathiou}, {Elsner},
  {En{\ss}lin}, {Eriksen}, {Farhang}, {Fergusson}, {Finelli}, {Forni},
  {Frailis}, {Fraisse}, {Franceschi}, {Frejsel}, {Galeotta}, {Galli}, {Ganga},
  {Gauthier}, {Gerbino}, {Ghosh}, {Giard}, {Giraud-H{\'e}raud}, {Giusarma},
  {Gjerl{\o}w}, {Gonz{\'a}lez-Nuevo}, {G{\'o}rski}, {Gratton}, {Gregorio},
  {Gruppuso}, {Gudmundsson}, {Hamann}, {Hansen}, {Hanson}, {Harrison}, {Helou},
  {Henrot-Versill{\'e}}, {Hern{\'a}ndez-Monteagudo}, {Herranz}, {Hildebrandt},
  {Hivon}, {Hobson}, {Holmes}, {Hornstrup}, {Hovest}, {Huang}, {Huffenberger},
  {Hurier}, {Jaffe}, {Jaffe}, {Jones}, {Juvela}, {Keih{\"a}nen}, {Keskitalo},
  {Kisner}, {Kneissl}, {Knoche}, {Knox}, {Kunz}, {Kurki-Suonio}, {Lagache},
  {L{\"a}hteenm{\"a}ki}, {Lamarre}, {Lasenby}, {Lattanzi}, {Lawrence}, {Leahy},
  {Leonardi}, {Lesgourgues}, {Levrier}, {Lewis}, {Liguori}, {Lilje},
  {Linden-V{\o}rnle}, {L{\'o}pez-Caniego}, {Lubin}, {Mac{\'\i}as-P{\'e}rez},
  {Maggio}, {Maino}, {Mandolesi}, {Mangilli}, {Marchini}, {Maris}, {Martin},
  {Martinelli}, {Mart{\'\i}nez-Gonz{\'a}lez}, {Masi}, {Matarrese}, {McGehee},
  {Meinhold}, {Melchiorri}, {Melin}, {Mendes}, {Mennella}, {Migliaccio},
  {Millea}, {Mitra}, {Miville-Desch{\^e}nes}, {Moneti}, {Montier}, {Morgante},
  {Mortlock}, {Moss}, {Munshi}, {Murphy}, {Naselsky}, {Nati}, {Natoli},
  {Netterfield}, {N{\o}rgaard-Nielsen}, {Noviello}, {Novikov}, {Novikov},
  {Oxborrow}, {Paci}, {Pagano}, {Pajot}, {Paladini}, {Paoletti}, {Partridge},
  {Pasian}, {Patanchon}, {Pearson}, {Perdereau}, {Perotto}, {Perrotta},
  {Pettorino}, {Piacentini}, {Piat}, {Pierpaoli}, {Pietrobon}, {Plaszczynski},
  {Pointecouteau}, {Polenta}, {Popa}, {Pratt}, \&
  {Pr{\'e}zeau}}]{2016A&A...594A..13P}
{Ade}, P.~A.~R., {Aghanim}, N., {Arnaud}, M., {et~al.} 2016, \aap, 594, A13,
  \dodoi{10.1051/0004-6361/201525830}

\bibitem[{Agathos {et~al.}(2015)Agathos, Meidam, Del~Pozzo, Li, Tompitak,
  Veitch, Vitale, \& Van Den~Broeck}]{Agathos:2015uaa}
Agathos, M., Meidam, J., Del~Pozzo, W., {et~al.} 2015, Phys. Rev. D, 92,
  023012, \dodoi{10.1103/PhysRevD.92.023012}

\bibitem[{Ajith {et~al.}(2007)}]{Ajith:2007qp}
Ajith, P., {et~al.} 2007, Class. Quant. Grav., 24, S689,
  \dodoi{10.1088/0264-9381/24/19/S31}

\bibitem[{{Ajith} {et~al.}(2011){Ajith}, {Hannam}, {Husa}, {Chen},
  {Br{\"u}gmann}, {Dorband}, {M{\"u}ller}, {Ohme}, {Pollney}, {Reisswig},
  {Santamar{\'\i}a}, \& {Seiler}}]{2011PhRvL.106x1101A}
{Ajith}, P., {Hannam}, M., {Husa}, S., {et~al.} 2011, \prl, 106, 241101,
  \dodoi{10.1103/PhysRevLett.106.241101}

\bibitem[{Akcay {et~al.}(2019)Akcay, Bernuzzi, Messina, Nagar, Ortiz, \&
  Rettegno}]{Akcay:2018yyh}
Akcay, S., Bernuzzi, S., Messina, F., {et~al.} 2019, Phys. Rev. D, 99, 044051,
  \dodoi{10.1103/PhysRevD.99.044051}

\bibitem[{{Akutsu} {et~al.}(2019){Akutsu}, {Ando}, {Arai}, {Arai}, {Araki},
  {Araya}, {Aritomi}, {Asada}, {Aso}, {Atsuta}, {Awai}, {Bae}, {Baiotti},
  {Barton}, {Cannon}, {Capocasa}, {Chen}, {Chiu}, {Cho}, {Chu}, {Craig},
  {Creus}, {Doi}, {Eda}, {Enomoto}, {Flaminio}, {Fujii}, {Fujimoto},
  {Fukunaga}, {Fukushima}, {Furuhata}, {Haino}, {Hasegawa}, {Hashino},
  {Hayama}, {Hirobayashi}, {Hirose}, {Hsieh}, {Huang}, {Ikenoue}, {Inoue},
  {Ioka}, {Itoh}, {Izumi}, {Kaji}, {Kajita}, {Kakizaki}, {Kamiizumi},
  {Kanbara}, {Kanda}, {Kanemura}, {Kaneyama}, {Kang}, {Kasuya}, {Kataoka},
  {Kawai}, {Kawamura}, {Kawasaki}, {Kim}, {Kim}, {Kim}, {Kim}, {Kim}, {Kimura},
  {Kinugawa}, {Kirii}, {Kitaoka}, {Kitazawa}, {Kojima}, {Kokeyama}, {Komori},
  {Kong}, {Kotake}, {Kozu}, {Kumar}, {Kuo}, {Kuroyanagi}, {Lee}, {Lee}, {Lee},
  {Leonardi}, {Lin}, {Lin}, {Liu}, {Liu}, {Majorana}, {Mano}, {Marchio},
  {Matsui}, {Matsushima}, {Michimura}, {Mio}, {Miyakawa}, {Miyamoto},
  {Miyamoto}, {Miyo}, {Miyoki}, {Morii}, {Morisaki}, {Moriwaki}, {Morozumi},
  {Musha}, {Nagano}, {Nagano}, {Nakamura}, {Nakamura}, {Nakano}, {Nakano},
  {Nakao}, {Narikawa}, {Naticchioni}, {Nguyen Quynh}, {Ni}, {Nishizawa},
  {Obuchi}, {Ochi}, {Oh}, {Oh}, {Ohashi}, {Ohishi}, {Ohkawa}, {Okutomi}, {Ono},
  {Oohara}, {Ooi}, {Pan}, {Park}, {Pe{\~n}a Arellano}, {Pinto}, {Sago},
  {Saijo}, {Saitou}, {Saito}, {Sakai}, {Sakai}, {Sakai}, {Sasai}, {Sasaki},
  {Sasaki}, {Sato}, {Sato}, {Sato}, {Sekiguchi}, {Seto}, {Shibata}, {Shimoda},
  {Shinkai}, {Shishido}, {Shoda}, {Somiya}, {Son}, {Suemasa}, {Suzuki},
  {Suzuki}, {Tagoshi}, {Tahara}, {Takahashi}, {Takahashi}, {Takamori},
  {Takeda}, {Tanaka}, {Tanaka}, {Tanaka}, {Tanioka}, {Tapia San Martin},
  {Tatsumi}, {Tomaru}, {Tomura}, {Travasso}, {Tsubono}, {Tsuchida}, {Uchikata},
  {Uchiyama}, {Uehara}, {Ueki}, {Ueno}, {Uraguchi}, {Ushiba}, {van Putten},
  {Vocca}, {Wada}, {Wakamatsu}, {Watanabe}, {Xu}, {Yamada}, {Yamamoto},
  {Yamamoto}, {Yamamoto}, {Yamamoto}, {Yamamoto}, {Yokogawa}, {Yokoyama},
  {Yokozawa}, {Yoon}, {Yoshioka}, {Yuzurihara}, {Zeidler}, \&
  {Zhu}}]{2019NatAs...3...35K}
{Akutsu}, T., {Ando}, M., {Arai}, K., {et~al.} 2019, Nature Astronomy, 3, 35,
  \dodoi{10.1038/s41550-018-0658-y}

\bibitem[{{Allen}(2005)}]{2005PhRvD..71f2001A}
{Allen}, B. 2005, \prd, 71, 062001, \dodoi{10.1103/PhysRevD.71.062001}

\bibitem[{{Allen} {et~al.}(2012){Allen}, {Anderson}, {Brady}, {Brown}, \&
  {Creighton}}]{2012PhRvD..85l2006A}
{Allen}, B., {Anderson}, W.~G., {Brady}, P.~R., {Brown}, D.~A., \& {Creighton},
  J. D.~E. 2012, \prd, 85, 122006, \dodoi{10.1103/PhysRevD.85.122006}

\bibitem[{{All{\'e}n{\'e}} {et~al.}(2025){All{\'e}n{\'e}}, {Aubin}, {Bentara},
  {Buskulic}, {Guidi}, {Juste}, {Lethuillier}, {Marion}, {Mobilia}, {Mours},
  {Ouzriat}, {Sainrat}, \& {Sordini}}]{2025CQGra..42j5009A}
{All{\'e}n{\'e}}, C., {Aubin}, F., {Bentara}, I., {et~al.} 2025, Classical and
  Quantum Gravity, 42, 105009, \dodoi{10.1088/1361-6382/add234}

\bibitem[{{{\'A}lvarez-L{\'o}pez} {et~al.}(2024){{\'A}lvarez-L{\'o}pez},
  {Liyanage}, {Ding}, {Ng}, \& {McIver}}]{2024CQGra..41h5007A}
{{\'A}lvarez-L{\'o}pez}, S., {Liyanage}, A., {Ding}, J., {Ng}, R., \& {McIver},
  J. 2024, Classical and Quantum Gravity, 41, 085007,
  \dodoi{10.1088/1361-6382/ad2194}

\bibitem[{{Andres} {et~al.}(2022){Andres}, {Assiduo}, {Aubin}, {Chierici},
  {Estevez}, {Faedi}, {Guidi}, {Juste}, {Marion}, {Mours}, {Nitoglia}, \&
  {Sordini}}]{2022CQGra..39e5002A}
{Andres}, N., {Assiduo}, M., {Aubin}, F., {et~al.} 2022, Classical and Quantum
  Gravity, 39, 055002, \dodoi{10.1088/1361-6382/ac482a}

\bibitem[{{Arnaud} {et~al.}(2026){Arnaud}, {Driggers}, {O'Reilly}, {Sawada},
  {The VIRGO Collaboration}, \& {The Kagra
  Collaboration}}]{2026JPhCS3177a2067A}
{Arnaud}, N., {Driggers}, J., {O'Reilly}, B., {et~al.} 2026, in Journal of
  Physics Conference Series, Vol. 3177, Journal of Physics Conference Series
  (IOP), 012067, \dodoi{10.1088/1742-6596/3177/1/012067}

\bibitem[{{Ashton} \& {Talbot}(2021)}]{2021MNRAS.507.2037A}
{Ashton}, G., \& {Talbot}, C. 2021, \mnras, 507, 2037,
  \dodoi{10.1093/mnras/stab2236}

\bibitem[{Ashton {et~al.}(2022)Ashton, Udall, \& Yarbrough}]{cbcflow}
Ashton, G., Udall, R., \& Yarbrough, Z. 2022, CBC Workflow.
\newblock \url{https://git.ligo.org/cbc/projects/cbcflow}

\bibitem[{{Ashton} {et~al.}(2019){Ashton}, {H{\"u}bner}, {Lasky}, {Talbot},
  {Ackley}, {Biscoveanu}, {Chu}, {Divakarla}, {Easter}, {Goncharov}, {Hernandez
  Vivanco}, {Harms}, {Lower}, {Meadors}, {Melchor}, {Payne}, {Pitkin},
  {Powell}, {Sarin}, {Smith}, \& {Thrane}}]{2019ApJS..241...27A}
{Ashton}, G., {H{\"u}bner}, M., {Lasky}, P.~D., {et~al.} 2019, \apjs, 241, 27,
  \dodoi{10.3847/1538-4365/ab06fc}

\bibitem[{{Aubin} {et~al.}(2021){Aubin}, {Brighenti}, {Chierici}, {Estevez},
  {Greco}, {Guidi}, {Juste}, {Marion}, {Mours}, {Nitoglia}, {Sauter}, \&
  {Sordini}}]{2021CQGra..38i5004A}
{Aubin}, F., {Brighenti}, F., {Chierici}, R., {et~al.} 2021, Classical and
  Quantum Gravity, 38, 095004, \dodoi{10.1088/1361-6382/abe913}

\bibitem[{Aubin {et~al.}(2025)Aubin, Bentara, Buskulic, Guidi, Juste,
  Lethuillier, Marion, Mobilia, Mours, Ouzriat, \& Sainrat}]{Aubin:2025TDS}
Aubin, F., Bentara, I., Buskulic, D., {et~al.} 2025, {The MBTA PSD calculation
  for the O4a offline all-sky search}, Tech. Rep. TDS-0722A.
\newblock \url{https://tds.virgo-gw.eu/?r=25024}

\bibitem[{{Babak} {et~al.}(2013){Babak}, {Biswas}, {Brady}, {Brown}, {Cannon},
  {Capano}, {Clayton}, {Cokelaer}, {Creighton}, {Dent}, {Dietz}, {Fairhurst},
  {Fotopoulos}, {Gonz{\'a}lez}, {Hanna}, {Harry}, {Jones}, {Keppel},
  {McKechan}, {Pekowsky}, {Privitera}, {Robinson}, {Rodriguez},
  {Sathyaprakash}, {Sengupta}, {Vallisneri}, {Vaulin}, \&
  {Weinstein}}]{2013PhRvD..87b4033B}
{Babak}, S., {Biswas}, R., {Brady}, P.~R., {et~al.} 2013, \prd, 87, 024033,
  \dodoi{10.1103/PhysRevD.87.024033}

\bibitem[{Baka {et~al.}(2025)Baka, Wright, Romero-Shaw, Berry, Haster, Hoy,
  Talbot, Pang, Raymond, Williams, Ashton, Bossilkov, Dartez, Manning, Veitch,
  Zimmerman, Sun, \& Farr}]{Baka:2025bbb}
Baka, T., Wright, M., Romero-Shaw, I., {et~al.} 2025, Correcting
  misspecification of calibration uncertainties in gravitational-wave data
  analysis with efficient reweighting, Tech. Rep. {LIGO}-T2500295, {LIGO}
  Project.
\newblock \url{https://dcc.ligo.org/LIGO-T2500295/public}

\bibitem[{Barausse {et~al.}(2012)Barausse, Buonanno, Hughes, Khanna,
  O'Sullivan, \& Pan}]{Barausse:2011kb}
Barausse, E., Buonanno, A., Hughes, S.~A., {et~al.} 2012, Phys. Rev. D, 85,
  024046, \dodoi{10.1103/PhysRevD.85.024046}

\bibitem[{{Bernuzzi} {et~al.}(2015){Bernuzzi}, {Nagar}, {Dietrich}, \&
  {Damour}}]{2015PhRvL.114p1103B}
{Bernuzzi}, S., {Nagar}, A., {Dietrich}, T., \& {Damour}, T. 2015, \prl, 114,
  161103, \dodoi{10.1103/PhysRevLett.114.161103}

\bibitem[{Bhaumik {et~al.}(2025)Bhaumik, Gayathri, Bartos, Anglin, Carullo,
  Healy, Klimenko, Lange, Lousto, Mishra, \& Szczepa\ifmmode~\acute{n}\else
  \'{n}\fi{}czyk}]{eBBH:2025sbgt}
Bhaumik, S., Gayathri, V., Bartos, I., {et~al.} 2025, Phys. Rev. D, 111,
  123032, \dodoi{10.1103/hwr5-scp4}

\bibitem[{{Biscoveanu} {et~al.}(2021){Biscoveanu}, {Isi}, {Varma}, \&
  {Vitale}}]{2021PhRvD.104j3018B}
{Biscoveanu}, S., {Isi}, M., {Varma}, V., \& {Vitale}, S. 2021, \prd, 104,
  103018, \dodoi{10.1103/PhysRevD.104.103018}

\bibitem[{{Blackman} {et~al.}(2017{\natexlab{a}}){Blackman}, {Field}, {Scheel},
  {Galley}, {Hemberger}, {Schmidt}, \& {Smith}}]{2017PhRvD..95j4023B}
{Blackman}, J., {Field}, S.~E., {Scheel}, M.~A., {et~al.} 2017{\natexlab{a}},
  \prd, 95, 104023, \dodoi{10.1103/PhysRevD.95.104023}

\bibitem[{{Blackman} {et~al.}(2017{\natexlab{b}}){Blackman}, {Field}, {Scheel},
  {Galley}, {Ott}, {Boyle}, {Kidder}, {Pfeiffer}, \&
  {Szil{\'a}gyi}}]{2017PhRvD..96b4058B}
---. 2017{\natexlab{b}}, \prd, 96, 024058, \dodoi{10.1103/PhysRevD.96.024058}

\bibitem[{{Blanchet}(2014)}]{2014LRR....17....2B}
{Blanchet}, L. 2014, Living Reviews in Relativity, 17, 2,
  \dodoi{10.12942/lrr-2014-2}

\bibitem[{Boh{\'e} {et~al.}(2016)Boh{\'e}, Hannam, Husa, Ohme, Puerrer, \&
  Schmidt}]{Bohe:PPv2}
Boh{\'e}, A., Hannam, M., Husa, S., {et~al.} 2016, PhenomPv2 - Technical Notes
  for LAL Implementation, Tech. Rep. {LIGO}-T1500602, {LIGO} Project.
\newblock \url{https://dcc.ligo.org/LIGO-T1500602}

\bibitem[{Bohe {et~al.}(2013)Bohe, Marsat, Faye, \& Blanchet}]{Bohe:2012mr}
Bohe, A., Marsat, S., Faye, G., \& Blanchet, L. 2013, Class. Quant. Grav., 30,
  075017, \dodoi{10.1088/0264-9381/30/7/075017}

\bibitem[{{Boh{\'e}} {et~al.}(2017){Boh{\'e}}, {Shao}, {Taracchini},
  {Buonanno}, {Babak}, {Harry}, {Hinder}, {Ossokine}, {P{\"u}rrer}, {Raymond},
  {Chu}, {Fong}, {Kumar}, {Pfeiffer}, {Boyle}, {Hemberger}, {Kidder},
  {Lovelace}, {Scheel}, \& {Szil{\'a}gyi}}]{2017PhRvD..95d4028B}
{Boh{\'e}}, A., {Shao}, L., {Taracchini}, A., {et~al.} 2017, \prd, 95, 044028,
  \dodoi{10.1103/PhysRevD.95.044028}

\bibitem[{Borchers {et~al.}(2024)Borchers, Ohme, Mielke, \&
  Ghosh}]{Borchers:2024tdi}
Borchers, A., Ohme, F., Mielke, J., \& Ghosh, S. 2024, Phys. Rev. D, 110,
  024037, \dodoi{10.1103/PhysRevD.110.024037}

\bibitem[{Bose {et~al.}(2011)Bose, Dayanga, Ghosh, \& Talukder}]{Bose:2011chs}
Bose, S., Dayanga, T., Ghosh, S., \& Talukder, D. 2011, Classical and Quantum
  Gravity, 28, 134009, \dodoi{10.1088/0264-9381/28/13/134009}

\bibitem[{{Boyle} {et~al.}(2011){Boyle}, {Owen}, \&
  {Pfeiffer}}]{2011PhRvD..84l4011B}
{Boyle}, M., {Owen}, R., \& {Pfeiffer}, H.~P. 2011, \prd, 84, 124011,
  \dodoi{10.1103/PhysRevD.84.124011}

\bibitem[{{Brown} {et~al.}(2013){Brown}, {Kumar}, \&
  {Nitz}}]{2013PhRvD..87h2004B}
{Brown}, D.~A., {Kumar}, P., \& {Nitz}, A.~H. 2013, \prd, 87, 082004,
  \dodoi{10.1103/PhysRevD.87.082004}

\bibitem[{{Buonanno} {et~al.}(2003){Buonanno}, {Chen}, \&
  {Vallisneri}}]{2003PhRvD..67j4025B}
{Buonanno}, A., {Chen}, Y., \& {Vallisneri}, M. 2003, \prd, 67, 104025,
  \dodoi{10.1103/PhysRevD.67.104025}

\bibitem[{{Buonanno} \& {Damour}(1999)}]{1999PhRvD..59h4006B}
{Buonanno}, A., \& {Damour}, T. 1999, \prd, 59, 084006,
  \dodoi{10.1103/PhysRevD.59.084006}

\bibitem[{{Buonanno} \& {Damour}(2000)}]{2000PhRvD..62f4015B}
---. 2000, \prd, 62, 064015, \dodoi{10.1103/PhysRevD.62.064015}

\bibitem[{{Buonanno} {et~al.}(2009){Buonanno}, {Iyer}, {Ochsner}, {Pan}, \&
  {Sathyaprakash}}]{2009PhRvD..80h4043B}
{Buonanno}, A., {Iyer}, B.~R., {Ochsner}, E., {Pan}, Y., \& {Sathyaprakash},
  B.~S. 2009, \prd, 80, 084043, \dodoi{10.1103/PhysRevD.80.084043}

\bibitem[{Buonanno {et~al.}(2007)Buonanno, Pan, Baker, Centrella, Kelly,
  McWilliams, \& van Meter}]{Buonanno:2007pf}
Buonanno, A., Pan, Y., Baker, J.~G., {et~al.} 2007, Phys. Rev. D, 76, 104049,
  \dodoi{10.1103/PhysRevD.76.104049}

\bibitem[{{Cabero} {et~al.}(2019){Cabero}, {Lundgren}, {Nitz}, {Dent},
  {Barker}, {Goetz}, {Kissel}, {Nuttall}, {Schale}, {Schofield}, \&
  {Davis}}]{2019CQGra..36o5010C}
{Cabero}, M., {Lundgren}, A., {Nitz}, A.~H., {et~al.} 2019, Classical and
  Quantum Gravity, 36, 155010, \dodoi{10.1088/1361-6382/ab2e14}

\bibitem[{{Cabourn Davies} \& {Harry}(2022)}]{2022CQGra..39u5012C}
{Cabourn Davies}, G.~S., \& {Harry}, I.~W. 2022, Classical and Quantum Gravity,
  39, 215012, \dodoi{10.1088/1361-6382/ac8862}

\bibitem[{{Cahillane} {et~al.}(2017){Cahillane}, {Betzwieser}, {Brown},
  {Goetz}, {Hall}, {Izumi}, {Kandhasamy}, {Karki}, {Kissel}, {Mendell},
  {Savage}, {Tuyenbayev}, {Urban}, {Viets}, {Wade}, \&
  {Weinstein}}]{2017PhRvD..96j2001C}
{Cahillane}, C., {Betzwieser}, J., {Brown}, D.~A., {et~al.} 2017, \prd, 96,
  102001, \dodoi{10.1103/PhysRevD.96.102001}

\bibitem[{{Calder{\'o}n Bustillo} {et~al.}(2016){Calder{\'o}n Bustillo},
  {Husa}, {Sintes}, \& {P{\"u}rrer}}]{2016PhRvD..93h4019C}
{Calder{\'o}n Bustillo}, J., {Husa}, S., {Sintes}, A.~M., \& {P{\"u}rrer}, M.
  2016, \prd, 93, 084019, \dodoi{10.1103/PhysRevD.93.084019}

\bibitem[{{Calder{\'o}n Bustillo} {et~al.}(2017){Calder{\'o}n Bustillo},
  {Laguna}, \& {Shoemaker}}]{2017PhRvD..95j4038C}
{Calder{\'o}n Bustillo}, J., {Laguna}, P., \& {Shoemaker}, D. 2017, \prd, 95,
  104038, \dodoi{10.1103/PhysRevD.95.104038}

\bibitem[{Calder\'on~Bustillo {et~al.}(2021)Calder\'on~Bustillo, Sanchis-Gual,
  Torres-Forn\'e, \& Font}]{CalderonBustillo:2020xms}
Calder\'on~Bustillo, J., Sanchis-Gual, N., Torres-Forn\'e, A., \& Font, J.~A.
  2021, Phys. Rev. Lett., 126, 201101, \dodoi{10.1103/PhysRevLett.126.201101}

\bibitem[{{Canizares} {et~al.}(2015){Canizares}, {Field}, {Gair}, {Raymond},
  {Smith}, \& {Tiglio}}]{2015PhRvL.114g1104C}
{Canizares}, P., {Field}, S.~E., {Gair}, J., {et~al.} 2015, \prl, 114, 071104,
  \dodoi{10.1103/PhysRevLett.114.071104}

\bibitem[{{Cannon} {et~al.}(2012{\natexlab{a}}){Cannon}, {Hanna}, \&
  {Keppel}}]{2012PhRvD..85h1504C}
{Cannon}, K., {Hanna}, C., \& {Keppel}, D. 2012{\natexlab{a}}, \prd, 85,
  081504, \dodoi{10.1103/PhysRevD.85.081504}

\bibitem[{{Cannon} {et~al.}(2013){Cannon}, {Hanna}, \&
  {Keppel}}]{2013PhRvD..88b4025C}
---. 2013, \prd, 88, 024025, \dodoi{10.1103/PhysRevD.88.024025}

\bibitem[{{Cannon} {et~al.}(2015){Cannon}, {Hanna}, \&
  {Peoples}}]{2015arXiv150404632C}
{Cannon}, K., {Hanna}, C., \& {Peoples}, J. 2015, arXiv e-prints,
  arXiv:1504.04632, \dodoi{10.48550/arXiv.1504.04632}

\bibitem[{{Cannon} {et~al.}(2012{\natexlab{b}}){Cannon}, {Cariou}, {Chapman},
  {Crispin-Ortuzar}, {Fotopoulos}, {Frei}, {Hanna}, {Kara}, {Keppel}, {Liao},
  {Privitera}, {Searle}, {Singer}, \& {Weinstein}}]{2012ApJ...748..136C}
{Cannon}, K., {Cariou}, R., {Chapman}, A., {et~al.} 2012{\natexlab{b}}, \apj,
  748, 136, \dodoi{10.1088/0004-637X/748/2/136}

\bibitem[{{Cannon} {et~al.}(2021){Cannon}, {Caudill}, {Chan}, {Cousins},
  {Creighton}, {Ewing}, {Fong}, {Godwin}, {Hanna}, {Hooper}, {Huxford},
  {Magee}, {Meacher}, {Messick}, {Morisaki}, {Mukherjee}, {Ohta}, {Pace},
  {Privitera}, {de Ruiter}, {Sachdev}, {Singer}, {Singh}, {Tapia}, {Tsukada},
  {Tsuna}, {Tsutsui}, {Ueno}, {Viets}, {Wade}, \& {Wade}}]{2021SoftX..1400680C}
{Cannon}, K., {Caudill}, S., {Chan}, C., {et~al.} 2021, SoftwareX, 14, 100680,
  \dodoi{10.1016/j.softx.2021.100680}

\bibitem[{Capano {et~al.}(2014)Capano, Pan, \& Buonanno}]{Capano:2013raa}
Capano, C., Pan, Y., \& Buonanno, A. 2014, Phys. Rev. D, 89, 102003,
  \dodoi{10.1103/PhysRevD.89.102003}

\bibitem[{{Chandra} {et~al.}(2021){Chandra}, {Villa-Ortega}, {Dent}, {McIsaac},
  {Pai}, {Harry}, {Davies}, \& {Soni}}]{2021PhRvD.104d2004C}
{Chandra}, K., {Villa-Ortega}, V., {Dent}, T., {et~al.} 2021, \prd, 104,
  042004, \dodoi{10.1103/PhysRevD.104.042004}

\bibitem[{Chatziioannou(2020)}]{Chatziioannou:2020pqz}
Chatziioannou, K. 2020, Gen. Rel. Grav., 52, 109,
  \dodoi{10.1007/s10714-020-02754-3}

\bibitem[{{Chatziioannou} {et~al.}(2015){Chatziioannou}, {Cornish}, {Klein}, \&
  {Yunes}}]{2015ApJ...798L..17C}
{Chatziioannou}, K., {Cornish}, N., {Klein}, A., \& {Yunes}, N. 2015, \apjl,
  798, L17, \dodoi{10.1088/2041-8205/798/1/L17}

\bibitem[{{Chatziioannou} {et~al.}(2021){Chatziioannou}, {Cornish},
  {Wijngaarden}, \& {Littenberg}}]{2021PhRvD.103d4013C}
{Chatziioannou}, K., {Cornish}, N., {Wijngaarden}, M., \& {Littenberg}, T.~B.
  2021, \prd, 103, 044013, \dodoi{10.1103/PhysRevD.103.044013}

\bibitem[{{Chatziioannou} {et~al.}(2019){Chatziioannou}, {Haster},
  {Littenberg}, {Farr}, {Ghonge}, {Millhouse}, {Clark}, \&
  {Cornish}}]{2019PhRvD.100j4004C}
{Chatziioannou}, K., {Haster}, C.-J., {Littenberg}, T.~B., {et~al.} 2019, \prd,
  100, 104004, \dodoi{10.1103/PhysRevD.100.104004}

\bibitem[{Chatziioannou {et~al.}(2013)Chatziioannou, Klein, Yunes, \&
  Cornish}]{Chatziioannou:2013dza}
Chatziioannou, K., Klein, A., Yunes, N., \& Cornish, N. 2013, Phys. Rev. D, 88,
  063011, \dodoi{10.1103/PhysRevD.88.063011}

\bibitem[{{Chatziioannou} {et~al.}(2017){Chatziioannou}, {Klein}, {Yunes}, \&
  {Cornish}}]{2017PhRvD..95j4004C}
{Chatziioannou}, K., {Klein}, A., {Yunes}, N., \& {Cornish}, N. 2017, \prd, 95,
  104004, \dodoi{10.1103/PhysRevD.95.104004}

\bibitem[{Chia {et~al.}(2024)Chia, Edwards, Wadekar, Zimmerman, Olsen, Roulet,
  Venumadhav, Zackay, \& Zaldarriaga}]{Chia:2023tle}
Chia, H.~S., Edwards, T. D.~P., Wadekar, D., {et~al.} 2024, Phys. Rev. D, 110,
  063007, \dodoi{10.1103/PhysRevD.110.063007}

\bibitem[{{Chu} {et~al.}(2022){Chu}, {Kovalam}, {Wen}, {Slaven-Blair},
  {Bosveld}, {Chen}, {Clearwater}, {Codoreanu}, {Du}, {Guo}, {Guo}, {Kim},
  {Li}, {Oloworaran}, {Panther}, {Powell}, {Sengupta}, {Wette}, \&
  {Zhu}}]{2022PhRvD.105b4023C}
{Chu}, Q., {Kovalam}, M., {Wen}, L., {et~al.} 2022, \prd, 105, 024023,
  \dodoi{10.1103/PhysRevD.105.024023}

\bibitem[{{Colleoni} {et~al.}(2025{\natexlab{a}}){Colleoni}, {Ramis Vidal},
  {Garc{\'\i}a-Quir{\'o}s}, {Ak{\c{c}}ay}, \& {Bera}}]{2025PhRvD.111j4019C}
{Colleoni}, M., {Ramis Vidal}, F.~A., {Garc{\'\i}a-Quir{\'o}s}, C.,
  {Ak{\c{c}}ay}, S., \& {Bera}, S. 2025{\natexlab{a}}, \prd, 111, 104019,
  \dodoi{10.1103/PhysRevD.111.104019}

\bibitem[{{Colleoni} {et~al.}(2025{\natexlab{b}}){Colleoni}, {Ramis Vidal},
  {Johnson-McDaniel}, {Dietrich}, {Haney}, \& {Pratten}}]{2025PhRvD.111f4025C}
{Colleoni}, M., {Ramis Vidal}, F.~A., {Johnson-McDaniel}, N.~K., {et~al.}
  2025{\natexlab{b}}, \prd, 111, 064025, \dodoi{10.1103/PhysRevD.111.064025}

\bibitem[{{Cornish} \& {Littenberg}(2015)}]{2015CQGra..32m5012C}
{Cornish}, N.~J., \& {Littenberg}, T.~B. 2015, Classical and Quantum Gravity,
  32, 135012, \dodoi{10.1088/0264-9381/32/13/135012}

\bibitem[{{Cornish} {et~al.}(2021){Cornish}, {Littenberg}, {B{\'e}csy},
  {Chatziioannou}, {Clark}, {Ghonge}, \& {Millhouse}}]{2021PhRvD.103d4006C}
{Cornish}, N.~J., {Littenberg}, T.~B., {B{\'e}csy}, B., {et~al.} 2021, \prd,
  103, 044006, \dodoi{10.1103/PhysRevD.103.044006}

\bibitem[{{Cotesta} {et~al.}(2018){Cotesta}, {Buonanno}, {Boh{\'e}},
  {Taracchini}, {Hinder}, \& {Ossokine}}]{2018PhRvD..98h4028C}
{Cotesta}, R., {Buonanno}, A., {Boh{\'e}}, A., {et~al.} 2018, \prd, 98, 084028,
  \dodoi{10.1103/PhysRevD.98.084028}

\bibitem[{{Cotesta} {et~al.}(2020){Cotesta}, {Marsat}, \&
  {P{\"u}rrer}}]{2020PhRvD.101l4040C}
{Cotesta}, R., {Marsat}, S., \& {P{\"u}rrer}, M. 2020, \prd, 101, 124040,
  \dodoi{10.1103/PhysRevD.101.124040}

\bibitem[{Creighton(2019)}]{Creighton:fgmc_identities}
Creighton, J. 2019, {Certain Identities in FGMC}, Tech. Rep. LIGO-T1700029-v2,
  {LIGO}.
\newblock \url{https://dcc.ligo.org/T1700029/public}

\bibitem[{{Cutler} \& {Flanagan}(1994)}]{1994PhRvD..49.2658C}
{Cutler}, C., \& {Flanagan}, {\'E}.~E. 1994, \prd, 49, 2658,
  \dodoi{10.1103/PhysRevD.49.2658}

\bibitem[{{Dal Canton} \& {Harry}(2017)}]{2017arXiv170501845D}
{Dal Canton}, T., \& {Harry}, I.~W. 2017, arXiv e-prints, arXiv:1705.01845,
  \dodoi{10.48550/arXiv.1705.01845}

\bibitem[{Dal~Canton {et~al.}(2015)Dal~Canton, Lundgren, \&
  Nielsen}]{DalCanton:2014qjd}
Dal~Canton, T., Lundgren, A.~P., \& Nielsen, A.~B. 2015, Phys. Rev. D, 91,
  062010, \dodoi{10.1103/PhysRevD.91.062010}

\bibitem[{{Dal Canton} {et~al.}(2021){Dal Canton}, {Nitz}, {Gadre}, {Cabourn
  Davies}, {Villa-Ortega}, {Dent}, {Harry}, \& {Xiao}}]{2021ApJ...923..254D}
{Dal Canton}, T., {Nitz}, A.~H., {Gadre}, B., {et~al.} 2021, \apj, 923, 254,
  \dodoi{10.3847/1538-4357/ac2f9a}

\bibitem[{{Dal Canton} {et~al.}(2014){Dal Canton}, {Nitz}, {Lundgren},
  {Nielsen}, {Brown}, {Dent}, {Harry}, {Krishnan}, {Miller}, {Wette},
  {Wiesner}, \& {Willis}}]{2014PhRvD..90h2004D}
{Dal Canton}, T., {Nitz}, A.~H., {Lundgren}, A.~P., {et~al.} 2014, \prd, 90,
  082004, \dodoi{10.1103/PhysRevD.90.082004}

\bibitem[{Damour {et~al.}(2001)Damour, Iyer, \& Sathyaprakash}]{Damour:2000zb}
Damour, T., Iyer, B.~R., \& Sathyaprakash, B.~S. 2001, Phys. Rev. D, 63,
  044023, \dodoi{10.1103/PhysRevD.63.044023}

\bibitem[{Damour \& Nagar(2014)}]{Damour:2014sva}
Damour, T., \& Nagar, A. 2014, Phys. Rev. D, 90, 044018,
  \dodoi{10.1103/PhysRevD.90.044018}

\bibitem[{{Damour} {et~al.}(2012){Damour}, {Nagar}, \&
  {Villain}}]{2012PhRvD..85l3007D}
{Damour}, T., {Nagar}, A., \& {Villain}, L. 2012, \prd, 85, 123007,
  \dodoi{10.1103/PhysRevD.85.123007}

\bibitem[{Dartez {et~al.}(2025)}]{O4aCalPaper}
Dartez, L., {et~al.} 2025, {Characterization of systematic error in Advanced
  LIGO calibration in the fourth observing run}, In preparation

\bibitem[{{Davies} {et~al.}(2020){Davies}, {Dent}, {T{\'a}pai}, {Harry},
  {McIsaac}, \& {Nitz}}]{2020PhRvD.102b2004D}
{Davies}, G.~S., {Dent}, T., {T{\'a}pai}, M., {et~al.} 2020, \prd, 102, 022004,
  \dodoi{10.1103/PhysRevD.102.022004}

\bibitem[{{Davis} {et~al.}(2022){Davis}, {Littenberg}, {Romero-Shaw},
  {Millhouse}, {McIver}, {Di Renzo}, \& {Ashton}}]{2022CQGra..39x5013D}
{Davis}, D., {Littenberg}, T.~B., {Romero-Shaw}, I.~M., {et~al.} 2022,
  Classical and Quantum Gravity, 39, 245013, \dodoi{10.1088/1361-6382/aca238}

\bibitem[{{Davis} {et~al.}(2026){Davis}, {Yarbrough}, {Areeda}, {Macas},
  {Arnaud}, {Helmling-Cornell}, {Doliva}, {Godwin}, {Yuzurihara}, {Mannix},
  {Alvarez-Lopez}, {Trevor}, {Huxford}, {Nguyen}, {Berger}, {Chatterjee}, {Di
  Renzo}, {Palomba}, {Sordini}, {Pesios}, {Walker}, {Ahuja}, {Chan}, {Ding},
  {Frey}, {Herbst}, {Lecoeuche}, {Liyanage}, {McIver}, {Ng}, {Perry},
  {Rawcliffe}, \& {Schofield}}]{2026arXiv260516183D}
{Davis}, D., {Yarbrough}, Z., {Areeda}, J., {et~al.} 2026, arXiv e-prints,
  arXiv:2605.16183.
\newblock \doarXiv{2605.16183}

\bibitem[{{Dax} {et~al.}(2021){Dax}, {Green}, {Gair}, {Macke}, {Buonanno}, \&
  {Sch{\"o}lkopf}}]{2021PhRvL.127x1103D}
{Dax}, M., {Green}, S.~R., {Gair}, J., {et~al.} 2021, \prl, 127, 241103,
  \dodoi{10.1103/PhysRevLett.127.241103}

\bibitem[{{Dax} {et~al.}(2023){Dax}, {Green}, {Gair}, {P{\"u}rrer},
  {Wildberger}, {Macke}, {Buonanno}, \& {Sch{\"o}lkopf}}]{2023PhRvL.130q1403D}
---. 2023, \prl, 130, 171403, \dodoi{10.1103/PhysRevLett.130.171403}

\bibitem[{Dent(2025)}]{Dent:multi_pastro}
Dent, T. 2025, {Extending the PyCBC pastro calculation to a global network},
  Tech. Rep. LIGO-T2100060-v3, {LIGO}.
\newblock \url{https://dcc.ligo.org/T2100060/public}

\bibitem[{Devine {et~al.}(2016)Devine, Etienne, \& McWilliams}]{Devine:2016ovp}
Devine, C., Etienne, Z.~B., \& McWilliams, S.~T. 2016, Class. Quant. Grav., 33,
  125025, \dodoi{10.1088/0264-9381/33/12/125025}

\bibitem[{Dhurkunde \& Nitz(2025)}]{Dhurkunde:2025}
Dhurkunde, R., \& Nitz, A.~H. 2025, Physical Review D, 111,
  \dodoi{10.1103/physrevd.111.103018}

\bibitem[{{Di Renzo} {et~al.}(2024){Di Renzo}, {Fidecaro}, {Razzano}, \&
  {Sorrentino}}]{2024arXiv240115392D}
{Di Renzo}, F., {Fidecaro}, F., {Razzano}, M., \& {Sorrentino}, N. 2024, arXiv
  e-prints, arXiv:2401.15392, \dodoi{10.48550/arXiv.2401.15392}

\bibitem[{{Di Renzo} \& {Virgo Collaboration}(2023)}]{2023chep.confE.110D}
{Di Renzo}, F., \& {Virgo Collaboration}. 2023, in 41st International
  Conference on High Energy Physics, 110

\bibitem[{{Dietrich} {et~al.}(2017){Dietrich}, {Bernuzzi}, \&
  {Tichy}}]{2017PhRvD..96l1501D}
{Dietrich}, T., {Bernuzzi}, S., \& {Tichy}, W. 2017, \prd, 96, 121501,
  \dodoi{10.1103/PhysRevD.96.121501}

\bibitem[{{Dietrich} {et~al.}(2019{\natexlab{a}}){Dietrich}, {Samajdar},
  {Khan}, {Johnson-McDaniel}, {Dudi}, \& {Tichy}}]{2019PhRvD.100d4003D}
{Dietrich}, T., {Samajdar}, A., {Khan}, S., {et~al.} 2019{\natexlab{a}}, \prd,
  100, 044003, \dodoi{10.1103/PhysRevD.100.044003}

\bibitem[{Dietrich {et~al.}(2018)Dietrich, Radice, Bernuzzi, Zappa, Perego,
  Br\"ugmann, Chaurasia, Dudi, Tichy, \& Ujevic}]{Dietrich:2018phi}
Dietrich, T., Radice, D., Bernuzzi, S., {et~al.} 2018, Class. Quant. Grav., 35,
  24LT01, \dodoi{10.1088/1361-6382/aaebc0}

\bibitem[{{Dietrich} {et~al.}(2019{\natexlab{b}}){Dietrich}, {Khan}, {Dudi},
  {Kapadia}, {Kumar}, {Nagar}, {Ohme}, {Pannarale}, {Samajdar}, {Bernuzzi},
  {Carullo}, {Del Pozzo}, {Haney}, {Markakis}, {P{\"u}rrer}, {Riemenschneider},
  {Setyawati}, {Tsang}, \& {Van Den Broeck}}]{2019PhRvD..99b4029D}
{Dietrich}, T., {Khan}, S., {Dudi}, R., {et~al.} 2019{\natexlab{b}}, \prd, 99,
  024029, \dodoi{10.1103/PhysRevD.99.024029}

\bibitem[{{Drago} {et~al.}(2021){Drago}, {Klimenko}, {Lazzaro}, {Milotti},
  {Mitselmakher}, {Necula}, {O'Brian}, {Prodi}, {Salemi}, {Szczepanczyk},
  {Tiwari}, {Tiwari}, {Gayathri}, {Vedovato}, \&
  {Yakushin}}]{2021SoftX..1400678D}
{Drago}, M., {Klimenko}, S., {Lazzaro}, C., {et~al.} 2021, SoftwareX, 14,
  100678, \dodoi{10.1016/j.softx.2021.100678}

\bibitem[{{Earl} \& {Deem}(2005)}]{2005PCCP....7.3910E}
{Earl}, D.~J., \& {Deem}, M.~W. 2005, Physical Chemistry Chemical Physics
  (Incorporating Faraday Transactions), 7, 3910, \dodoi{10.1039/B509983H}

\bibitem[{Essick {et~al.}(2020)Essick, Godwin, Hanna, Blackburn, \&
  Katsavounidis}]{Essick:2020qpo}
Essick, R., Godwin, P., Hanna, C., Blackburn, L., \& Katsavounidis, E. 2020,
  Machine Learning: Science and Technology, 2, 015004,
  \dodoi{10.1088/2632-2153/abab5f}

\bibitem[{{Essick} {et~al.}(2020){Essick}, {Godwin}, {Hanna}, {Blackburn}, \&
  {Katsavounidis}}]{2020arXiv200512761E}
{Essick}, R., {Godwin}, P., {Hanna}, C., {Blackburn}, L., \& {Katsavounidis},
  E. 2020, arXiv e-prints, arXiv:2005.12761, \dodoi{10.48550/arXiv.2005.12761}

\bibitem[{{Essick} {et~al.}(2025){Essick}, {Coughlin}, {Zevin}, {Chatterjee},
  {Clarke}, {Colloms}, {Mali}, {Miller}, {Steinle}, {Baral}, {Baylor}, {Cabourn
  Davies}, {Dent}, {Joshi}, {Kumar}, {Messick}, {Mishra}, {Ouzriat}, {Phukon},
  {Piccari}, {Pillas}, {Trevor}, {Callister}, \&
  {Fishbach}}]{2025PhRvD.112j2001E}
{Essick}, R., {Coughlin}, M.~W., {Zevin}, M., {et~al.} 2025, \prd, 112, 102001,
  \dodoi{10.1103/44x3-hv3y}

\bibitem[{{Estell{\'e}s} {et~al.}(2026){Estell{\'e}s}, {Buonanno}, {Enficiaud},
  {Foo}, \& {Pompili}}]{2026PhRvD.113d4049E}
{Estell{\'e}s}, H., {Buonanno}, A., {Enficiaud}, R., {Foo}, C., \& {Pompili},
  L. 2026, \prd, 113, 044049, \dodoi{10.1103/pjbd-pjxn}

\bibitem[{{Estell{\'e}s} {et~al.}(2022{\natexlab{a}}){Estell{\'e}s},
  {Colleoni}, {Garc{\'\i}a-Quir{\'o}s}, {Husa}, {Keitel}, {Mateu-Lucena},
  {Planas}, \& {Ramos-Buades}}]{2022PhRvD.105h4040E}
{Estell{\'e}s}, H., {Colleoni}, M., {Garc{\'\i}a-Quir{\'o}s}, C., {et~al.}
  2022{\natexlab{a}}, \prd, 105, 084040, \dodoi{10.1103/PhysRevD.105.084040}

\bibitem[{{Estell{\'e}s} {et~al.}(2022{\natexlab{b}}){Estell{\'e}s}, {Husa},
  {Colleoni}, {Keitel}, {Mateu-Lucena}, {Garc{\'\i}a-Quir{\'o}s},
  {Ramos-Buades}, \& {Borchers}}]{2022PhRvD.105h4039E}
{Estell{\'e}s}, H., {Husa}, S., {Colleoni}, M., {et~al.} 2022{\natexlab{b}},
  \prd, 105, 084039, \dodoi{10.1103/PhysRevD.105.084039}

\bibitem[{{Estell{\'e}s} {et~al.}(2021){Estell{\'e}s}, {Ramos-Buades}, {Husa},
  {Garc{\'\i}a-Quir{\'o}s}, {Colleoni}, {Haegel}, \&
  {Jaume}}]{2021PhRvD.103l4060E}
{Estell{\'e}s}, H., {Ramos-Buades}, A., {Husa}, S., {et~al.} 2021, \prd, 103,
  124060, \dodoi{10.1103/PhysRevD.103.124060}

\bibitem[{{Ewing} {et~al.}(2024){Ewing}, {Huxford}, {Singh}, {Tsukada},
  {Hanna}, {Huang}, {Joshi}, {Li}, {Magee}, {Messick}, {Pace}, {Ray},
  {Sachdev}, {Sakon}, {Tapia}, {Adhicary}, {Baral}, {Baylor}, {Cannon},
  {Caudill}, {Chaudhary}, {Coughlin}, {Cousins}, {Creighton}, {Essick}, {Fong},
  {George}, {Godwin}, {Harada}, {Kennington}, {Kuwahara}, {Meacher},
  {Morisaki}, {Mukherjee}, {Niu}, {Posnansky}, {Toivonen}, {Tsutsui}, {Ueno},
  {Viets}, {Wade}, {Wade}, \& {Waratkar}}]{2024PhRvD.109d2008E}
{Ewing}, B., {Huxford}, R., {Singh}, D., {et~al.} 2024, \prd, 109, 042008,
  \dodoi{10.1103/PhysRevD.109.042008}

\bibitem[{Farr(2014)}]{Farr:2014aaa}
Farr, W. 2014, {Marginalisation of the time and phase parameters in CBC
  parameter estimation}, Tech. Rep. DCC-T1400460, {LIGO}.
\newblock \url{https://dcc.ligo.org/LIGO-T1400460/public}

\bibitem[{Farr {et~al.}(2014)Farr, Farr, \& Littenberg}]{Farr:2014aab}
Farr, W., Farr, B., \& Littenberg, T. 2014, {Modelling calibration errors in
  CBC waveforms}, Tech. Rep. DCC-T1400682, {LIGO}.
\newblock \url{https://dcc.ligo.org/LIGO-T1400682/public}

\bibitem[{{Farr} {et~al.}(2015){Farr}, {Gair}, {Mandel}, \&
  {Cutler}}]{2015PhRvD..91b3005F}
{Farr}, W.~M., {Gair}, J.~R., {Mandel}, I., \& {Cutler}, C. 2015, \prd, 91,
  023005, \dodoi{10.1103/PhysRevD.91.023005}

\bibitem[{{Favata} {et~al.}(2022){Favata}, {Kim}, {Arun}, {Kim}, \&
  {Lee}}]{2022PhRvD.105b3003F}
{Favata}, M., {Kim}, C., {Arun}, K.~G., {Kim}, J., \& {Lee}, H.~W. 2022, \prd,
  105, 023003, \dodoi{10.1103/PhysRevD.105.023003}

\bibitem[{{Finn}(1992)}]{1992PhRvD..46.5236F}
{Finn}, L.~S. 1992, \prd, 46, 5236, \dodoi{10.1103/PhysRevD.46.5236}

\bibitem[{{Flanagan} \& {Hinderer}(2008)}]{2008PhRvD..77b1502F}
{Flanagan}, {\'E}.~{\'E}., \& {Hinderer}, T. 2008, \prd, 77, 021502,
  \dodoi{10.1103/PhysRevD.77.021502}

\bibitem[{Fong(2018)}]{Fong:2018elx}
Fong, H. K.~Y. 2018, PhD thesis, Toronto U.

\bibitem[{Foucart {et~al.}(2014)Foucart, Deaton, Duez, O'Connor, Ott, Haas,
  Kidder, Pfeiffer, Scheel, \& Szilagyi}]{Foucart:2014nda}
Foucart, F., Deaton, M.~B., Duez, M.~D., {et~al.} 2014, Phys. Rev. D, 90,
  024026, \dodoi{10.1103/PhysRevD.90.024026}

\bibitem[{Gamba {et~al.}(2022)Gamba, Ak{\c{c}}ay, Bernuzzi, \&
  Williams}]{Gamba:2021ydi}
Gamba, R., Ak{\c{c}}ay, S., Bernuzzi, S., \& Williams, J. 2022, Phys. Rev. D,
  106, 024020, \dodoi{10.1103/PhysRevD.106.024020}

\bibitem[{Gamba {et~al.}(2021)Gamba, Bernuzzi, \& Nagar}]{Gamba:2020ljo}
Gamba, R., Bernuzzi, S., \& Nagar, A. 2021, Phys. Rev. D, 104, 084058,
  \dodoi{10.1103/PhysRevD.104.084058}

\bibitem[{Gamba {et~al.}(2023{\natexlab{a}})Gamba, Breschi, Carullo, Albanesi,
  Rettegno, Bernuzzi, \& Nagar}]{Gamba:2021gap}
Gamba, R., Breschi, M., Carullo, G., {et~al.} 2023{\natexlab{a}}, Nature
  Astron., 7, 11, \dodoi{10.1038/s41550-022-01813-w}

\bibitem[{Gamba {et~al.}(2023{\natexlab{b}})}]{Gamba:2023mww}
Gamba, R., {et~al.} 2023{\natexlab{b}}, {Analytically improved and
  numerical-relativity informed effective-one-body model for coalescing binary
  neutron stars}.
\newblock \doarXiv{2307.15125}

\bibitem[{{Garc{\'\i}a-Quir{\'o}s} {et~al.}(2020){Garc{\'\i}a-Quir{\'o}s},
  {Colleoni}, {Husa}, {Estell{\'e}s}, {Pratten}, {Ramos-Buades},
  {Mateu-Lucena}, \& {Jaume}}]{2020PhRvD.102f4002G}
{Garc{\'\i}a-Quir{\'o}s}, C., {Colleoni}, M., {Husa}, S., {et~al.} 2020, \prd,
  102, 064002, \dodoi{10.1103/PhysRevD.102.064002}

\bibitem[{{Garc{\'\i}a-Quir{\'o}s} {et~al.}(2021){Garc{\'\i}a-Quir{\'o}s},
  {Husa}, {Mateu-Lucena}, \& {Borchers}}]{2021CQGra..38a5006G}
{Garc{\'\i}a-Quir{\'o}s}, C., {Husa}, S., {Mateu-Lucena}, M., \& {Borchers}, A.
  2021, Classical and Quantum Gravity, 38, 015006,
  \dodoi{10.1088/1361-6382/abc36e}

\bibitem[{Gayathri {et~al.}(2022)Gayathri, Healy, Lange, O'Brien, Szczepanczyk,
  Bartos, Campanelli, Klimenko, Lousto, \& O'Shaughnessy}]{Gayathri:2020coq}
Gayathri, V., Healy, J., Lange, J., {et~al.} 2022, Nature Astron., 6, 344,
  \dodoi{10.1038/s41550-021-01568-w}

\bibitem[{Gerosa {et~al.}(2023)Gerosa, Fumagalli, Mould, Cavallotto, Monroy,
  Gangardt, \& De~Renzis}]{Gerosa:2023xsx}
Gerosa, D., Fumagalli, G., Mould, M., {et~al.} 2023, Phys. Rev. D, 108, 024042,
  \dodoi{10.1103/PhysRevD.108.024042}

\bibitem[{{Gerosa} {et~al.}(2015){Gerosa}, {Kesden}, {Sperhake}, {Berti}, \&
  {O'Shaughnessy}}]{2015PhRvD..92f4016G}
{Gerosa}, D., {Kesden}, M., {Sperhake}, U., {Berti}, E., \& {O'Shaughnessy}, R.
  2015, \prd, 92, 064016, \dodoi{10.1103/PhysRevD.92.064016}

\bibitem[{{Ghonge} {et~al.}(2020){Ghonge}, {Chatziioannou}, {Clark},
  {Littenberg}, {Millhouse}, {Cadonati}, \& {Cornish}}]{2020PhRvD.102f4056G}
{Ghonge}, S., {Chatziioannou}, K., {Clark}, J.~A., {et~al.} 2020, \prd, 102,
  064056, \dodoi{10.1103/PhysRevD.102.064056}

\bibitem[{Ghonge {et~al.}(2024)Ghonge, Brandt, Sullivan, Millhouse,
  Chatziioannou, Clark, Littenberg, Cornish, Hourihane, \&
  Cadonati}]{Ghonge:2023ksb}
Ghonge, S., Brandt, J., Sullivan, J.~M., {et~al.} 2024, Phys. Rev. D, 110,
  122002, \dodoi{10.1103/PhysRevD.110.122002}

\bibitem[{Ghosh {et~al.}(2024)Ghosh, Kolitsidou, \& Hannam}]{Ghosh:2023mhc}
Ghosh, S., Kolitsidou, P., \& Hannam, M. 2024, Phys. Rev. D, 109, 024061,
  \dodoi{10.1103/PhysRevD.109.024061}

\bibitem[{{Glanzer} {et~al.}(2023){Glanzer}, {Banagiri}, {Coughlin}, {Soni},
  {Zevin}, {Berry}, {Patane}, {Bahaadini}, {Rohani}, {Crowston}, {Kalogera},
  {{\O}sterlund}, {Trouille}, \& {Katsaggelos}}]{2023CQGra..40f5004G}
{Glanzer}, J., {Banagiri}, S., {Coughlin}, S.~B., {et~al.} 2023, Classical and
  Quantum Gravity, 40, 065004, \dodoi{10.1088/1361-6382/acb633}

\bibitem[{{Godwin}(2020)}]{2020PhDT........28G}
{Godwin}, P. 2020, PhD thesis, Pennsylvania State University

\bibitem[{{Godwin} {et~al.}(2020){Godwin}, {Essick}, {Hanna}, {Cannon},
  {Caudill}, {Chan}, {Creighton}, {Fong}, {Katsavounidis}, {Magee}, {Meacher},
  {Messick}, {Morisaki}, {Mukherjee}, {Ohta}, {Pace}, {de Ruiter}, {Sachdev},
  {Tsukada}, {Tsutsui}, {Ueno}, {Wade}, \& {Wade}}]{2020arXiv201015282G}
{Godwin}, P., {Essick}, R., {Hanna}, C., {et~al.} 2020, arXiv e-prints,
  arXiv:2010.15282, \dodoi{10.48550/arXiv.2010.15282}

\bibitem[{{Goggans} \& {Chi}(2004)}]{2004AIPC..707...59G}
{Goggans}, P.~M., \& {Chi}, Y. 2004, in American Institute of Physics
  Conference Series, Vol. 707, Bayesian Inference and Maximum Entropy Methods
  in Science and Engineering, ed. G.~J. {Erickson} \& Y.~{Zhai} (AIP), 59--66,
  \dodoi{10.1063/1.1751356}

\bibitem[{{Gonzalez} {et~al.}(2025){Gonzalez}, {Bernuzzi}, {Rashti},
  {Brandoli}, \& {Gamba}}]{2025arXiv250700113G}
{Gonzalez}, A., {Bernuzzi}, S., {Rashti}, A., {Brandoli}, F., \& {Gamba}, R.
  2025, arXiv e-prints, arXiv:2507.00113, \dodoi{10.48550/arXiv.2507.00113}

\bibitem[{Gonzalez {et~al.}(2023{\natexlab{a}})Gonzalez, Gamba, Breschi, Zappa,
  Carullo, Bernuzzi, \& Nagar}]{Gonzalez:2022prs}
Gonzalez, A., Gamba, R., Breschi, M., {et~al.} 2023{\natexlab{a}}, Phys. Rev.
  D, 107, 084026, \dodoi{10.1103/PhysRevD.107.084026}

\bibitem[{Gonzalez {et~al.}(2023{\natexlab{b}})}]{Gonzalez:2022mgo}
Gonzalez, A., {et~al.} 2023{\natexlab{b}}, Class. Quant. Grav., 40, 085011,
  \dodoi{10.1088/1361-6382/acc231}

\bibitem[{{Graff} {et~al.}(2015){Graff}, {Buonanno}, \&
  {Sathyaprakash}}]{2015PhRvD..92b2002G}
{Graff}, P.~B., {Buonanno}, A., \& {Sathyaprakash}, B.~S. 2015, \prd, 92,
  022002, \dodoi{10.1103/PhysRevD.92.022002}

\bibitem[{{Green} {et~al.}(2020){Green}, {Simpson}, \&
  {Gair}}]{2020PhRvD.102j4057G}
{Green}, S.~R., {Simpson}, C., \& {Gair}, J. 2020, \prd, 102, 104057,
  \dodoi{10.1103/PhysRevD.102.104057}

\bibitem[{{Guo} {et~al.}(2018){Guo}, {Chu}, {Chung}, {Du}, {Wen}, \&
  {Gu}}]{2018CoPhC.231...62G}
{Guo}, X., {Chu}, Q., {Chung}, S.~K., {et~al.} 2018, Computer Physics
  Communications, 231, 62, \dodoi{10.1016/j.cpc.2018.05.002}

\bibitem[{{Gupta} \& {Cornish}(2024)}]{2024PhRvD.109f4040G}
{Gupta}, T., \& {Cornish}, N.~J. 2024, \prd, 109, 064040,
  \dodoi{10.1103/PhysRevD.109.064040}

\bibitem[{Gupte {et~al.}(2024)}]{Gupte:2024jfe}
Gupte, N., {et~al.} 2024, {Evidence for eccentricity in the population of
  binary black holes observed by LIGO-Virgo-KAGRA}.
\newblock \doarXiv{2404.14286}

\bibitem[{{Hamilton} {et~al.}(2021){Hamilton}, {London}, {Thompson},
  {Fauchon-Jones}, {Hannam}, {Kalaghatgi}, {Khan}, {Pannarale}, \&
  {Vano-Vinuales}}]{2021PhRvD.104l4027H}
{Hamilton}, E., {London}, L., {Thompson}, J.~E., {et~al.} 2021, \prd, 104,
  124027, \dodoi{10.1103/PhysRevD.104.124027}

\bibitem[{{Hamilton} {et~al.}(2026){Hamilton}, {Colleoni}, {Thompson}, {Hoy},
  {Heffernan}, {Kinnear}, {Valencia}, {Ramis Vidal}, {Garc{\'\i}a-Quir{\'o}s},
  {Ghosh}, {London}, {Hannam}, \& {Husa}}]{2026PhRvD.113h4055H}
{Hamilton}, E., {Colleoni}, M., {Thompson}, J.~E., {et~al.} 2026, \prd, 113,
  084055, \dodoi{10.1103/kxsf-23rr}

\bibitem[{{Hanna} {et~al.}(2020){Hanna}, {Caudill}, {Messick}, {Reza},
  {Sachdev}, {Tsukada}, {Cannon}, {Blackburn}, {Creighton}, {Fong}, {Godwin},
  {Kapadia}, {Li}, {Magee}, {Meacher}, {Mukherjee}, {Pace}, {Privitera}, {Lo},
  \& {Wade}}]{2020PhRvD.101b2003H}
{Hanna}, C., {Caudill}, S., {Messick}, C., {et~al.} 2020, \prd, 101, 022003,
  \dodoi{10.1103/PhysRevD.101.022003}

\bibitem[{Hanna {et~al.}(2022)}]{Hanna:2021luk}
Hanna, C., {et~al.} 2022, Phys. Rev. D, 106, 084033,
  \dodoi{10.1103/PhysRevD.106.084033}

\bibitem[{{Hanna} {et~al.}(2023){Hanna}, {Kennington}, {Sakon}, {Privitera},
  {Fernandez}, {Wang}, {Messick}, {Pace}, {Cannon}, {Joshi}, {Huxford},
  {Caudill}, {Chan}, {Cousins}, {Creighton}, {Ewing}, {Fong}, {Godwin},
  {Magee}, {Meacher}, {Morisaki}, {Mukherjee}, {Ohta}, {Sachdev}, {Singh},
  {Tapia}, {Tsukada}, {Tsuna}, {Tsutsui}, {Ueno}, {Viets}, {Wade}, \&
  {Wade}}]{2023PhRvD.108d2003H}
{Hanna}, C., {Kennington}, J., {Sakon}, S., {et~al.} 2023, \prd, 108, 042003,
  \dodoi{10.1103/PhysRevD.108.042003}

\bibitem[{{Hannam} {et~al.}(2014){Hannam}, {Schmidt}, {Boh{\'e}}, {Haegel},
  {Husa}, {Ohme}, {Pratten}, \& {P{\"u}rrer}}]{2014PhRvL.113o1101H}
{Hannam}, M., {Schmidt}, P., {Boh{\'e}}, A., {et~al.} 2014, \prl, 113, 151101,
  \dodoi{10.1103/PhysRevLett.113.151101}

\bibitem[{{Harris} {et~al.}(2020){Harris}, {Millman}, {van der Walt},
  {Gommers}, {Virtanen}, {Cournapeau}, {Wieser}, {Taylor}, {Berg}, {Smith},
  {Kern}, {Picus}, {Hoyer}, {van Kerkwijk}, {Brett}, {Haldane}, {del R{\'\i}o},
  {Wiebe}, {Peterson}, {G{\'e}rard-Marchant}, {Sheppard}, {Reddy}, {Weckesser},
  {Abbasi}, {Gohlke}, \& {Oliphant}}]{2020Natur.585..357H}
{Harris}, C.~R., {Millman}, K.~J., {van der Walt}, S.~J., {et~al.} 2020, \nat,
  585, 357, \dodoi{10.1038/s41586-020-2649-2}

\bibitem[{Harris(1978)}]{Harris:1978}
Harris, F. 1978, Proceedings of the IEEE, 66, 51,
  \dodoi{10.1109/PROC.1978.10837}

\bibitem[{Harry \& Hinderer(2018)}]{Harry:2018hke}
Harry, I., \& Hinderer, T. 2018, Class. Quant. Grav., 35, 145010,
  \dodoi{10.1088/1361-6382/aac7e3}

\bibitem[{Harry {et~al.}(2016)Harry, Privitera, Boh{\'e}, \&
  Buonanno}]{Harry:2016ijz}
Harry, I., Privitera, S., Boh{\'e}, A., \& Buonanno, A. 2016, Phys. Rev. D, 94,
  024012, \dodoi{10.1103/PhysRevD.94.024012}

\bibitem[{{Harry} {et~al.}(2009){Harry}, {Allen}, \&
  {Sathyaprakash}}]{2009PhRvD..80j4014H}
{Harry}, I.~W., {Allen}, B., \& {Sathyaprakash}, B.~S. 2009, \prd, 80, 104014,
  \dodoi{10.1103/PhysRevD.80.104014}

\bibitem[{Harry \& Fairhurst(2011)}]{Harry:2011chs}
Harry, I.~W., \& Fairhurst, S. 2011, Physical Review D, 83,
  \dodoi{10.1103/physrevd.83.084002}

\bibitem[{{Hastings}(1970)}]{1970Bimka..57...97H}
{Hastings}, W.~K. 1970, Biometrika, 57, 97, \dodoi{10.1093/biomet/57.1.97}

\bibitem[{{Healy} \& {Lousto}(2017)}]{2017PhRvD..95b4037H}
{Healy}, J., \& {Lousto}, C.~O. 2017, \prd, 95, 024037,
  \dodoi{10.1103/PhysRevD.95.024037}

\bibitem[{{Helmling-Cornell} {et~al.}(2024){Helmling-Cornell}, {Nguyen},
  {Schofield}, \& {Frey}}]{2024CQGra..41n5003H}
{Helmling-Cornell}, A.~F., {Nguyen}, P., {Schofield}, R.~M.~S., \& {Frey}, R.
  2024, Classical and Quantum Gravity, 41, 145003,
  \dodoi{10.1088/1361-6382/ad5139}

\bibitem[{Henry(2023)}]{Henry:2022ccf}
Henry, Q. 2023, Phys. Rev. D, 107, 044057, \dodoi{10.1103/PhysRevD.107.044057}

\bibitem[{Henry {et~al.}(2020)Henry, Faye, \& Blanchet}]{Henry:2020ski}
Henry, Q., Faye, G., \& Blanchet, L. 2020, Phys. Rev. D, 102, 044033,
  \dodoi{10.1103/PhysRevD.102.044033}

\bibitem[{Hinderer(2008)}]{Hinderer:2007mb}
Hinderer, T. 2008, Astrophys. J., 677, 1216, \dodoi{10.1086/533487}

\bibitem[{{Hinderer} {et~al.}(2016){Hinderer}, {Taracchini}, {Foucart},
  {Buonanno}, {Steinhoff}, {Duez}, {Kidder}, {Pfeiffer}, {Scheel}, {Szilagyi},
  {Hotokezaka}, {Kyutoku}, {Shibata}, \& {Carpenter}}]{2016PhRvL.116r1101H}
{Hinderer}, T., {Taracchini}, A., {Foucart}, F., {et~al.} 2016, \prl, 116,
  181101, \dodoi{10.1103/PhysRevLett.116.181101}

\bibitem[{{Hofmann} {et~al.}(2016){Hofmann}, {Barausse}, \&
  {Rezzolla}}]{2016ApJ...825L..19H}
{Hofmann}, F., {Barausse}, E., \& {Rezzolla}, L. 2016, \apjl, 825, L19,
  \dodoi{10.3847/2041-8205/825/2/L19}

\bibitem[{{Hooper} {et~al.}(2012){Hooper}, {Chung}, {Luan}, {Blair}, {Chen}, \&
  {Wen}}]{2012PhRvD..86b4012H}
{Hooper}, S., {Chung}, S.~K., {Luan}, J., {et~al.} 2012, \prd, 86, 024012,
  \dodoi{10.1103/PhysRevD.86.024012}

\bibitem[{{Hourihane} {et~al.}(2022){Hourihane}, {Chatziioannou},
  {Wijngaarden}, {Davis}, {Littenberg}, \& {Cornish}}]{2022PhRvD.106d2006H}
{Hourihane}, S., {Chatziioannou}, K., {Wijngaarden}, M., {et~al.} 2022, \prd,
  106, 042006, \dodoi{10.1103/PhysRevD.106.042006}

\bibitem[{{Hoy} \& {Raymond}(2021)}]{2021SoftX..1500765H}
{Hoy}, C., \& {Raymond}, V. 2021, SoftwareX, 15, 100765,
  \dodoi{10.1016/j.softx.2021.100765}

\bibitem[{{Hunter}(2007)}]{2007CSE.....9...90H}
{Hunter}, J.~D. 2007, Computing in Science and Engineering, 9, 90,
  \dodoi{10.1109/MCSE.2007.55}

\bibitem[{{Husa} {et~al.}(2016){Husa}, {Khan}, {Hannam}, {P{\"u}rrer}, {Ohme},
  {Forteza}, \& {Boh{\'e}}}]{2016PhRvD..93d4006H}
{Husa}, S., {Khan}, S., {Hannam}, M., {et~al.} 2016, \prd, 93, 044006,
  \dodoi{10.1103/PhysRevD.93.044006}

\bibitem[{Islam {et~al.}(2023)Islam, Field, \& Khanna}]{Islam:2023mob}
Islam, T., Field, S.~E., \& Khanna, G. 2023, Phys. Rev. D, 108, 064048,
  \dodoi{10.1103/PhysRevD.108.064048}

\bibitem[{{Isogai} {et~al.}(2010){Isogai}, {LIGO Scientific Collaboration}, \&
  {Virgo Collaboration}}]{2010JPhCS.243a2005I}
{Isogai}, T., {LIGO Scientific Collaboration}, \& {Virgo Collaboration}. 2010,
  in Journal of Physics Conference Series, Vol. 243, Journal of Physics
  Conference Series (IOP), 012005, \dodoi{10.1088/1742-6596/243/1/012005}

\bibitem[{Isoyama {et~al.}(2020)Isoyama, Sturani, \& Nakano}]{Isoyama:2020lls}
Isoyama, S., Sturani, R., \& Nakano, H. 2020,
  \dodoi{10.1007/978-981-15-4702-7_31-1}

\bibitem[{Jan {et~al.}(2025)Jan, Tsao, O'Shaughnessy, Shoemaker, \&
  Laguna}]{Jan:2025fps}
Jan, A., Tsao, B.-J., O'Shaughnessy, R., Shoemaker, D., \& Laguna, P. 2025.
\newblock \doarXiv{2508.12460}

\bibitem[{{Jim{\'e}nez-Forteza} {et~al.}(2017){Jim{\'e}nez-Forteza}, {Keitel},
  {Husa}, {Hannam}, {Khan}, \& {P{\"u}rrer}}]{2017PhRvD..95f4024J}
{Jim{\'e}nez-Forteza}, X., {Keitel}, D., {Husa}, S., {et~al.} 2017, \prd, 95,
  064024, \dodoi{10.1103/PhysRevD.95.064024}

\bibitem[{{Johnson-McDaniel} {et~al.}(2022{\natexlab{a}}){Johnson-McDaniel},
  {Ghosh}, {Ghonge}, {Saleem}, {Krishnendu}, \& {Clark}}]{2022PhRvD.105d4020J}
{Johnson-McDaniel}, N.~K., {Ghosh}, A., {Ghonge}, S., {et~al.}
  2022{\natexlab{a}}, \prd, 105, 044020, \dodoi{10.1103/PhysRevD.105.044020}

\bibitem[{Johnson-McDaniel {et~al.}(2016)Johnson-McDaniel, Gupta, Ajith,
  Keitel, Birnholtz, \& Ohme}]{JohnsonMcDaniel:2016aaa}
Johnson-McDaniel, N.~K., Gupta, A., Ajith, P., {et~al.} 2016, {Determining the
  final spin of a binary black hole system including in-plane spins: Method and
  checks of accuracy}, Tech. Rep. DCC-T1600168, {LIGO}.
\newblock \url{https://dcc.ligo.org/LIGO-T1600168/public}

\bibitem[{{Johnson-McDaniel} {et~al.}(2022{\natexlab{b}}){Johnson-McDaniel},
  {Kulkarni}, \& {Gupta}}]{2022PhRvD.106b3001J}
{Johnson-McDaniel}, N.~K., {Kulkarni}, S., \& {Gupta}, A. 2022{\natexlab{b}},
  \prd, 106, 023001, \dodoi{10.1103/PhysRevD.106.023001}

\bibitem[{{Joshi} {et~al.}(2023){Joshi}, {Tsukada}, \&
  {Hanna}}]{2023PhRvD.108h4032J}
{Joshi}, P., {Tsukada}, L., \& {Hanna}, C. 2023, \prd, 108, 084032,
  \dodoi{10.1103/PhysRevD.108.084032}

\bibitem[{{Joshi} {et~al.}(2025{\natexlab{a}}){Joshi}, {Tsukada}, {Hanna},
  {Adhicary}, {Mukherjee}, {Niu}, {Sakon}, {Singh}, {Baral}, {Baylor},
  {Cannon}, {Caudill}, {Cousins}, {Creighton}, {Ewing}, {Fong}, {George},
  {Godwin}, {Harada}, {Huang}, {Huxford}, {Kennington}, {Kuwahara}, {Li},
  {Magee}, {Meacher}, {Messick}, {Morisaki}, {Pace}, {Posnansky}, {Ray},
  {Sachdev}, {Schmidt}, {Shah}, {Tapia}, {Ueno}, {Viets}, {Wade}, {Wade},
  {Yarbrough}, \& {Zhang}}]{2025arXiv250606497J}
{Joshi}, P., {Tsukada}, L., {Hanna}, C., {et~al.} 2025{\natexlab{a}}, arXiv
  e-prints, arXiv:2506.06497, \dodoi{10.48550/arXiv.2506.06497}

\bibitem[{{Joshi} {et~al.}(2025{\natexlab{b}}){Joshi}, {Niu}, {Hanna},
  {Huxford}, {Singh}, {Tsukada}, {Adhicary}, {Baral}, {Baylor}, {Cannon},
  {Caudill}, {Coughlin}, {Cousins}, {Creighton}, {Ewing}, {Fong}, {George},
  {Ghosh}, {Godwin}, {Harada}, {Huang}, {Messick}, {Morisaki}, {Mukherjee},
  {Pace}, {Posnansky}, {Ray}, {Sachdev}, {Sakon}, {Shah}, {Tapia}, {Ueno},
  {Viets}, {Wade}, {Wade}, {Yarbrough}, \& {Zhang}}]{2025arXiv250523959J}
{Joshi}, P., {Niu}, W., {Hanna}, C., {et~al.} 2025{\natexlab{b}}, arXiv
  e-prints, arXiv:2505.23959, \dodoi{10.48550/arXiv.2505.23959}

\bibitem[{Kacanja {et~al.}(2025)Kacanja, Soni, \& Nitz}]{Kacanja:2025kpr}
Kacanja, K., Soni, K., \& Nitz, A.~H. 2025.
\newblock \doarXiv{2508.00179}

\bibitem[{{Kapadia} {et~al.}(2020){Kapadia}, {Caudill}, {Creighton}, {Farr},
  {Mendell}, {Weinstein}, {Cannon}, {Fong}, {Godwin}, {Lo}, {Magee}, {Meacher},
  {Messick}, {Mohite}, {Mukherjee}, \& {Sachdev}}]{2020CQGra..37d5007K}
{Kapadia}, S.~J., {Caudill}, S., {Creighton}, J. D.~E., {et~al.} 2020,
  Classical and Quantum Gravity, 37, 045007, \dodoi{10.1088/1361-6382/ab5f2d}

\bibitem[{Kawaguchi {et~al.}(2015)Kawaguchi, Kyutoku, Nakano, Okawa, Shibata,
  \& Taniguchi}]{Kawaguchi:2015bwa}
Kawaguchi, K., Kyutoku, K., Nakano, H., {et~al.} 2015, Phys. Rev. D, 92,
  024014, \dodoi{10.1103/PhysRevD.92.024014}

\bibitem[{{Keitel} {et~al.}(2017){Keitel}, {Forteza}, {Husa}, {London},
  {Bernuzzi}, {Harms}, {Nagar}, {Hannam}, {Khan}, {P{\"u}rrer}, {Pratten}, \&
  {Chaurasia}}]{2017PhRvD..96b4006K}
{Keitel}, D., {Forteza}, X.~J., {Husa}, S., {et~al.} 2017, \prd, 96, 024006,
  \dodoi{10.1103/PhysRevD.96.024006}

\bibitem[{Khalil {et~al.}(2023)Khalil, Buonanno, Estelles, Mihaylov, Ossokine,
  Pompili, \& Ramos-Buades}]{Khalil:2023kep}
Khalil, M., Buonanno, A., Estelles, H., {et~al.} 2023, Phys. Rev. D, 108,
  124036, \dodoi{10.1103/PhysRevD.108.124036}

\bibitem[{{Khan} {et~al.}(2016){Khan}, {Husa}, {Hannam}, {Ohme}, {P{\"u}rrer},
  {Forteza}, \& {Boh{\'e}}}]{2016PhRvD..93d4007K}
{Khan}, S., {Husa}, S., {Hannam}, M., {et~al.} 2016, \prd, 93, 044007,
  \dodoi{10.1103/PhysRevD.93.044007}

\bibitem[{{Khan} {et~al.}(2020){Khan}, {Ohme}, {Chatziioannou}, \&
  {Hannam}}]{2020PhRvD.101b4056K}
{Khan}, S., {Ohme}, F., {Chatziioannou}, K., \& {Hannam}, M. 2020, \prd, 101,
  024056, \dodoi{10.1103/PhysRevD.101.024056}

\bibitem[{Kiuchi {et~al.}(2017)Kiuchi, Kawaguchi, Kyutoku, Sekiguchi, Shibata,
  \& Taniguchi}]{Kiuchi:2017pte}
Kiuchi, K., Kawaguchi, K., Kyutoku, K., {et~al.} 2017, Phys. Rev. D, 96,
  084060, \dodoi{10.1103/PhysRevD.96.084060}

\bibitem[{Klein {et~al.}(2013)Klein, Cornish, \& Yunes}]{Klein:2013qda}
Klein, A., Cornish, N., \& Yunes, N. 2013, Phys. Rev. D, 88, 124015,
  \dodoi{10.1103/PhysRevD.88.124015}

\bibitem[{Klein {et~al.}(2014)Klein, Cornish, \& Yunes}]{Klein:2014bua}
---. 2014, Phys. Rev. D, 90, 124029, \dodoi{10.1103/PhysRevD.90.124029}

\bibitem[{{Klimenko}(2022)}]{2022arXiv220101096K}
{Klimenko}, S. 2022, arXiv e-prints, arXiv:2201.01096,
  \dodoi{10.48550/arXiv.2201.01096}

\bibitem[{{Klimenko} {et~al.}(2005){Klimenko}, {Mohanty}, {Rakhmanov}, \&
  {Mitselmakher}}]{2005PhRvD..72l2002K}
{Klimenko}, S., {Mohanty}, S., {Rakhmanov}, M., \& {Mitselmakher}, G. 2005,
  \prd, 72, 122002, \dodoi{10.1103/PhysRevD.72.122002}

\bibitem[{{Klimenko} {et~al.}(2008){Klimenko}, {Yakushin}, {Mercer}, \&
  {Mitselmakher}}]{2008CQGra..25k4029K}
{Klimenko}, S., {Yakushin}, I., {Mercer}, A., \& {Mitselmakher}, G. 2008,
  Classical and Quantum Gravity, 25, 114029,
  \dodoi{10.1088/0264-9381/25/11/114029}

\bibitem[{{Klimenko} {et~al.}(2016){Klimenko}, {Vedovato}, {Drago}, {Salemi},
  {Tiwari}, {Prodi}, {Lazzaro}, {Ackley}, {Tiwari}, {Da Silva}, \&
  {Mitselmakher}}]{2016PhRvD..93d2004K}
{Klimenko}, S., {Vedovato}, G., {Drago}, M., {et~al.} 2016, \prd, 93, 042004,
  \dodoi{10.1103/PhysRevD.93.042004}

\bibitem[{Kolitsidou {et~al.}(2024)Kolitsidou, Thompson, \&
  Hannam}]{Kolitsidou:2024vub}
Kolitsidou, P., Thompson, J.~E., \& Hannam, M. 2024, {Impact of anti-symmetric
  contributions to signal multipoles in the measurement of black-hole spins}.
\newblock \doarXiv{2402.00813}

\bibitem[{{Kovalam} {et~al.}(2022){Kovalam}, {Kaium Patwary}, {Sreekumar},
  {Wen}, {Panther}, \& {Chu}}]{2022ApJ...927L...9K}
{Kovalam}, M., {Kaium Patwary}, M.~A., {Sreekumar}, A.~K., {et~al.} 2022,
  \apjl, 927, L9, \dodoi{10.3847/2041-8213/ac5687}

\bibitem[{{Krishnendu} \& {Ohme}(2022)}]{2022PhRvD.105f4012K}
{Krishnendu}, N.~V., \& {Ohme}, F. 2022, \prd, 105, 064012,
  \dodoi{10.1103/PhysRevD.105.064012}

\bibitem[{{Krolak} \& {Schutz}(1987)}]{1987GReGr..19.1163K}
{Krolak}, A., \& {Schutz}, B.~F. 1987, General Relativity and Gravitation, 19,
  1163, \dodoi{10.1007/BF00759095}

\bibitem[{{Kumar} \& {Dent}(2024)}]{2024PhRvD.110d3036K}
{Kumar}, P., \& {Dent}, T. 2024, \prd, 110, 043036,
  \dodoi{10.1103/PhysRevD.110.043036}

\bibitem[{Kumar {et~al.}(2025)Kumar, Nitz, \& Forteza}]{Kumar:2025}
Kumar, S., Nitz, A.~H., \& Forteza, X.~J. 2025, The Astrophysical Journal, 982,
  67, \dodoi{10.3847/1538-4357/adb973}

\bibitem[{Kyutoku {et~al.}(2011)Kyutoku, Okawa, Shibata, \&
  Taniguchi}]{Kyutoku:2011vz}
Kyutoku, K., Okawa, H., Shibata, M., \& Taniguchi, K. 2011, Phys. Rev. D, 84,
  064018, \dodoi{10.1103/PhysRevD.84.064018}

\bibitem[{{Lackey} {et~al.}(2019){Lackey}, {P{\"u}rrer}, {Taracchini}, \&
  {Marsat}}]{2019PhRvD.100b4002L}
{Lackey}, B.~D., {P{\"u}rrer}, M., {Taracchini}, A., \& {Marsat}, S. 2019,
  \prd, 100, 024002, \dodoi{10.1103/PhysRevD.100.024002}

\bibitem[{Lang \& Hughes(2006)}]{Lang:2006bsg}
Lang, R.~N., \& Hughes, S.~A. 2006, Phys. Rev. D, 74, 122001,
  \dodoi{10.1103/PhysRevD.75.089902}

\bibitem[{{Lange} {et~al.}(2017){Lange}, {O'Shaughnessy}, {Boyle},
  {Calder{\'o}n Bustillo}, {Campanelli}, {Chu}, {Clark}, {Demos}, {Fong},
  {Healy}, {Hemberger}, {Hinder}, {Jani}, {Khamesra}, {Kidder}, {Kumar},
  {Laguna}, {Lousto}, {Lovelace}, {Ossokine}, {Pfeiffer}, {Scheel},
  {Shoemaker}, {Szilagyi}, {Teukolsky}, \& {Zlochower}}]{2017PhRvD..96j4041L}
{Lange}, J., {O'Shaughnessy}, R., {Boyle}, M., {et~al.} 2017, \prd, 96, 104041,
  \dodoi{10.1103/PhysRevD.96.104041}

\bibitem[{{Lenon} {et~al.}(2020){Lenon}, {Nitz}, \&
  {Brown}}]{2020MNRAS.497.1966L}
{Lenon}, A.~K., {Nitz}, A.~H., \& {Brown}, D.~A. 2020, \mnras, 497, 1966,
  \dodoi{10.1093/mnras/staa2120}

\bibitem[{{LIGO--Virgo--KAGRA Collaboration}(2018)}]{lalsuite}
{LIGO--Virgo--KAGRA Collaboration}. 2018, {LVK} {A}lgorithm {L}ibrary -
  {LALS}uite, Free software (GPL), \dodoi{10.7935/GT1W-FZ16}

\bibitem[{{LIGO--Virgo--KAGRA
  Collaboration}(2025{\natexlab{a}})}]{GWTC-4.0:Search}
---. 2025{\natexlab{a}}, GWTC-4.0: Candidate Data Release,  Zenodo,
  \dodoi{10.5281/zenodo.17014083}

\bibitem[{{LIGO--Virgo--KAGRA Collaboration}(2025{\natexlab{b}})}]{OPA}
---. 2025{\natexlab{b}}, {LIGO/Virgo/KAGRA Public Alerts User Guide}.
\newblock \url{https://emfollow.docs.ligo.org/userguide}

\bibitem[{{LIGO--Virgo--KAGRA Collaboration}(2025{\natexlab{c}})}]{GWTC-4.0:DQ}
---. 2025{\natexlab{c}}, GWTC-4.0: Data Quality Products for Transient
  Gravitational Wave Searches,  Zenodo, \dodoi{10.5281/zenodo.16856919}

\bibitem[{{LIGO--Virgo--KAGRA Collaboration}(2025{\natexlab{d}})}]{GWTC-4.0:PE}
---. 2025{\natexlab{d}}, GWTC-4.0: Parameter Estimation Data Release,  Zenodo,
  \dodoi{10.5281/zenodo.17014085}

\bibitem[{{LIGO--Virgo--KAGRA Collaboration}(2026)}]{GWTC-5.0:Deglitch}
---. 2026, GWTC-5.0: Glitch Modelling for Events,  In preparation

\bibitem[{{LIGO Scientific Collaboration} {et~al.}(2022){LIGO Scientific
  Collaboration}, {Virgo Collaboration}, \& {KAGRA
  collaboration}}]{LIGO-T2000012}
{LIGO Scientific Collaboration}, {Virgo Collaboration}, \& {KAGRA
  collaboration}. 2022, {Noise curves used for Simulations in the update of the
  Observing Scenarios Paper}, Tech. Rep. LIGO-T2000012.
\newblock \url{https://dcc.ligo.org/LIGO-T2000012/public}

\bibitem[{{Lin}(1991)}]{1991ITIT...37..145L}
{Lin}, J. 1991, IEEE Transactions on Information Theory, 37, 145,
  \dodoi{10.1109/18.61115}

\bibitem[{{Littenberg} \& {Cornish}(2009)}]{2009PhRvD..80f3007L}
{Littenberg}, T.~B., \& {Cornish}, N.~J. 2009, \prd, 80, 063007,
  \dodoi{10.1103/PhysRevD.80.063007}

\bibitem[{{Littenberg} \& {Cornish}(2015)}]{2015PhRvD..91h4034L}
---. 2015, \prd, 91, 084034, \dodoi{10.1103/PhysRevD.91.084034}

\bibitem[{{Liu} {et~al.}(2012){Liu}, {Du}, {Chung}, {Hooper}, {Blair}, \&
  {Wen}}]{2012CQGra..29w5018L}
{Liu}, Y., {Du}, Z., {Chung}, S.~K., {et~al.} 2012, Classical and Quantum
  Gravity, 29, 235018, \dodoi{10.1088/0264-9381/29/23/235018}

\bibitem[{{London} {et~al.}(2018){London}, {Khan}, {Fauchon-Jones},
  {Garc{\'\i}a}, {Hannam}, {Husa}, {Jim{\'e}nez-Forteza}, {Kalaghatgi}, {Ohme},
  \& {Pannarale}}]{2018PhRvL.120p1102L}
{London}, L., {Khan}, S., {Fauchon-Jones}, E., {et~al.} 2018, \prl, 120,
  161102, \dodoi{10.1103/PhysRevLett.120.161102}

\bibitem[{{Lower} {et~al.}(2018){Lower}, {Thrane}, {Lasky}, \&
  {Smith}}]{2018PhRvD..98h3028L}
{Lower}, M.~E., {Thrane}, E., {Lasky}, P.~D., \& {Smith}, R. 2018, \prd, 98,
  083028, \dodoi{10.1103/PhysRevD.98.083028}

\bibitem[{{Luan} {et~al.}(2012){Luan}, {Hooper}, {Wen}, \&
  {Chen}}]{2012PhRvD..85j2002L}
{Luan}, J., {Hooper}, S., {Wen}, L., \& {Chen}, Y. 2012, \prd, 85, 102002,
  \dodoi{10.1103/PhysRevD.85.102002}

\bibitem[{{Maggiore}(2007)}]{2007gwte.book.....M}
{Maggiore}, M. 2007, {Gravitational Waves: Volume 1: Theory and Experiments},
  \dodoi{10.1093/acprof:oso/9780198570745.001.0001}

\bibitem[{Marsat {et~al.}(2013)Marsat, Blanchet, Bohe, \&
  Faye}]{Marsat:2013wwa}
Marsat, S., Blanchet, L., Bohe, A., \& Faye, G. 2013, {Gravitational waves from
  spinning compact object binaries: New post-Newtonian results}.
\newblock \doarXiv{1312.5375}

\bibitem[{{Martel} \& {Poisson}(1999)}]{1999PhRvD..60l4008M}
{Martel}, K., \& {Poisson}, E. 1999, \prd, 60, 124008,
  \dodoi{10.1103/PhysRevD.60.124008}

\bibitem[{{Matas} {et~al.}(2020){Matas}, {Dietrich}, {Buonanno}, {Hinderer},
  {P{\"u}rrer}, {Foucart}, {Boyle}, {Duez}, {Kidder}, {Pfeiffer}, \&
  {Scheel}}]{2020PhRvD.102d3023M}
{Matas}, A., {Dietrich}, T., {Buonanno}, A., {et~al.} 2020, \prd, 102, 043023,
  \dodoi{10.1103/PhysRevD.102.043023}

\bibitem[{Mehta {et~al.}(2025)Mehta, Wadekar, Roulet, Anantpurkar, Venumadhav,
  Mushkin, Zackay, Zaldarriaga, \& Islam}]{Mehta:2025jiq}
Mehta, A.~K., Wadekar, D., Roulet, J., {et~al.} 2025.
\newblock \doarXiv{2501.17939}

\bibitem[{{Messick} {et~al.}(2017){Messick}, {Blackburn}, {Brady}, {Brockill},
  {Cannon}, {Cariou}, {Caudill}, {Chamberlin}, {Creighton}, {Everett}, {Hanna},
  {Keppel}, {Lang}, {Li}, {Meacher}, {Nielsen}, {Pankow}, {Privitera}, {Qi},
  {Sachdev}, {Sadeghian}, {Singer}, {Thomas}, {Wade}, {Wade}, {Weinstein}, \&
  {Wiesner}}]{2017PhRvD..95d2001M}
{Messick}, C., {Blackburn}, K., {Brady}, P., {et~al.} 2017, \prd, 95, 042001,
  \dodoi{10.1103/PhysRevD.95.042001}

\bibitem[{Mihaylov {et~al.}(2023)Mihaylov, Ossokine, Buonanno, Estelles,
  Pompili, P\"urrer, \& Ramos-Buades}]{Mihaylov:2023bkc}
Mihaylov, D.~P., Ossokine, S., Buonanno, A., {et~al.} 2023, {pySEOBNR: a
  software package for the next generation of effective-one-body multipolar
  waveform models}.
\newblock \doarXiv{2303.18203}

\bibitem[{Mihaylov {et~al.}(2021)Mihaylov, Ossokine, Buonanno, \&
  Ghosh}]{Mihaylov:2021bpf}
Mihaylov, D.~P., Ossokine, S., Buonanno, A., \& Ghosh, A. 2021, Phys. Rev. D,
  104, 124087, \dodoi{10.1103/PhysRevD.104.124087}

\bibitem[{{Mills} \& {Fairhurst}(2021)}]{2021PhRvD.103b4042M}
{Mills}, C., \& {Fairhurst}, S. 2021, \prd, 103, 024042,
  \dodoi{10.1103/PhysRevD.103.024042}

\bibitem[{{Mishra} {et~al.}(2025){Mishra}, {Bhaumik}, {Gayathri},
  {Szczepa{\'n}czyk}, {Bartos}, \& {Klimenko}}]{2025PhRvD.111b3054M}
{Mishra}, T., {Bhaumik}, S., {Gayathri}, V., {et~al.} 2025, \prd, 111, 023054,
  \dodoi{10.1103/PhysRevD.111.023054}

\bibitem[{Mishra {et~al.}(2021)Mishra, O'Brien, Gayathri,
  Szczepa\ifmmode~\acute{n}\else \'{n}\fi{}czyk, Bhaumik, Bartos, \&
  Klimenko}]{PhysRevD.104.023014}
Mishra, T., O'Brien, B., Gayathri, V., {et~al.} 2021, Phys. Rev. D, 104,
  023014, \dodoi{10.1103/PhysRevD.104.023014}

\bibitem[{{Mishra} {et~al.}(2022){Mishra}, {O'Brien}, {Szczepa{\'n}czyk},
  {Vedovato}, {Bhaumik}, {Gayathri}, {Prodi}, {Salemi}, {Milotti}, {Bartos}, \&
  {Klimenko}}]{2022PhRvD.105h3018M}
{Mishra}, T., {O'Brien}, B., {Szczepa{\'n}czyk}, M., {et~al.} 2022, \prd, 105,
  083018, \dodoi{10.1103/PhysRevD.105.083018}

\bibitem[{Moe {et~al.}(2014)Moe, Brady, Stephens, Katsavounidis, Williams,
  Zhang, {et~al.}}]{GraceDB2014}
Moe, B., Brady, P., Stephens, B., {et~al.} 2014, {GraceDB: A Gravitational Wave
  Candidate Event Database}, Tech. Rep. LIGO-T1400365-v5, LIGO Scientific
  Collaboration.
\newblock \url{https://dcc.ligo.org/LIGO-T1400365/public}

\bibitem[{{Morisaki} {et~al.}(2023){Morisaki}, {Smith}, {Tsukada}, {Sachdev},
  {Stevenson}, {Talbot}, \& {Zimmerman}}]{2023PhRvD.108l3040M}
{Morisaki}, S., {Smith}, R., {Tsukada}, L., {et~al.} 2023, \prd, 108, 123040,
  \dodoi{10.1103/PhysRevD.108.123040}

\bibitem[{Morras {et~al.}(2025{\natexlab{a}})Morras, Pratten, \&
  Schmidt}]{Morras:2025nlp}
Morras, G., Pratten, G., \& Schmidt, P. 2025{\natexlab{a}}, Phys. Rev. D, 111,
  084052, \dodoi{10.1103/PhysRevD.111.084052}

\bibitem[{Morras {et~al.}(2025{\natexlab{b}})Morras, Pratten, \&
  Schmidt}]{Morras:2025xfu}
---. 2025{\natexlab{b}}.
\newblock \doarXiv{2503.15393}

\bibitem[{Mould \& Gerosa(2022)}]{Mould:2021xst}
Mould, M., \& Gerosa, D. 2022, Phys. Rev. D, 105, 024076,
  \dodoi{10.1103/PhysRevD.105.024076}

\bibitem[{Mozzon {et~al.}(2022)Mozzon, Ashton, Nuttall, \&
  Williamson}]{Mozzon:2022}
Mozzon, S., Ashton, G., Nuttall, L.~K., \& Williamson, A.~R. 2022, Phys. Rev.
  D, 106, 043504, \dodoi{10.1103/PhysRevD.106.043504}

\bibitem[{{Mozzon} {et~al.}(2020){Mozzon}, {Nuttall}, {Lundgren}, {Dent},
  {Kumar}, \& {Nitz}}]{2020CQGra..37u5014M}
{Mozzon}, S., {Nuttall}, L.~K., {Lundgren}, A., {et~al.} 2020, Classical and
  Quantum Gravity, 37, 215014, \dodoi{10.1088/1361-6382/abac6c}

\bibitem[{{Mukherjee} {et~al.}(2021){Mukherjee}, {Caudill}, {Magee}, {Messick},
  {Privitera}, {Sachdev}, {Blackburn}, {Brady}, {Brockill}, {Cannon},
  {Chamberlin}, {Chatterjee}, {Creighton}, {Fong}, {Godwin}, {Hanna},
  {Kapadia}, {Lang}, {Li}, {Lo}, {Meacher}, {Pace}, {Sadeghian}, {Tsukada},
  {Wade}, {Wade}, {Weinstein}, \& {Xiao}}]{2021PhRvD.103h4047M}
{Mukherjee}, D., {Caudill}, S., {Magee}, R., {et~al.} 2021, \prd, 103, 084047,
  \dodoi{10.1103/PhysRevD.103.084047}

\bibitem[{Nagar \& Rettegno(2019)}]{Nagar:2018gnk}
Nagar, A., \& Rettegno, P. 2019, Phys. Rev. D, 99, 021501,
  \dodoi{10.1103/PhysRevD.99.021501}

\bibitem[{Nagar \& Shah(2016)}]{Nagar:2016ayt}
Nagar, A., \& Shah, A. 2016, Phys. Rev. D, 94, 104017,
  \dodoi{10.1103/PhysRevD.94.104017}

\bibitem[{Nagar {et~al.}(2018)}]{Nagar:2018zoe}
Nagar, A., {et~al.} 2018, Phys. Rev. D, 98, 104052,
  \dodoi{10.1103/PhysRevD.98.104052}

\bibitem[{{Necula} {et~al.}(2012){Necula}, {Klimenko}, \&
  {Mitselmakher}}]{2012JPhCS.363a2032N}
{Necula}, V., {Klimenko}, S., \& {Mitselmakher}, G. 2012, in Journal of Physics
  Conference Series, Vol. 363, Journal of Physics Conference Series (IOP),
  012032, \dodoi{10.1088/1742-6596/363/1/012032}

\bibitem[{{Nitz}(2018)}]{2018CQGra..35c5016N}
{Nitz}, A.~H. 2018, Classical and Quantum Gravity, 35, 035016,
  \dodoi{10.1088/1361-6382/aaa13d}

\bibitem[{{Nitz} {et~al.}(2019){Nitz}, {Capano}, {Nielsen}, {Reyes}, {White},
  {Brown}, \& {Krishnan}}]{2019ApJ...872..195N}
{Nitz}, A.~H., {Capano}, C., {Nielsen}, A.~B., {et~al.} 2019, \apj, 872, 195,
  \dodoi{10.3847/1538-4357/ab0108}

\bibitem[{{Nitz} {et~al.}(2018){Nitz}, {Dal Canton}, {Davis}, \&
  {Reyes}}]{2018PhRvD..98b4050N}
{Nitz}, A.~H., {Dal Canton}, T., {Davis}, D., \& {Reyes}, S. 2018, \prd, 98,
  024050, \dodoi{10.1103/PhysRevD.98.024050}

\bibitem[{{Nitz} {et~al.}(2017){Nitz}, {Dent}, {Dal Canton}, {Fairhurst}, \&
  {Brown}}]{2017ApJ...849..118N}
{Nitz}, A.~H., {Dent}, T., {Dal Canton}, T., {Fairhurst}, S., \& {Brown}, D.~A.
  2017, \apj, 849, 118, \dodoi{10.3847/1538-4357/aa8f50}

\bibitem[{{Nitz} {et~al.}(2020){Nitz}, {Dent}, {Davies}, {Kumar}, {Capano},
  {Harry}, {Mozzon}, {Nuttall}, {Lundgren}, \&
  {T{\'a}pai}}]{2020ApJ...891..123N}
{Nitz}, A.~H., {Dent}, T., {Davies}, G.~S., {et~al.} 2020, \apj, 891, 123,
  \dodoi{10.3847/1538-4357/ab733f}

\bibitem[{{Nuttall}(2018)}]{2018RSPTA.37670286N}
{Nuttall}, L.~K. 2018, Philosophical Transactions of the Royal Society of
  London Series A, 376, 20170286, \dodoi{10.1098/rsta.2017.0286}

\bibitem[{O'Shaughnessy {et~al.}(2012)O'Shaughnessy, Healy, London, Meeks, \&
  Shoemaker}]{OShaughnessy:2012lay}
O'Shaughnessy, R., Healy, J., London, L., Meeks, Z., \& Shoemaker, D. 2012,
  Phys. Rev. D, 85, 084003, \dodoi{10.1103/PhysRevD.85.084003}

\bibitem[{{O'Shea} \& {Kumar}(2023)}]{2023PhRvD.108j4018O}
{O'Shea}, E., \& {Kumar}, P. 2023, \prd, 108, 104018,
  \dodoi{10.1103/PhysRevD.108.104018}

\bibitem[{{Ossokine} {et~al.}(2020){Ossokine}, {Buonanno}, {Marsat}, {Cotesta},
  {Babak}, {Dietrich}, {Haas}, {Hinder}, {Pfeiffer}, {P{\"u}rrer}, {Woodford},
  {Boyle}, {Kidder}, {Scheel}, \& {Szil{\'a}gyi}}]{2020PhRvD.102d4055O}
{Ossokine}, S., {Buonanno}, A., {Marsat}, S., {et~al.} 2020, \prd, 102, 044055,
  \dodoi{10.1103/PhysRevD.102.044055}

\bibitem[{{Owen}(1996)}]{1996PhRvD..53.6749O}
{Owen}, B.~J. 1996, \prd, 53, 6749, \dodoi{10.1103/PhysRevD.53.6749}

\bibitem[{{Pan} {et~al.}(2014){Pan}, {Buonanno}, {Taracchini}, {Kidder},
  {Mrou{\'e}}, {Pfeiffer}, {Scheel}, \& {Szil{\'a}gyi}}]{2014PhRvD..89h4006P}
{Pan}, Y., {Buonanno}, A., {Taracchini}, A., {et~al.} 2014, \prd, 89, 084006,
  \dodoi{10.1103/PhysRevD.89.084006}

\bibitem[{{Pankow} {et~al.}(2015){Pankow}, {Brady}, {Ochsner}, \&
  {O'Shaughnessy}}]{2015PhRvD..92b3002P}
{Pankow}, C., {Brady}, P., {Ochsner}, E., \& {O'Shaughnessy}, R. 2015, \prd,
  92, 023002, \dodoi{10.1103/PhysRevD.92.023002}

\bibitem[{{Pankow} {et~al.}(2018){Pankow}, {Chatziioannou}, {Chase},
  {Littenberg}, {Evans}, {McIver}, {Cornish}, {Haster}, {Kanner}, {Raymond},
  {Vitale}, \& {Zimmerman}}]{2018PhRvD..98h4016P}
{Pankow}, C., {Chatziioannou}, K., {Chase}, E.~A., {et~al.} 2018, \prd, 98,
  084016, \dodoi{10.1103/PhysRevD.98.084016}

\bibitem[{Pannarale {et~al.}(2015)Pannarale, Berti, Kyutoku, Lackey, \&
  Shibata}]{Pannarale:2015jka}
Pannarale, F., Berti, E., Kyutoku, K., Lackey, B.~D., \& Shibata, M. 2015,
  Phys. Rev. D, 92, 084050, \dodoi{10.1103/PhysRevD.92.084050}

\bibitem[{Pannarale {et~al.}(2013)Pannarale, Berti, Kyutoku, \&
  Shibata}]{Pannarale:2013uoa}
Pannarale, F., Berti, E., Kyutoku, K., \& Shibata, M. 2013, Phys. Rev. D, 88,
  084011, \dodoi{10.1103/PhysRevD.88.084011}

\bibitem[{{Pannarale} {et~al.}(2011){Pannarale}, {Rezzolla}, {Ohme}, \&
  {Read}}]{2011PhRvD..84j4017P}
{Pannarale}, F., {Rezzolla}, L., {Ohme}, F., \& {Read}, J.~S. 2011, \prd, 84,
  104017, \dodoi{10.1103/PhysRevD.84.104017}

\bibitem[{{Payne} {et~al.}(2020){Payne}, {Talbot}, {Lasky}, {Thrane}, \&
  {Kissel}}]{2020PhRvD.102l2004P}
{Payne}, E., {Talbot}, C., {Lasky}, P.~D., {Thrane}, E., \& {Kissel}, J.~S.
  2020, \prd, 102, 122004, \dodoi{10.1103/PhysRevD.102.122004}

\bibitem[{{Peters}(1964)}]{1964PhRv..136.1224P}
{Peters}, P.~C. 1964, Physical Review, 136, 1224,
  \dodoi{10.1103/PhysRev.136.B1224}

\bibitem[{{Peters} \& {Mathews}(1963)}]{1963PhRv..131..435P}
{Peters}, P.~C., \& {Mathews}, J. 1963, Physical Review, 131, 435,
  \dodoi{10.1103/PhysRev.131.435}

\bibitem[{Phukon {et~al.}(2025)Phukon, Schmidt, \& Pratten}]{Phukon:2024amh}
Phukon, K.~S., Schmidt, P., \& Pratten, G. 2025, Phys. Rev. D, 111, 043040,
  \dodoi{10.1103/PhysRevD.111.043040}

\bibitem[{Planas {et~al.}(2025{\natexlab{a}})Planas, Husa, Ramos-Buades, \&
  Valencia}]{Planas:2025plq}
Planas, M. d.~L., Husa, S., Ramos-Buades, A., \& Valencia, J.
  2025{\natexlab{a}}.
\newblock \doarXiv{2506.01760}

\bibitem[{Planas {et~al.}(2025{\natexlab{b}})Planas, Ramos-Buades,
  Garc\'\i{}a-Quir\'os, Estell\'es, Husa, \& Haney}]{Planas:2025jny}
Planas, M. d.~L., Ramos-Buades, A., Garc\'\i{}a-Quir\'os, C., {et~al.}
  2025{\natexlab{b}}, {Eccentric or circular? A reanalysis of binary black hole
  gravitational wave events for orbital eccentricity signatures}.
\newblock \doarXiv{2504.15833}

\bibitem[{{Poisson}(1998)}]{1998PhRvD..57.5287P}
{Poisson}, E. 1998, \prd, 57, 5287, \dodoi{10.1103/PhysRevD.57.5287}

\bibitem[{{Pompili} {et~al.}(2023){Pompili}, {Buonanno}, {Estell{\'e}s},
  {Khalil}, {van de Meent}, {Mihaylov}, {Ossokine}, {P{\"u}rrer},
  {Ramos-Buades}, {Mehta}, {Cotesta}, {Marsat}, {Boyle}, {Kidder}, {Pfeiffer},
  {Scheel}, {R{\"u}ter}, {Vu}, {Dudi}, {Ma}, {Mitman}, {Melchor}, {Thomas}, \&
  {Sanchez}}]{2023PhRvD.108l4035P}
{Pompili}, L., {Buonanno}, A., {Estell{\'e}s}, H., {et~al.} 2023, \prd, 108,
  124035, \dodoi{10.1103/PhysRevD.108.124035}

\bibitem[{{Powell}(2018)}]{2018CQGra..35o5017P}
{Powell}, J. 2018, Classical and Quantum Gravity, 35, 155017,
  \dodoi{10.1088/1361-6382/aacf18}

\bibitem[{{Pratten} {et~al.}(2020{\natexlab{a}}){Pratten}, {Husa},
  {Garc{\'\i}a-Quir{\'o}s}, {Colleoni}, {Ramos-Buades}, {Estell{\'e}s}, \&
  {Jaume}}]{2020PhRvD.102f4001P}
{Pratten}, G., {Husa}, S., {Garc{\'\i}a-Quir{\'o}s}, C., {et~al.}
  2020{\natexlab{a}}, \prd, 102, 064001, \dodoi{10.1103/PhysRevD.102.064001}

\bibitem[{{Pratten} {et~al.}(2020{\natexlab{b}}){Pratten}, {Schmidt},
  {Buscicchio}, \& {Thomas}}]{2020PhRvR...2d3096P}
{Pratten}, G., {Schmidt}, P., {Buscicchio}, R., \& {Thomas}, L.~M.
  2020{\natexlab{b}}, Physical Review Research, 2, 043096,
  \dodoi{10.1103/PhysRevResearch.2.043096}

\bibitem[{{Pratten} {et~al.}(2021){Pratten}, {Garc{\'\i}a-Quir{\'o}s},
  {Colleoni}, {Ramos-Buades}, {Estell{\'e}s}, {Mateu-Lucena}, {Jaume}, {Haney},
  {Keitel}, {Thompson}, \& {Husa}}]{2021PhRvD.103j4056P}
{Pratten}, G., {Garc{\'\i}a-Quir{\'o}s}, C., {Colleoni}, M., {et~al.} 2021,
  \prd, 103, 104056, \dodoi{10.1103/PhysRevD.103.104056}

\bibitem[{{Privitera} {et~al.}(2014){Privitera}, {Mohapatra}, {Ajith},
  {Cannon}, {Fotopoulos}, {Frei}, {Hanna}, {Weinstein}, \&
  {Whelan}}]{2014PhRvD..89b4003P}
{Privitera}, S., {Mohapatra}, S. R.~P., {Ajith}, P., {et~al.} 2014, \prd, 89,
  024003, \dodoi{10.1103/PhysRevD.89.024003}

\bibitem[{{Ramos-Buades} {et~al.}(2023){Ramos-Buades}, {Buonanno},
  {Estell{\'e}s}, {Khalil}, {Mihaylov}, {Ossokine}, {Pompili}, \&
  {Shiferaw}}]{2023PhRvD.108l4037R}
{Ramos-Buades}, A., {Buonanno}, A., {Estell{\'e}s}, H., {et~al.} 2023, \prd,
  108, 124037, \dodoi{10.1103/PhysRevD.108.124037}

\bibitem[{Ramos-Buades {et~al.}(2020)Ramos-Buades, Husa, Pratten, Estell{\'e}s,
  Garc{\'\i}a-Quir{\'o}s, Mateu-Lucena, Colleoni, \&
  Jaume}]{Ramos-Buades:2019uvh}
Ramos-Buades, A., Husa, S., Pratten, G., {et~al.} 2020, Phys. Rev. D, 101,
  083015, \dodoi{10.1103/PhysRevD.101.083015}

\bibitem[{{Ramos-Buades} {et~al.}(2020{\natexlab{a}}){Ramos-Buades}, {Schmidt},
  {Pratten}, \& {Husa}}]{2020PhRvD.101j3014R}
{Ramos-Buades}, A., {Schmidt}, P., {Pratten}, G., \& {Husa}, S.
  2020{\natexlab{a}}, \prd, 101, 103014, \dodoi{10.1103/PhysRevD.101.103014}

\bibitem[{{Ramos-Buades} {et~al.}(2020{\natexlab{b}}){Ramos-Buades}, {Tiwari},
  {Haney}, \& {Husa}}]{2020PhRvD.102d3005R}
{Ramos-Buades}, A., {Tiwari}, S., {Haney}, M., \& {Husa}, S.
  2020{\natexlab{b}}, \prd, 102, 043005, \dodoi{10.1103/PhysRevD.102.043005}

\bibitem[{{Ray} {et~al.}(2023){Ray}, {Niu}, {Sakon}, {Ewing}, {Creighton},
  {Hanna}, {Adhicary}, {Baral}, {Baylor}, {Cannon}, {Caudill}, {Cousins},
  {Fong}, {George}, {Godwin}, {Harada}, {Huang}, {Huxford}, {Joshi}, {Kapadia},
  {Kennington}, {Kuwahara}, {Li}, {Magee}, {Meacher}, {Messick}, {Morisaki},
  {Mukherjee}, {Pace}, {Posnansky}, {Sachdev}, {Singh}, {Tapia}, {Tsukada},
  {Tsutsui}, {Ueno}, {Viets}, {Wade}, \& {Wade}}]{2023arXiv230607190R}
{Ray}, A., {Niu}, W., {Sakon}, S., {et~al.} 2023, arXiv e-prints,
  arXiv:2306.07190, \dodoi{10.48550/arXiv.2306.07190}

\bibitem[{{Rodriguez} {et~al.}(2016){Rodriguez}, {Zevin}, {Pankow}, {Kalogera},
  \& {Rasio}}]{2016ApJ...832L...2R}
{Rodriguez}, C.~L., {Zevin}, M., {Pankow}, C., {Kalogera}, V., \& {Rasio},
  F.~A. 2016, \apjl, 832, L2, \dodoi{10.3847/2041-8205/832/1/L2}

\bibitem[{{Romero-Shaw} {et~al.}(2020{\natexlab{a}}){Romero-Shaw}, {Lasky},
  {Thrane}, \& {Calder{\'o}n Bustillo}}]{2020ApJ...903L...5R}
{Romero-Shaw}, I., {Lasky}, P.~D., {Thrane}, E., \& {Calder{\'o}n Bustillo}, J.
  2020{\natexlab{a}}, \apjl, 903, L5, \dodoi{10.3847/2041-8213/abbe26}

\bibitem[{Romero-Shaw {et~al.}(2025)Romero-Shaw, Stegmann, Tagawa, Gerosa,
  Samsing, Gupte, \& Green}]{Romero-Shaw:2025vbc}
Romero-Shaw, I., Stegmann, J., Tagawa, H., {et~al.} 2025, Phys. Rev. D, 112,
  063052, \dodoi{10.1103/jj7m-x66y}

\bibitem[{Romero-Shaw {et~al.}(2023)Romero-Shaw, Gerosa, \&
  Loutrel}]{Romero-Shaw:2022fbf}
Romero-Shaw, I.~M., Gerosa, D., \& Loutrel, N. 2023, Mon. Not. Roy. Astron.
  Soc., 519, 5352, \dodoi{10.1093/mnras/stad031}

\bibitem[{{Romero-Shaw} {et~al.}(2020{\natexlab{b}}){Romero-Shaw}, {Talbot},
  {Biscoveanu}, {D'Emilio}, {Ashton}, {Berry}, {Coughlin}, {Galaudage}, {Hoy},
  {H{\"u}bner}, {Phukon}, {Pitkin}, {Rizzo}, {Sarin}, {Smith}, {Stevenson},
  {Vajpeyi}, {Ar{\`e}ne}, {Athar}, {Banagiri}, {Bose}, {Carney},
  {Chatziioannou}, {Clark}, {Colleoni}, {Cotesta}, {Edelman}, {Estell{\'e}s},
  {Garc{\'\i}a-Quir{\'o}s}, {Ghosh}, {Green}, {Haster}, {Husa}, {Keitel},
  {Kim}, {Hernandez-Vivanco}, {Maga{\~n}a Hernandez}, {Karathanasis}, {Lasky},
  {De Lillo}, {Lower}, {Macleod}, {Mateu-Lucena}, {Miller}, {Millhouse},
  {Morisaki}, {Oh}, {Ossokine}, {Payne}, {Powell}, {Pratten}, {P{\"u}rrer},
  {Ramos-Buades}, {Raymond}, {Thrane}, {Veitch}, {Williams}, {Williams}, \&
  {Xiao}}]{2020MNRAS.499.3295R}
{Romero-Shaw}, I.~M., {Talbot}, C., {Biscoveanu}, S., {et~al.}
  2020{\natexlab{b}}, \mnras, 499, 3295, \dodoi{10.1093/mnras/staa2850}

\bibitem[{{Roy} {et~al.}(2019){Roy}, {Sengupta}, \&
  {Ajith}}]{2019PhRvD..99b4048R}
{Roy}, S., {Sengupta}, A.~S., \& {Ajith}, P. 2019, \prd, 99, 024048,
  \dodoi{10.1103/PhysRevD.99.024048}

\bibitem[{{Roy} {et~al.}(2017){Roy}, {Sengupta}, \&
  {Thakor}}]{2017PhRvD..95j4045R}
{Roy}, S., {Sengupta}, A.~S., \& {Thakor}, N. 2017, \prd, 95, 104045,
  \dodoi{10.1103/PhysRevD.95.104045}

\bibitem[{{Sachdev} {et~al.}(2019){Sachdev}, {Caudill}, {Fong}, {Lo},
  {Messick}, {Mukherjee}, {Magee}, {Tsukada}, {Blackburn}, {Brady}, {Brockill},
  {Cannon}, {Chamberlin}, {Chatterjee}, {Creighton}, {Godwin}, {Gupta},
  {Hanna}, {Kapadia}, {Lang}, {Li}, {Meacher}, {Pace}, {Privitera},
  {Sadeghian}, {Wade}, {Wade}, {Weinstein}, \& {Liting
  Xiao}}]{2019arXiv190108580S}
{Sachdev}, S., {Caudill}, S., {Fong}, H., {et~al.} 2019, arXiv e-prints,
  arXiv:1901.08580, \dodoi{10.48550/arXiv.1901.08580}

\bibitem[{{Sakon} {et~al.}(2024){Sakon}, {Tsukada}, {Fong}, {Kennington},
  {Niu}, {Hanna}, {Adhicary}, {Baral}, {Baylor}, {Cannon}, {Caudill},
  {Cousins}, {Creighton}, {Ewing}, {George}, {Godwin}, {Harada}, {Huang},
  {Huxford}, {Joshi}, {Kuwahara}, {Li}, {Magee}, {Meacher}, {Messick},
  {Morisaki}, {Mukherjee}, {Pace}, {Posnansky}, {Ray}, {Sachdev}, {Singh},
  {Tapia}, {Tsutsui}, {Ueno}, {Viets}, {Wade}, {Wade}, \&
  {Wang}}]{2024PhRvD.109d4066S}
{Sakon}, S., {Tsukada}, L., {Fong}, H., {et~al.} 2024, \prd, 109, 044066,
  \dodoi{10.1103/PhysRevD.109.044066}

\bibitem[{{Salemi} {et~al.}(2019){Salemi}, {Milotti}, {Prodi}, {Vedovato},
  {Lazzaro}, {Tiwari}, {Vinciguerra}, {Drago}, \&
  {Klimenko}}]{2019PhRvD.100d2003S}
{Salemi}, F., {Milotti}, E., {Prodi}, G.~A., {et~al.} 2019, \prd, 100, 042003,
  \dodoi{10.1103/PhysRevD.100.042003}

\bibitem[{Salpeter(1955)}]{Salpeter:1955it}
Salpeter, E.~E. 1955, Astrophys. J., 121, 161, \dodoi{10.1086/145971}

\bibitem[{Samajdar \& Dietrich(2018)}]{Samajdar:2018dcx}
Samajdar, A., \& Dietrich, T. 2018, Phys. Rev. D, 98, 124030,
  \dodoi{10.1103/PhysRevD.98.124030}

\bibitem[{Samsing \& Ramirez-Ruiz(2017)}]{Samsing:2017rat}
Samsing, J., \& Ramirez-Ruiz, E. 2017, Astrophys. J. Lett., 840, L14,
  \dodoi{10.3847/2041-8213/aa6f0b}

\bibitem[{{Santamar{\'\i}a} {et~al.}(2010){Santamar{\'\i}a}, {Ohme}, {Ajith},
  {Br{\"u}gmann}, {Dorband}, {Hannam}, {Husa}, {M{\"o}sta}, {Pollney},
  {Reisswig}, {Robinson}, {Seiler}, \& {Krishnan}}]{2010PhRvD..82f4016S}
{Santamar{\'\i}a}, L., {Ohme}, F., {Ajith}, P., {et~al.} 2010, \prd, 82,
  064016, \dodoi{10.1103/PhysRevD.82.064016}

\bibitem[{{Schmidt} {et~al.}(2012){Schmidt}, {Hannam}, \&
  {Husa}}]{2012PhRvD..86j4063S}
{Schmidt}, P., {Hannam}, M., \& {Husa}, S. 2012, \prd, 86, 104063,
  \dodoi{10.1103/PhysRevD.86.104063}

\bibitem[{{Schmidt} {et~al.}(2011){Schmidt}, {Hannam}, {Husa}, \&
  {Ajith}}]{2011PhRvD..84b4046S}
{Schmidt}, P., {Hannam}, M., {Husa}, S., \& {Ajith}, P. 2011, \prd, 84, 024046,
  \dodoi{10.1103/PhysRevD.84.024046}

\bibitem[{{Schmidt} {et~al.}(2015){Schmidt}, {Ohme}, \&
  {Hannam}}]{2015PhRvD..91b4043S}
{Schmidt}, P., {Ohme}, F., \& {Hannam}, M. 2015, \prd, 91, 024043,
  \dodoi{10.1103/PhysRevD.91.024043}

\bibitem[{{Singer} \& {Price}(2016)}]{2016PhRvD..93b4013S}
{Singer}, L.~P., \& {Price}, L.~R. 2016, \prd, 93, 024013,
  \dodoi{10.1103/PhysRevD.93.024013}

\bibitem[{{Singer} {et~al.}(2016){Singer}, {Chen}, {Holz}, {Farr}, {Price},
  {Raymond}, {Cenko}, {Gehrels}, {Cannizzo}, {Kasliwal}, {Nissanke},
  {Coughlin}, {Farr}, {Urban}, {Vitale}, {Veitch}, {Graff}, {Berry},
  {Mohapatra}, \& {Mandel}}]{2016ApJ...829L..15S}
{Singer}, L.~P., {Chen}, H.-Y., {Holz}, D.~E., {et~al.} 2016, \apjl, 829, L15,
  \dodoi{10.3847/2041-8205/829/1/L15}

\bibitem[{Singer {et~al.}(2016)}]{Singer:2016erz}
Singer, L.~P., {et~al.} 2016, Astrophys. J. Suppl., 226, 10,
  \dodoi{10.3847/0067-0049/226/1/10}

\bibitem[{Skilling(2006)}]{Skilling:2006gxv}
Skilling, J. 2006, Bayesian Analysis, 1, 833, \dodoi{10.1214/06-BA127}

\bibitem[{{Smith} {et~al.}(2011){Smith}, {Abbott}, {Hirose}, {Leroy},
  {MacLeod}, {McIver}, {Saulson}, \& {Shawhan}}]{2011CQGra..28w5005S}
{Smith}, J.~R., {Abbott}, T., {Hirose}, E., {et~al.} 2011, Classical and
  Quantum Gravity, 28, 235005, \dodoi{10.1088/0264-9381/28/23/235005}

\bibitem[{{Smith} {et~al.}(2016){Smith}, {Field}, {Blackburn}, {Haster},
  {P{\"u}rrer}, {Raymond}, \& {Schmidt}}]{2016PhRvD..94d4031S}
{Smith}, R., {Field}, S.~E., {Blackburn}, K., {et~al.} 2016, \prd, 94, 044031,
  \dodoi{10.1103/PhysRevD.94.044031}

\bibitem[{{Smith} {et~al.}(2020){Smith}, {Ashton}, {Vajpeyi}, \&
  {Talbot}}]{2020MNRAS.498.4492S}
{Smith}, R. J.~E., {Ashton}, G., {Vajpeyi}, A., \& {Talbot}, C. 2020, \mnras,
  498, 4492, \dodoi{10.1093/mnras/staa2483}

\bibitem[{Soni {et~al.}(2024)Soni, Glanzer, Effler, Frolov, Gonz{\'a}lez, Pele,
  \& Schofield}]{Soni:2023kqq}
Soni, S., Glanzer, J., Effler, A., {et~al.} 2024, Class. Quant. Grav., 41,
  135015, \dodoi{10.1088/1361-6382/ad494a}

\bibitem[{{Soni} {et~al.}(2021){Soni}, {Austin}, {Effler}, {Schofield},
  {Gonz{\'a}lez}, {Frolov}, {Driggers}, {Pele}, {Urban}, {Valdes}, {Abbott},
  {Adams}, {Adhikari}, {Ananyeva}, {Appert}, {Arai}, {Areeda}, {Asali},
  {Aston}, {Baer}, {Ball}, {Ballmer}, {Banagiri}, {Barker}, {Barsotti},
  {Bartlett}, {Berger}, {Betzwieser}, {Bhattacharjee}, {Billingsley},
  {Biscans}, {Blair}, {Blair}, {Bode}, {Booker}, {Bork}, {Bramley}, {Brooks},
  {Brown}, {Buikema}, {Cahillane}, {Cannon}, {Chen}, {Ciobanu}, {Clara},
  {Cooper}, {Corley}, {Countryman}, {Covas}, {Coyne}, {Datrier}, {Davis},
  {Fronzo}, {Dooley}, {Dupej}, {Dwyer}, {Etzel}, {Evans}, {Evans}, {Feicht},
  {Fernandez-Galiana}, {Fritschel}, {Fulda}, {Fyffe}, {Giaime}, {Giardina},
  {Godwin}, {Goetz}, {Gras}, {Gray}, {Gray}, {Green}, {Gustafson}, {Gustafson},
  {Hanks}, {Hanson}, {Hardwick}, {Hasskew}, {Heintze}, {Helmling-Cornell},
  {Holland}, {Jones}, {Kandhasamy}, {Karki}, {Kasprzack}, {Kawabe},
  {Kijbunchoo}, {King}, {Kissel}, {Kumar}, {Landry}, {Lane}, {Lantz}, {Laxen},
  {Lecoeuche}, {Leviton}, {Liu}, {Lormand}, {Lundgren}, {Macas}, {MacInnis},
  {Macleod}, {Mansell}, {M{\'a}rka}, {M{\'a}rka}, {Martynov}, {Mason},
  {Massinger}, {Matichard}, {Mavalvala}, {McCarthy}, {McClelland}, {McCormick},
  {McCuller}, {McIver}, {McRae}, {Mendell}, {Merfeld}, {Merilh}, {Meylahn},
  {Mistry}, {Mittleman}, {Moreno}, {Mow-Lowry}, {Mozzon}, {Mullavey}, {Nelson},
  {Nguyen}, {Nuttall}, {Oberling}, {Oram}, {Osthelder}, {Ottaway}, {Overmier},
  {Palamos}, {Parker}, {Payne}, {Penhorwood}, {Perez}, {Pirello}, {Radkins},
  {Ramirez}, {Richardson}, {Riles}, {Robertson}, {Rollins}, {Romel}, {Romie},
  {Ross}, {Ryan}, {Sadecki}, {Sanchez}, {Sanchez}, {Saravanan}, {Savage},
  {Schaetzl}, {Schnabel}, {Schwartz}, {Sellers}, {Shaffer}, {Sigg},
  {Slagmolen}, {Smith}, {Sorazu}, {Spencer}, {Strain}, {Sun},
  {Szczepa{\'n}czyk}, {Thomas}, {Thomas}, {Thorne}, {Toland}, {Torrie},
  {Traylor}, {Tse}, {Vajente}, {Vander-Hyde}, {Veitch}, {Venkateswara},
  {Venugopalan}, {Viets}, {Vo}, {Vorvick}, {Wade}, {Ward}, {Warner}, {Weaver},
  {Weiss}, {Whittle}, {Willke}, {Wipf}, {Xiao}, {Yamamoto}, {Yu}, {Yu},
  {Zhang}, {Zucker}, {Zweizig}, \& {LIGO Scientific
  Collaboration}}]{2021CQGra..38b5016S}
{Soni}, S., {Austin}, C., {Effler}, A., {et~al.} 2021, Classical and Quantum
  Gravity, 38, 025016, \dodoi{10.1088/1361-6382/abc906}

\bibitem[{{Soni} {et~al.}(2025){Soni}, {Berger}, {Davis}, {Di Renzo}, {Effler},
  {Ferreira}, {Glanzer}, {Goetz}, {Gonz{\'a}lez}, {Helmling-Cornell}, {Hughey},
  {Huxford}, {Mannix}, {Mo}, {Nandi}, {Neunzert}, {Nichols}, {Pham}, {Renzini},
  {Schofield}, {Stuver}, {Trevor}, {{\'A}lvarez-L{\'o}pez}, {Beda}, {Berry},
  {Bhuiyan}, {Blagg}, {Bruntz}, {Callos}, {Chan}, {Charlton}, {Christensen},
  {Connolly}, {Dhatri}, {Ding}, {Garg}, {Holley-Bockelmann}, {Hourihane},
  {Jani}, {Janssens}, {Jarov}, {Knee}, {Lattal}, {Lecoeuche}, {Littenberg},
  {Liyanage}, {Lott}, {Macas}, {Malakar}, {McGowan}, {McIver}, {Millhouse},
  {Nuttall}, {Nykamp}, {Ota}, {Rawcliffe}, {Scully}, {Tasson}, {Tejera},
  {Thiele}, {Udall}, {Winborn}, {Yarbrough}, {Zhang}, {Zheng}, {Abbott},
  {Abouelfettouh}, {Adhikari}, {Ananyeva}, {Appert}, {Arai}, {Aritomi},
  {Aston}, {Ball}, {Ballmer}, {Barker}, {Barsotti}, {Betzwieser},
  {Billingsley}, {Biscans}, {Bode}, {Bonilla}, {Bossilkov}, {Branch}, {Brooks},
  {Brown}, {Bryant}, {Cahillane}, {Cao}, {Capote}, {Clara}, {Collins},
  {Compton}, {Cottingham}, {Coyne}, {Crouch}, {Csizmazia}, {Cullen}, {Dartez},
  {Demos}, {Dohmen}, {Driggers}, {Dwyer}, {Ejlli}, {Etzel}, {Evans}, {Feicht},
  {Frey}, {Frischhertz}, {Fritschel}, {Frolov}, {Fulda}, {Fyffe}, {Ganapathy},
  {Gateley}, {Giaime}, {Giardina}, {Goetz}, {Goodwin-Jones}, {Gras}, {Gray},
  {Griffith}, {Grote}, {Guidry}, {Hall}, {Hanks}, {Hanson}, {Heintze},
  {Holland}, {Hoyland}, {Huang}, {Inoue}, {James}, {Jennings}, {Jia}, {Karat},
  {Karki}, {Kasprzack}, {Kawabe}, {Kijbunchoo}, {King}, {Kissel}, {Komori},
  {Kontos}, {Kumar}, {Kuns}, {Landry}, {Lantz}, {Laxen}, {Lee}, {Lesovsky},
  {Llamas}, {Lormand}, {Loughlin}, {MacInnis}, {Makarem}, {Mansell}, {Martin},
  {Mason}, {Matichard}, {Mavalvala}, {Maxwell}, {McCarrol}, {McCarthy},
  {McClelland}, {McCormick}, {McCuller}, {McRae}, {Mera}, {Merilh}, {Meylahn},
  {Mittleman}, {Moraru}, {Moreno}, {Mullavey}, {Nakano}, {Nelson}, {Notte},
  {Oberling}, {O'Hanlon}, {Osthelder}, {Ottaway}, {Overmier}, {Parker}, {Pele},
  {Pham}, {Pirello}, {Quetschke}, {Ramirez}, {Reyes}, {Richardson}, {Robinson},
  {Rollins}, {Romel}, {Romie}, {Ross}, {Ryan}, {Sadecki}, {Sanchez}, \&
  {Sanchez}}]{2025CQGra..42h5016S}
{Soni}, S., {Berger}, B.~K., {Davis}, D., {et~al.} 2025, Classical and Quantum
  Gravity, 42, 085016, \dodoi{10.1088/1361-6382/adc4b6}

\bibitem[{{Speagle}(2020)}]{2020MNRAS.493.3132S}
{Speagle}, J.~S. 2020, \mnras, 493, 3132, \dodoi{10.1093/mnras/staa278}

\bibitem[{{Steinhoff} {et~al.}(2016){Steinhoff}, {Hinderer}, {Buonanno}, \&
  {Taracchini}}]{2016PhRvD..94j4028S}
{Steinhoff}, J., {Hinderer}, T., {Buonanno}, A., \& {Taracchini}, A. 2016,
  \prd, 94, 104028, \dodoi{10.1103/PhysRevD.94.104028}

\bibitem[{Steinle \& Kesden(2022)}]{Steinle:2022rhj}
Steinle, N., \& Kesden, M. 2022, Phys. Rev. D, 106, 063028,
  \dodoi{10.1103/PhysRevD.106.063028}

\bibitem[{{Stevenson} {et~al.}(2017){Stevenson}, {Berry}, \&
  {Mandel}}]{2017MNRAS.471.2801S}
{Stevenson}, S., {Berry}, C. P.~L., \& {Mandel}, I. 2017, \mnras, 471, 2801,
  \dodoi{10.1093/mnras/stx1764}

\bibitem[{{Stovall} {et~al.}(2018){Stovall}, {Freire}, {Chatterjee},
  {Demorest}, {Lorimer}, {McLaughlin}, {Pol}, {van Leeuwen}, {Wharton},
  {Allen}, {Boyce}, {Brazier}, {Caballero}, {Camilo}, {Camuccio}, {Cordes},
  {Crawford}, {Deneva}, {Ferdman}, {Hessels}, {Jenet}, {Kaspi}, {Knispel},
  {Lazarus}, {Lynch}, {Parent}, {Patel}, {Pleunis}, {Ransom}, {Scholz},
  {Seymour}, {Siemens}, {Stairs}, {Swiggum}, \& {Zhu}}]{2018ApJ...854L..22S}
{Stovall}, K., {Freire}, P.~C.~C., {Chatterjee}, S., {et~al.} 2018, \apjl, 854,
  L22, \dodoi{10.3847/2041-8213/aaad06}

\bibitem[{Sturani(2015)}]{Sturani:2015aaa}
Sturani, R. 2015, {Note on the derivation of the angular momentum and spin
  precessing equations in SpinTaylor codes}, Tech. Rep. DCC-T1500554, {LIGO}.
\newblock \url{https://dcc.ligo.org/LIGO-T1500554/public}

\bibitem[{{Sun} {et~al.}(2020){Sun}, {Goetz}, {Kissel}, {Betzwieser}, {Karki},
  {Viets}, {Wade}, {Bhattacharjee}, {Bossilkov}, {Covas}, {Datrier}, {Gray},
  {Kandhasamy}, {Lecoeuche}, {Mendell}, {Mistry}, {Payne}, {Savage},
  {Weinstein}, {Aston}, {Buikema}, {Cahillane}, {Driggers}, {Dwyer}, {Kumar},
  \& {Urban}}]{2020CQGra..37v5008S}
{Sun}, L., {Goetz}, E., {Kissel}, J.~S., {et~al.} 2020, Classical and Quantum
  Gravity, 37, 225008, \dodoi{10.1088/1361-6382/abb14e}

\bibitem[{{Sun} {et~al.}(2021){Sun}, {Goetz}, {Kissel}, {Betzwieser}, {Karki},
  {Bhattacharjee}, {Covas}, {Datrier}, {Kandhasamy}, {Lecoeuche}, {Mendell},
  {Mistry}, {Payne}, {Savage}, {Viets}, {Wade}, {Weinstein}, {Aston},
  {Cahillane}, {Driggers}, {Dwyer}, \& {Urban}}]{2021arXiv210700129S}
---. 2021, arXiv e-prints, arXiv:2107.00129, \dodoi{10.48550/arXiv.2107.00129}

\bibitem[{{Talbot} \& {Thrane}(2017)}]{2017PhRvD..96b3012T}
{Talbot}, C., \& {Thrane}, E. 2017, \prd, 96, 023012,
  \dodoi{10.1103/PhysRevD.96.023012}

\bibitem[{{Talbot} {et~al.}(2025){Talbot}, {Biscoveanu}, {Zimmerman}, {Baka},
  {Farr}, {Golomb}, {Hoy}, {Lundgren}, {Tissino}, {Veitch}, {Vijaykumar}, \&
  {Williams}}]{2025CQGra..42w5023T}
{Talbot}, C., {Biscoveanu}, S., {Zimmerman}, A., {et~al.} 2025, Classical and
  Quantum Gravity, 42, 235023, \dodoi{10.1088/1361-6382/ae1ac7}

\bibitem[{Tantau(2023)}]{tantau:2023}
Tantau, T. 2023, The TikZ and PGF Packages.
\newblock \url{https://github.com/pgf-tikz/pgf}

\bibitem[{Taracchini {et~al.}(2014{\natexlab{a}})Taracchini, Buonanno, Khanna,
  \& Hughes}]{Taracchini:2014zpa}
Taracchini, A., Buonanno, A., Khanna, G., \& Hughes, S.~A. 2014{\natexlab{a}},
  Phys. Rev. D, 90, 084025, \dodoi{10.1103/PhysRevD.90.084025}

\bibitem[{Taracchini {et~al.}(2014{\natexlab{b}})}]{Taracchini:2013rva}
Taracchini, A., {et~al.} 2014{\natexlab{b}}, Phys. Rev. D, 89, 061502,
  \dodoi{10.1103/PhysRevD.89.061502}

\bibitem[{{Thompson} {et~al.}(2020){Thompson}, {Fauchon-Jones}, {Khan},
  {Nitoglia}, {Pannarale}, {Dietrich}, \& {Hannam}}]{2020PhRvD.101l4059T}
{Thompson}, J.~E., {Fauchon-Jones}, E., {Khan}, S., {et~al.} 2020, \prd, 101,
  124059, \dodoi{10.1103/PhysRevD.101.124059}

\bibitem[{{Thompson} {et~al.}(2024){Thompson}, {Hamilton}, {London}, {Ghosh},
  {Kolitsidou}, {Hoy}, \& {Hannam}}]{2024PhRvD.109f3012T}
{Thompson}, J.~E., {Hamilton}, E., {London}, L., {et~al.} 2024, \prd, 109,
  063012, \dodoi{10.1103/PhysRevD.109.063012}

\bibitem[{{Thrane} \& {Talbot}(2019)}]{2019PASA...36...10T}
{Thrane}, E., \& {Talbot}, C. 2019, \pasa, 36, e010,
  \dodoi{10.1017/pasa.2019.2}

\bibitem[{{Tiwari} {et~al.}(2016){Tiwari}, {Klimenko}, {Necula}, \&
  {Mitselmakher}}]{2016CQGra..33aLT01T}
{Tiwari}, V., {Klimenko}, S., {Necula}, V., \& {Mitselmakher}, G. 2016,
  Classical and Quantum Gravity, 33, 01LT01,
  \dodoi{10.1088/0264-9381/33/1/01LT01}

\bibitem[{{Tsukada} {et~al.}(2023){Tsukada}, {Joshi}, {Adhicary}, {George},
  {Guimaraes}, {Hanna}, {Magee}, {Zimmerman}, {Baral}, {Baylor}, {Cannon},
  {Caudill}, {Cousins}, {Creighton}, {Ewing}, {Fong}, {Godwin}, {Harada},
  {Huang}, {Huxford}, {Kennington}, {Kuwahara}, {Li}, {Meacher}, {Messick},
  {Morisaki}, {Mukherjee}, {Niu}, {Pace}, {Posnansky}, {Ray}, {Sachdev},
  {Sakon}, {Singh}, {Tapia}, {Tsutsui}, {Ueno}, {Viets}, {Wade}, \&
  {Wade}}]{2023PhRvD.108d3004T}
{Tsukada}, L., {Joshi}, P., {Adhicary}, S., {et~al.} 2023, \prd, 108, 043004,
  \dodoi{10.1103/PhysRevD.108.043004}

\bibitem[{{Tucker} \& {Will}(2021)}]{2021PhRvD.104j4023T}
{Tucker}, A., \& {Will}, C.~M. 2021, \prd, 104, 104023,
  \dodoi{10.1103/PhysRevD.104.104023}

\bibitem[{{Usman} {et~al.}(2016){Usman}, {Nitz}, {Harry}, {Biwer}, {Brown},
  {Cabero}, {Capano}, {Dal Canton}, {Dent}, {Fairhurst}, {Kehl}, {Keppel},
  {Krishnan}, {Lenon}, {Lundgren}, {Nielsen}, {Pekowsky}, {Pfeiffer},
  {Saulson}, {West}, \& {Willis}}]{2016CQGra..33u5004U}
{Usman}, S.~A., {Nitz}, A.~H., {Harry}, I.~W., {et~al.} 2016, Classical and
  Quantum Gravity, 33, 215004, \dodoi{10.1088/0264-9381/33/21/215004}

\bibitem[{{Vajente}(2022)}]{2022PhRvD.105j2005V}
{Vajente}, G. 2022, \prd, 105, 102005, \dodoi{10.1103/PhysRevD.105.102005}

\bibitem[{van~de Meent {et~al.}(2023)van~de Meent, Buonanno, Mihaylov,
  Ossokine, Pompili, Warburton, Pound, Wardell, Durkan, \&
  Miller}]{vandeMeent:2023ols}
van~de Meent, M., Buonanno, A., Mihaylov, D.~P., {et~al.} 2023, Phys. Rev. D,
  108, 124038, \dodoi{10.1103/PhysRevD.108.124038}

\bibitem[{{van der Sluys} {et~al.}(2008){van der Sluys}, {Raymond}, {Mandel},
  {R{\"o}ver}, {Christensen}, {Kalogera}, {Meyer}, \&
  {Vecchio}}]{2008CQGra..25r4011V}
{van der Sluys}, M., {Raymond}, V., {Mandel}, I., {et~al.} 2008, Classical and
  Quantum Gravity, 25, 184011, \dodoi{10.1088/0264-9381/25/18/184011}

\bibitem[{{Varma} {et~al.}(2019){Varma}, {Field}, {Scheel}, {Blackman},
  {Gerosa}, {Stein}, {Kidder}, \& {Pfeiffer}}]{2019PhRvR...1c3015V}
{Varma}, V., {Field}, S.~E., {Scheel}, M.~A., {et~al.} 2019, Physical Review
  Research, 1, 033015, \dodoi{10.1103/PhysRevResearch.1.033015}

\bibitem[{Varma {et~al.}(2019)Varma, Gerosa, Stein, H\'ebert, \&
  Zhang}]{Varma:2018aht}
Varma, V., Gerosa, D., Stein, L.~C., H\'ebert, F., \& Zhang, H. 2019, Phys.
  Rev. Lett., 122, 011101, \dodoi{10.1103/PhysRevLett.122.011101}

\bibitem[{Varma {et~al.}(2020)Varma, Isi, \& Biscoveanu}]{Varma:2020nbm}
Varma, V., Isi, M., \& Biscoveanu, S. 2020, Phys. Rev. Lett., 124, 101104,
  \dodoi{10.1103/PhysRevLett.124.101104}

\bibitem[{{Vazsonyi} \& {Davis}(2023)}]{2023CQGra..40c5008V}
{Vazsonyi}, L., \& {Davis}, D. 2023, Classical and Quantum Gravity, 40, 035008,
  \dodoi{10.1088/1361-6382/acafd2}

\bibitem[{Vecchio(2004)}]{Vecchio:2003tn}
Vecchio, A. 2004, Phys. Rev. D, 70, 042001, \dodoi{10.1103/PhysRevD.70.042001}

\bibitem[{{Veitch} {et~al.}(2015{\natexlab{a}}){Veitch}, {P{\"u}rrer}, \&
  {Mandel}}]{2015PhRvL.115n1101V}
{Veitch}, J., {P{\"u}rrer}, M., \& {Mandel}, I. 2015{\natexlab{a}}, \prl, 115,
  141101, \dodoi{10.1103/PhysRevLett.115.141101}

\bibitem[{{Veitch} {et~al.}(2015{\natexlab{b}}){Veitch}, {Raymond}, {Farr},
  {Farr}, {Graff}, {Vitale}, {Aylott}, {Blackburn}, {Christensen}, {Coughlin},
  {Del Pozzo}, {Feroz}, {Gair}, {Haster}, {Kalogera}, {Littenberg}, {Mandel},
  {O'Shaughnessy}, {Pitkin}, {Rodriguez}, {R{\"o}ver}, {Sidery}, {Smith}, {Van
  Der Sluys}, {Vecchio}, {Vousden}, \& {Wade}}]{2015PhRvD..91d2003V}
{Veitch}, J., {Raymond}, V., {Farr}, B., {et~al.} 2015{\natexlab{b}}, \prd, 91,
  042003, \dodoi{10.1103/PhysRevD.91.042003}

\bibitem[{Villa-Ortega {et~al.}(2022)Villa-Ortega, Dent, \&
  Barroso}]{Villa-Ortega:2022qdo}
Villa-Ortega, V., Dent, T., \& Barroso, A.~C. 2022, Mon. Not. Roy. Astron.
  Soc., 515, 5718, \dodoi{10.1093/mnras/stac2120}

\bibitem[{Vinciguerra {et~al.}(2017)Vinciguerra, Veitch, \&
  Mandel}]{Vinciguerra:2017ngf}
Vinciguerra, S., Veitch, J., \& Mandel, I. 2017, Class. Quant. Grav., 34,
  115006, \dodoi{10.1088/1361-6382/aa6d44}

\bibitem[{{Vines} {et~al.}(2011){Vines}, {Flanagan}, \&
  {Hinderer}}]{2011PhRvD..83h4051V}
{Vines}, J., {Flanagan}, {\'E}.~{\'E}., \& {Hinderer}, T. 2011, \prd, 83,
  084051, \dodoi{10.1103/PhysRevD.83.084051}

\bibitem[{{Vitale} {et~al.}(2014){Vitale}, {Lynch}, {Veitch}, {Raymond}, \&
  {Sturani}}]{2014PhRvL.112y1101V}
{Vitale}, S., {Lynch}, R., {Veitch}, J., {Raymond}, V., \& {Sturani}, R. 2014,
  \prl, 112, 251101, \dodoi{10.1103/PhysRevLett.112.251101}

\bibitem[{{Wadekar} {et~al.}(2024){Wadekar}, {Venumadhav}, {Roulet}, {Mehta},
  {Zackay}, {Mushkin}, \& {Zaldarriaga}}]{2024PhRvD.110d4063W}
{Wadekar}, D., {Venumadhav}, T., {Roulet}, J., {et~al.} 2024, \prd, 110,
  044063, \dodoi{10.1103/PhysRevD.110.044063}

\bibitem[{Warburton {et~al.}(2021)Warburton, Pound, Wardell, Miller, \&
  Durkan}]{Warburton:2021kwk}
Warburton, N., Pound, A., Wardell, B., Miller, J., \& Durkan, L. 2021, Phys.
  Rev. Lett., 127, 151102, \dodoi{10.1103/PhysRevLett.127.151102}

\bibitem[{Wen(2003)}]{Wen:2002km}
Wen, L. 2003, Astrophys. J., 598, 419, \dodoi{10.1086/378794}

\bibitem[{Wen(2008)}]{Wen:2008chs}
---. 2008, International Journal of Modern Physics D, 17, 1095–1104,
  \dodoi{10.1142/s0218271808012723}

\bibitem[{{Wette}(2020)}]{2020SoftX..1200634W}
{Wette}, K. 2020, SoftwareX, 12, 100634, \dodoi{10.1016/j.softx.2020.100634}

\bibitem[{{Williams} {et~al.}(2023){Williams}, {Veitch}, {Chiofalo}, {Schmidt},
  {Udall}, {Vajpeji}, \& {Hoy}}]{2023JOSS....8.4170W}
{Williams}, D., {Veitch}, J., {Chiofalo}, M., {et~al.} 2023, The Journal of
  Open Source Software, 8, 4170, \dodoi{10.21105/joss.04170}

\bibitem[{{Wysocki} {et~al.}(2019){Wysocki}, {O'Shaughnessy}, {Lange}, \&
  {Fang}}]{2019PhRvD..99h4026W}
{Wysocki}, D., {O'Shaughnessy}, R., {Lange}, J., \& {Fang}, Y.-L.~L. 2019,
  \prd, 99, 084026, \dodoi{10.1103/PhysRevD.99.084026}

\bibitem[{Yu {et~al.}(2023)Yu, Roulet, Venumadhav, Zackay, \&
  Zaldarriaga}]{Yu:2023lml}
Yu, H., Roulet, J., Venumadhav, T., Zackay, B., \& Zaldarriaga, M. 2023, Phys.
  Rev. D, 108, 064059, \dodoi{10.1103/PhysRevD.108.064059}

\bibitem[{{Zackay} {et~al.}(2021){Zackay}, {Venumadhav}, {Roulet}, {Dai}, \&
  {Zaldarriaga}}]{2021PhRvD.104f3034Z}
{Zackay}, B., {Venumadhav}, T., {Roulet}, J., {Dai}, L., \& {Zaldarriaga}, M.
  2021, \prd, 104, 063034, \dodoi{10.1103/PhysRevD.104.063034}

\bibitem[{{Zappa} {et~al.}(2019){Zappa}, {Bernuzzi}, {Pannarale}, {Mapelli}, \&
  {Giacobbo}}]{2019PhRvL.123d1102Z}
{Zappa}, F., {Bernuzzi}, S., {Pannarale}, F., {Mapelli}, M., \& {Giacobbo}, N.
  2019, \prl, 123, 041102, \dodoi{10.1103/PhysRevLett.123.041102}

\bibitem[{Zhu {et~al.}(2018)Zhu, Thrane, Oslowski, Levin, \&
  Lasky}]{Zhu:2017znf}
Zhu, X., Thrane, E., Oslowski, S., Levin, Y., \& Lasky, P.~D. 2018, Phys. Rev.
  D, 98, 043002, \dodoi{10.1103/PhysRevD.98.043002}

\end{thebibliography}

\ifprintauthors
\expandafter\gdef\csname
currCollabName\the\allentries\endcsname{\LVKcollaboration}
\allauthors
\fi

\end{document}